%% file: ypp_survey_results_long.tex
\documentstyle[12pt,epsf,color,colordvi,epsfig,rotate]{article}

\newlength{\reducedwidth}
\begin{document}
\newlength{\headboxline}
\setlength{\headboxline}{2mm}
\renewcommand{\fboxrule}{\headboxline}
\renewcommand{\labelitemii}{$\triangleright$}
\lefthyphenmin=3
\righthyphenmin=3

%%%%%%%%%%%%%%%%%%%%%%%%%%%%%%%%%%%%%%%%%%%%%%%%%%%%%%%%%%%%%%%%%%%%%%
\input{title}
\newpage

\input{abstract}

\newpage

\tableofcontents
\newpage

\listoffigures
\newpage
%%%%%%%%%%%%%%%%%%%%%%%%%%%%%%%%%%%%%%%%%%%%%%%%%%%%%%%%%%%%%%%%%%%%%%

\input{authorlist.tex}
\input{intro.tex}
\input{method.tex}

\input{demographics.tex}

\input{balance.tex}

\input{globalization.tex}
\input{outreach.tex}
\input{building.tex}

\input{physics.tex}
\input{pick.tex}

\input{conclusions.tex}
\input{acknowledge.tex}

\input{survey.tex}

%%%%%%%%%%%%%%%%%%%%%%%%%%%%%%%%%%%%%%%%%%%%%%%%%%%%%%%%%%%%%%%%%%%%%%%%%%%%%
\newpage
 \thispagestyle{plain}
   
%%%%%%%%%%%%%%%%%%%%%%%%%%%%%%%%%%%%%%%%%%%%%%%%%%%%%%%%%%%%%%%%%%%%%%%%%%%%%
\newpage
\input{comments.tex}
%%%%%%%%%%%%%%%%%%%%%%%%%%%%%%%%%%%%%%%%%%%%%%%%%%%%%%%%%%%%%%%%%%%%%
\end{document}

%% file: title.tex
\begin{center}
 \Large{\verb+http://ypp.hep.net+}
\end{center}

\begin{centering}
 \vspace{0.8in}
 \textbf{\Huge The Young Physicists Panel (YPP)}\\
 \vspace{0.3in}
 \textbf{\Large presents}\\
 \vspace{0.3in}
 \textbf{\Huge Results of the Survey on the Future of HEP}\\
\vspace{0.8in}

\begin{center}
   \framebox[5.5in]{\epsfxsize=5.0in\epsfbox{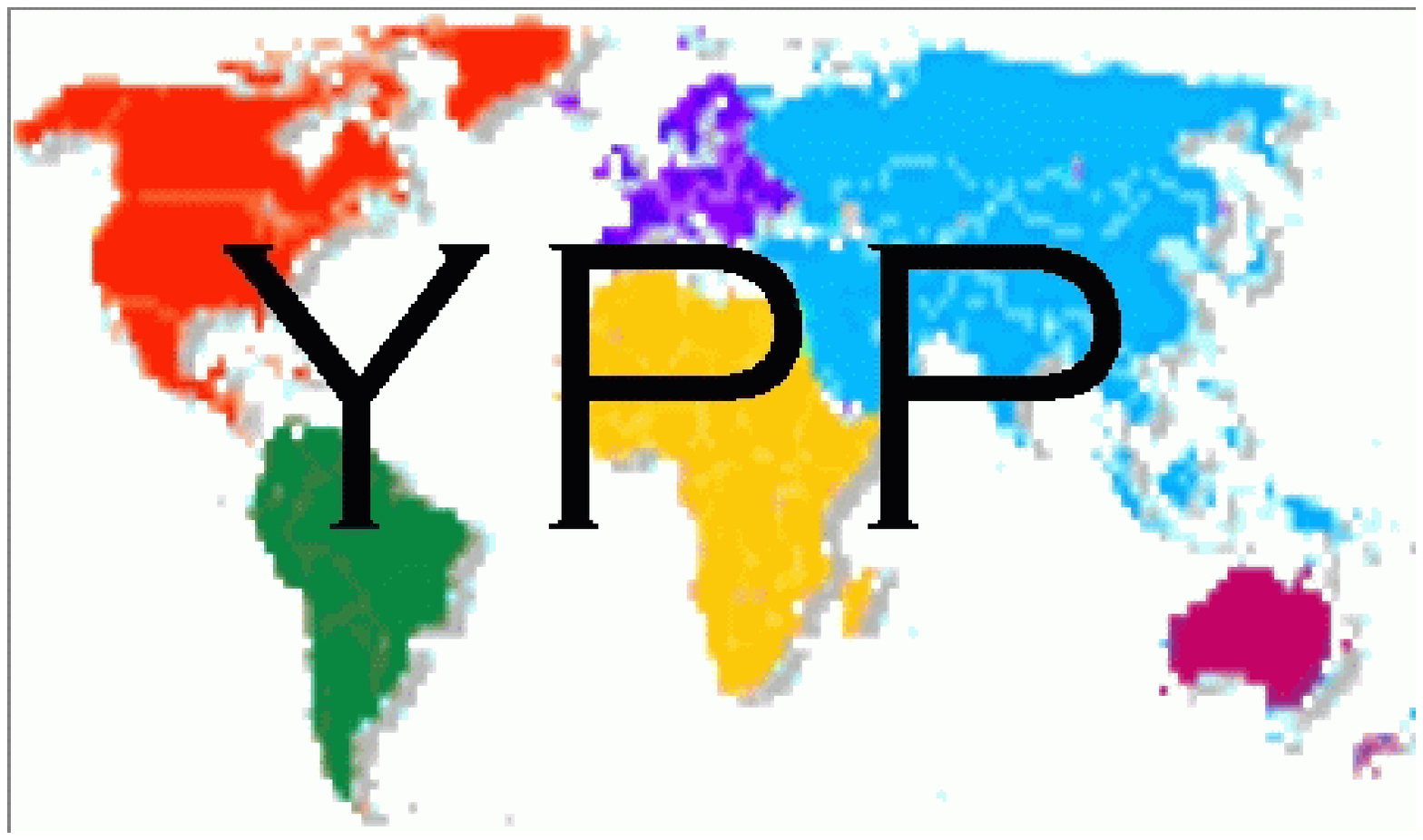}}
\end{center}

\vspace{0.8in}
submitted to the DOE/NSF Subpanel on Long Range Planning \\
for U.S. High Energy Physics

\vspace{0.5in}
August 23, 2001

\end{centering}

%% file: abstract.tex
\begin{center}
\vspace*{.2 in}
{\large \bf ABSTRACT\\}
\vspace*{.5 in}
\end{center}

\vspace{0.5 in}
The Young Physicists Panel (YPP) summarizes results from
their \textit{Survey on the Future of High Energy Physics}, which was conducted
from July 2 to July 15, 2001. Over 1500 physicists from around the world, both
young and tenured, responded to the survey. We first describe the origin, design, 
and advertisement of the survey, and then present the demographics of the
respondents. Following sections analyze the collected opinions on the desire 
for balance in the field, the 
impact of globalization, the necessity of outreach, how to build the field, 
the physics that drives the field, and the selection of the next big machine. 
Respondents' comments follow the survey analysis in an appendix.

%% file: authorlist.tex
\begin{center}
\vspace*{.2 in}
{\large \bf AUTHOR LIST\\}
\vspace*{.1 in}
\end{center}

\begin{center}
  \large{\verb+ypp2000@fnal.gov+}

\vspace{0.8in}
B.~T.~Fleming$^{1}$, J.~Krane$^{2}$, G.~P.~Zeller$^{3}$,
F.~Canelli$^{4}$, B.~Connolly$^{5}$, R.~Erbacher$^{6}$, G.~T.~Fleming$^{7}$,
D.~A.~Harris$^{6}$, E.~Hawker$^{8}$, M.~Hildreth$^{9}$, B.~Kilminster$^{4}$,
S.~Lammers$^{10}$, D.~Medoff$^{11}$, J.~Monroe$^{1}$, D.~Toback$^{12}$, 
M.~Sorel$^{1}$, A.~Turcot$^{13}$, M.~Velasco$^{3}$, M.~O.~Wascko$^{14}$,
G.~Watts$^{15}$, and~E.~D.~Zimmerman$^{16}$.

\vspace{0.5in}
\noindent
$^1$Columbia University \\
$^2$Iowa State University \\
$^3$Northwestern University \\
$^4$University of Rochester \\
$^5$Florida State University \\
$^6$Fermi National Accelerator Laboratory\\
$^7$The Ohio State University\\
$^8$University of Cincinnati\\
$^9$University of Notre Dame\\
$^{10}$University of Wisconsin\\
$^{11}$University of Maryland\\
$^{12}$Texas A\&M University\\
$^{13}$Brookhaven National Laboratory\\
$^{14}$Louisiana State University\\
$^{15}$University of Washington\\
$^{16}$University of Colorado at Boulder\\

\end{center}

\newpage

%% file: intro.tex
\section{Introduction}

The Young Physicists' Panel (YPP) was founded in May of 2000 by Bonnie
Fleming, John Krane, and Sam Zeller  
to provide young particle physicists with a forum in which to discuss the 
future of our field.  Our first goal was to directly involve young physicists 
in the planning of the future of HEP, and to act as their conduit to 
the DOE/NSF HEPAP Subpanel on Long Range Planning for U.S. High Energy
Physics \cite{hepap}. Originally, the YPP intended to present the subpanel 
with a simple and brief document that summarized our views.  We found that 
the members of our group, and we think physicists in general, did not have a
single opinion to convey.  Our vision grew from the need to poll our own
members for their opinions, and culminated in the creation of a \textit{Survey
on the Future of HEP}, a tool to quantify the opinions of all HEP physicists 
around the globe. \\

To be most useful to the HEPAP subpanel, we created a survey with several
questions that were specific to the United States.  In keeping with our
mission, we also asked questions of particular relevance to young high
energy physicists.  Rather than limit the survey to a particular 
subsample of physicists, we opened it to all physicists to see if opinions 
differed by career stage, by laboratory, by nationality, or any other metric. 
We believe the final product provides a uniquely useful ``snapshot'' of the
field.\\

What is a ``young'' physicist? We have defined a young physicist as any 
researcher who has neither tenure nor any other permanent position in the 
field. This classification includes undergraduates, graduate 
students, postdocs, untenured faculty, and term staff; thus, ``young'' refers 
to one's sense of career security more than to one's age.

\newpage

%% file: method.tex
\section{Methodology}

The survey questions were designed to extract maximum information with
the minimum possible bias in as short a time as possible.  Although
essay questions might seem ideal, many people are daunted by the idea
of spending hours on a well-constructed response.  We believed that
multiple choice questions would achieve a much higher yield of 
responses, and spark brief but useful essay responses as well.\\

The YPP survey questions suffered many months of invention, selection,
modification, and rejection. Essential contributions were made by Dr. Deborah 
Medoff, a statistician from the University of Maryland's School of Medicine, 
who specializes in the creation and review of surveys for psychiatric studies.
She advised us to carefully identify our event sample and refine our question 
content, in particular helping us to remove bias from leading questions in the
survey. Eventually the survey was released to external reviewers 
including several senior scientists and members of the HEPAP subpanel. \\

The survey opened on July 2, 2001 to coincide with the Snowmass
workshop -- as large a concentration of high energy physicists as one
could hope for -- and ran for two weeks.  We offered both a web-based 
version and an identical paper version. 
Advertisement at Snowmass consisted of flyers posted in the 
common areas, mention in the pamphlets for young physicists (available 
outside the main conference room), and note cards distributed by hand and 
placed in front of the common PCs in the computing center.  A single mass 
e-mail was sent to a list of young Snowmass participants compiled by the 
Young Physicists Forum steering committee \cite{ypf}.  Advertisement 
outside Snowmass was entirely electronic and targeted individual labs and the 
home institutions of YPP members.  Notices were also placed in 
D\O\ news, CDF news, and mailed to collaboration lists at SLAC, BNL, DESY, 
KEK, and several university HEP groups.\\

More than 99$\%$ of the surveys were received via the web. The
approximately 10 paper copies collected at Snowmass were entered 
into the web interface by hand. At two points during the course of the survey
we refined our advertising focus after examining demographic data: 
the fraction of young respondents and the individual lab response rates.
None of the other survey data were reviewed until three days
before the close of the survey. This practice minimized the chances 
of accidentally biasing the survey's sample pool and still allowed time to 
prepare plots and talks for presentation at the Snowmass Young Physicists'
Forum on July 17th, 2001 \cite{townmeet}. There were no additional 
advertisements after the full review. The results did not change to any 
significant degree in the final three days of the survey.\\

Responses were screened for possible duplicate submissions with a simple
script that located identical responses received within minutes of
each other from the same IP address.  The script flagged several occurrences,
in which case only a single copy of the response was retained.  With the 
exception of this filter, we relied on our colleagues to answer the survey 
only once.\\

A combination of \verb+tcsh+ and \verb+awk+ scripts converted the web 
responses into raw data files.  The raw data are available from 
\verb+http://ypp.hep.net+ 
in three formats: a .csv file for use with Excel, ntuples for {\sc Paw}, 
and a Root-tuple for the {\sc Root} analysis system.  All histograms 
in this document were generated with a standard macro file and the {\sc Paw} 
program.  This macro and its {\sc Root} analogue are available with the 
aforementioned data.  Our goal is to make these results as accessible 
as possible, so please contact us \cite{ypp} if you have special needs.\\

Most of the histograms in this document are normalized 
to yield the fraction of physicists who selected a given
response.  When two lines appear in the same histogram, the solid line 
indicates responses from young physicists, the dotted line indicates 
responses from tenured physicists.
Often, the samples are subdivided based on geography (see next section), 
yielding multiple-panel histograms.
Several histograms are not normalized, in which case the scale at the 
left is absolute.  In one instance, the sample size is dramatically reduced 
so we include the statistical error bars.\\

The figure captions contain the survey questions, either accurately reproduced
or faithfully paraphrased.  The response options, listed on the x-axis of the
histograms, are accurately reproduced where space allows and
paraphrased elsewhere.  Please refer to the original survey (included as
section 12 of this document) for the exact form of any questions 
or response options. \\

All comments from the survey are included in the long form of this document.  
They were edited for spelling and to remove names in some cases, but
are otherwise intact.  The comments, reordered into sections
by topic, should provide a good complement to the numerical
data provided by the survey. 

\newpage

%% file: demographics.tex
\section{Demographics}
This section summarizes the personal information of the respondents
to the YPP survey.  It was not the intention of YPP to gather
a representative sample of HEP physicists; rather, we wanted to
gather sufficient statistics from each existing demographic.
The interested analyst can re-weight the samples if desired.\\

\begin{figure}[ht]
\centering
     \epsfysize=3.0in \epsfbox{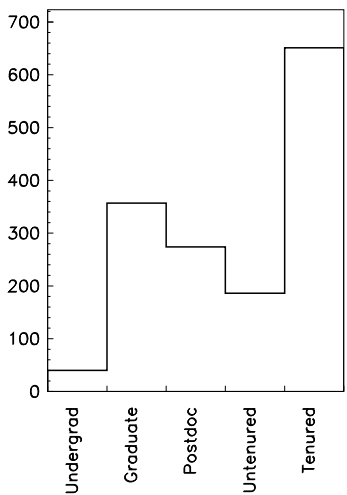}\vspace{-0.3in}
     \caption{Current career stage of survey respondents.}
  \label{fig:careerstage}
\end{figure}

A total of 1508 survey responses were received before the deadline. 
Another 30 responses came in after deadline, and were excluded from the sample.
More than half of the respondents (857) met the YPP definition of a young
physicist. Figure~\ref{fig:careerstage} shows the total number of
responses from each of the career stage groups.  More than 650
responses came from tenured physicists, compared to roughly 350 from
graduate students.

\begin{figure}[ht]
\centering
     \epsfysize=3.0in \epsfbox{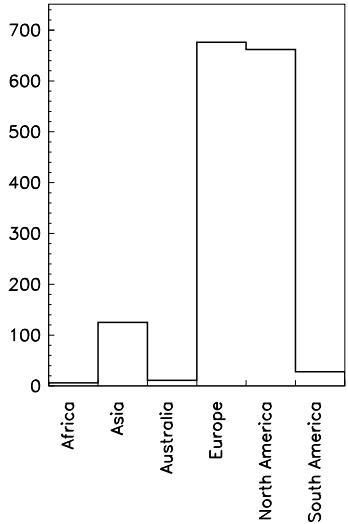}\vspace{-0.2in}
     \caption{Continent where survey respondents grew up.}
  \label{fig:contrais}
\end{figure}

\vspace{0.2in}
Because survey responses seemed to correlate more strongly with geography 
than with career stage, we subdivided the samples based on the respondents' 
continent of origin.  Respondents stating that they had ``grown
up'' in North America were classified as \textit{Americans}, while all others were 
classified as \textit{non-Americans} (Figure~\ref{fig:contrais}). 
To illustrate these differences, the data sample is often divided
into 4 subsets: young Americans, young non-Americans, tenured
Americans, and tenured non-Americans.  Separating the samples based on where
one currently works yields similar results, except in questions that relate most
specifically to the future of HEP in the United States.  In these instances,
separate histograms show the responses subdivided in both ways, first by
continent of origin, and then by continent on which one works.
The largest single subclass, \textit{young non-Americans}, is approximately 
$50\%$ larger than the three other subclasses (Table~\ref{table:population}).\\

\begin{table}[ht]
\centering
   \begin{tabular}{|l|c|c|c|}\hline
                             &  Young  & Tenured  & Total  \\ \hline
 {Americans}                 & {334}   & {328}    & {662}  \\ \hline
 {Non-Americans}             & {523}   & {323}    & {846}  \\ \hline\hline
 {Total}                     & {857}   & {651}    & {1508} \\ 
\hline
   \end{tabular}
   \caption{Number of survey respondents by career stage
            and continent of origin.} 
   \label{table:population}
\end{table} 
\vspace{0.2in}

\begin{figure}[ht]
\mbox{
\begin{minipage}{0.5\textwidth}
\centerline{\epsfxsize 3.0 truein \epsfbox{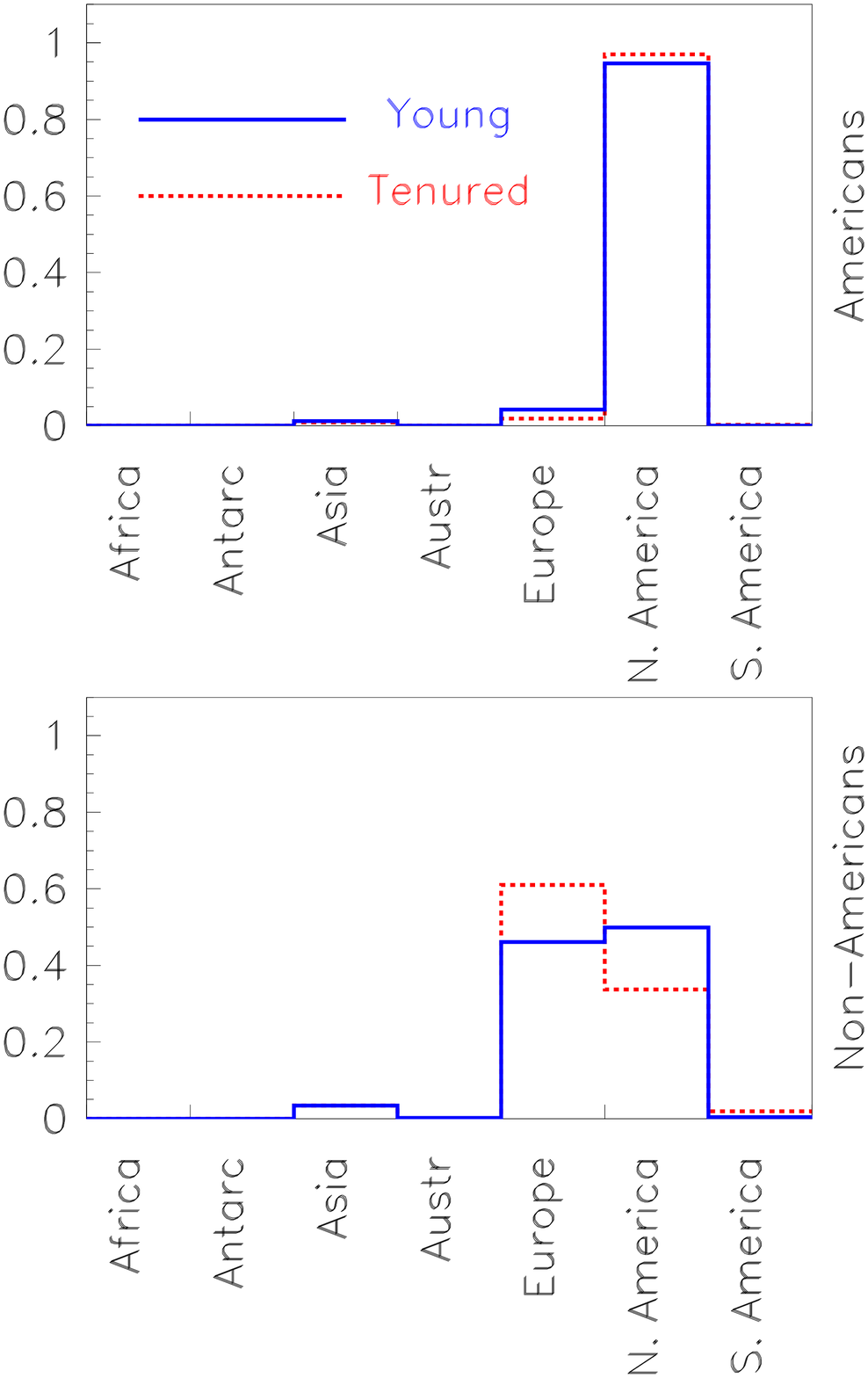}}   
\vspace{-0.2in}
\caption{On which continent do you currently work (greatest part of the 
         year)?}
\label{fig:contwork}
\end{minipage}\hspace*{0.02\textwidth}
\begin{minipage}{0.5\textwidth}
\centerline{\epsfxsize 3.0 truein \epsfbox{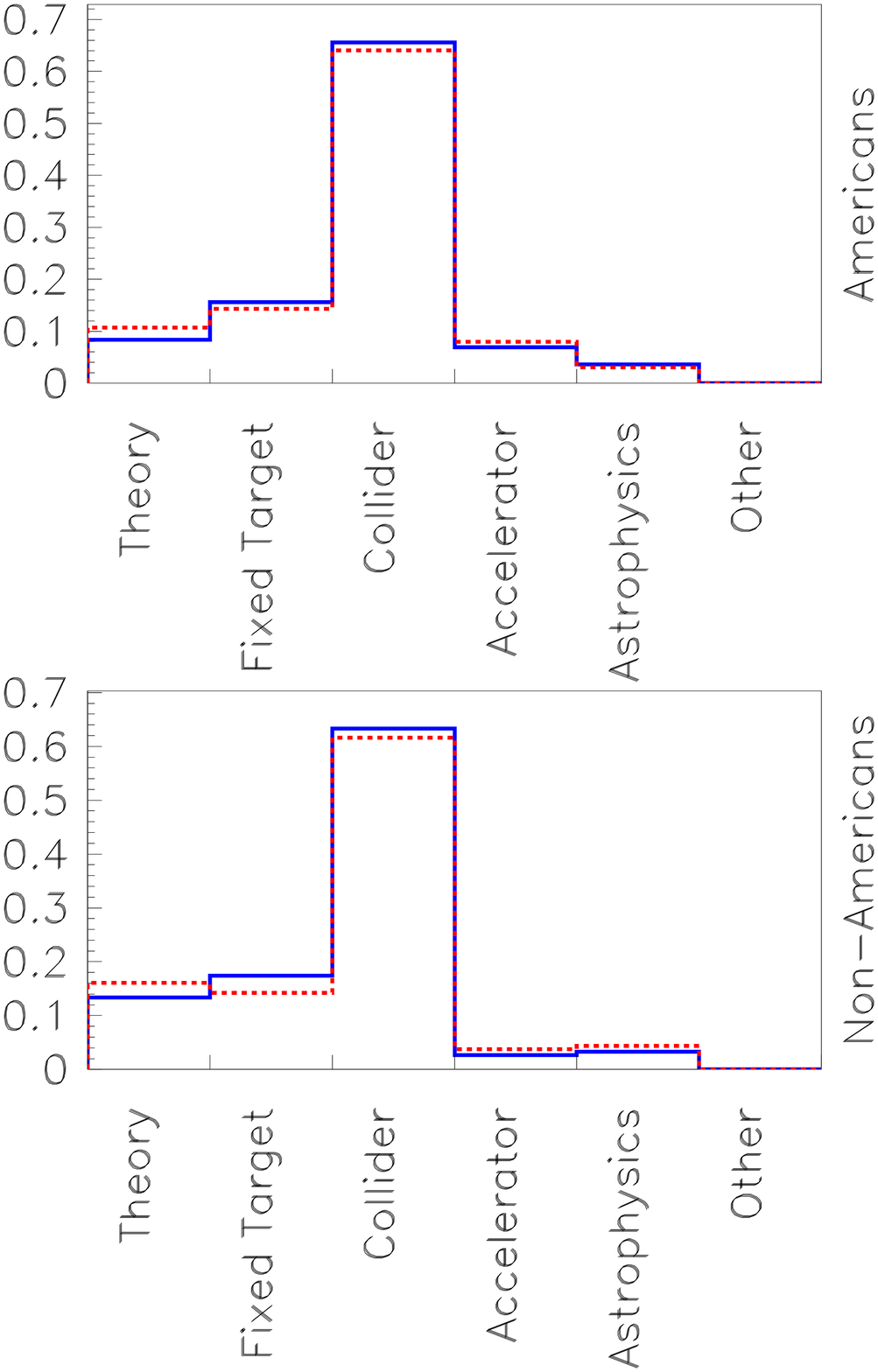}}   
\vspace{-0.2in}
\caption{What type of physics have you been working on this year?}
\label{fig:physwork}
\end{minipage}
}
\end{figure}

Of the 662 respondents who grew up in North America, almost all of them still 
work in North America (Figure~\ref{fig:contwork}).  In contrast, roughly 
half of the 676 European and 125 Asian respondents (including Russia and the 
Indian subcontinent) have traveled to North America to work 
(Figure~\ref{fig:contwork}).\\

\begin{figure}[hb]
\mbox{
\begin{minipage}{0.5\textwidth}
\centerline{\epsfxsize 3.0 truein \epsfbox{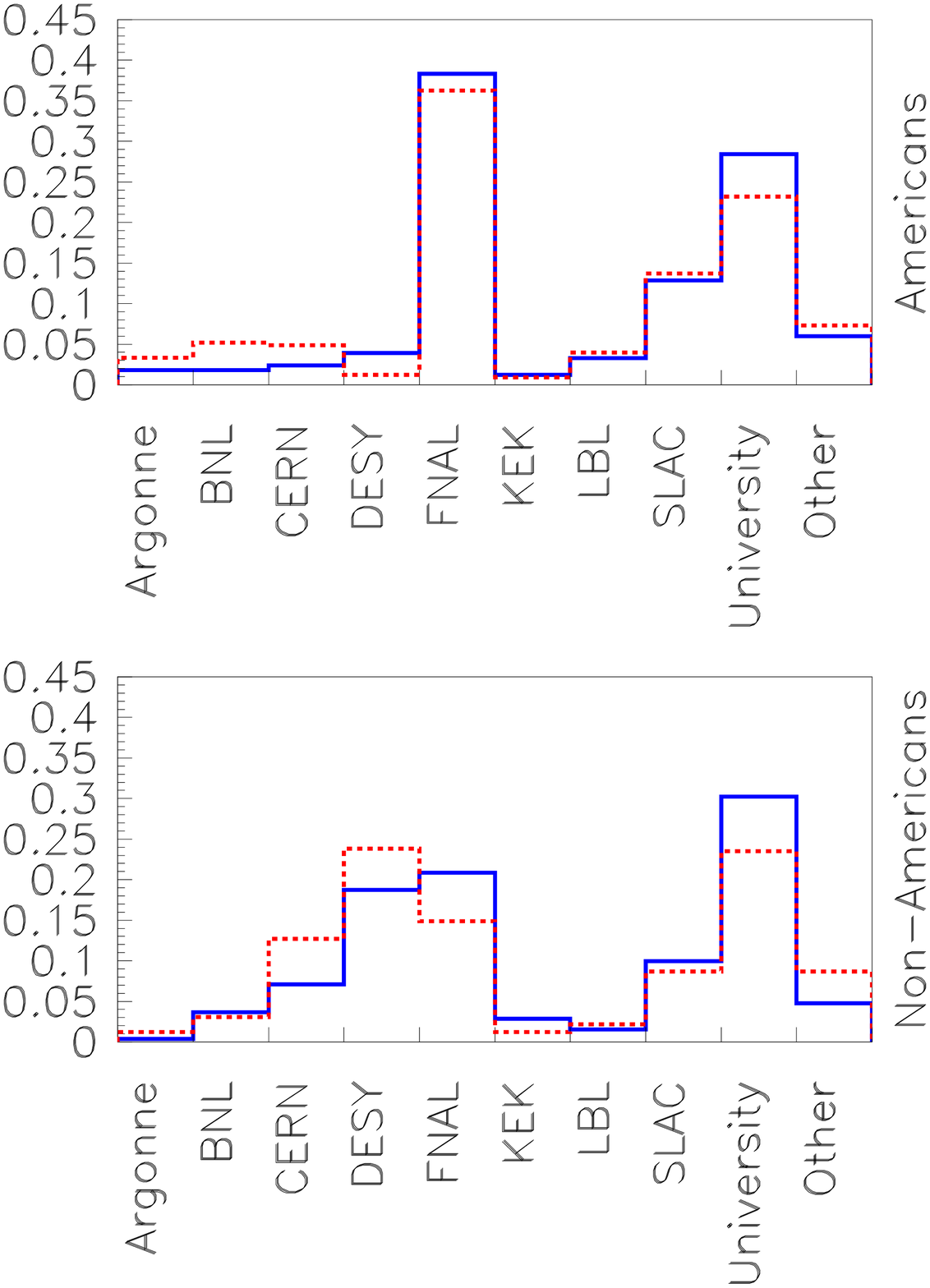}}   
\vspace{-0.2in}
\caption{Where do you do your research (greatest part of year)?}
\label{fig:labwork}
\end{minipage}\hspace*{0.02\textwidth}
\begin{minipage}{0.5\textwidth}
\centerline{\epsfxsize 3.0 truein \epsfbox{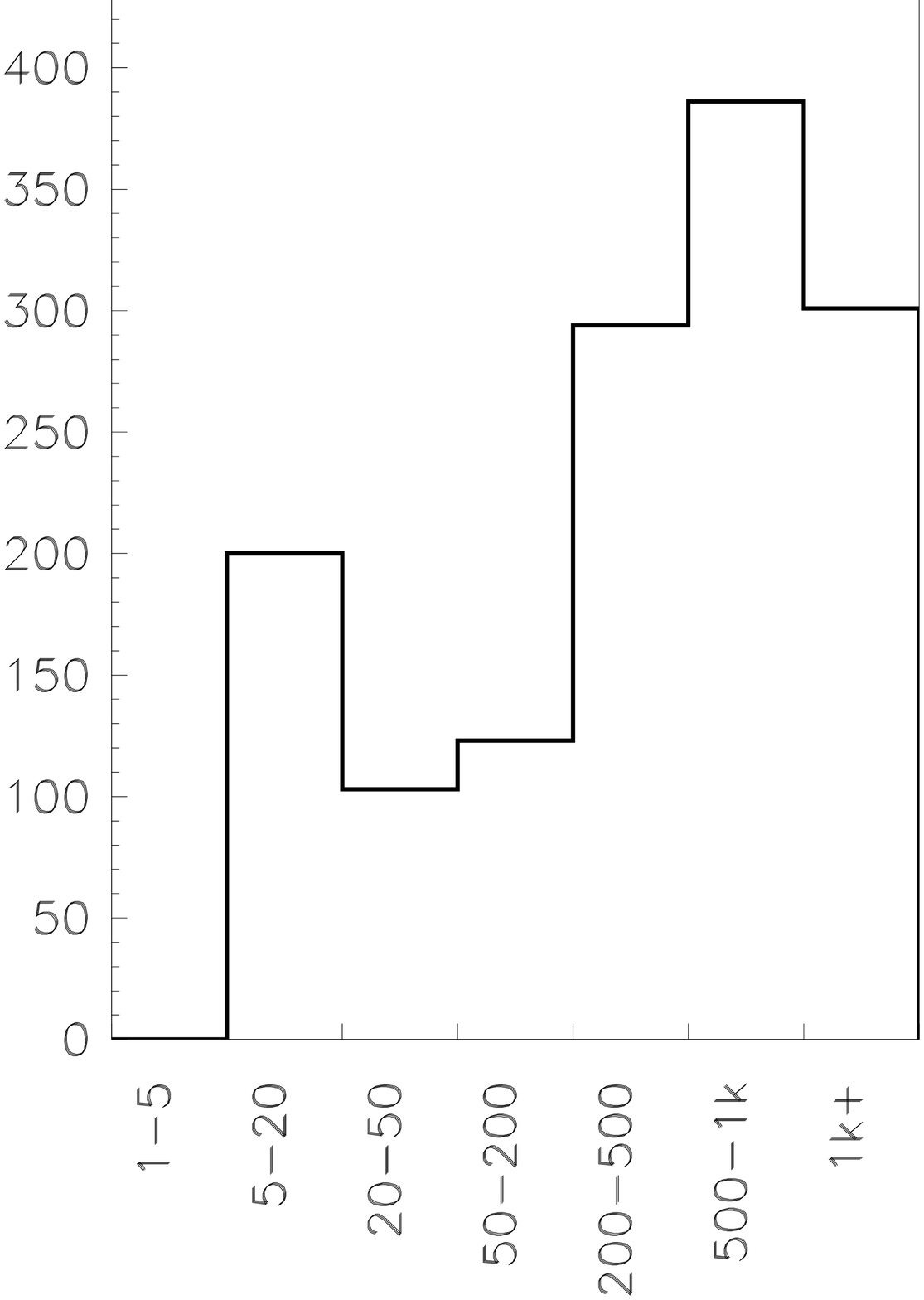}}   
\vspace{-0.2in}
\caption{How large is your current collaboration (in number of people)?}
\label{fig:collsize}
\end{minipage}
}
\end{figure}

Most responses came from experimentalists working in collider physics 
(Figure~\ref{fig:physwork}) at FNAL (404 responses), SLAC (168), 
DESY (192), CERN (102), or a university group (405), as depicted by 
Figure~\ref{fig:labwork}.  Responses in the ``Other'' category in Figure 5 
include 26 people from Cornell/LNS, 5 from LANL, 4 from JLAB, 3 from Kamland 
and the rest from a wide variety of institutes and research centers around the 
world. Most responses indicated work on large collaborations of more than 
200 participants (Figure~\ref{fig:collsize}).  In the survey analysis, the 
opinions of all respondents generally follow those specific to this major
constituency of collider experimentalists on large collaborations.\\

\begin{figure}[ht]
\centering
     \epsfysize=4.0in \epsfbox{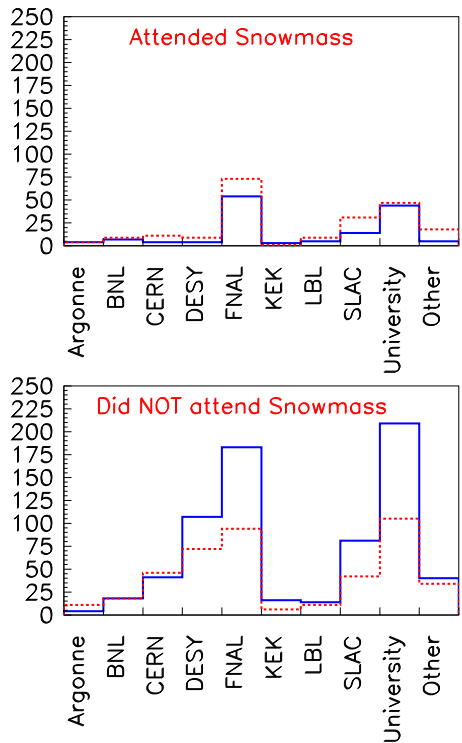}\vspace{-0.2in}
     \caption[Place where respondents currently work]
             {Place where respondents currently work separated by those who
              attended Snowmass (top) versus those who did not (bottom).}
  \label{fig:labwork-atsmass}
\end{figure}

Of the 1508 respondents, only 24$\%$ (356)\footnote{In the presentation
at Snowmass, the fractions of responses from inside and outside
Snowmass were inadvertently reversed.} reported that they attended the 
Snowmass 2001 workshop, for which 1162 scientists were registered. As a 
result, this survey is predominantly representative of those who were not in
attendance at Snowmass, and hence reflects a slightly wider global distribution
of physicists (Figure~\ref{fig:labwork-atsmass}).

\clearpage
\newpage

%% file: balance.tex
\section{Balance vs. Focus}

A balanced physics program at a laboratory can be achieved through several 
mechanisms; the YPP survey explores four of them.  First, an accelerator 
complex could host a range of accelerator experiments, including fixed-target
experiments, specialized collider experiments, and large multi-purpose
collider detectors. Second, two or more redundant detectors at one
facility can be built to allow several measurements of the same
quantities.  Third, more attempts could be made to balance the number
of experimentalists, theorists, and phenomenologists in the field.
Finally, facilities and resources can be shared with astrophysics experiments.
This section of the survey attempts to determine what HEP physicists are 
willing to sacrifice if diversity cannot be maintained in the face of major 
new construction and a finite budget.\\

\begin{figure}[ht]
\mbox{
\begin{minipage}{0.5\textwidth}
\centerline{\epsfxsize 3.0 truein \epsfbox{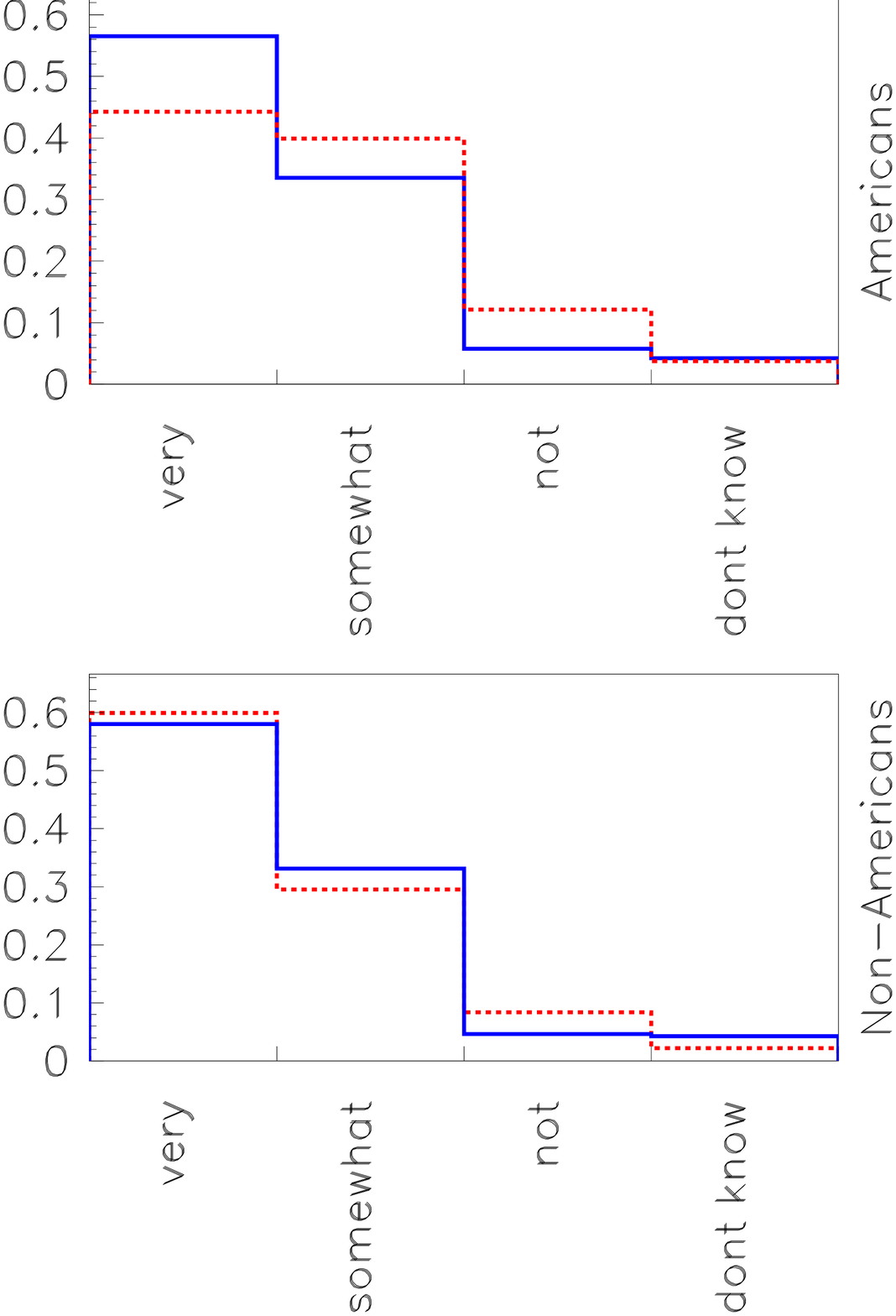}}
\vspace{-0.2in}
\caption[Should the ``next machine'' host diverse experiments and collaborations?]
        {Is it important that the ``next machine'' host a diverse (multi-physics) 
         range of experiments and collaborations?
         Recall that the solid lines indicate young physicists, and 
         dashed lines indicates tenured physicists.}
\label{fig:diversity}
\end{minipage}\hspace*{0.02\textwidth}
\begin{minipage}{0.5\textwidth}
\centerline{\epsfxsize 3.0 truein \epsfbox{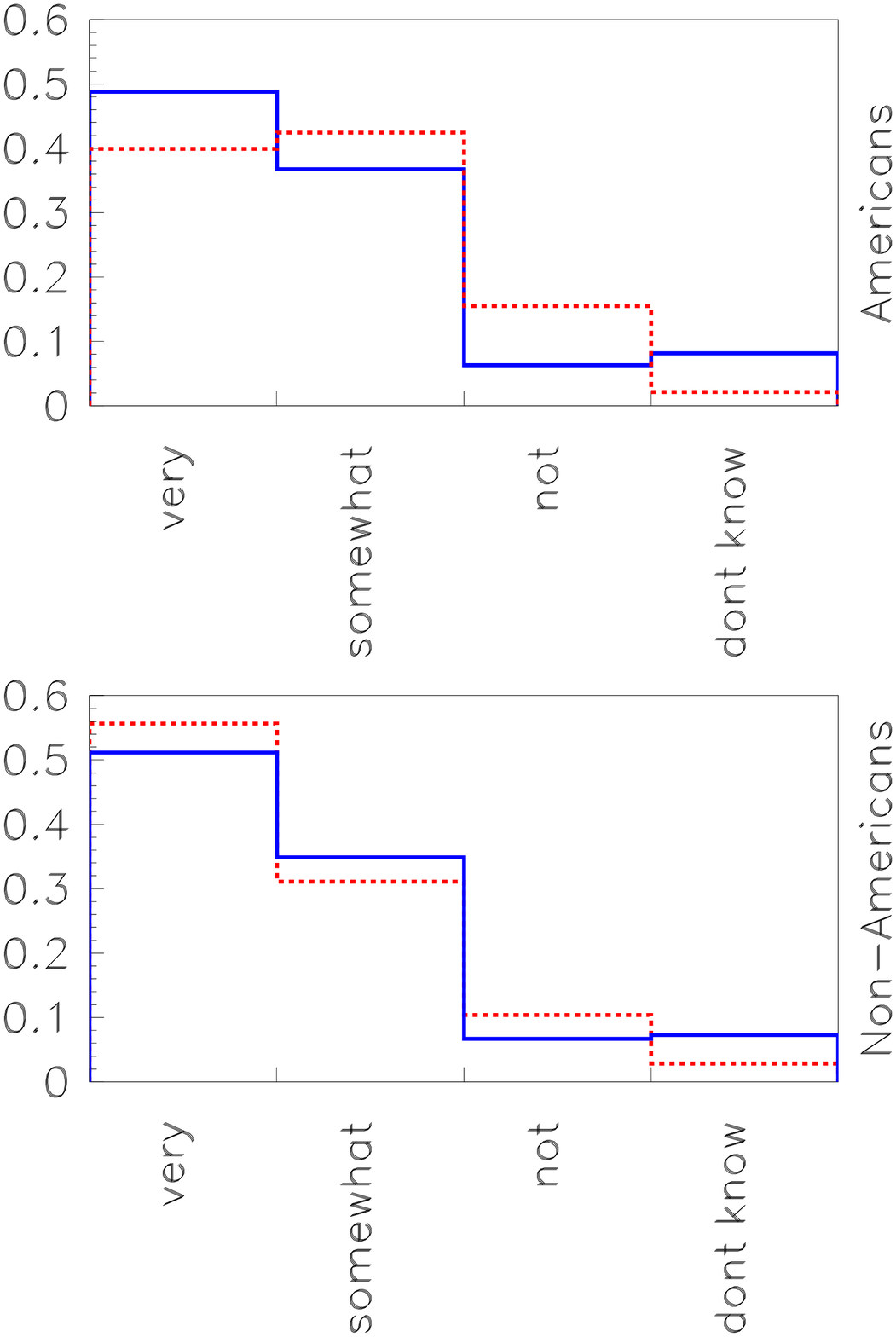}}   
\vspace{-0.2in}
     \caption[Is it important to have at least two detectors?]
             {If the ``next machine'' is a collider ($e^{+}e^{-},p
              \overline{p},\mu^{+}\mu^{-}$), is it important to have at 
              least two large, multipurpose detectors versus a single 
              detector?}
  \label{fig:redundancy}
\end{minipage}
}
\end{figure}

As indicated in Figure~\ref{fig:diversity}, a majority of all
scientists believe diversity is very important in terms of the
different types of physics results they expect from the next facility.
A fixed-target program has been a natural way to achieve this
diversity at a $p\overline{p}$ machine.  A linear collider might achieve this
by having a low-energy interaction point and/or an x-ray facility,
and a muon collider by having a neutrino factory.\\

In terms of redundancy, very few physicists seem willing to rely on
a single large detector for any given analysis (Figure~\ref{fig:redundancy}). 
Despite the presumed high cost of a second multi-purpose detector, it seems
we believe the ability to verify experimental measurements is worth the 
additional expense.\\

%apparently we believe we can't afford to let inaccurate results stand 
%unchallenged.\\

\begin{figure}[ht]
\mbox{
\begin{minipage}{0.5\textwidth}
\centerline{\epsfxsize 3.0 truein \epsfbox{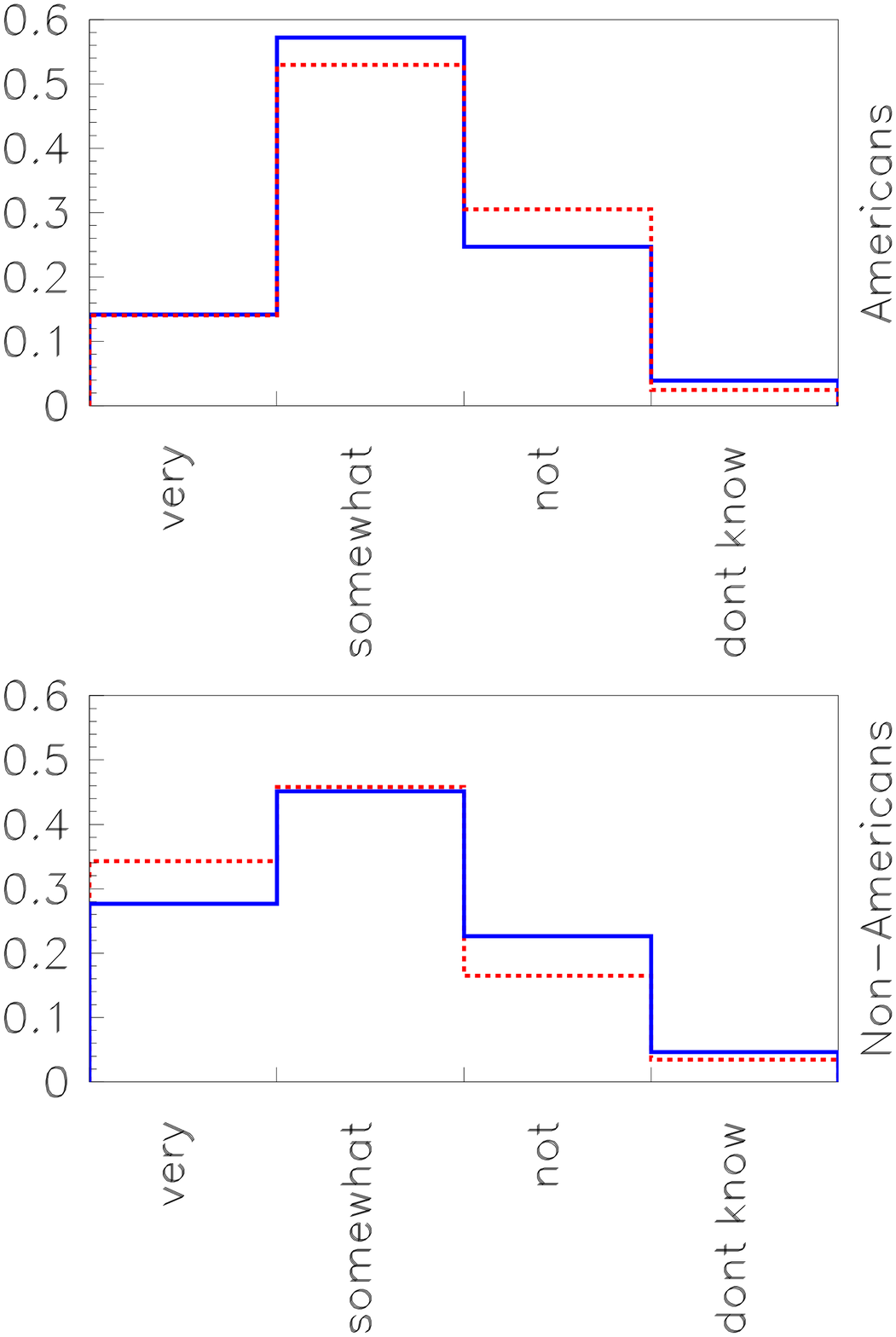}}
\vspace{-0.2in}
\caption[Is it important to balance the numbers of 
         theorists and experimentalists?]
        {Is it important to promote a balance between the numbers of 
         theorists, phenomenologists, and experimentalists?}
\label{fig:theorists}
\end{minipage}\hspace*{0.02\textwidth}
\begin{minipage}{0.5\textwidth}
\centerline{\epsfxsize 3.0 truein \epsfbox{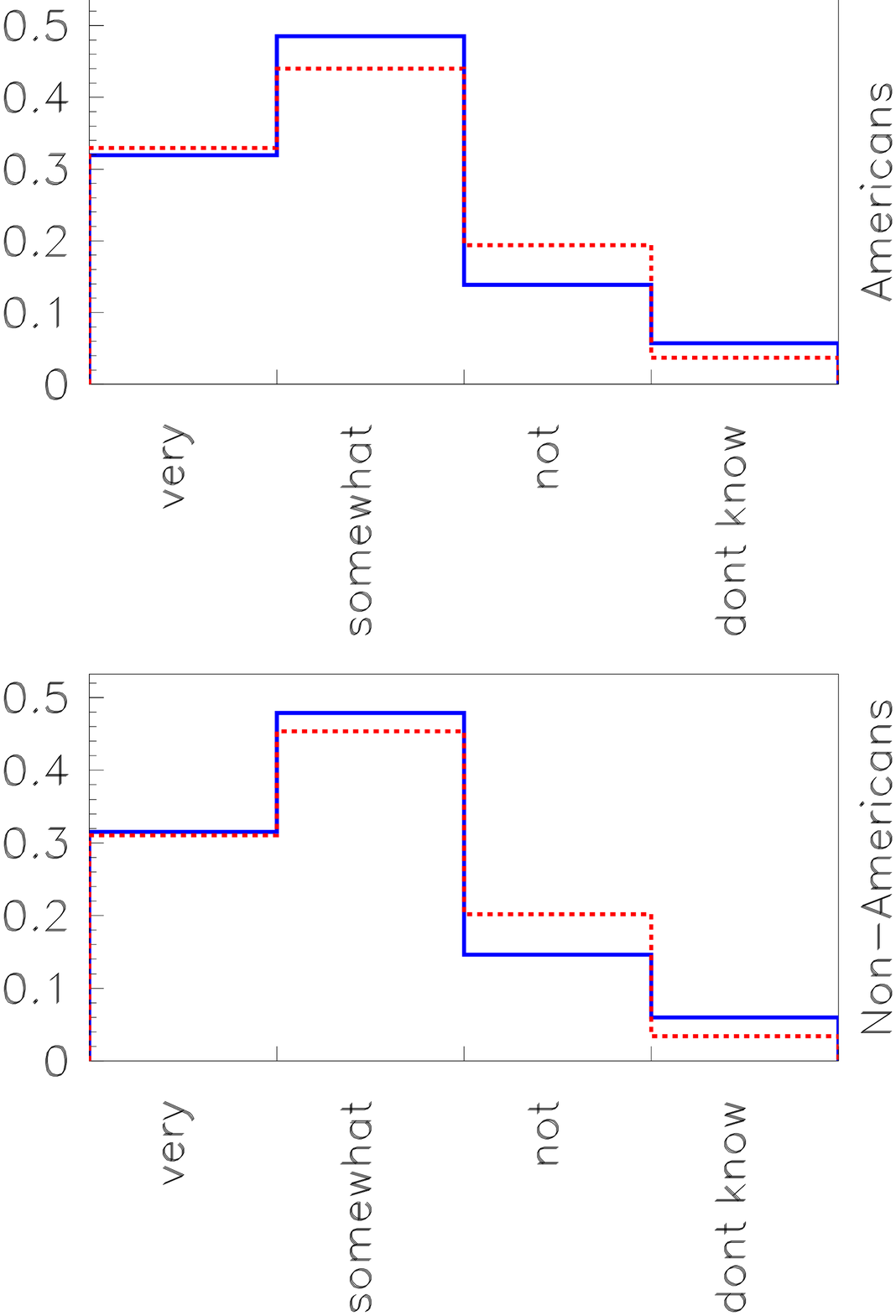}}   
\vspace{-0.2in}
\caption{Is it important that the HEP labs host astrophysics efforts?}
\label{fig:astro}
\end{minipage}
}
\end{figure}

\vspace{0.1in}
Some believe that few permanent positions exist for phenomenologists relative 
to experimentalists. Comments from the survey and town meeting \cite{townmeet}
suggest a perception that undue resources are going to particular branches
of theory (e.g. string theory) at the cost of other theoretical research.
One could imagine trying to ``correct'' each perceived imbalance, or 
to continue the current free-market policy. Figure~\ref{fig:theorists} shows 
that most Americans, both young and tenured, believe it is somewhat 
important to promote a particular mix; non-Americans appear to favor 
incentives more strongly.\\

\vspace{0.1in}
HEP laboratories host several astrophysics efforts, including the Sloan 
Digital Sky Survey, GLAST, and the Pierre Auger project.  Scientists feel 
it is somewhat important for this to continue, as shown in 
Figure~\ref{fig:astro}.\\

One might have expected a willingness to sacrifice redundant detectors
and small experiments in the quest for ever higher energies.  In this section,
high energy physicists have clearly stated the importance of a full physics
program at the site of the next major facility, not just the bare minimum 
required to operate the lab with a single detector at the frontier energy.  

\clearpage
\newpage

%% file: globalization.tex
\section{The Impact of Globalization}

This section of the survey asked questions that become most relevant
if an experiment and accelerator are located far from the home
institution.  For many respondents, the questions asked here might not
draw from direct experience.  To minimize responses drawn from
conjecture, several of the questions focus on the respondents'
current situation.\\

\begin{figure}[htp]
\mbox{
\begin{minipage}{0.5\textwidth}
\centerline{\epsfxsize 3.0 truein \epsfbox{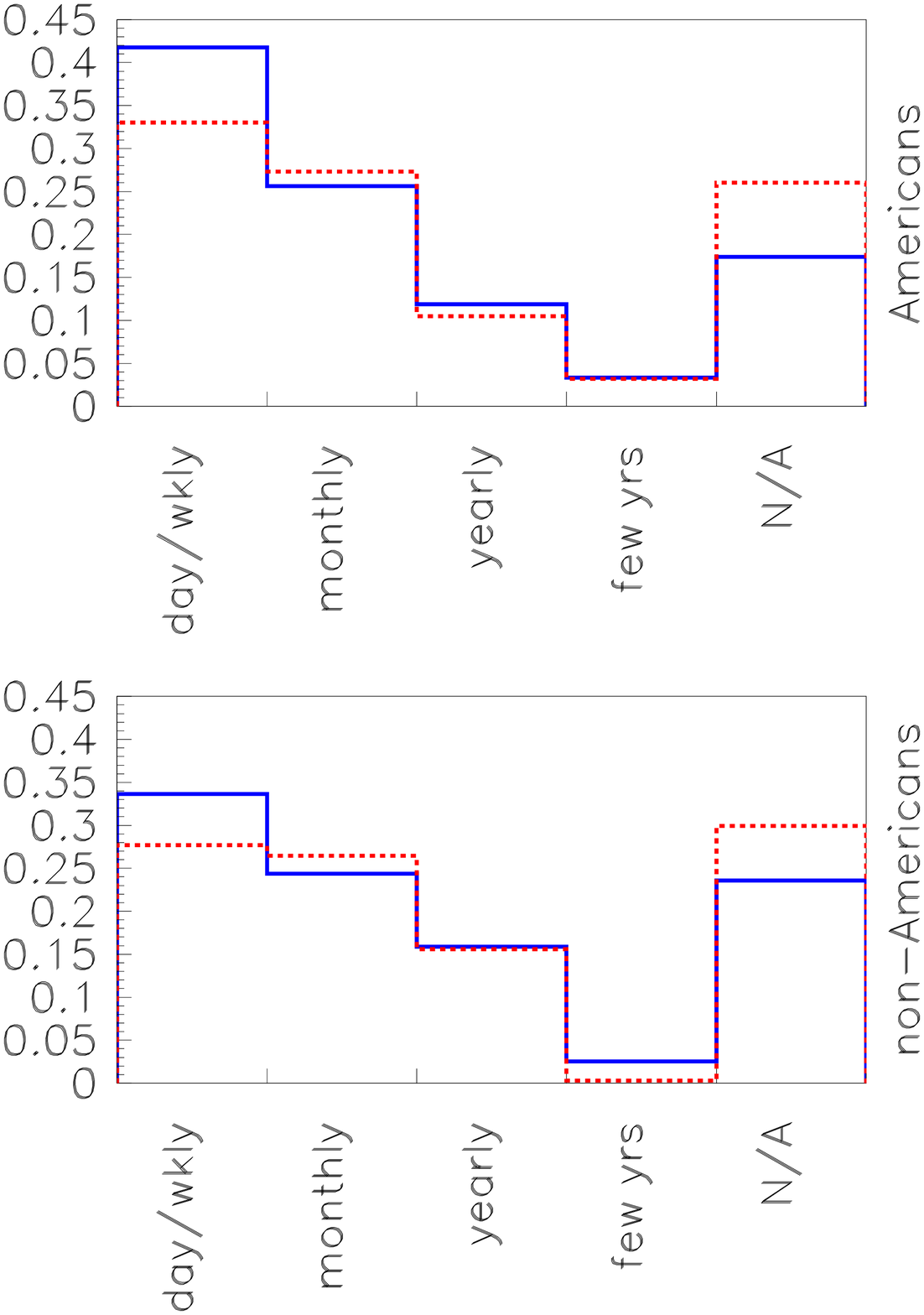}}   
\caption{How frequently do you currently see or work on your detector?}
\label{fig:seeDet}
\end{minipage}\hspace*{0.02\textwidth}

\begin{minipage}{0.5\textwidth}
\centerline{\epsfxsize 3.0 truein \epsfbox{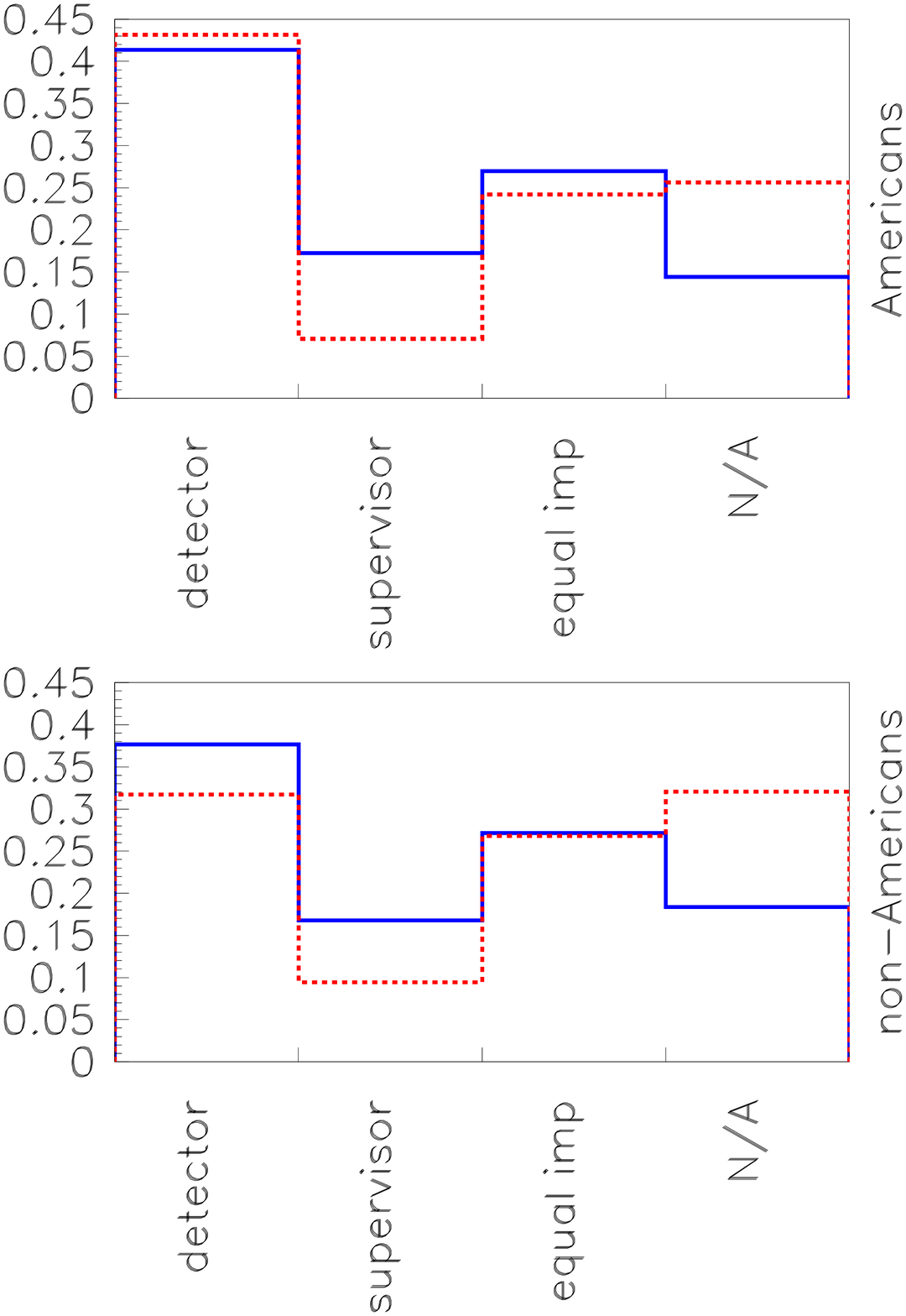}} 
\caption[How do you rate being near your detector versus being near your 
              supervisor?]
{In your current situation, given the choice, how do you 
              rate being near your detector versus being near your 
              advisor/supervisor?}
\label{fig:advisDet}
\end{minipage}
}
\end{figure}

The first question asked how frequently the respondent saw or worked
on their detector (Figure~\ref{fig:seeDet}).  Is it, in fact,
important to see your detector?  The frequency
people reported in this first question
seems quite high and might have been more informative had we asked,
``How often do you see or work on your detector for more than two hours
at a time?''  In the complementary question on the importance of
being near the detector versus near their supervisor, most selected the
detector as having the primary importance (Figure~\ref{fig:advisDet}).
We conclude that, given the choice, few physicists would prefer to work
exclusively at their home institution if they could instead work at 
the lab for a significant fraction of their time.

\begin{figure}[ht]
\mbox{
\begin{minipage}{0.5\textwidth}
\centerline{\epsfxsize 3.0 truein \epsfbox{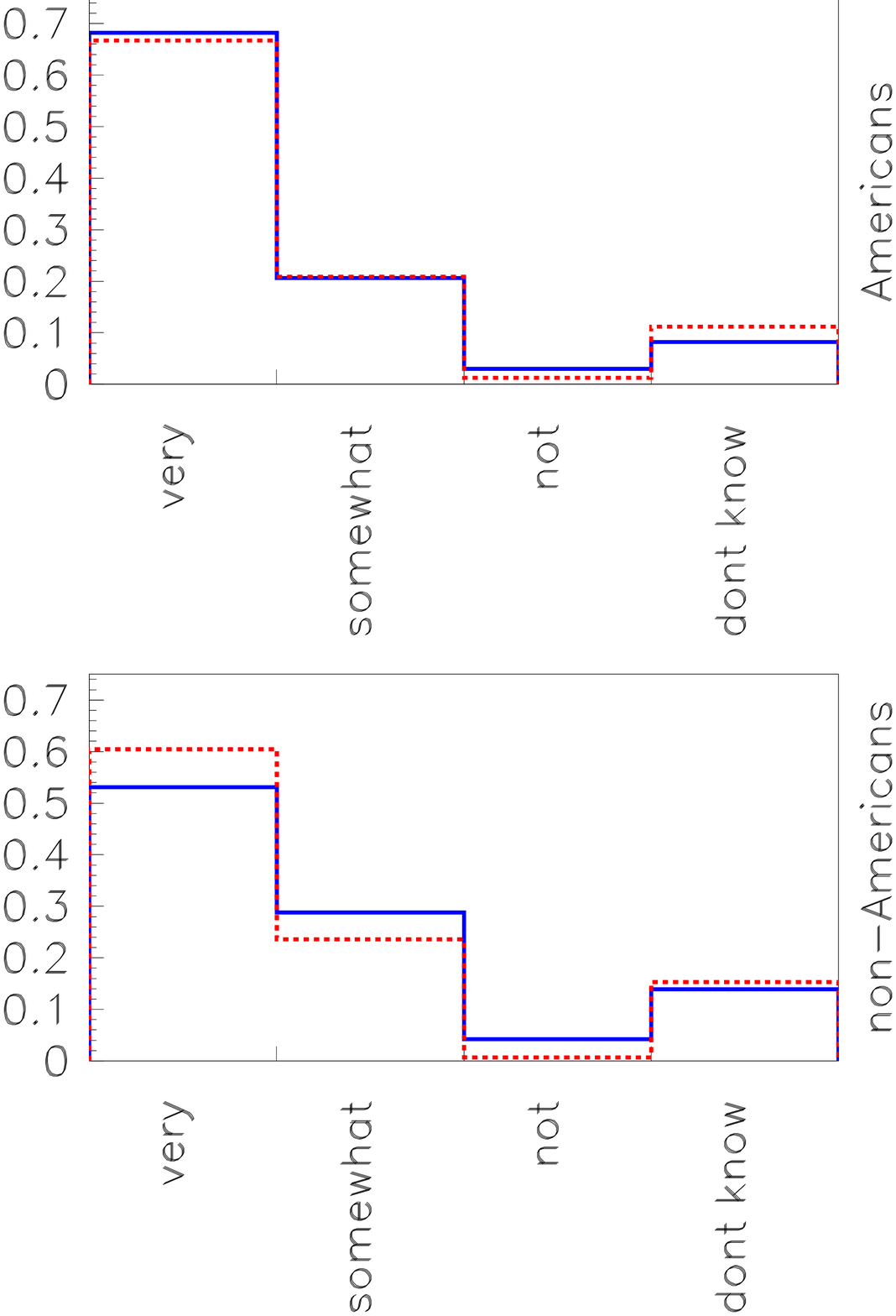}}   
\caption{How important do you feel it is to have hands-on
              hardware experience?}
\label{fig:handsDet}
\end{minipage}\hspace*{0.02\textwidth}

\begin{minipage}{0.5\textwidth}
\centerline{\epsfxsize 3.0 truein \epsfbox{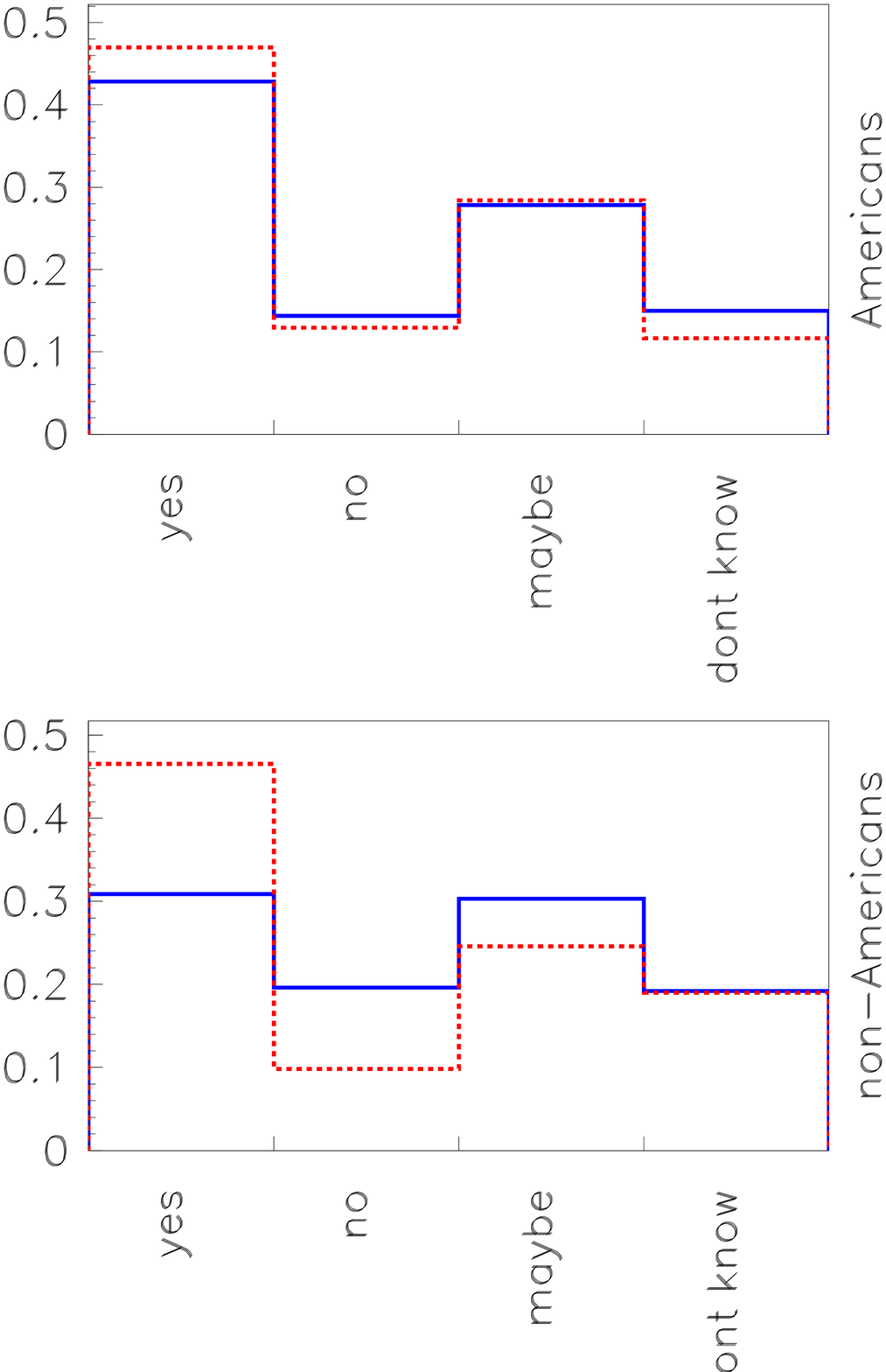}}
\vspace{0.1in}   
\caption[Should we have ``regional centers'' in the U.S.?]
{If the next machine is not located in the U.S., will the 
              quality of beams and physics results improve if we have
  	      ``regional centers'' in the U.S.?}
\label{fig:rceasy}
\end{minipage}
}
\vspace{0.2in}
\end{figure}

The prevailing opinion is that hands-on hardware experience is
very important (Figure~\ref{fig:handsDet}).  In a future survey, 
it might be interesting to learn how much hardware work is 
enough to satisfy the need.  Likewise, one could ask 
how much analysis work is sufficient to make a well-rounded physicist.\\

\begin{figure}[htp]
\mbox{
\begin{minipage}{0.5\textwidth}
\centerline{\epsfxsize 3.0 truein \epsfbox{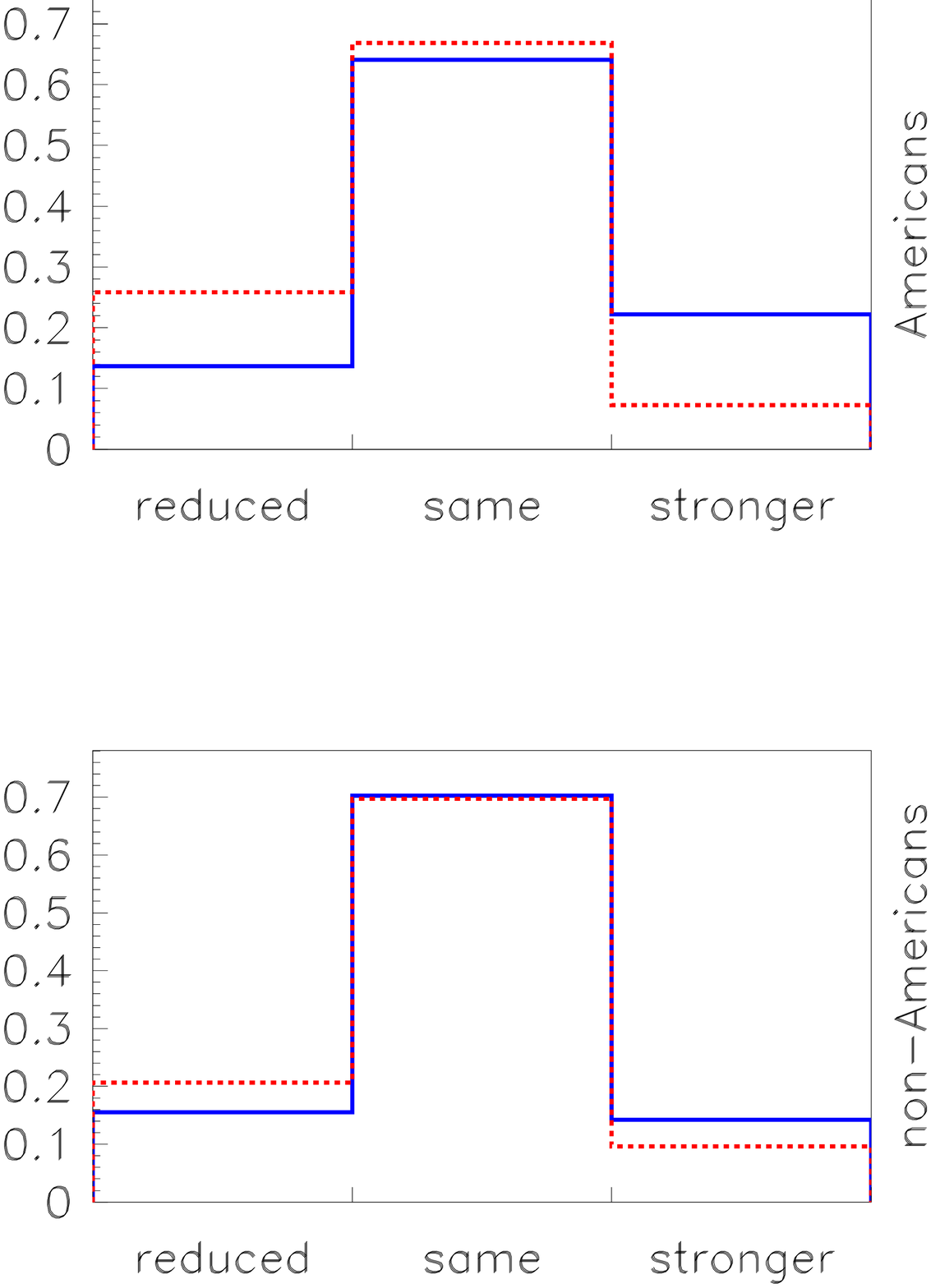}}   
\caption{What should be the role of the U.S. national labs 
              in the next 10-25 years?}
\label{fig:uslabs}
\end{minipage}\hspace*{0.02\textwidth}

\begin{minipage}{0.5\textwidth}
\centerline{\epsfxsize 3.0 truein \epsfbox{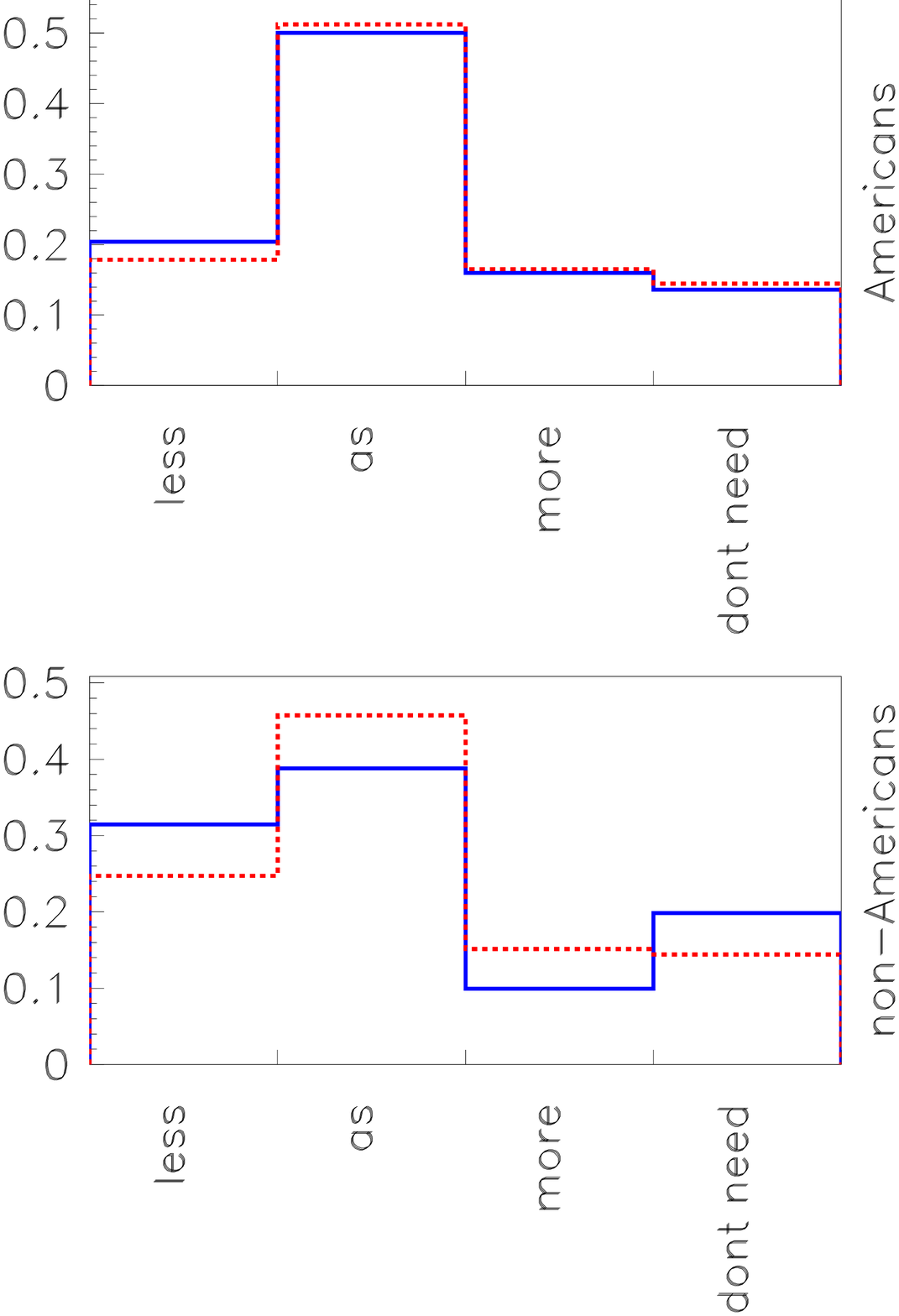}}   
\caption[How frequently should these regional centers be distributed?]
        {How frequently should these regional centers be distributed
         relative to the current national labs?}
\label{fig:rclocate}
\end{minipage}
}
\end{figure}

The next three questions dealt, directly or indirectly, with the
concept of regional centers.  These proposed structures consist of a
facility with remote control rooms, excellent (perhaps permanent) video
conferencing, and data processing ability.  The concept of a regional
center is very new and complicated; the survey questions were not able
to explore the idea as deeply as the town meeting \cite{townmeet} at
Snowmass 2001.  Although most physicists
appear to be interested in the concept of regional centers 
(Figure~\ref{fig:rceasy}), they need to see the concept demonstrated before
they can evaluate it.  The majority of physicists want to
see national labs keep their strong roles; so if a single new frontier
facility is built, a regional center could be placed at each national
lab in the countries that do not host the facility (Figures~\ref{fig:rclocate}
and \ref{fig:uslabs}).

\clearpage
\newpage

%% file: outreach.tex
\section{Outreach}

\begin{figure}[ht]
\mbox{
\begin{minipage}{0.5\textwidth}
\centerline{\epsfxsize 3.0 truein \epsfbox{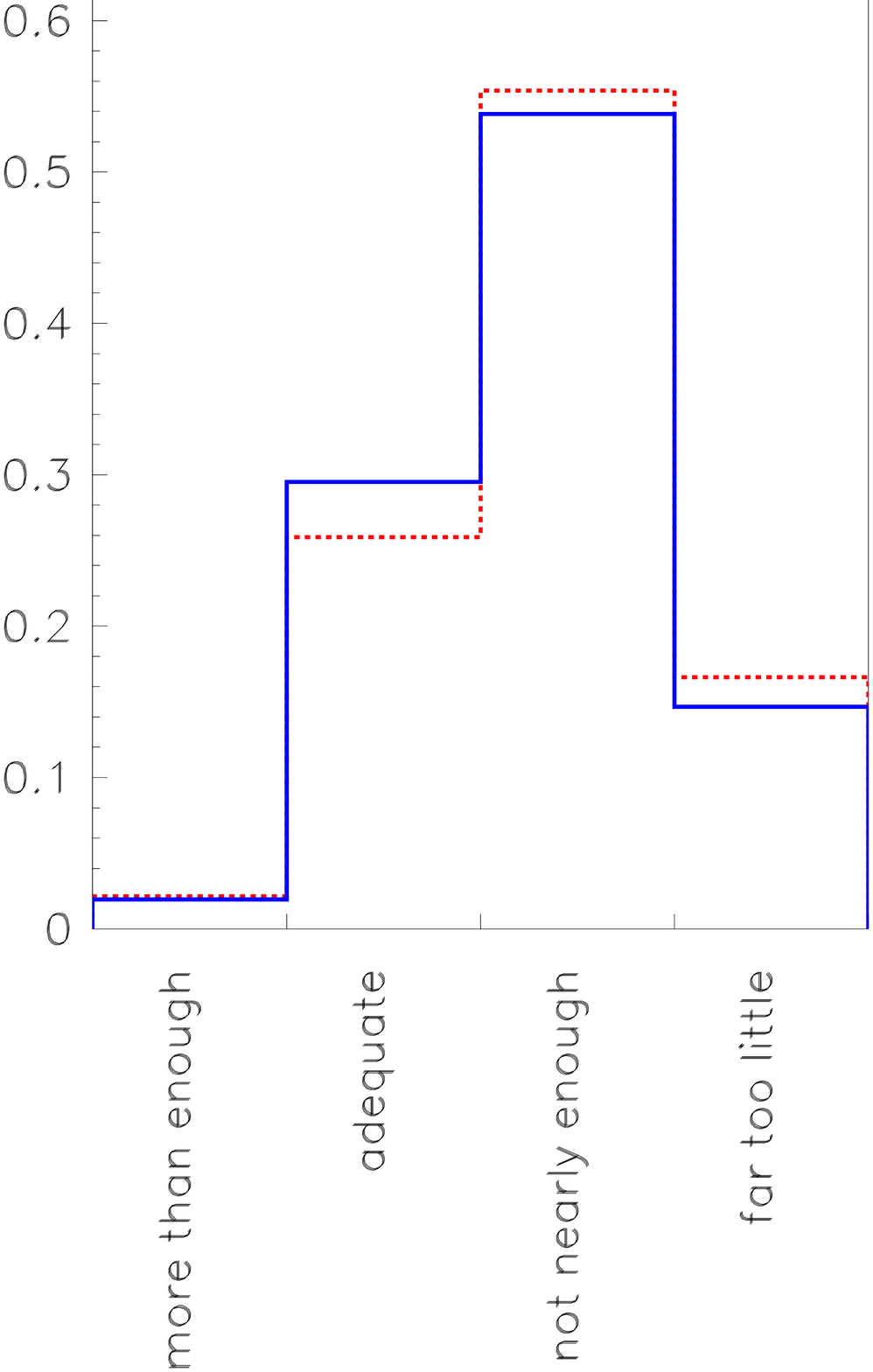}}   
\vspace{-0.2in}
     \caption{Are we currently doing enough outreach to funding agencies?}
\label{fig:outreach-congress}
\end{minipage}\hspace*{0.02\textwidth}
\begin{minipage}{0.5\textwidth}
\centerline{\epsfxsize 3.0 truein \epsfbox{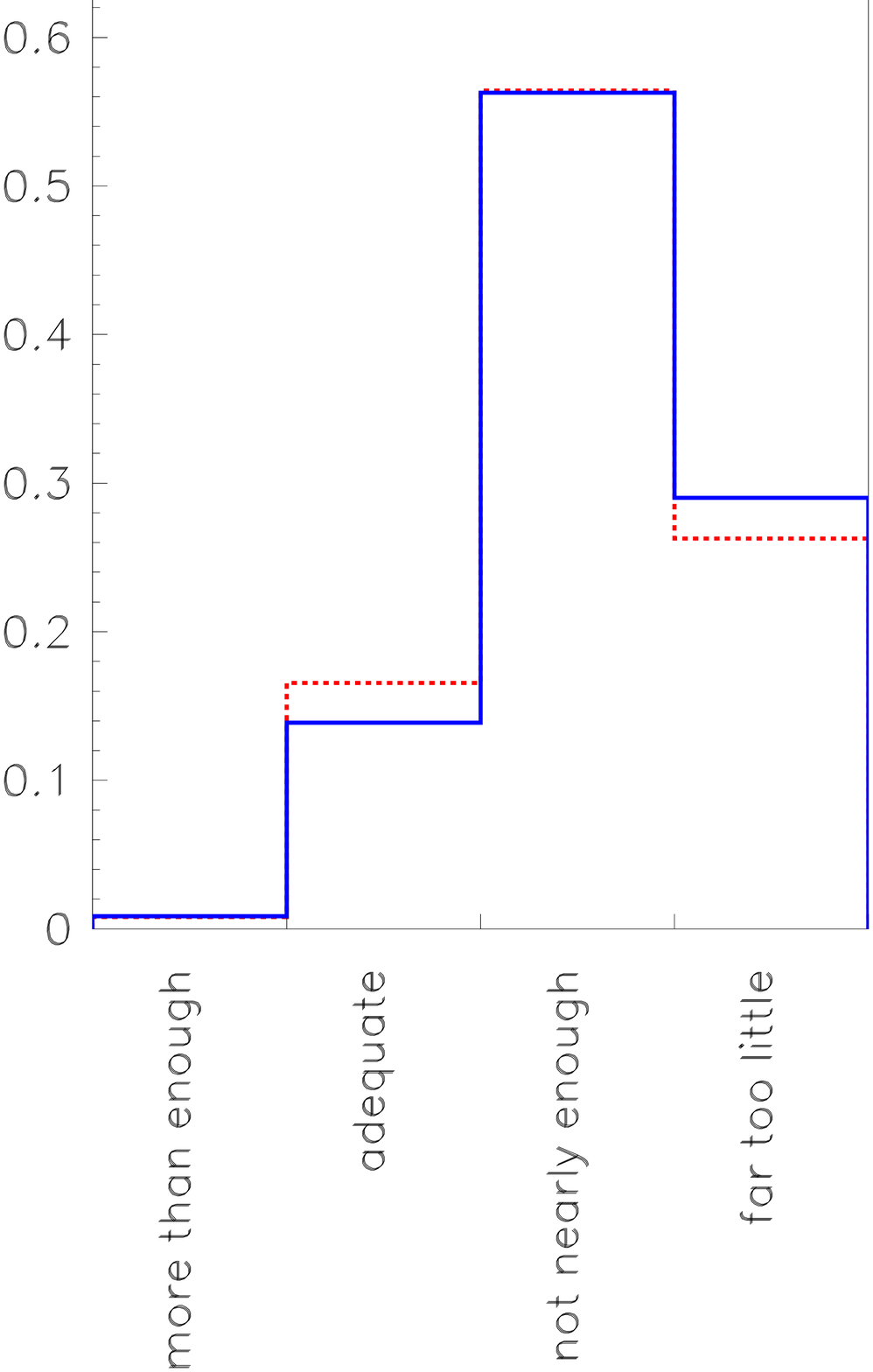}}   
\vspace{-0.2in}
\caption{Are we currently doing enough outreach to the general public?}
\label{fig:outreach-public}
\end{minipage}
}
\end{figure}

In the survey results, a clear majority believe
we are currently not doing enough outreach to either the funding agencies
(Figure~\ref{fig:outreach-congress}) or the general public
(Figure~\ref{fig:outreach-public}). In fact, physicists believe we
are doing a slightly worse job with the general public than with Congress.
These opinions on outreach hold whether you are a 
young or tenured physicist, an American or a non-American. \\

\vspace{0.2in}
In the survey, the options should have read, ``more than enough, adequate,
not enough, or far too little.''  The third option came out as ``not nearly 
enough,'' a phrase which is not easily distinguishable from the fourth option.
From the observed results, we believe that most respondents 
interpreted the third option to lie somewhere in between options 2 and 4.\\

The survey contained a third question in the section, asking if the 
respondent would be willing to commit more time to outreach 
than they are currently.  Due to a technical error, the second and third
response options were accidentally combined, rendering the results 
less meaningful. Nonetheless, 49$\%$ (743) of all respondents said 
``yes'', they would personally be willing to dedicate more time to outreach;
the rest replied ``maybe'' or ``no''.

\clearpage
\newpage

%% file: building.tex
\section{Building the Field}

For purposes of the survey, ``building the field'' refers to increasing the
numbers and quality of physicists in HEP.  This can be accomplished
by attracting larger numbers of talented graduate students, by
retaining larger numbers of talented graduates, or both.  In this
section, we explore the reasons people come to physics, why they
might feel compelled to leave and if we are retaining enough
talent to accomplish the goals of the field.\\

\begin{figure}[ht]
\mbox{
\begin{minipage}{0.5\textwidth}
\centerline{\epsfxsize 3.0 truein \epsfbox{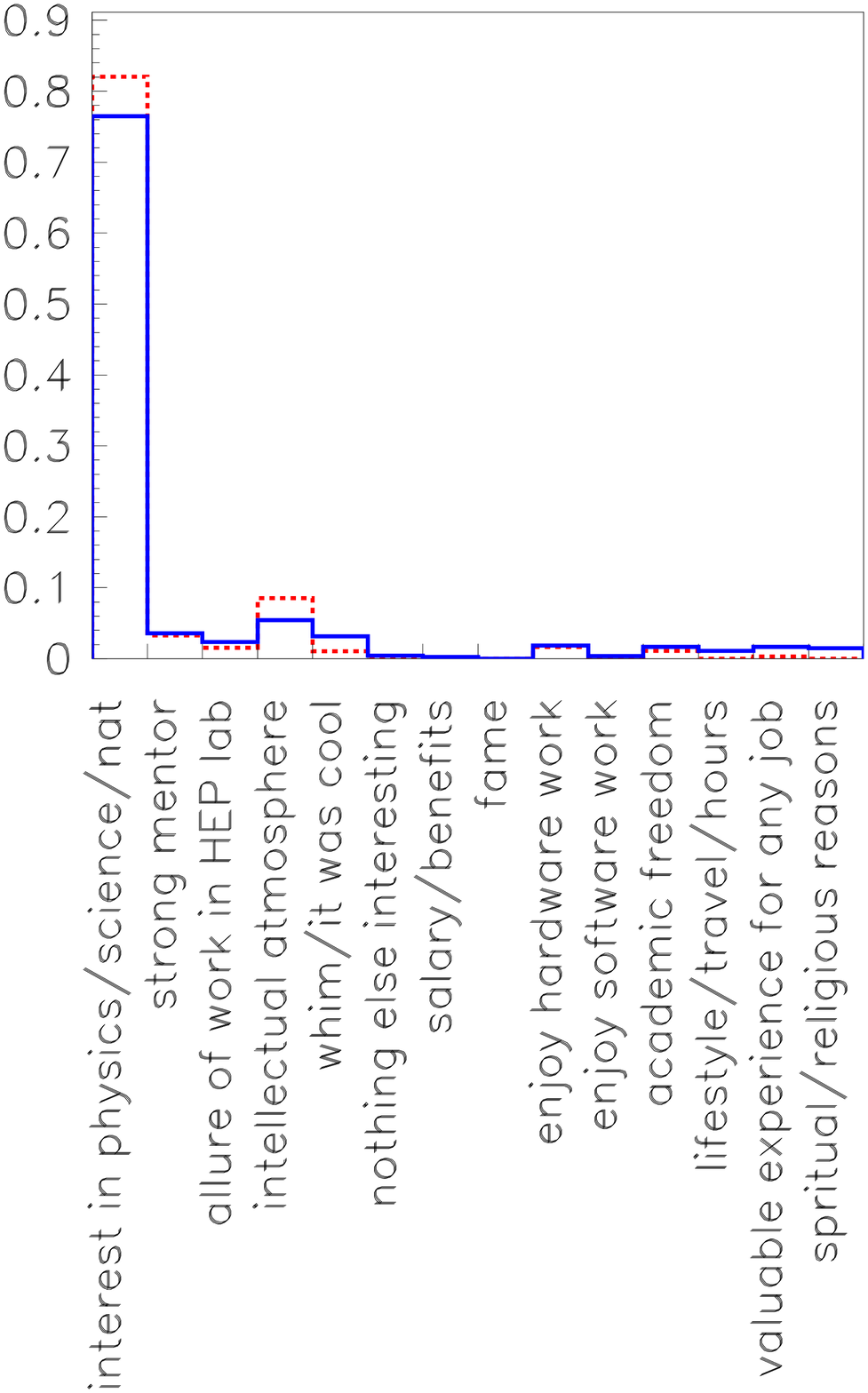}}
\vspace{-0.2in}
\caption{What is it that \textbf{most} attracted you to HEP?}
\label{fig:attracted}
\end{minipage}\hspace*{0.02\textwidth}
\begin{minipage}{0.5\textwidth}
\centerline{\epsfxsize 3.0 truein \epsfbox{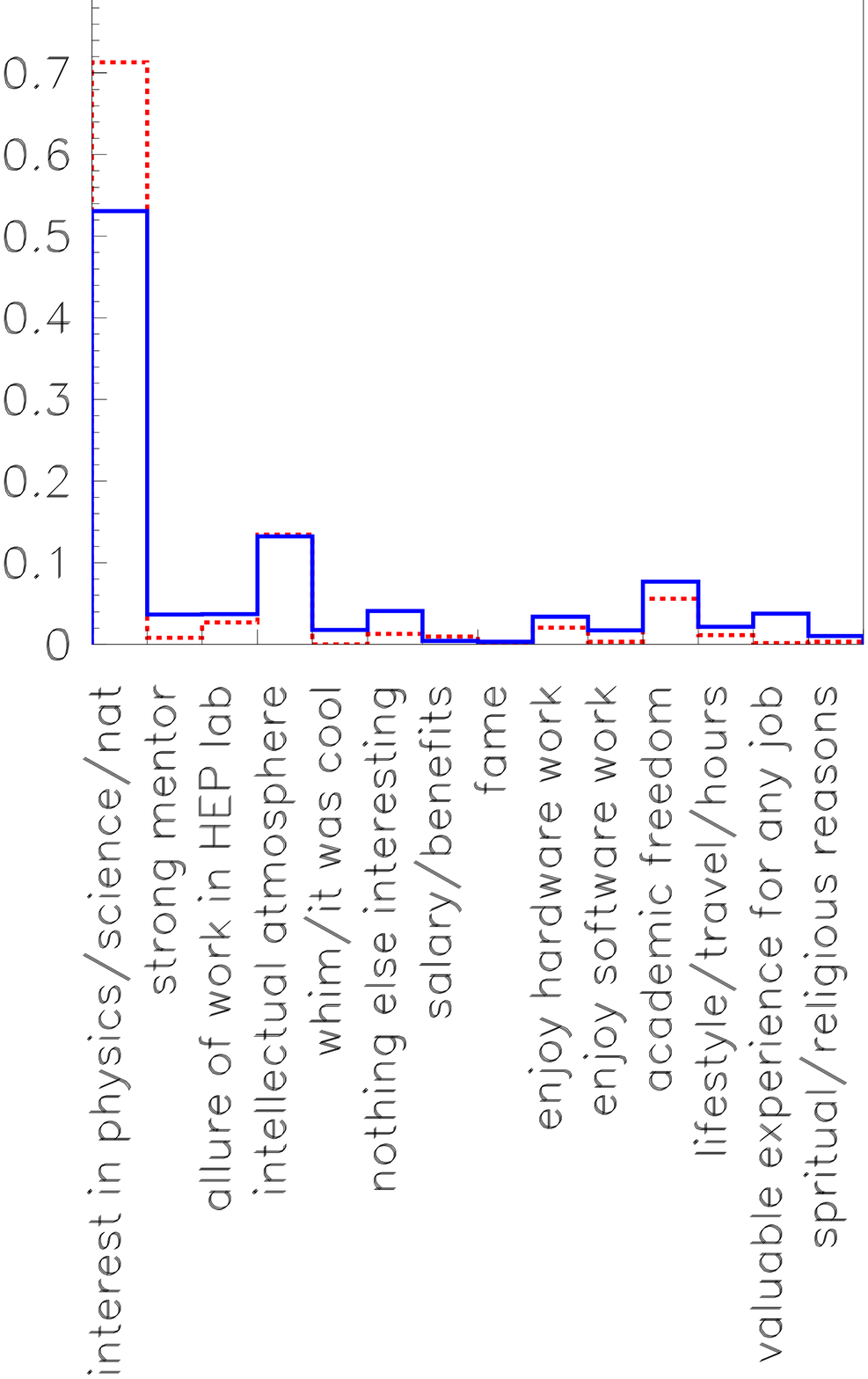}}   
\vspace{-0.2in}
\caption{What is the main reason you have stayed in the field so far?}
\label{fig:stay}
\end{minipage}
}
\end{figure}

In answer to the question of what most attracted them to the field,
most respondents selected ``interest in physics/science/nature'' as
their reason (Figure~\ref{fig:attracted}).  This was also the most
popular reason people have stayed (Figure~\ref{fig:stay}). 
The intellectual atmosphere and academic freedom options placed a
distant second and third, respectively. Only a few percent were attracted 
because it was ``good experience for any career path'' (i.e. fewer than the 
number that joined because they simply ``thought it was cool''), so this seems 
to be a weak selling point.

\vspace{-0.1in}

\begin{figure}[ht]
\mbox{
\begin{minipage}{0.5\textwidth}
\centerline{\epsfxsize 3.0 truein \epsfbox{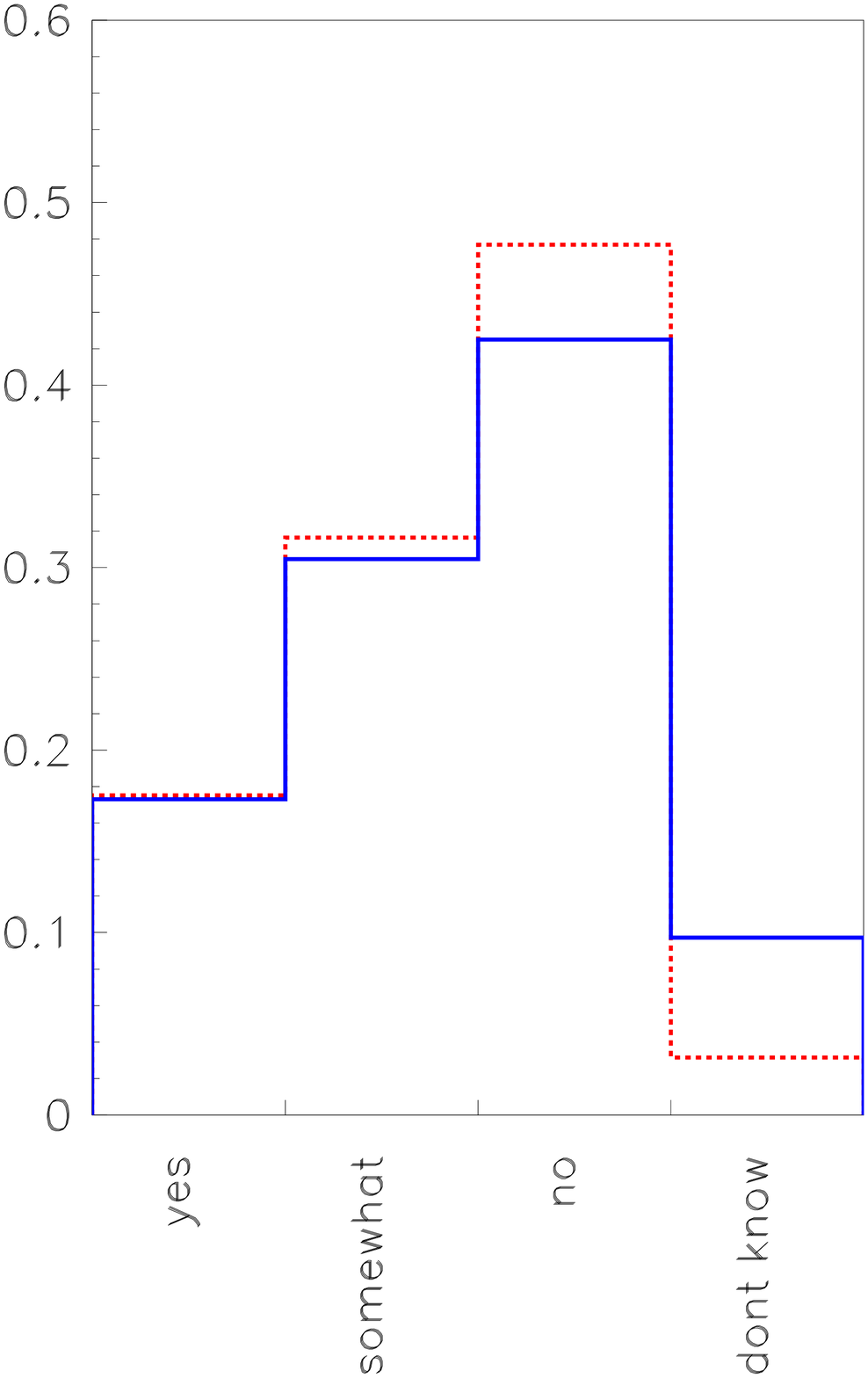}}
\vspace{-0.2in}
\caption[Are we retaining adequate numbers of talented physicists in HEP?]
        {Do you think we are currently retaining adequate numbers of 
         talented physicists in HEP?}
\label{fig:talent}
\end{minipage}\hspace*{0.02\textwidth}
\begin{minipage}{0.5\textwidth}
\centerline{\epsfxsize 3.0 truein \epsfbox{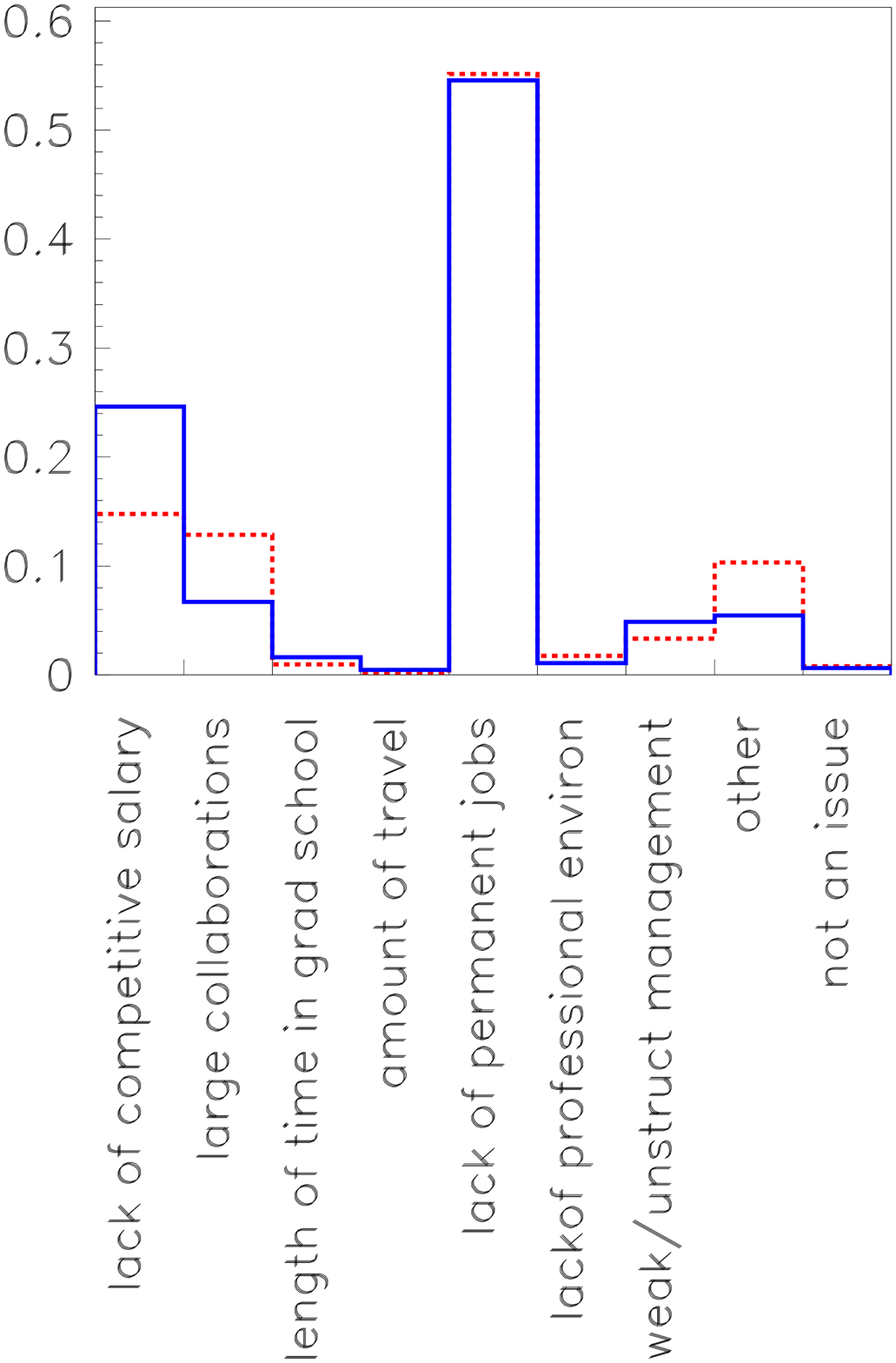}}   
\vspace{-0.2in}
\caption[What most
         influences young people to leave the field?]
{What feature of the field do you think might \textbf{most} 
         influence young people to leave the field?}
\label{fig:leave}
\end{minipage}
}
\end{figure}

\vspace{0.1in}
As shown in Figure~\ref{fig:talent}, less than 20$\%$ of all
respondents are certain that we are retaining adequate numbers of talented
physicists.  If we do indeed need to build the field, then we must
address the outward pressures that are causing HEP physicists to leave
their posts.  As shown in Figure~\ref{fig:leave}, most physicists 
believe that people leave primarily because of the lack of permanent 
positions.  Because universities employ approximately two-thirds of tenured 
scientists in the field, increased funding for university HEP might place
them in a better position to address this shortfall than the labs.\\

\vspace{0.2in}
We emphasize that Figure~\ref{fig:careerstage} in the Demographics section
indicates ample numbers of graduate students answering the survey; if we 
retain them all there certainly would be no lack of scientists for our work.  
If it is believed that the field will face a talent shortage in the future, 
then the survey results and many of the comments support one particular solution: 
create more permanent positions.

\clearpage
\newpage

%% file: physics.tex
\section{Physics}

This section attempts to identify the physics that is driving HEP today.
The answer to this question should shape our decisions not only on the
next big facility but also on smaller programs, on our theoretical focus,
and on our strategies for the next 20 years. \\

\begin{figure}[ht]
\mbox{
\begin{minipage}{0.5\textwidth}
\centerline{\epsfxsize 3.0 truein \epsfbox{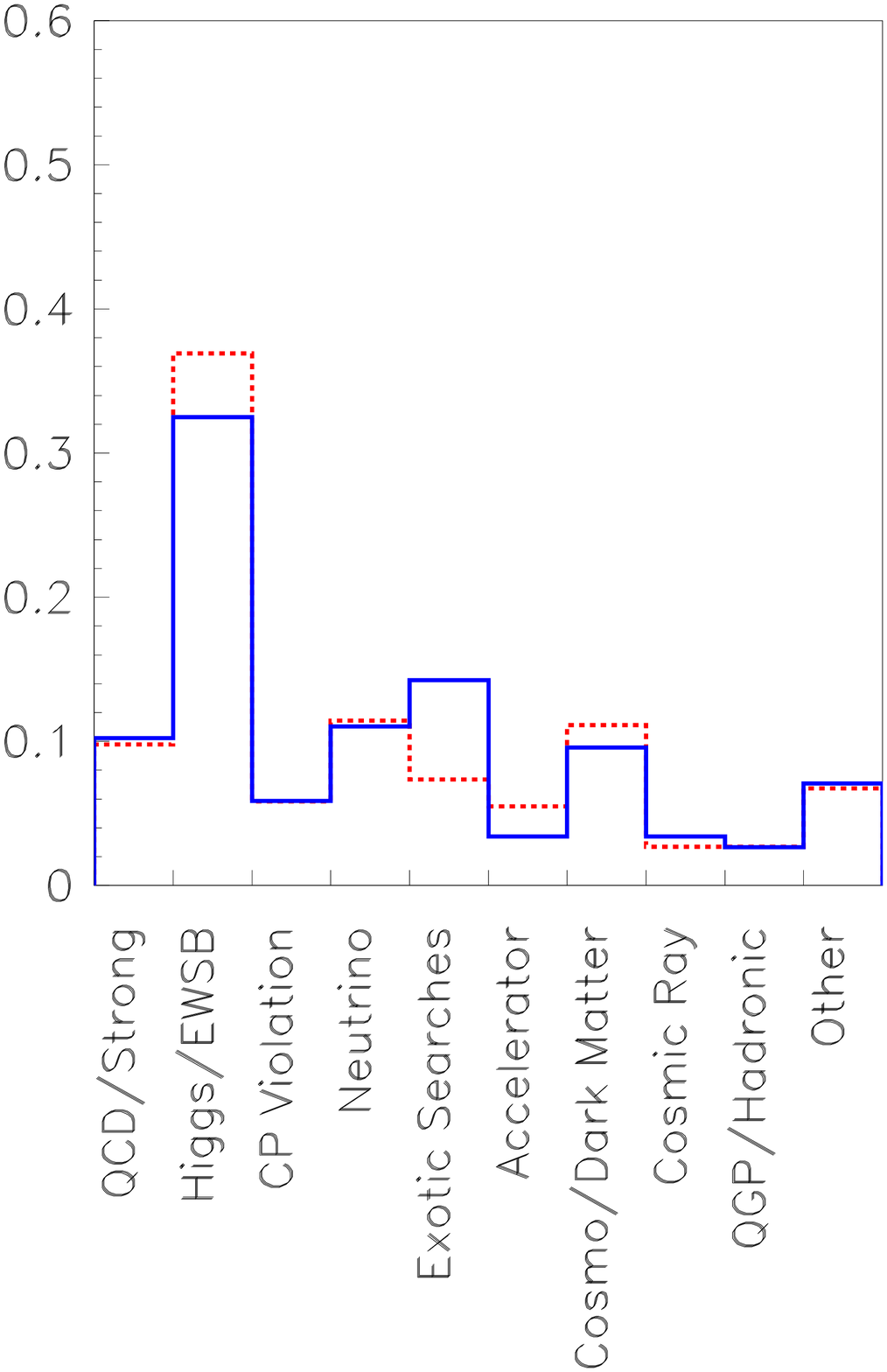}}
\vspace{-0.2in}
\caption[What physics would \textbf{you} personally find the most compelling in the future]
        {What physics would \textbf{you} personally find the most compelling 
         in the next 10-25 years?}
\label{fig:you}
\end{minipage}\hspace*{0.02\textwidth}
\begin{minipage}{0.5\textwidth}
\centerline{\epsfxsize 3.0 truein \epsfbox{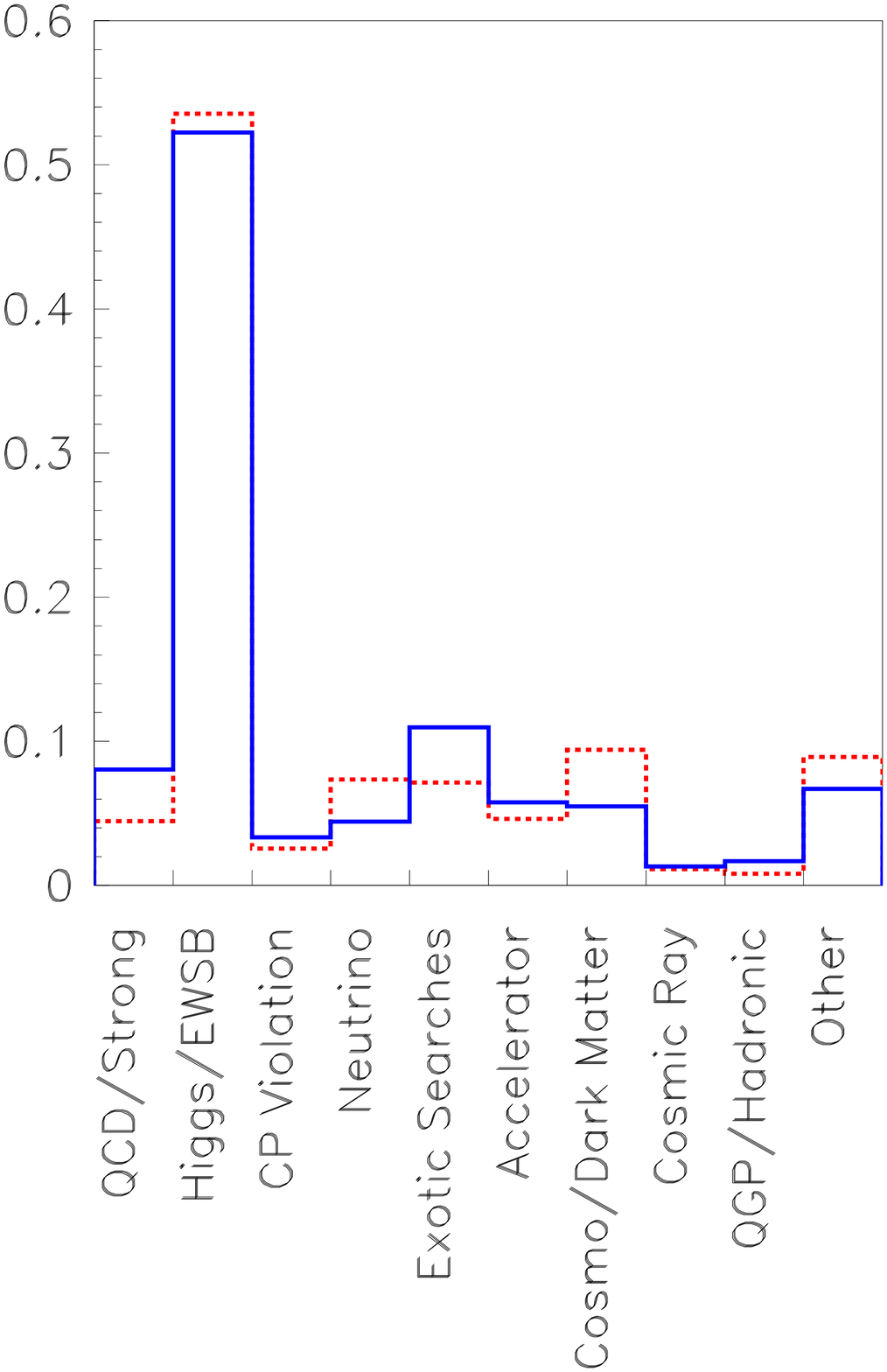}}   
\vspace{-0.2in}
\caption[What do you think is the most important physics for 
         the field in the future]
        {What do you think is the most important physics for 
         \textbf{the field} over the next 10-25 years?}
\label{fig:field}
\end{minipage}
}
\end{figure}

Along with the clear preference for studies of Higgs and electroweak
symmetry breaking (EWSB), Figure~\ref{fig:you} also shows a broad interest 
in many other areas of physics.  Interestingly, younger scientists
find exotic searches to be more compelling than do tenured scientists.  
There are no significant variations in this plot as a function of geography.
In comparing Figures~\ref{fig:you} and \ref{fig:field}, it is clear
that despite their varied personal interests, the majority of physicists
believe Higgs/EWSB is the most important physics for the field 
to study in the coming years.\\

\begin{figure}[ht]
\centering
   \epsfxsize 3.0 truein \epsfbox{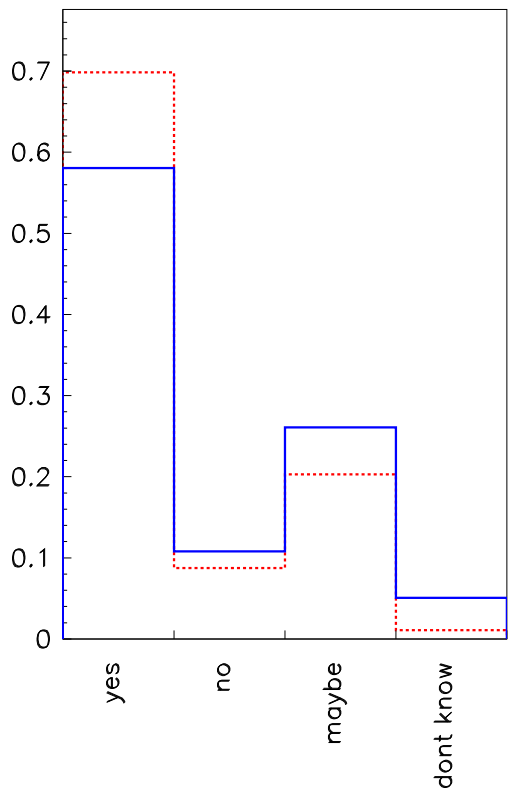}   
   \vspace{-0.2in}
   \caption{Does this science you selected require a major new facility?}
   \label{fig:newfacility}
\end{figure}

\vspace{0.2in}
Most respondents believe the most important physics does indeed require 
a major new facility (Figure~\ref{fig:newfacility}). Tenured scientists 
appear to believe this more strongly than their younger colleagues. 
Despite this perceived need, only 5$\%$ of all respondents 
listed accelerator physics as the most important physics
for the field in the future.\\

\newpage

\begin{figure}[ht]
\mbox{
\begin{minipage}{0.5\textwidth}
\centerline{\epsfxsize 3.0 truein \epsfbox{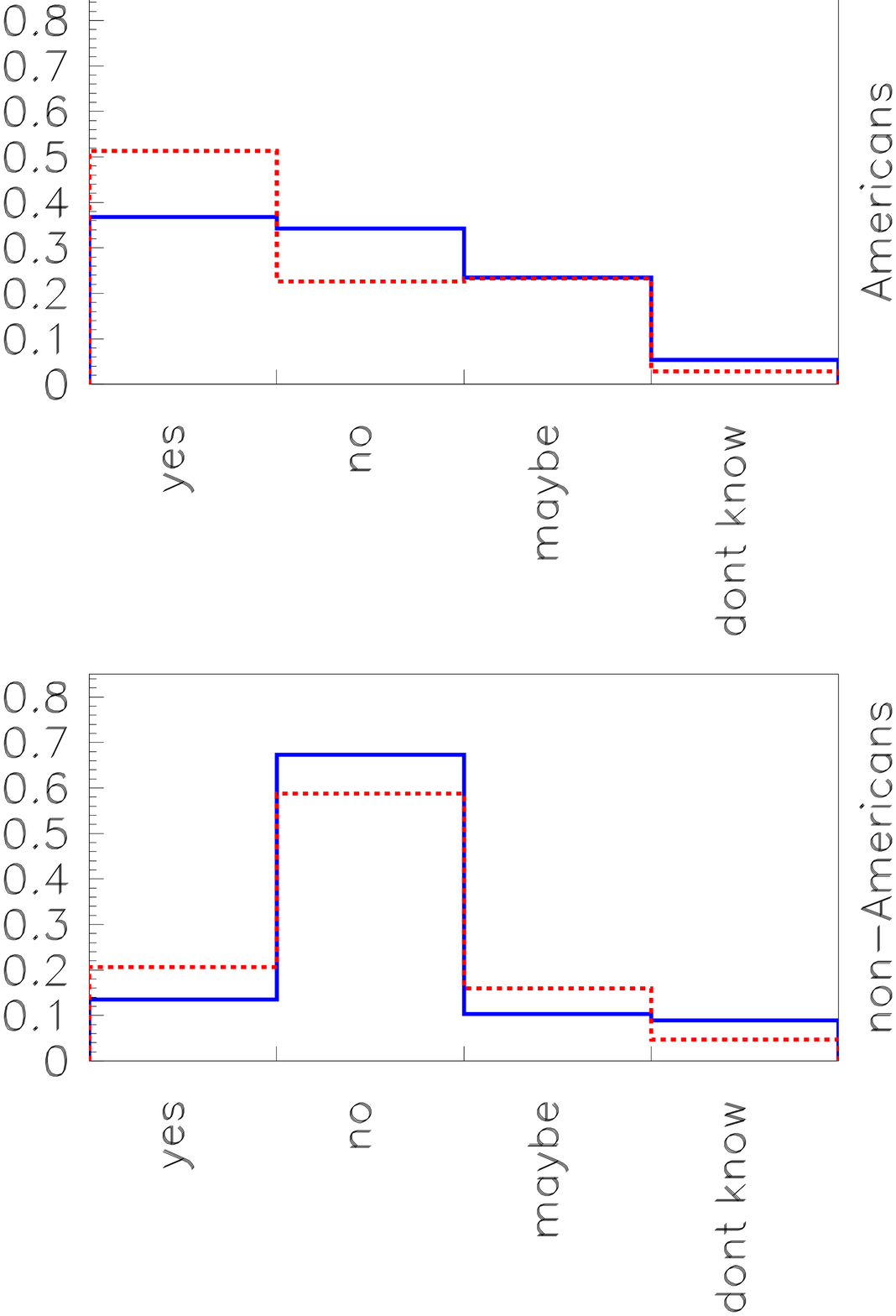}}
\vspace{-0.2in}
\caption{Do you think it is important that the next new facility be 
         in the U.S.?}
\label{fig:newfacilityus}
\end{minipage}\hspace*{0.02\textwidth}
\begin{minipage}{0.5\textwidth}\vspace{0.9in}
\centerline{\epsfxsize 3.0 truein \epsfbox{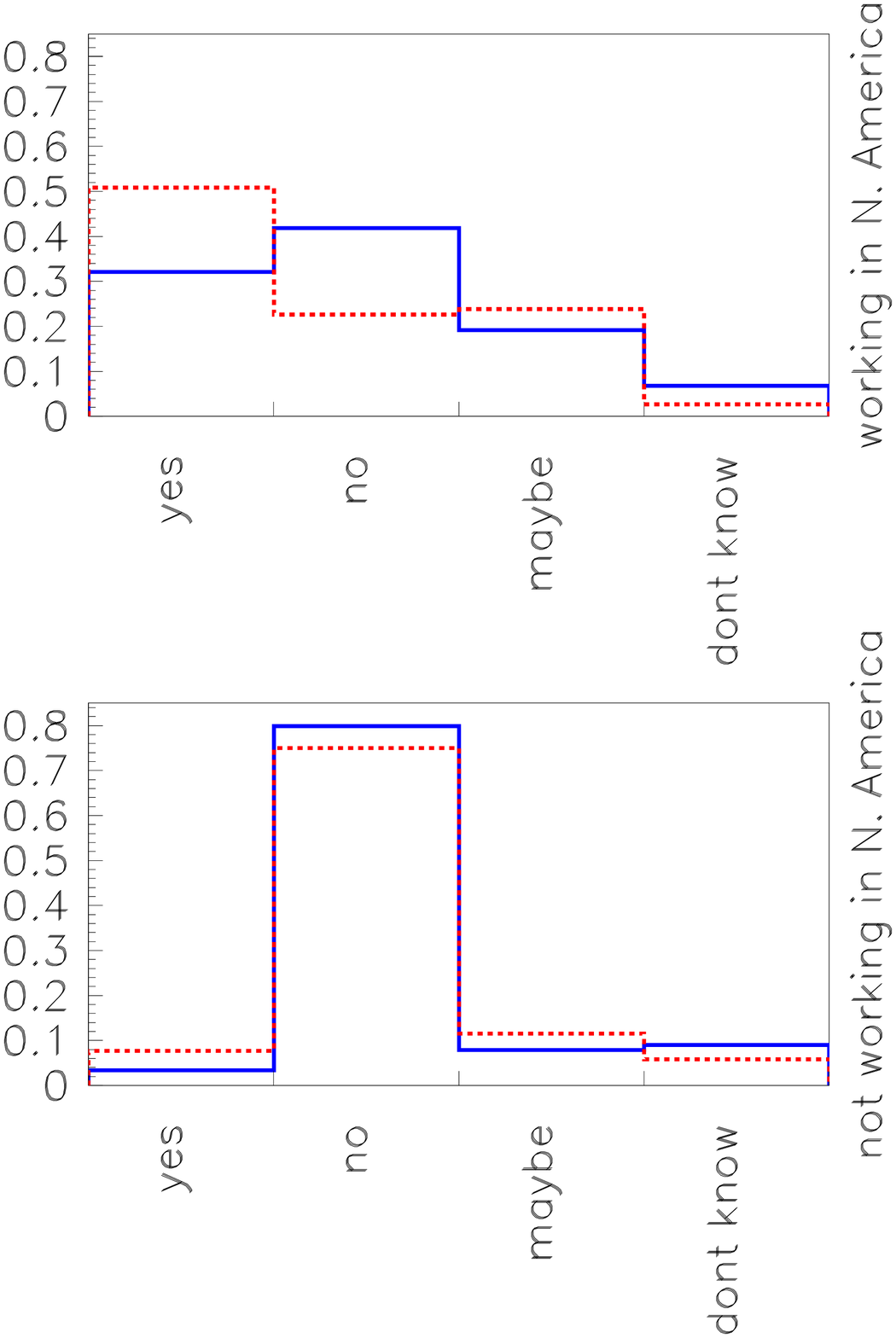}}   
\vspace{-0.2in}
\caption[Do you think it is important that the next new facility be 
         in the U.S.?]
        {Do you think it is important that the next new facility be 
         in the U.S.? Responses are now separated based on where the 
         respondent currently works and not where they grew up. Top plot 
         are those currently working in North America, bottom plot are those 
         currently not working in North America.}
\label{fig:newfacilityus-work-in-us}
\end{minipage}
}
\end{figure}

\vspace{0.2in}
Proposals for future frontier facilities exist in a number of countries
around the world.  Given the high cost to build and operate any of the proposed
large-scale machines, only one such facility (of a given kind) is 
likely to be built.  So how does one select a host country?
We specifically asked how important it is that the next new facility be in 
the U.S. (Figure~\ref{fig:newfacilityus}). While young American scientists 
appear to be more divided, most tenured American scientists think it is 
important to have a new facility in the U.S.  The overwhelming majority of 
non-Americans believe it is not important that a new facility be in the U.S.  
These results change slightly if they are subdivided based on where 
the respondent currently works rather than where they grew up 
(Figure~\ref{fig:newfacilityus-work-in-us}). In light of these responses,
the HEP community must take great care to select the location of the next 
machine in a way that results in global cooperation and shared funding,
rather than in duplicate, separate efforts.

\clearpage
\newpage

%% file: pick.tex
\section{Finding Consensus and Picking a Plan}

This section attempted to determine if the HEP community should
commit to a particular course within the next few years or wait
for more information.  In addition to plainly asking for a favored machine 
plan, this section tried to gauge the respondents' knowledge 
of each of the proposed machines, and to identify the most 
effective learning media.\\

\begin{figure}[ht]
\centering
     \epsfysize=4.0in \epsfbox{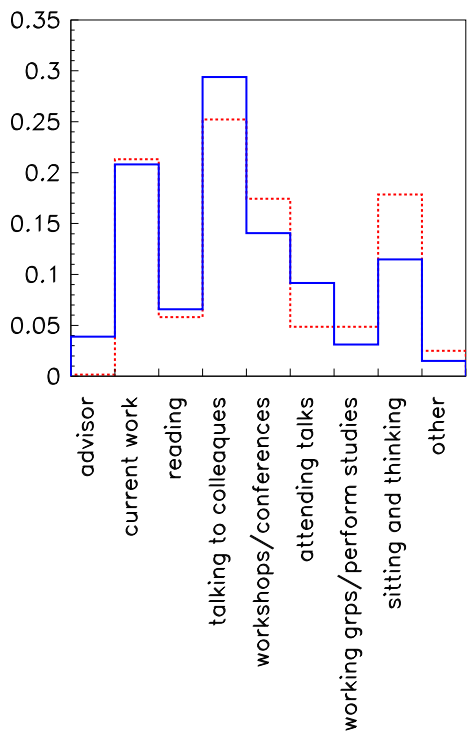}\vspace{-0.1in}
     \caption[What most affected your current opinion on the future options?]
             {What has most affected your current opinion(s) on the options 
              for the future of HEP?}
  \label{fig:affectopinion}
\vspace{0.2in}
\end{figure}

Figure~\ref{fig:affectopinion} shows that a number of factors influenced
people's current opinions. Conversations with colleagues ranked first,
followed by current work, and workshop attendance.  Young physicists
are much less likely than tenured ones to report ``sitting and thinking'' 
as their major influence.

\begin{figure}[h]
\mbox{
\begin{minipage}{0.5\textwidth}
\centerline{\epsfxsize 3.0 truein \epsfbox{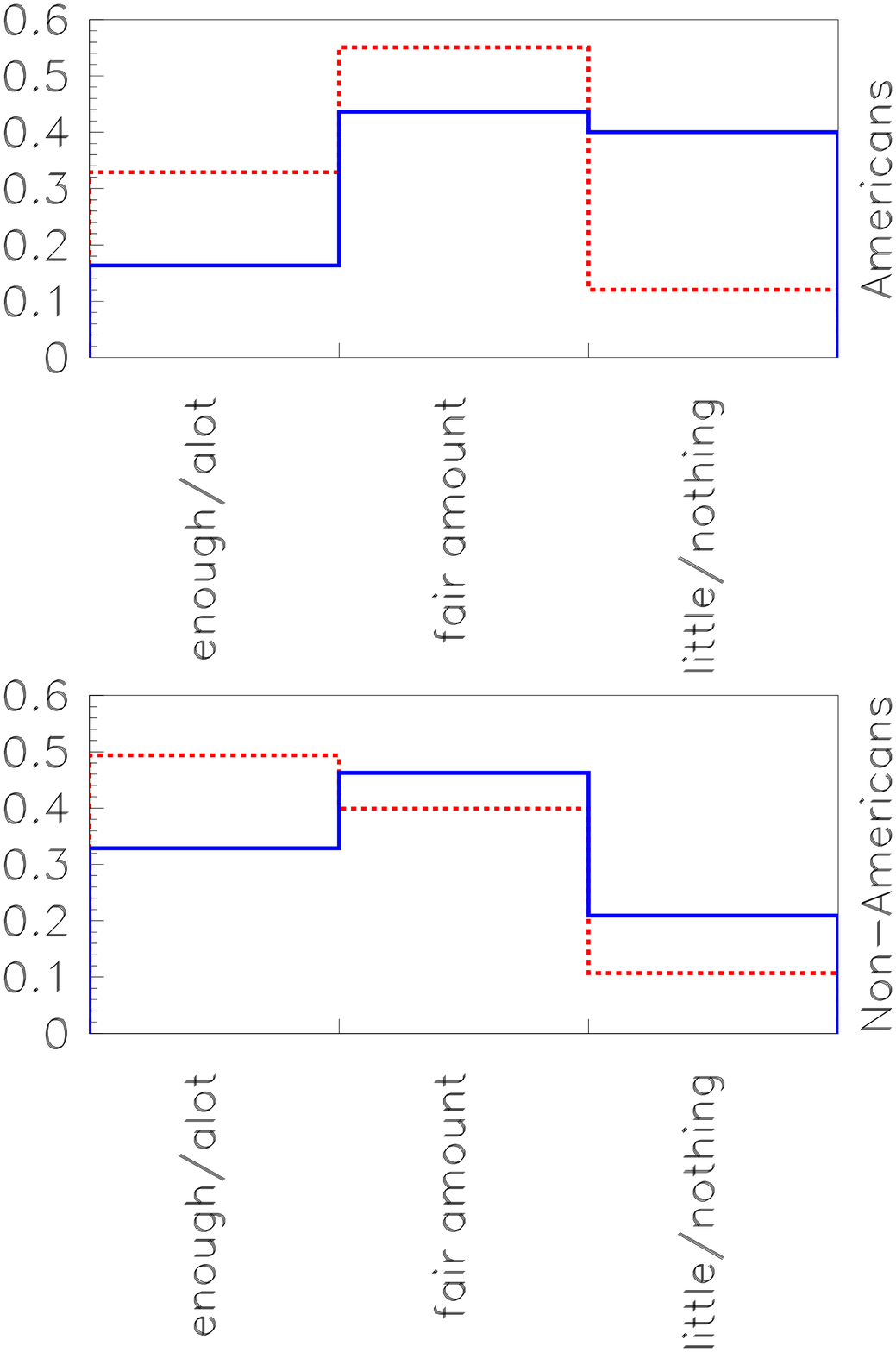}}   
\vspace{-0.2in}
\caption{How much do you know about TESLA?}
\label{fig:tesla}
\end{minipage}\hspace*{0.02\textwidth}
\begin{minipage}{0.5\textwidth}
\centerline{\epsfxsize 3.0 truein \epsfbox{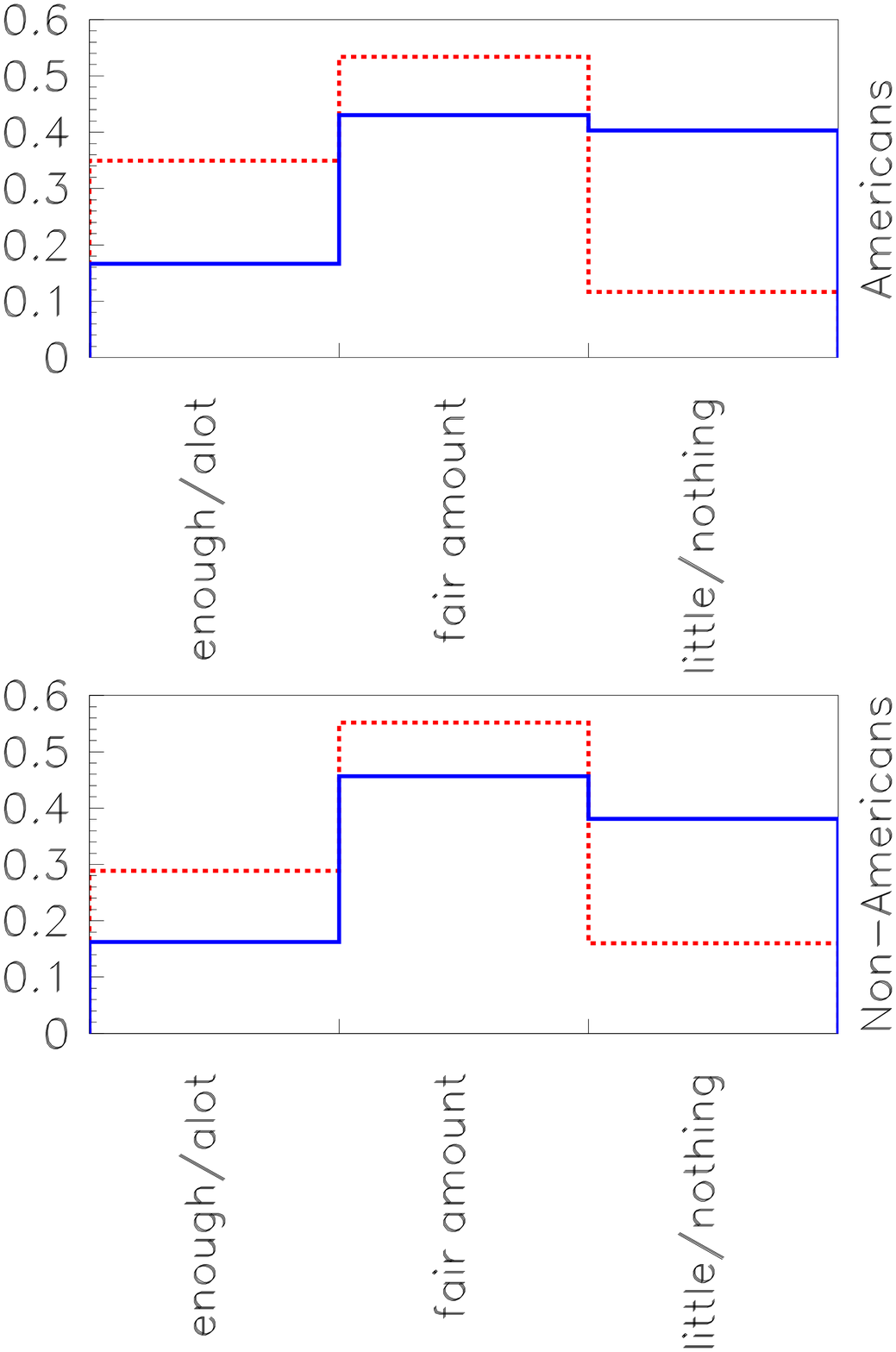}}   
\vspace{-0.2in}
\caption{How much do you know about LC non-TESLA versions (i.e. NLC, CLIC)?}
\label{fig:nlc}
\end{minipage}
}
\end{figure}

\begin{figure}[h]
\centering
     \epsfysize=4.0in \epsfbox{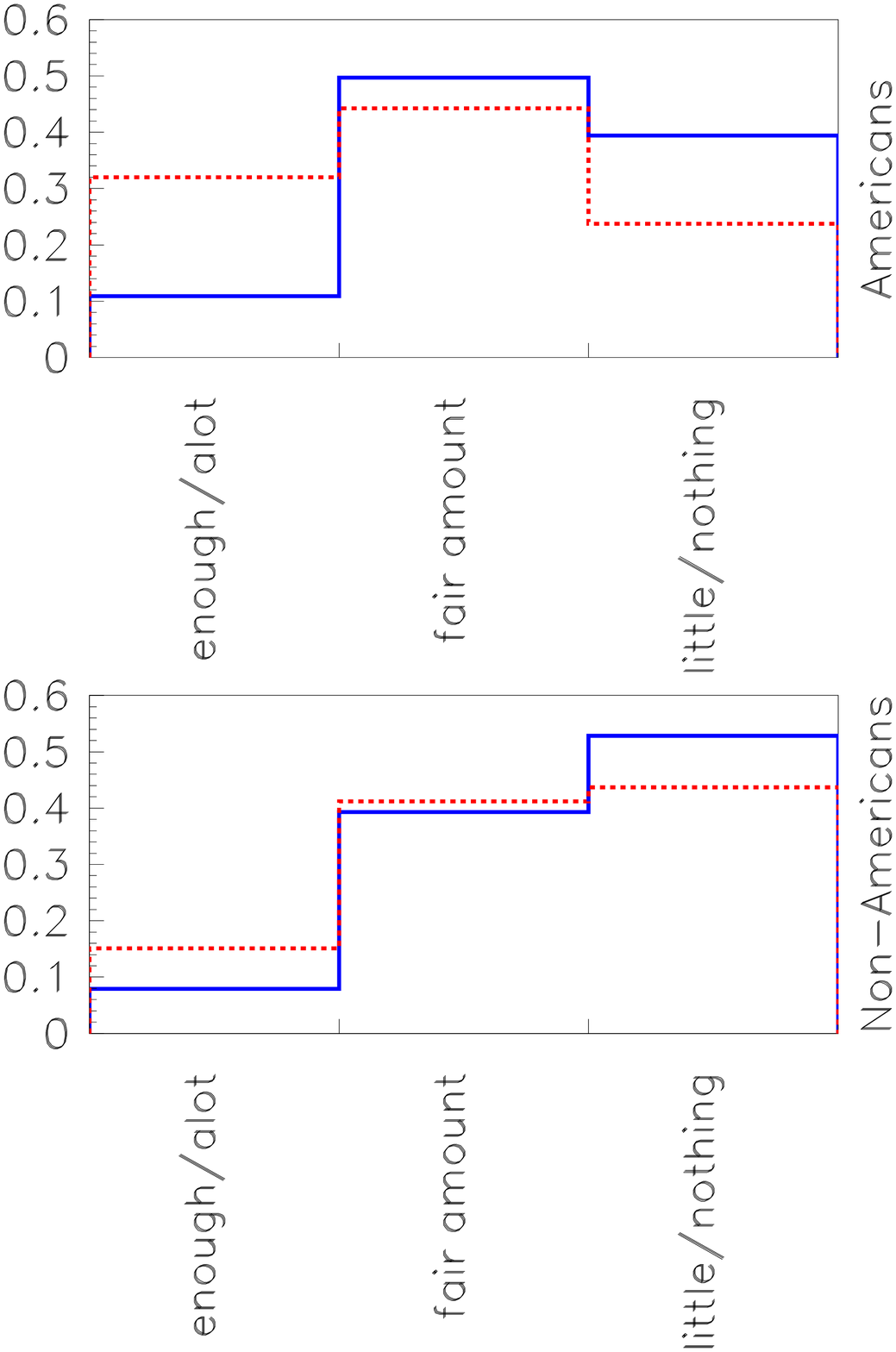}\vspace{-0.2in}
     \caption{How much do you know about the VLHC?}
  \label{fig:vlhc}
\end{figure}

\begin{figure}[h]
\mbox{
\begin{minipage}{0.5\textwidth}
\centerline{\epsfxsize 3.0 truein \epsfbox{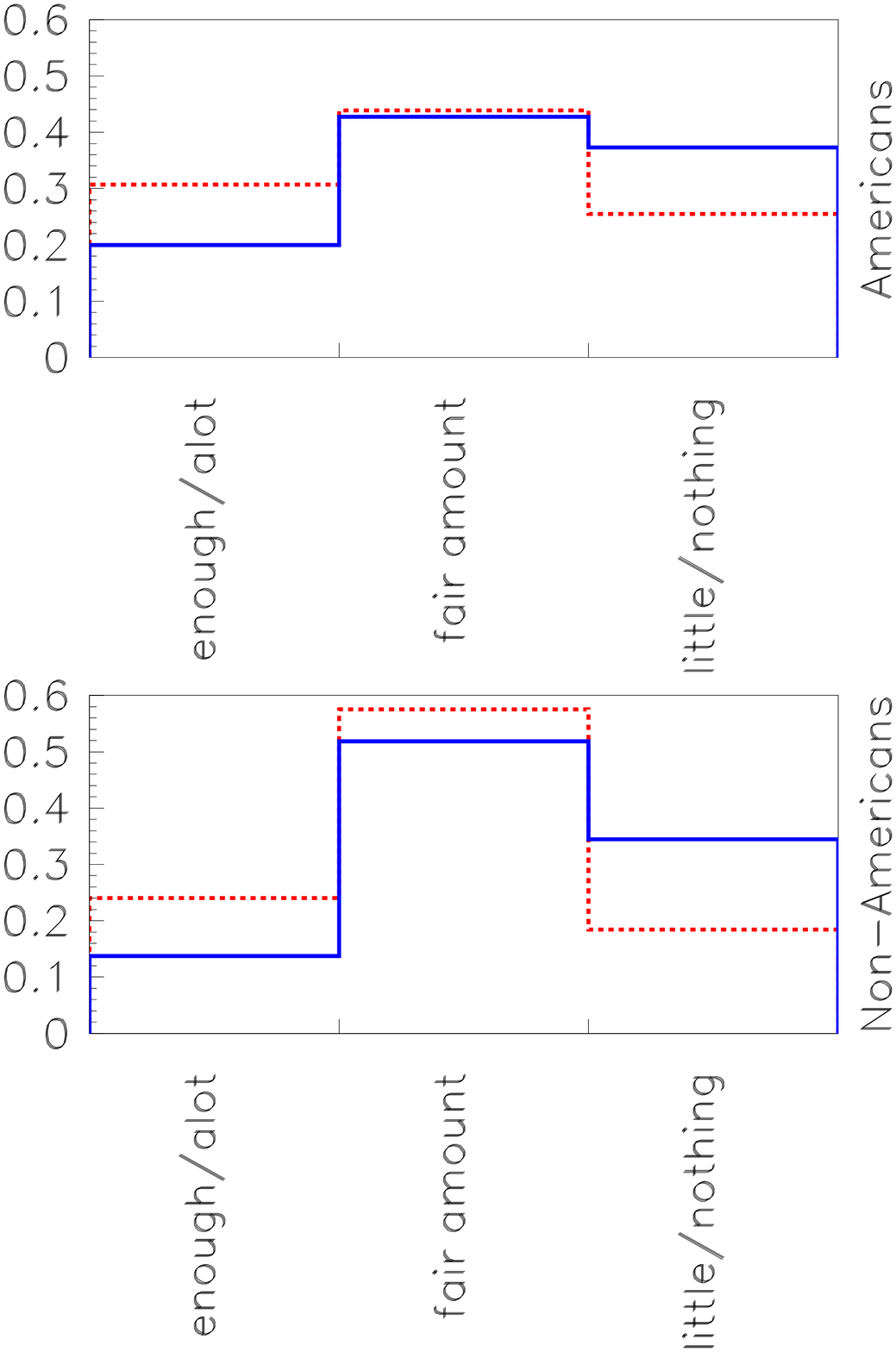}}   
\vspace{-0.2in}
\caption{How much do you know about the Neutrino Factory?}
\label{fig:nufact}
\end{minipage}\hspace*{0.02\textwidth}
\begin{minipage}{0.5\textwidth}
\centerline{\epsfxsize 3.0 truein \epsfbox{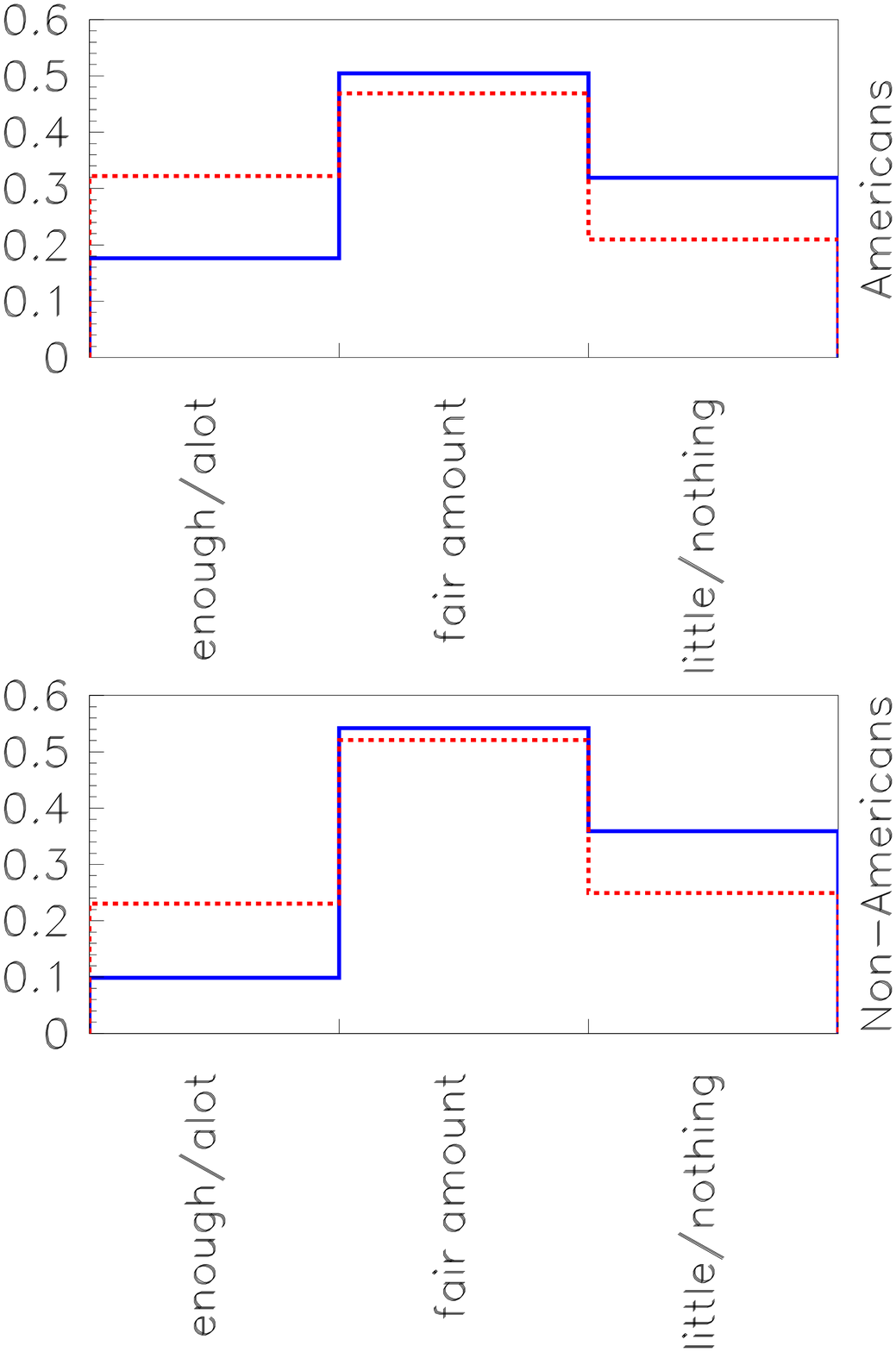}}   
\vspace{-0.2in}
\caption{How much do you know about the Muon Collider?}
\label{fig:mumu}
\end{minipage}
}
\end{figure}

We asked respondents to rate their knowledge of each of the various
machine options, including TESLA (Figure~\ref{fig:tesla}), NLC and
CLIC (Figure~\ref{fig:nlc}), the VLHC (Figure~\ref{fig:vlhc}), the
neutrino factory (Figure~\ref{fig:nufact}), and the muon collider
(Figure~\ref{fig:mumu}).  Non-American physicists are more
knowledgeable about TESLA than their American counterparts, and
significantly less knowledgeable about the VLHC.\\

\begin{figure}[h]
\mbox{
\begin{minipage}{0.5\textwidth}
\centerline{\epsfxsize 3.0 truein \epsfbox{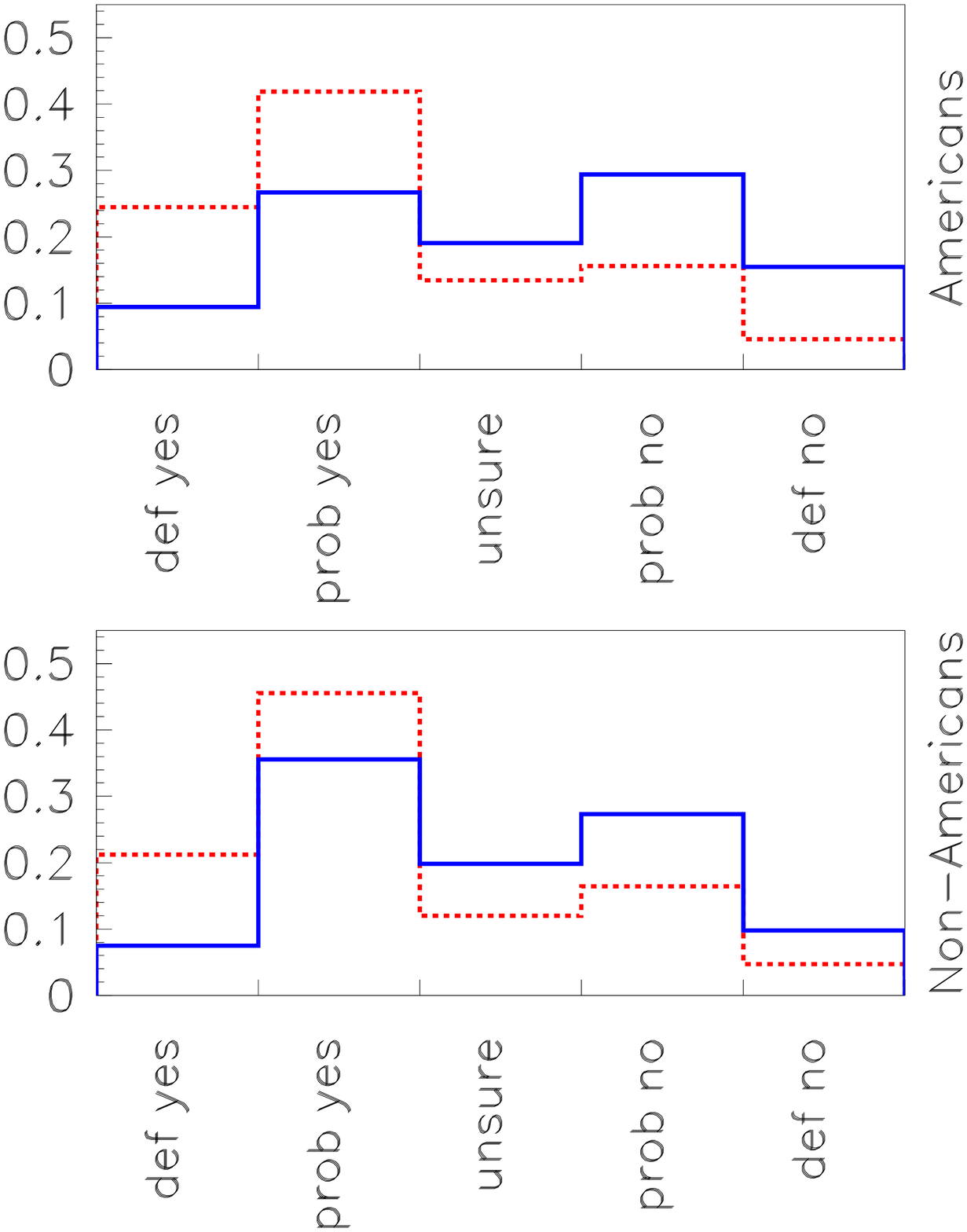}}   
\vspace{-0.3in}
\caption[Do you think you know enough to decide?]
{Do you think you know enough to form a decision and choose 
         from the various options?}
\label{fig:decide}
\end{minipage}\hspace*{0.02\textwidth}
\begin{minipage}{0.5\textwidth}
\centerline{\epsfxsize 3.0 truein \epsfbox{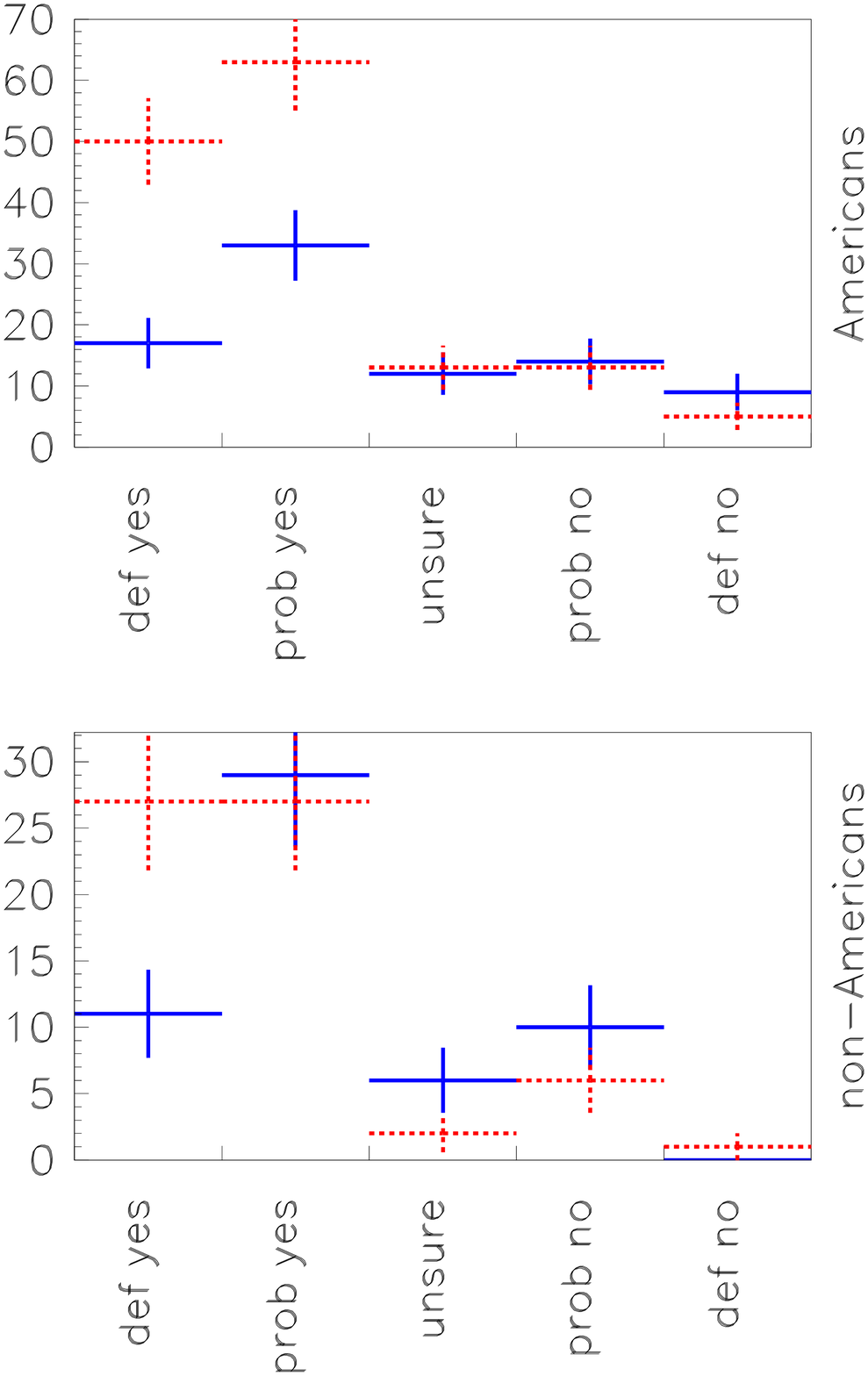}}   
\vspace{-0.1in}
\caption[Snowmass: Do you think you know enough to decide?]
 {Respondents \textbf{at Snowmass}: ``Do you think you know enough to 
         form a decision and chose 
         from the various options''. Errors are statistical only; here, the
	 total number of responses is 356 for all four histograms combined.}
\label{fig:decide_notsmass}
\end{minipage}
}
\end{figure}

We also asked respondents if they felt they currently knew enough to form 
a decision. Figure~\ref{fig:decide} shows tenured physicists are slightly more
confident in their ability to decide than young physicists. The physicists who
attended Snowmass were far more confident that they had sufficient information
to decide (Figure~\ref{fig:decide_notsmass}).

\clearpage

Finally, the respondents were asked to choose between several options
for a future flagship facility.  Using the online materials provided by the
proponents of each of the machine efforts, we identified the 
time-scales of the proposals and determined that only a few options
could begin construction within three to five years.\footnote{If one of 
the listed options requires more than
``several'' years to begin construction, or if an unlisted option becomes
viable in that time-frame, then the results would have to be interpreted
differently.}
We reasoned that
``committing'' to a plan that didn't begin construction within that
time was ostensibly the same as waiting and continuing research 
along the many branches.  The survey asked,
``if you had to choose from the following options, which one would you
pick:

\begin{itemize}
     \item A new e$^{+}$e$^{-}$ collider (TESLA,NLC,CLIC) in the U.S. soon,
           with continued research in VLHC, and muon storage ring/muon collider
           technologies.
     \item TESLA in Germany soon, with the goal of a post-LHC higher energy
           VLHC in the U.S.
     \item TESLA in Germany soon, with the goal of an eventual muon storage
           ring/muon collider in the U.S.
     \item Reserve judgment for several years, continue research.
     \item Other (specify in the comments section).''
\end{itemize}

The first option was intended to denote a linear collider that used 
unspecified technology \footnote{In hindsight, we wish we had listed JLC 
instead of CLIC, which is a longer-term project.}. The TESLA option was 
split in an attempt to glean 
additional information about machines in the U.S., resulting in options 2 
and 3. We only listed options that precluded all other options. For instance,
a $\sim$\$150M proton driver effort could exist concurrently with a 
multi-billion dollar flagship machine initiative. \\

\begin{figure}[pt]
\centering
   \epsfxsize 2.6 truein \epsfbox{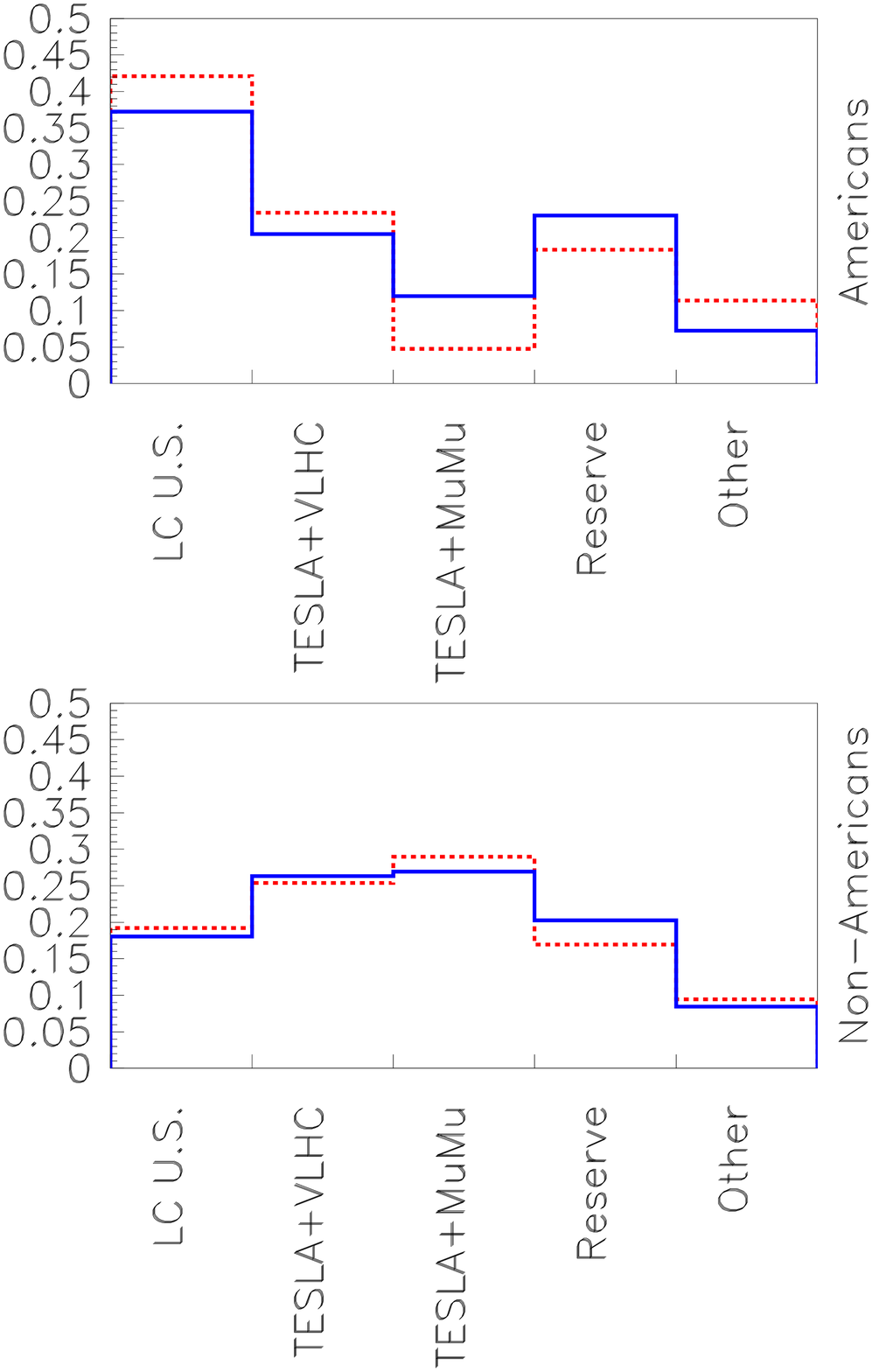}
   \vspace{-0.1in}
   \caption{Which of the following machine options would you select?}
   \label{fig:pick}
\end{figure}

\begin{figure}[pb]
\mbox{
\begin{minipage}{0.5\textwidth}
\centerline{\epsfxsize 3.0 truein \epsfbox{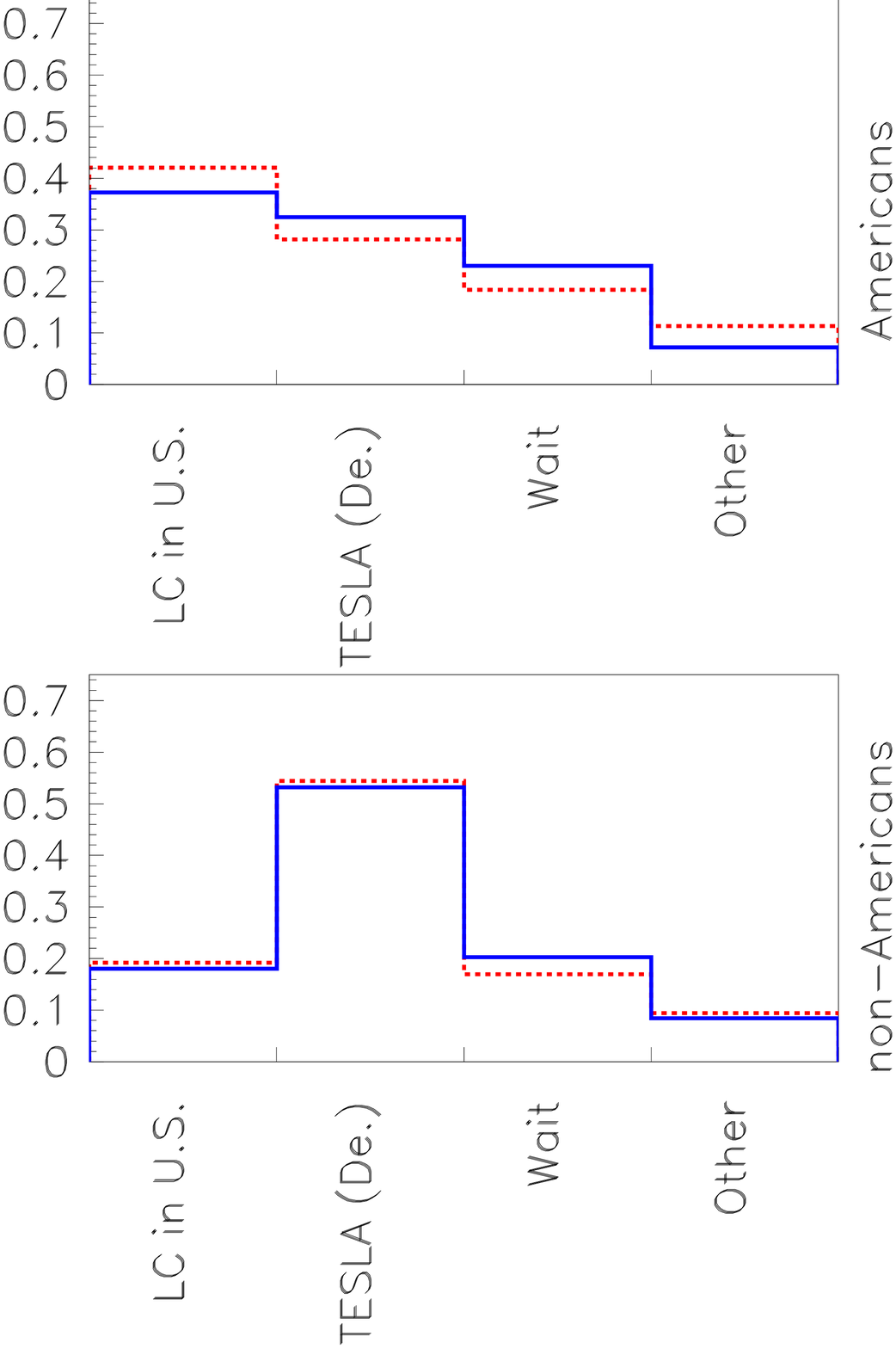}}   
\vspace{-0.1in}
\caption{Which would you select, after summing TESLA options.}
\label{fig:pick-sumtesla}
\end{minipage}\hspace*{0.02\textwidth}
\begin{minipage}{0.5\textwidth}
   \epsfxsize 3.0 truein \epsfbox{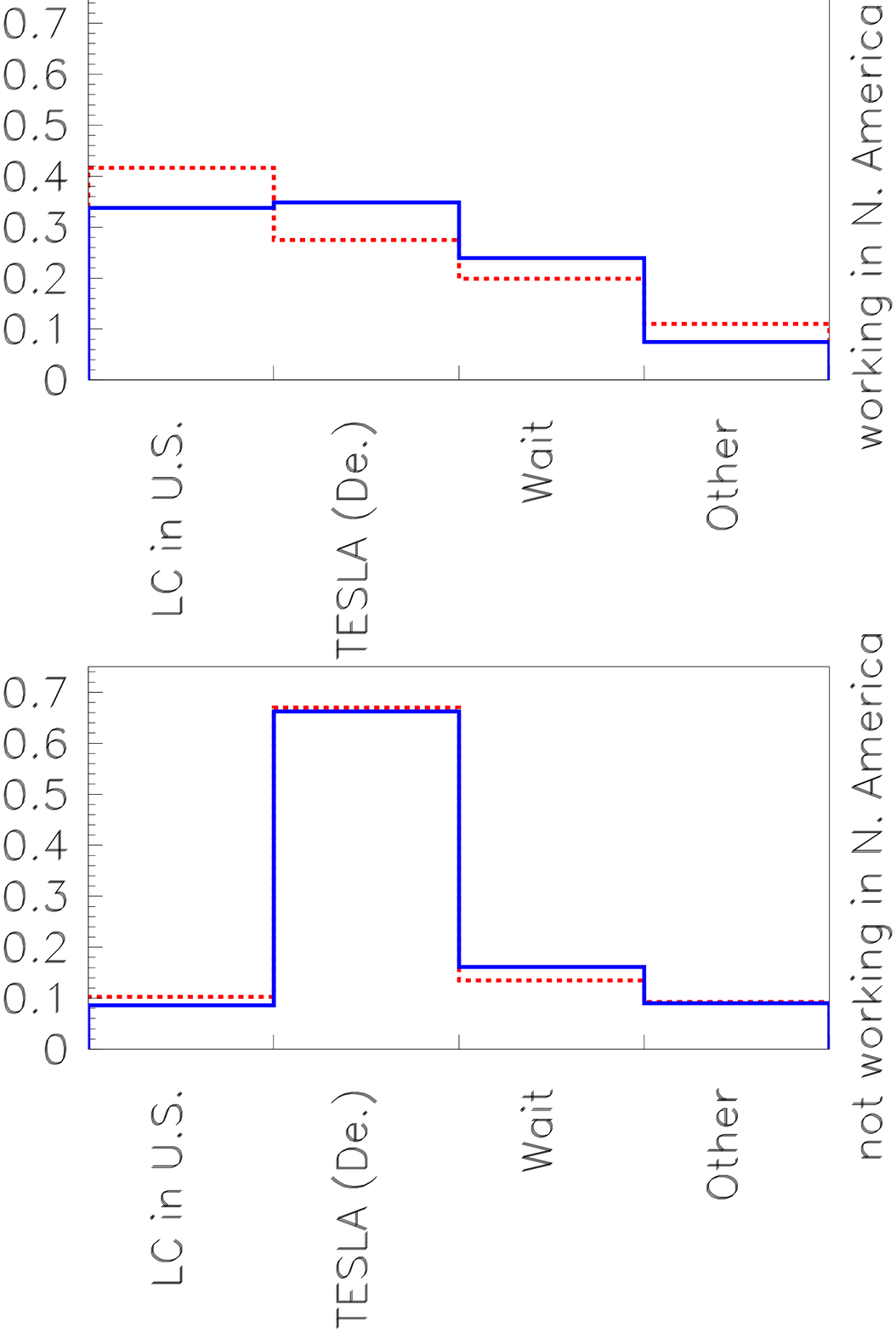}   
   \vspace{-0.1in}
   \caption[Which would you select, separated by where the respondent 
            works]
{Which would you select, with responses separated by where the respondent 
            currently works instead of where they grew up.}
   \label{fig:pick-workus}
\end{minipage}
}
\end{figure}

Figure~\ref{fig:pick} shows the results.  To directly compare the location 
options, the two responses to build TESLA in Germany were summed and the 
result shown in Figure~\ref{fig:pick-sumtesla}.  Note the striking
difference between the viewpoints of Americans and non-Americans. This
difference becomes slightly more pronounced if we subdivide the responses 
based on where one currently works (Figure~\ref{fig:pick-workus}). In
the American responses, tenured scientists favor a linear 
collider in the U.S., while younger scientists are split on whether to build 
a linear collider here or TESLA in Germany. \\

\begin{figure}[h]
\mbox{
\begin{minipage}{0.5\textwidth}
\centerline{\epsfxsize 3.0 truein \epsfbox{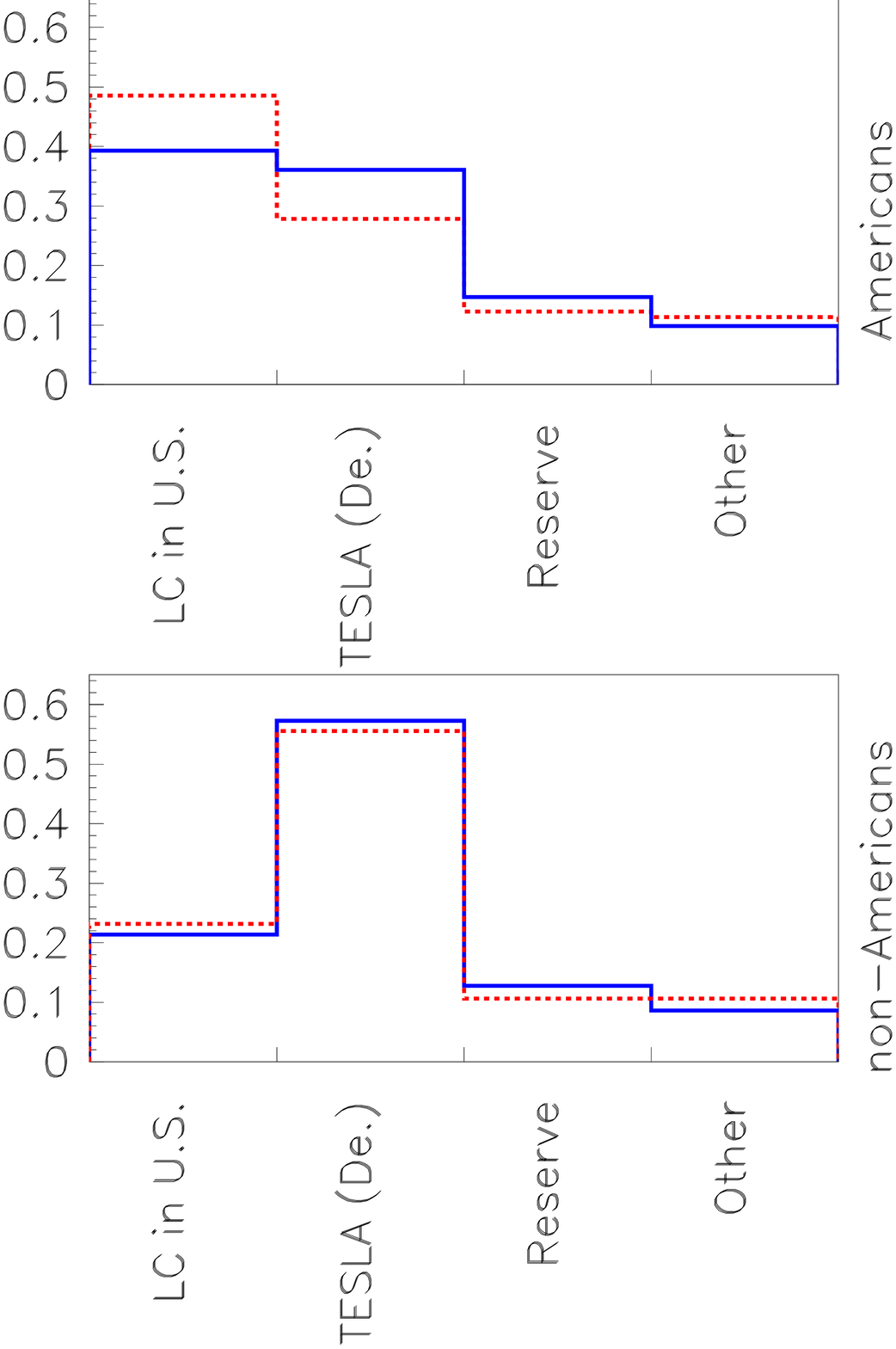}}   
\vspace{-0.2in}
\caption{Selections of people claiming to know enough to make a decision.}
\label{fig:finale_knoenou}
\end{minipage}\hspace*{0.02\textwidth}
\begin{minipage}{0.5\textwidth}
\centerline{\epsfxsize 3.0 truein \epsfbox{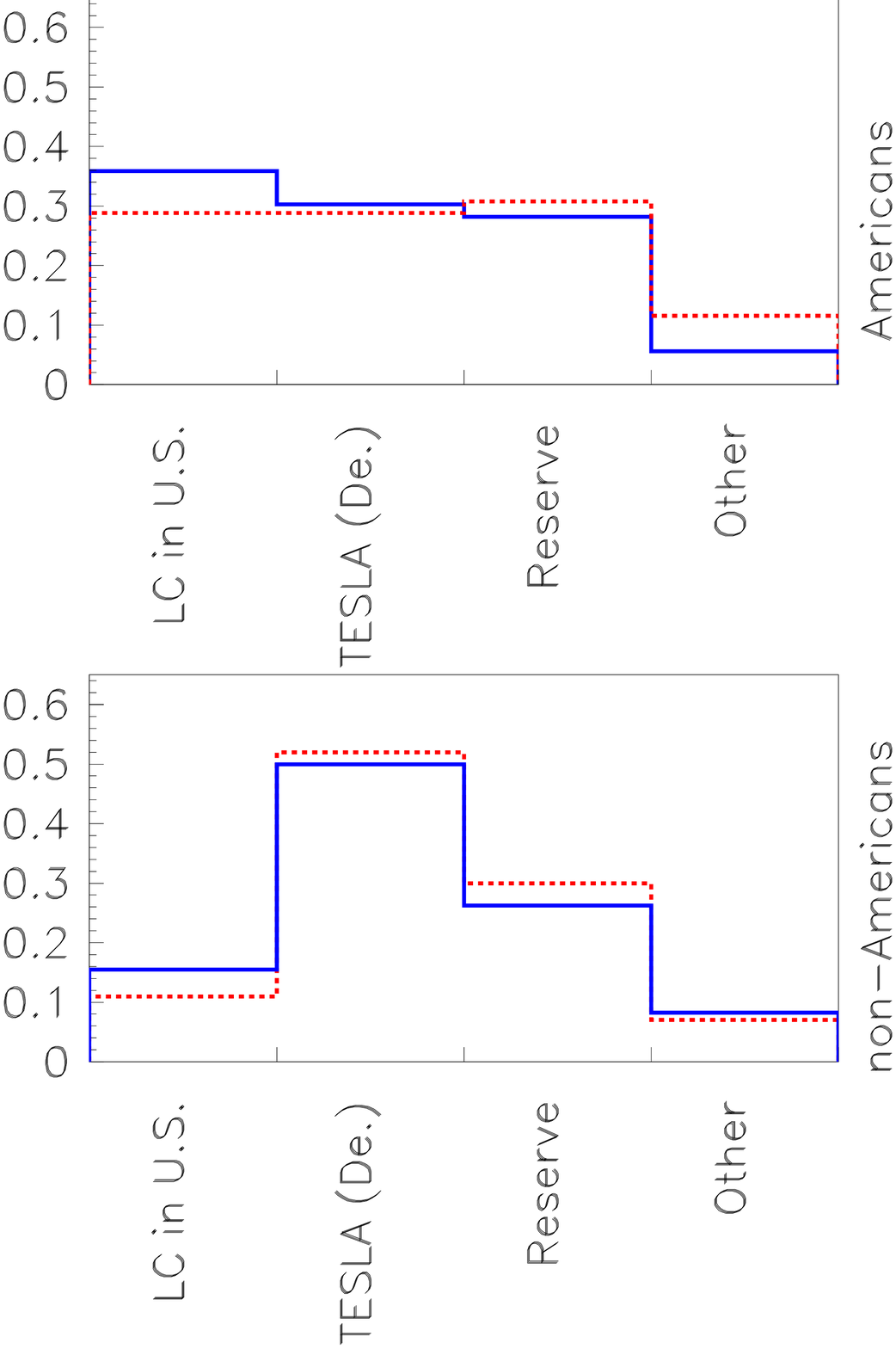}}   
\vspace{-0.2in}
\caption[Selections of people who were unsure or claimed they do not know enough]
{Selections of people who were unsure or claimed they do not know enough to make a decision.}
\label{fig:finale_not_knoenou}
\end{minipage}
}
\vspace{0.2in}
\end{figure}

Finally, we present the selections of the people that claimed to know
enough to make a decision (Figure~\ref{fig:finale_knoenou}) and those 
who were unsure or claimed not to know enough to make a decision 
(Figure~\ref{fig:finale_not_knoenou}).  It is interesting that many people
who claimed they did not know enough to make a decision still selected
an option other than ``reserve judgment''.\\

This section summarized respondents' opinions of the current machine choices.
Most physicists knew a ``fair amount'' about the current options, and most 
selected a linear collider somewhere in the world if the alternative was 
waiting for several more years to decide.  Americans tended to prefer a 
U.S. option and non-Americans strongly favored the German option.\\

\clearpage
\newpage

%% file: conclusions.tex
\section{Conclusions}

First and foremost, both young and tenured physicists appear to 
have similar opinions on the future course of high energy 
physics. There are, however, strong correlations 
depending on geography.\\

The survey results indicate accelerator based science is the driving force 
in HEP.  Maintaining diversity and 
balance in their work is seen as very important, but support for 
astrophysics at HEP labs is only somewhat supported.  Higgs and 
electroweak symmetry breaking are clear 
priorities, and this science requires a new facility.\\

Physicists indicated that they value 
the kind of hands-on experience that comes from being stationed at a 
laboratory.  If lab proximity becomes difficult, regional centers
might become a necessary alternative.\\

An overwhelming majority of physicists believe we are not doing enough
outreach to the funding agencies or to the general public. At least
half of these physicists would be willing to dedicate more of their
time to this effort. \\

There is some agreement that we are currently not retaining
enough young talent in the field.   The primary reason for this
is the lack of permanent positions, indicating that a lack of 
career security is the perceived force driving away young physicists.\\

Given the choice of building a linear collider (TESLA, NLC, or CLIC), 
reserving judgement for several years, or creating their own option, 
80\% of Americans and non-Americans prefer a linear collider.
Site location does not exhibit the same clear agreement: 

\begin{itemize}
     \item Non-Americans clearly do not think it is important 
           that the next machine be built in the U.S., and they
           selected the German option three to one over a U.S. option
     \item Young Americans are almost equally divided on whether there 
           should be a new linear collider in the U.S. or TESLA in Germany
     \item Tenured Americans select the U.S. option over the German
           option in a ratio of three to two
\end{itemize}

The results cannot determine how acceptable the respondents would 
find their second choice to be.\\

Given current construction and operation cost estimates, 
many countries will have to cooperate to build a new flagship facility,
and support of high energy physicists within these countries will need 
to be strong. The differences in opinion regarding the location of the 
future machine might make this necessary collaboration difficult.  More 
work is needed to forge a truly global consensus on the next machine.

\newpage

%% file: acknowledge.tex
\section{Acknowledgments}

We thank all who have helped us in the preparation and advertisement 
of the YPP survey, including Jonathan 
Bagger, Ken Bloom, Liubo Borissov, Janet Conrad, Kevin McFarland, Paul 
Nienaber, Marv Goldberg, and Chris Quigg;  special thanks to Diana Canzone 
and Fred Ullrich of Fermilab Visual Media Services.

%% file: survey.tex
\section{Survey}

The following is a complete listing of the survey questions and
possible responses as it appeared on the web. As a guide for the PAW ntuple 
and ROOT-tree (available on the web at \verb+http://ypp.hep.net+), variable 
names are listed in brackets at the end of each survey question.

\newpage

%%%%%%%%%%%%%%%%%%%%%%%%%%%%%%%%%%%%%%%%%%%%%%%%%%%%%%%%%%%%%%%%%%%%%%%%%%%%%
% First Page
\vspace{-0.6in}
\begin{center}
  \Huge{Survey on the Future of HEP}
\end{center}

\rule{\textwidth}{0.02in} 

\vspace{0.3in}
This survey is geared toward members of the high energy physics community.
Please give us your opinions on the future direction of high energy
physics. Please answer honestly, as many questions as you can.\\

The survey consists of only 36 questions, mostly multiple choice.
We ask that you first fill out a brief demographics section to help
us correlate responses. All entries will remain strictly anonymous. \\
 
We kindly ask that you fill out either the web-based version of this survey
(available at \verb+http://ypp.hep.net/survey.html+) or 
the following paper version, but not both. You can place your completed survey
in any of the orange ``YPP Survey'' boxes located around Snowmass.

\vspace{0.5in}

\OliveGreen{\framebox[6.5in]{
\parbox{\reducedwidth}{\begin{center}
{\epsfxsize=6.0in \epsfbox{plots/ypp-logo3.eps}}\end{center}}}}

\vspace{0.5in}
\begin{center}
 This survey was prepared by the Young Physicists Panel (YPP). More
 information can be found at the following website:\\

 \verb+http://ypp.hep.net+
\end{center}

%%%%%%%%%%%%%%%%%%%%%%%%%%%%%%%%%%%%%%%%%%%%%%%%%%%%%%%%%%%%%%%%%%%%%%%%%%%%%
% Demographics 
\newpage

\twocolumn
\begin{center}
  \Huge{Demographic Information}
\end{center}
\vspace{-0.1in}
\rule{3.3in}{0.02in} 

\renewcommand{\theenumi}{\arabic{enumi}}
\renewcommand{\theenumii}{\roman{enumii}}
\begin{enumerate}
   \item At what stage are you currently in your career [\verb+career+]?
     \begin{enumerate}
        \item Undergraduate 
        \item Graduate student 
        \item Postdoc
        \item Untenured faculty or term staff
        \item Tenured faculty or permanent staff
     \end{enumerate}

   \item On which continent did you grow up [\verb+contrais+]?
     \begin{enumerate}
        \item Africa
        \item Asia
        \item Australia
        \item Europe
        \item North America
        \item South America
     \end{enumerate}

   \item On which continent do you currently work greatest part of year
         [\verb+contwork+]?
     \begin{enumerate}
        \item Africa
        \item Antarctica
        \item Asia
        \item Australia
        \item Europe
        \item North America
        \item South America
     \end{enumerate}

\newpage
    \item What type of physics have you been working on this year
          [\verb+worktype+]?
     \begin{enumerate}
        \item theory
        \item fixed target experiment
        \item collider experiment ($p\overline{p}$, e$^+$e$^-$, etc.)
        \item beam research/accelerator physics
        \item astrophysics
        \item other (please specify below)
     \end{enumerate}

   \item Where do you do your research \\ greatest part of year
         [\verb+labwork+]?
     \begin{enumerate}
        \item Argonne
        \item BNL
        \item CERN
        \item DESY
        \item FNAL
        \item KEK
        \item LBL
        \item SLAC
        \item University
        \item Other (please specify in comment section below)
     \end{enumerate}

    \item How large is your current collaboration [\verb+collsize+]?
     \begin{enumerate}
        \item 1-5 people
        \item 5-20 people
        \item 20-50 people
        \item 50-200 people
        \item 200-500 people
        \item 500-1000 people
        \item more than 1000 people
     \end{enumerate}

    \item Are you attending Snowmass [\verb+atsmass+]?
     \begin{enumerate}
        \item yes
        \item no
     \end{enumerate}

    \item Additional comments:

%%%%%%%%%%%%%%%%%%%%%%%%%%%%%%%%%%%%%%%%%%%%%%%%%%%%%%%%%%%%%%%%%%%%%%%%%%%%%
% Balance
\cleardoublepage

\begin{center}
  \Huge{Balance vs. Focus}
\end{center}

For the following questions, we define "next machine" as the next energy \\
frontier or flagship machine. Please \\ circle one response for each of the
\\ following questions.

\rule{3.0in}{0.02in} 
   \item Is it important that the "next machine" host a diverse (multi-physics)
         range of experiments and collaborations [\verb+divnext+]?
         \begin{enumerate}
            \item very important
            \item somewhat important
            \item not important
            \item don't know/no opinion
         \end{enumerate}

   \item Is it important that the HEP labs host astrophysics efforts
         [\verb+astrnext+]?
         \begin{enumerate}
            \item very important
            \item somewhat important
            \item not important
            \item don't know/no opinion
         \end{enumerate}

   \item If the "next machine" is a collider \\ (e$^+$e$^-$, $p\overline{p}$,
         $\mu^+\mu^-$), is it important to have at least two large, 
         multi-purpose detectors versus a single detector 
         [\verb+multnext+]?
         \begin{enumerate}
            \item redundancy is very important
            \item redundancy is somewhat important
            \item redundancy is not important
            \item don't know/no opinion
         \end{enumerate}

\newpage
   \item Is it important to promote a balance \\ between the numbers of
         theorists, \\ phenomenologists, and experimentalists [\verb+prombal+]?
         \begin{enumerate}
            \item very important/mandate the mixture
            \item somewhat important/encourage a particular mixture
            \item not important/let the market decide
            \item don't know/no opinion
         \end{enumerate}

\vspace{0.2in}
   \item Additional comments on this section \\ including what HEP can support 
         given a limited budget:

%%%%%%%%%%%%%%%%%%%%%%%%%%%%%%%%%%%%%%%%%%%%%%%%%%%%%%%%%%%%%%%%%%%%%%%%%%%%%
% Globalization
\cleardoublepage

\begin{center}
  \Huge{The Impact of Globalization}
\end{center}

In order to best understand your \\ responses, we ask that you answer a 
number of questions on your present \\ situation. By "regional centers" 
we mean facilities that function as remote \\ virtual control rooms with 
experts present (monitoring, data acquisition, voltages, setting parameters, 
etc), with \\exceedingly good video conferencing and data processing 
capability. Please circle one response for each of the following questions.

\rule{3.0in}{0.02in} 
   \item How frequently do you currently see or work on your detector
         [\verb+seedet+]?
         \begin{enumerate}
            \item daily/weekly
            \item monthly
            \item yearly
            \item every few years
            \item N/A - don't have a detector
         \end{enumerate}
      
   \item How important do you feel it is to have hands-on hardware
         experience [\verb+handsdet+]?
         \begin{enumerate}
            \item very important
            \item somewhat important
            \item not at all important
            \item N/A - don't have a detector
         \end{enumerate}
\newpage
   \item It may be the case that your supervisor is in a different location 
         than your \\ detector. In your current situation, given the choice, 
         how do you rate being near your detector versus being near your 
         advisor/supervisor [\verb+advisdet+]?
         \begin{enumerate}
            \item detector proximity more important
            \item advisor proximity more important
            \item roughly equal in importance
            \item N/A - don't have a detector
         \end{enumerate}

   \item If the next machine is not located in the U.S., will the
         quality of beams and physics results improve if we have\\
         "regional centers" in the U.S. [\verb+rceasy+]?
         \begin{enumerate}
            \item yes
            \item no
            \item maybe
            \item don't know/no opinion
         \end{enumerate}

   \item  How frequently should these regional \\ centers be distributed
          [\verb+rclocate+]?
         \begin{enumerate}
            \item less frequently distributed than the national labs
            \item as numerous as the national labs
            \item more frequently than the national labs
            \item don't need them
         \end{enumerate}

   \item What should be the importance of the U.S. national labs 
         in 10-25 years [\verb+uslabs+]?
         \begin{enumerate}
            \item reduced role/some labs should close
            \item much the same role/current labs should persist
            \item strong role/additional labs should open
         \end{enumerate}

\vspace{0.1in}
   \item Your comments on this section, \\ including thoughts on how you might 
         envision global collaboration working:

%%%%%%%%%%%%%%%%%%%%%%%%%%%%%%%%%%%%%%%%%%%%%%%%%%%%%%%%%%%%%%%%%%%%%%%%%%%%%
% Outreach
\cleardoublepage

\begin{center}
  \Huge{Outreach to Influence Funding Agencies and/or the Public}
\end{center}

How important is outreach to \\ continued support of our work? Please circle 
one response for each of the \\ following questions.

\rule{3.0in}{0.02in} 
    \item Are we doing enough outreach to funding agencies, i.e. Congress
          [\verb+congoutr+]?
         \begin{enumerate}
            \item more than enough
            \item adequate
            \item not nearly enough
            \item far too little
         \end{enumerate}

    \item Are we doing enough outreach to the general public [\verb+puboutr+]?
         \begin{enumerate}
            \item more than enough
            \item adequate
            \item not nearly enough
            \item far too little
         \end{enumerate}

    \item  Would you personally be willing to \\ dedicate more of your time
           to outreach [\verb+persoutr+]?
         \begin{enumerate}
            \item yes
            \item no
            \item maybe
         \end{enumerate}

\vspace{0.2in}
    \item Your brief comments on this section and thoughts on what we
          might do \\ differently. What we might do better:

%%%%%%%%%%%%%%%%%%%%%%%%%%%%%%%%%%%%%%%%%%%%%%%%%%%%%%%%%%%%%%%%%%%%%%%%%%%%%
% Building the Field
\cleardoublepage

\begin{center}
  \Huge{Building the Field}
\end{center}

Your personal experience can be insightful. Please circle one response for 
each question.

\rule{2.8in}{0.02in} 
   \item What is it that \textbf{most} attracted you to HEP [\verb+attrhep+]?
      \begin{enumerate}
        \item interest in physics/science/nature
        \item strong mentor in the field 
        \item allure of working at a HEP lab
        \item intellectual atmosphere
        \item whim/it was cool
        \item nothing else is interesting
        \item salary and benefits
        \item possibility of fame
        \item enjoy hardware work
        \item enjoy software work
        \item academic freedom
        \item lifestyle/travel/hours
        \item gain valuable experience for any career choice
        \item spiritual/religious reasons
      \end{enumerate}

   \item What is main reason you have stayed in the field so far 
         [\verb+stayhep+]?
      \begin{enumerate}
        \item interest in physics/science/nature
        \item strong mentor in the field 
        \item allure of working at a HEP lab
        \item intellectual atmosphere
        \item whim/it was cool
        \item nothing else is interesting
        \item salary and benefits
        \item possibility of fame
        \item enjoy hardware work
        \item enjoy software work
        \item academic freedom
        \item lifestyle/travel/hours
        \item gain valuable experience for any career choice
        \item spiritual/religious reasons
      \end{enumerate}

   \item Do you think we are currently retaining adequate numbers 
         of talented physicists in HEP [\verb+talenhep+]?
      \begin{enumerate}
        \item yes
        \item somewhat
        \item no
        \item don't know/no opinion
      \end{enumerate}

   \item What feature of the field do you think might \textbf{most} 
         influence young physicists to leave the field [\verb+neghep+]?
      \begin{enumerate}
        \item lack of competitive salary
        \item large collaboration sizes
        \item length of time spent in grad school
        \item amount of travel required
        \item lack of permanent job \\ opportunities
        \item lack of professional environment
        \item weak or unstructured management
        \item other (please specify in comments)
        \item not an issue (don't think many will leave)
      \end{enumerate}
    
\vspace{0.2in}    
   \item Additional comments are welcome. How do we encourage talented young 
         people to do graduate work in HEP, and then stay in the field? What 
         issues facing young physicists in the future concern you the most?

%%%%%%%%%%%%%%%%%%%%%%%%%%%%%%%%%%%%%%%%%%%%%%%%%%%%%%%%%%%%%%%%%%%%%%%%%%%%%
% Physics
\cleardoublepage

\begin{center}
  \Huge{Physics}
\end{center}

The direction you select for the field is largely driven by the 
fundamental \\ questions you find most compelling. \\ Please circle one
response for each of the following questions.

\rule{2.8in}{0.02in} 
   \item What physics would \textbf{you} personally find most compelling in 
         the next 10-25 years [\verb+persphys+]?
      \begin{enumerate}
         \item QCD/strong interactions
         \item Higgs physics/EW symmetry \\
               breaking
         \item CP violation
         \item neutrino physics
         \item exotic particle searches
         \item accelerator physics
         \item cosmological constant/dark matter
         \item cosmic ray physics
         \item quark gluon plasma/hadronic physics
         \item other (please specify below)
      \end{enumerate}

   \item What do you think is the most important physics for \textbf{the field}
         over the next 10-25 years [\verb+fieldphy+]? 
      \begin{enumerate}
         \item QCD/strong interactions
         \item Higgs physics/EW symmetry \\
               breaking
         \item CP violation
         \item neutrino physics
         \item exotic particle searches
         \item accelerator physics
         \item cosmological constant/dark matter
         \item cosmic ray physics
         \item quark gluon plasma/hadronic physics
         \item other (please specify below)
      \end{enumerate}
\newpage
   \item Does the science you selected above \\
         require a major new facility [\verb+facphy+]?
      \begin{enumerate}
         \item yes
         \item no
         \item maybe
         \item don't know/no opinion
      \end{enumerate}
   \item If so, do you think it is important that the new facility be in
         the U.S. [\verb+facusphy+]?
      \begin{enumerate}
         \item yes
         \item no
         \item maybe
         \item don't know/no opinion
      \end{enumerate}

\vspace{0.1in}    
   \item Your additional comments on this \\ section are welcome:

%%%%%%%%%%%%%%%%%%%%%%%%%%%%%%%%%%%%%%%%%%%%%%%%%%%%%%%%%%%%%%%%%%%%%%%%%%%%%
% Finding Consensus
\newpage

\begin{center}
  \Huge{Finding Consensus, Picking a Plan}
\end{center}

Please circle one response for each.

\rule{2.8in}{0.02in} 

   \item What has most affected your current opinion(s) on the
         options for the future of HEP [\verb+affopin+]?
         \begin{enumerate}
              \item your advisor
              \item your current work
              \item reading
              \item talking to colleagues
              \item attending workshops/conferences
              \item attending talks
              \item joining working groups/ \\
                    performing studies
              \item sitting and thinking
              \item other
         \end{enumerate}
 
   \item How much do you know about TESLA [\verb+knowtes+]?
          \begin{enumerate}
             \item enough or alot
             \item fair amount, but need to know more           
             \item very little or nothing  
          \end{enumerate}  
   \item How much do you know about LC non-TESLA versions, i.e. NLC, CLIC
             [\verb+knowolc+]?
          \begin{enumerate}  
             \item enough or alot
             \item fair amount, but need to know more           
             \item very little or nothing  
          \end{enumerate}  
   \item  How much do you know about the VLHC [\verb+knowvlhc+]? 
          \begin{enumerate}  
             \item enough or alot
             \item fair amount, but need to know more           
             \item very little or nothing  
          \end{enumerate}  
\vspace{0.5in}
   \item  How much do you know about the \\Neutrino Factory [\verb+knownfac+]? 
          \begin{enumerate}  
             \item enough or alot
             \item fair amount, but need to know more           
             \item very little or nothing  
          \end{enumerate}
   \item  How much do you know about the Muon Collider [\verb+knowmumu+]?
          \begin{enumerate}  
             \item enough or alot
             \item fair amount, but need to know more           
             \item very little or nothing  
          \end{enumerate}  

   \item Do you think you know enough to form a decision and chose from
         the various options [\verb+knowenou+]?
         \begin{enumerate}
             \item definitely yes
             \item probably yes
             \item unsure
             \item probably no
             \item definitely no
         \end{enumerate}

   \item If you had to choose from the following options, which one would
         you pick [\verb+choose1+]?
         \begin{enumerate}
             \item a new e+e- collider (TESLA,NLC,CLIC) in the U.S. soon, 
               with continued research in VLHC and muon storage ring/muon 
               collider technologies
             \item TESLA in Germany soon, with the goal of a post-LHC
               higher energy VLHC in the U.S.
             \item TESLA in Germany soon, with the goal of an eventual muon
               storage ring/ muon collider in the U.S.
             \item reserve judgment for several years, continue research
             \item other (please specify in comment section below) 
         \end{enumerate}

\vspace{0.1in}
   \item Your comments here, please amplify on your response to the above:

\end{enumerate}

%%%%%%%%%%%%%%%%%%%%%%%%%%%%%%%%%%%%%%%%%%%%%%%%%%%%%%%%%%%%%%%%%%%%%%%%%%%%%
\onecolumn
\newpage

%% file: comments.tex
\section{Sound Bites}

The following is a complete listing of \textbf{all} of the comments received 
with the survey responses as well as letters addressed to YPP. The comments 
are categorized into the six survey subsections.

\input{comments-letters.tex}

\newpage

\twocolumn
\input{comments-demographics.tex}

\newpage

\input{comments-balance.tex}

\newpage

\input{comments-globalization.tex}
\newpage

\input{comments-outreach.tex}
\newpage

\input{comments-building.tex}
\newpage

\input{comments-physics.tex}
\newpage

\input{comments-pick.tex}
\newpage

%% file: comments-letters.tex
\pagestyle{myheadings}
\markboth{Letters to YPP}
         {Letters to YPP} 
\subsection{Separate Letters Sent to YPP}
   \begin{itemize}

    \item I must first applaud your efforts to actively involve the 
          community and provide many opportunities for individuals and groups 
          to voice their opinions. Thank you.

          I am nearing completion of the graduate student life cycle on the 
          BaBar experiment at SLAC, and want to share a few thoughts on the 
          U.S. high energy physics (HEP) community. 
          Our spokesperson, Stewart Smith, recently introduced BaBar 
          (before the announcement of a new result) with the words, "Our 
          experiment has two products:  new science, and highly trained 
          people."  This is, I think, the key issue. The field of high energy 
          physics, be it experimental or theoretical (or perhaps not even high
          energy), has TWO products, two purposes:  discoveries about our 
          Universe and the relationships that govern it (exciting enough to be
          pretty heady); and the rigorous training of individuals in the 
          skills of critical thinking, clear communication, and objective 
          observation (a rewarding and necessary challenge). This 
          duality is not divisive, rather it is inclusive -- it asserts 
          that there is more to HEP than merely a derived understanding of the 
          Universe; there is a practice, a paradigm, a set of skills that are 
          part of high energy physics. 

          Some argue that the educational aspect of HEP training is most 
          well carried out by having a machine here in the U.S.  Though I 
          might be inclined to agree, having lived through the SSC experience,
          and not seeing currently a much better political climate for science
          (if not a worse one) I personally worry about whether we as a 
          community will be able to make our leaders understand why building a
          large-scale machine such as an ILC is important for so many reasons.
          So for now -- I am completely willing to support all efforts for 
          getting Congress to fund a machine within the U.S.  However -- 
          before working on it myself, I personally would like more 
          assurance if it ever began that the project was going to be carried 
          through to completion. 

          Also, having recently seen a talk by Albrecht Wagner, I am very 
          impressed with how far DESY has progressed on the TESLA concept, and
          if we as a nation decide *against* funding such a machine here in 
          the U.S., I would be *most* happy to support them and send funds 
          over in the model of how we have for LHC.  I think it would be 
          another grievous blow for science in this country, but again -- HEP 
          is always becoming more global (we did invent the Web, after all), 
          and I would be quite content to live abroad for many years to do the
          necessary research.  I do like the idea of regional educational 
          centers, akin to the National Labs, but serving the primary purpose 
          of disseminating the culture, knowledge, and excitement inherent in 
          the HEP quest.  Further, Wagner has stated that he expects 
          future generation experiments will have *much* higher 
          maintainability and operability from afar.  (Clearly not the 
          hardware, but the software he's envisioning would be much superior 
          to EPICS in how well the machine can be run remotely, from what I 
          gather).  Given the limited resources in the world, and the many 
          other pressing problems society needs to address, I do feel that 
          there should only be *one* large world lab doing e+e- at the 500 
          GeV range, and I would offer arguments similar to those of Shelly 
          Glashow's who said much the same thing many years ago while the 
          LHC and SSC were both being built. 

          Further, the emphasis of high energy physics on education is 
          critical to the fitness, indeed, even the existence, of a successful
          U.S. HEP future. The strong attitude of elitism within our field is,
          I think, contrary to that purpose (and contributed strongly to the 
          demise of the SSC).  Many high energy physicists consider HEP as 
          self-evident, the only fundamental science; other fields are 
          devalued as mere large statistic studies of ensemble behavior. 
          Additionally, the physics institution still maintains that most 
          HEP-trained physicists should be doing research in academia; 
          failure to do so is considered an admission of inadequate talent. 
          These messages are all utterly ridiculous; high energy physics is no 
          more fundamental than psychology, which studies the perceptive 
          apparatus by which we make all of our observations.  These pervasive
          attitudes have two consequences: they alienate practioners of other 
          sciences, and they prevent open dialogue between those in the 
          community and those that have left it. I point again to the two 
          "products" of HEP: we're producing new science, and we're producing
          highly trained individuals; we should welcome customers of the 
          latter. Having HEP "graduates" dispersed across the 
          professional world is clearly advantageous to the field. As HEP 
          "alumni," these people not only bring the strength of 
          cross-disciplinary training to other fields, but they also 
          enthusiastically share and teach what they have learned from the 
          high energy community.  They are an under-utilized resource, an 
          inactive alumni network that still has passion for the "alma mater."
          We don't need to "keep young people in the field;" we need to 
          send more trained high energy physicists out into the world!  I feel 
          this very strongly, and feel that the Young Physicists Panel 
          questionnaire is fundamentally slanted towards the more common view 
          mentioned above when it talks about "how can we keep more HEP 
          researchers in the field?". 

          I strongly feel that the HEP community is not sufficiently focused 
          on outreach and education efforts. It is clear that the dividends of
          our work need to go back to the general public, again in terms of 
          our two products: new knowledge and trained individuals. But the 
          difficulty in organizing outreach efforts is in defining the target 
          audiences and the principal actors.  I think this is where we need 
          more focus. At present, the formula seems to involve press releases
          which are then distilled into exciting narratives by trained 
          journalists. Is this the most appropriate interaction?  Should we 
          instead encourage more of the principal researchers to work on the 
          front lines?  But we cannot expect all successful researchers to be 
          renaissance-persons, capable of submitting a PRL draft and a New 
          York Times science article in the same evening. We can, however, ask
          a greater number of active HEP researchers to engage the public in 
          workshops and tours, rather than retreating behind the cloak of 
          traditional media. 

          Lastly, I feel that there is a large sector of the general public 
          that is greatly dissatisfied with (overlooked by?) HEP outreach -- 
          the well-educated, middle-class professional. Some are business 
          professionals, some are scientists in other fields, some are 
          motivated amateurs; they are capable and enthusiastic, and want 
          something between Scientific American and physics graduate school.  
          HEP education often targets children, because these teaching 
          situations are often easier and more successful because the material
          can be correctly represented and controlled.  I say 
          that we must tackle the harder audiences as well, the ones with 
          intellectual training and critical thinking skills; they can 
          become our best customers, and our strongest advocates.

          I think it's quite reasonable that we should spend say even 1$\%$ of 
          our budget on educational outreach.  Currently, it's probably about 
          ... 1$\%$, if *that* -- at least at SLAC, I'm not sure how FNAL 
          compares, I know CERN does a better job and, as many have observed, 
          on the whole, NASA puts us all to shame with their incredible public
          outreach machine, and I think it's very short-sighted for us to 
          continue in the mode of not reaching out to the public more. 
          As far as a few vague ideas (some a little more SLAC-specific) 
          that would need filling out much more before being serious 
          proposals, how about: making IMAX movies, multimedia CD's, 
          traveling roadshows with demos, art $+$ science exhibitions (CERN 
          does these well), more fun knick-knack merchandise we can give 
          away/sell on the web for cheap, open houses for 
          all neighbors at the Venture Capital firms to come see what we do, 
          giving all lab users/employees a half day seminar on occasion to 
          educate them on all the amazing things that have come out of HEP 
          (e.g. first U.S. website, medical technologies from SSRL, Silicon 
          Valley connections etc.), some kind of economic estimational 
          analysis of the contribution of SLAC per year to the U.S. economy 
          through these things (similar to what environmental economists do 
          when they estimate the economic impact of externalizing 
          things like pollution) etc. 

   \item  I would like to add some extra comments to the survey form 
          I just submitted. Just to let you put my comments in context 
          I include at the end of the message my answers to the form. 
          I share the general concern for the future of HEP, and I think 
          it may be useful to look at how the whole thing started. Until 
          the 40s, nuclear physics was "small science", done in 
          University labs with very little funding. The amount of new 
          discoveries done in the early part of the century was amazing, 
          although still with relatively little impact on society. 
          The bomb changed it all, showing how this abstruse research 
          field could change the world, forever. It created a strong 
          political consensus to fund the HEP field, which really started 
          in the 50s. The discoveries of the 60s and 70s showed rapid, 
          albeit not truly understood, advancement in knowledge, with 
          an acceptable increase in funding, especially considering the 
          level of economic expansion that all developed countries where 
          then experiencing. But the mood changed in the 80s, for a number 
          of reasons. Discoveries have become rare, advancement slow and 
          increasingly obscure, funding enormous, the economy has been 
          slowing down making it harder and harder to defend this kind 
          of pure science project. In the U.S. this culminated 
          with the termination of the SSC project; in Europe the larger 
          continental inertia allowed LHC, but with delays and enormous 
          and still lingering funding problems (Atlas alone lacks 45MCHF 
          at this day). 

          Looking at this from a distance, I can't help wondering whether 
          the "next big machine" is real necessity for the advancement of 
          science, or instead is dictated by the sociological need of 
          keeping the HEP community together, of preventing the accumulated 
          knowledge from vanishing into retirement. I'm not a small science 
          champion, I've always been working in big experiments and I 
          appreciate what a big project can do. Still ... How can we attract 
          young brilliant scientists, when the prospect is to work for years 
          to give a small specialized contribution to a huge experiment one 
          may not be able to see running? There is something wrong in the 
          model. 

          Of course I have no magic solution. The direction I think we should 
          be moving on, though, is to violently promote the unity of science. 
          What I mean by this is that not only we should foster a much 
          higher level of communication between theoreticians, 
          phenomenologists and experimentalists, but also increase our 
          knowledge of, and outreach to other physics fields. As a field, we 
          know very little of what happens in solid state physics, or material
          physics, or even in astrophysics. The example of the relatively 
          recent astroparticle physics field is illuminating. Studying subtle 
          effects at low energies (say "Higgs boson corrections to Cooper
          pairs energy levels", just to invent one) may be more effective, 
          more rapid and in the end more rewarding than walking the relatively
          clear road of the next energy frontier. 

          Thanks for the effort of taking this survey, I think it 
          can be very useful.

    \item HEP has not and is not losing its vitality and will remain the 
          most "noble" and defining field of physics and other natural 
          sciences for decades to come. My primary concerns are the erosion 
          for the public's respect and support for science as a whole and 
          dismal job security for young people in the field. Follow Europe 
          and give a permanent job to every Ph.D. who get any job in 
          the field upon getting a degree!

  \item I just sent in my answers and comments.  You all are doing a very
        important job.  To get to a sensible policy for the future will take a
        great deal of prodding -- your group can be a major factor.

   \end{itemize}

%% file: comments-demographics.tex
\pagestyle{myheadings}
\markboth{Comments on Demographics}
         {Comments on Demographics} 
\subsection{Demographics}
   \begin{itemize}
    \item "Graduate students" are fictional characters invented by the 
          American system of higher education. Once past the M.S. level, they 
          are mature researchers who are paid below the federal minimum salary
          levels.     
    \item I am working in the field of nucleon nucleon interactions at the 
          Research Center Juelich, Cooler Synchrotron 3.2 GeV/c. A few of your
          questions can also be presented to and answered by a nuclear 
          physicist as well and I feel free to do so.     
    \item I have problems answering the question about the size of my current 
          collaboration since I collaborate with different ensembles of people
          on different projects (as most theorists do I suppose). 
    \item I do 50$\%$ accelerator research and 50$\%$ collider physics data 
          analysis (50/50 was not an option in above questions). 
    \item I'm Canadian; for demographic purposes Canada probably shouldn't be 
          lumped in with USA since we have distinct funding agencies, aren't 
          particularly affected by DOE budgets etc. 
    \item For question 4, I find complete inattention to non-accelerator 
          particle physics. I work in neutrino physics, which could be called 
          ``fixed target'' at some level. I also work on atmospheric 
          neutrinos. This is NOT astrophysics. 
    \item This survey represents a very narrow view of HEP. The committee 
          seems to believe that any work not being done at an accelerator 
          (fixed target, or collider) qualifies as astrophysics. Is the 
          committee unaware of the exciting physics taking place away from 
          accelerators (SNO, Super-K, MACRO, Monolith, Soudan-2, just to name 
          a few)?
    \item Previously in neutrino physics. Biggest shock for me was the very 
          limited (and narrow-minded) focus of the "big physics" community;
          re: other HEP subfields. For CDF people, if you're not at LEP, CLEO,
          SLAC, FNAL or DESY, you're nowhere. This close-mindedness is bad 
          for the field as a whole. 
    \item 1) I grew up on an island, not a continent! (and not even near 
          one!!) 2) Need to allow for people who work in more than one 
          category of physics (in my case, pbar p colliders and astrophysics).
    \item Very trivial, I grew up on an island that is not part of any 
          "continent".
   \item  I grew up in the middle east, which is not usually called Asia. 
    \item Too bad the type of research doesn't specify the type of physics 
          because correlations in bias and opinions could have been made with 
          that info. People always seem to vote for what they are/ have worked 
          on. 
    \item This survey doesn't seem to be geared towards "members of the HEP 
          community" but towards "members of the HEP-experiment community who 
          work in the USA" - and it's not the same thing.
    \item These questions seem to be geared towards experimentalists. I am 
          a theorist. 
    \item A year ago I formally retired from active research by resigning 
          from E871. I have been active in Science Education for the past 10 
          or so years. I am 79 in a few weeks.    
    \item Many of us, including myself, work equally on more than one 
          experiment. You have not taken this into account in your questions. 
    \item Maybe I could have said 'fixed target' but I put 'other' because I
          am in neutrino oscillation physics. 
\end{itemize}

%% file: comments-balance.tex
\pagestyle{myheadings}
\markboth{Comments on Balance vs. Focus}
         {Comments on Balance vs. Focus}
\subsection{Balance vs. Focus}
\subsubsection{Timelines:}
\begin{itemize}
    \item NLC VLHC CLIC
    \item Design the next machine to be staged, with each stage component 
          being able to yield interesting physics while ultimately 
          contributing to the final collider physics program. This way one 
          avoids an "all or nothing" scenario, similar to what happened with 
          the SSC.
    \item I strongly support incrementally upgrading Fermilab to a) proton 
          Driver b) Collecting and cooling muons c) Neutrino factory 
          d)Muon collider. This produces diverse physics. Each stage is under 
          \$1B. Neutrinos oscillate and provide the first evidence of physics 
          beyond the standard model.
    \item Whatever course is chosen toward the frontier (and I personally 
          lean toward VLHC), the mantra ought to be "staged development". This 
          approach should smooth funding profile, provide for more sustained 
          experimental program, and permit strategic reassessment at 
          appropriate times.
    \item One accelerator in the world next. Either LC in Germany or US. If 
          Germany, after construction, then go for VLHC or muon collider in 
          the U.S. (VLHC I prefer).
    \item Often a staged approach might be useful: start out at a new 
          machine with one experiment, but add a second one later. However 
          competition and redundancy are very important in the long run.

\subsubsection{Budget and Funding Issues:}
    \item It is important NOT to commit more than 15-20$\%$ per year from 
          the base budget to a long-term project that will not supply 
          research results for 10 years. The bulk of U.S. support for such a 
          new facility must come from NEW MONEY. If this principle is 
          violated, the actual science activities and achievements from U.S. 
          HEP will wither and die in the meantime.
    \item The assumption of a fixed budget is plain wrong and should not come 
          from young physicists. Young physicists, together with their senior
          colleagues should work hard on expanding the budget. Look at 
          biosciences! Mere whining about lack of money, so popular these 
          days, is the best way to ensure that we won't get any expansion. 
          Go to local schools, do PR, talk to journalists, talk to your 
          neighbors, for God's sake, but don't tell me that the richest 
          country in the world can not afford broad science programs after 
          5 years of bullish market! 
    \item It is essential that we expend sufficient resources to properly 
          analyze the data we work so hard and spend so much money to acquire.
          To do otherwise, we may as well just skip doing the experiment.
    \item In times of constant/decreasing budget, it is important to invest 
          in efforts to keep the cost of new accelerators down. There is also 
          a tendency within the HEP community to inflate the cost of new 
          accelerators to saturate the current budget.
    \item I do not accept the premise that our budget is limited. It is true 
          that it has been flat (or declining, if you count inflation). But 
          the future depends on our enthusiasm and willingness to pull for 
          each other. Right now we have too many sour pusses, harping on the 
          imperfections that any project will have.
    \item Over the next 20 years, world HEP can support three machines at 
          roughly \$5B each: An e+e- collider, a neutrino factory and stage 
          one of a VLHC. This will require a construction add-on of about 
          \$0.8B a year shared around the world...say half in the US. It 
          should permit a very active program of support of existing 
          facilities, niche experiments some astrophysics while we are 
          constructing in some phased way. If VLHC is in the U.S., we will 
          have a tunnel to exploit over the next 20 years.
    \item Economics and politics mandate that the next machine be a low 
          energy e+e- collider. We must do what is economically feasible. 
          A program that optimizes the physics potential but can never be 
          carried out is a disaster for the field.
    \item Preferred next choice: Linear e+ e- collider with one experiment 
          for e+ e- and one for e-e, e-gamma and gamma-gamma collisions. A 
          significant attempt should be made towards a multi-science centre 
          to attract and reach support by a wide scientific community - this 
          is most probably the only way that the necessary resources can be 
          found.
    \item Probably need to close some centres to fund new initiatives. A 
          nasty choice.
    \item For the next 20 year we can support only LHC and the linear collider.
    \item Some standardization to save money.
    \item In an era of limited resources and publicity/personality driven 
          political decisions regarding both short and long range 
          discretionary budgetary policy, the first thing HEP should support 
          is a strong public and governmental relations program it would 
          perhaps be prudent for an umbrella organization such as APS (or 
          a consortium of such organizations), to consult with 
          public/ governmental/ media relations experts from the business 
          world HEP has an advocacy and image problem in a world which, often,
          bases its decisions on images, perceived realities and received 
          wisdom.
    \item We can afford a frontier ee and a frontier pp machine as well as 
          smaller machines. Of course, the construction of the two big 
          machines will be sequenced. 
    \item We actually have a limited number of ideas; a new machine is ten 
          years away to its impact at least; A budget that can contemplate 
          \$10B is not very limited; WE FAILED TO USE THAT ORDER OF MONEY WITH 
          THE SSC AND WITH ISABELLE ... run flown into the ground with our 
          eyes wide open with learned committees run by theorists and 
          organizational scientists; budget reviews and slogans, management 
          teams oversight committees, Harvard Directors, (firing the 
          experienced director); examine how Bob Wilson DID it and with whom. 
          I find the NLC program more like one experiment which would be very 
          good if it could be done now; a lot like the SLC but a one shot deal
          very expensive no future and makes an oxymoron of "high energy 
          physics" mantra: rather is physics at 1/2 TeV when we can see 
          easily how to do physics at 100-200 TeV ... and this has to be where 
          we want to be in 10 years ... There is no Z pole glory hole to make 
          e+e- the known physics of choice; NLC physics is very likely that 
          of the Tristan, SLC and LeP II era: damn few events ... We would 
          have had hints something useful at a 1/2 TeV attackable with e+e- 
          is just around the corner in hadron physics; putting the future of 
          U.S. particle physics for the next generation on the e+e- option is 
          suicide for our field in the U.S. 
    \item Put off very expensive experiments. Maybe in the near future new 
          technologies will help to greatly reduce the cost. Just keep the 
          minimum going. 
    \item It's not at all clear we can afford any new machines. 
    \item Given the SSC debacle, without consensus inside HEP it doesn't 
          matter what budget we have.
    \item As to question 11, the number of detectors, and indeed even the 
          size of the machine, can't be just decided by fiat or even physics 
          desires; we need to seriously consider doing what most other fields 
          do, and figure out what we can do AFTER we have some idea of how 
          much money we can get, not BEFORE. Doing so requires us to fight 
          against critics, rather than working with them to figure out how 
          to get the most bang for the buck.
    \item Keep funding for accelerator physics R\&D at a generous level to 
          encourage young people who would like to enter the field.
    \item Value for money is more important now than it ever has been. 
          Providing for other disciplines to work at our big toys as well 
          is the way to go.
    \item Support the scientists and the field will survive. HEP budgets tend 
          to demand more clever ideas from its proponents. There are limits, 
          but we must maximize our resources and work to increase those 
          resources.
    \item More effort must be expended to restore a budget sized to our 
          needs and expectations.
    \item There is much work to be done with current major facilities. It 
          also is important that the community find a way to break the cycle 
          of increasing capital expenditure and lengthening development times.
          Advanced accelerator R\&D is one approach. 
    \item We should ask for the machine which allows for good and frontier
          science. Then we should make the case for it and justify the cost. 
          We shouldn't expect any money without justification and shouldn't 
          decide on the machine based on our understanding of the limited 
          budget. 
    \item What do you mean by limited budget? One should not assume that the 
          budget is necessarily all that limited. How limited it is depends 
          majorly on how well the HEP community can convince the funding 
          organizations and Congress that a particular facility is important 
          enough to fund. This in turn depends on how united the HEP community
          is in supporting one specific major facility. A continually split 
          community may very well cause the demise of the field. 
    \item Dispose of a large fraction of staff which merely functions as 
          unwanted by-product of funding; this will free resources and open 
          positions. 
    \item Good research needs enthusiasm and adequate support including 
          financial support. 
    \item Limited budgets restrict options for new machines. More important 
          than ever to base decisions on best science. When a new facility 
          comes on is much less important than that it be able to address a 
          range of important topics. 
    \item The HEP budget is quite large compared to many other areas of 
          science. We can support many disciplines and be very diverse, if we 
          avoid placing too many of our resources in one activity.
    \item Money will always be limited in research, especially in fundamental 
          research. In my point of view the combination of a HEP machine 
          (really fundamental research) with a more applied subject (like e.g.
          the use of a free electron laser beam) gives unique arguments in 
          our hands in order to convince politicians to give us the money to 
          build such a machine. It is also very good if we can tell our 
          children that in our lifetime we did not only study the properties 
          of the Higgs but also better understood the functionality of 
          proteins in our body.
     \item Lab budget is now 90$\%$ of HEP funding. Universities, only 
           10$\%$. This imbalance is bad for long term future of HEP 
           creativity and funding. 

\subsubsection{Universities:}
    \item The University Research, including equipment development, must be 
          more strongly supported at Universities. Universities are the
          traditional source of ideas, and young people. National Labs. 
          (e.g. BNL) were started as places where Universities did together 
          what they are too small to do separately. The National Labs should
          NOT be in control. Therefore the funding should come through 
          Universities whenever possible.
    \item Must always have a component that supports undergraduate as well 
          as graduate student involvement. Otherwise we have no future. 
    \item To attract young physicists, University support must improve.
    \item A balance is important and students should not be given false 
          encouragement, but given the time lag between when a student decides
          what to study and when she/he is looks for a potentially permanent
          position, I don't see what could reasonably be done. At the faculty
          level, the interplay between government funding and what a 
          department/ university wants is complicated. 
    \item A small number of universities eat up the majority of the HEP 
          budget. This needs to be better distributed. A minimum of one 
          post-doc or technician and one grad student per university project 
          would be a good start.

\subsubsection{Importance of Using Existing Resources:}
     \item Building a new accelerator at an already existing lab would be the 
           most cost-efficient thing to do. 
    \item The energy frontier is the holy grail. This is where the important 
          discoveries will come. Pushing accelerator and detector technology 
          is extremely important. I also believe that building on to an 
          existing facility is a wise move. That is, using a facility with a
          trained staff, infrastructure, existing research program, and 
          running machines would overcome many of the problems the SSC 
          encountered. 
     \item The infrastructure for HEP is extensive and expensive. It is most 
           productive when concentrated at few (national) labs. We must be 
           very cautious when proposing a new lab rather that expanding an 
           existing lab. The SSC is the poster child for inadequate planning. 
    \item Put money into increasing Tevatron Luminosity.
    \item The next collider should not be a "start from scratch" effort, 
          since I believe it would lead to another SSC fiasco. It should 
          rather be viewed as an upgrade to an (existing?) facility. Each 
          upgrade should be associated with particular physics goals to 
          justify funding and instigate the necessity for the next one.
    \item Support research at HEP facilities.
    \item In order of priority : 0 - Base program at FNAL 1 - 
          Superconducting Linear Collider 2 - LHC participation 3 - VLHC 
          Magnet R\&D 4 - Superconducting RF technology.
    \item We should be upgrading current machines to maximize their physics 
          potential and doing R\&D for the next big machine. It is premature 
          to be building the next big machine now in the US, or to be 
          proposing to do so.
    \item There is a difference between the "next machine" and a new facility.
          National laboratories are natural places for a range of experiments,
          because they already have a critical mass to build an 
          infrastructure. Smaller experiments can run parasitically off of 
          this. Also, while for my personal career it would be better if 
          phenomenology support was "mandated," I do think that the community 
          responds to needs. Just not necessarily on the time scale of our 
          careers.

\subsubsection{Importance of Redundacy:}
    \item It is obscene to spend \$8B on a machine and get 1 or max 2 
          interaction regions. Can you imagine collaborations with greater 
          than 2000 physicists! Diversity is important. Only neutrino factory 
          approach will guarantee this. VLHC timescale is after LHC, since 
          its physics case will only be made by the LHC. 
    \item Two cheap detectors are probably not better than a single, good 
          detector.
    \item It is not clear that we can afford multiple detectors at the next 
          facility. If two detectors means the death of everything else 
          (which is almost certainly the case), then the cost is too high. 
          While the next big machine is very important to the future of the 
          field, we cannot kill everything else for it. Whatever the facility 
          is, it will not answer all the important questions. 
    \item May not be feasible cost-wise to have 2 multipurpose detectors at 
          next collider - far better to put sufficient money into 1 detector. 
          This is not happening at the moment, where redundancy is favored 
          over adequate investment in each detector. If budgets allow, would 
          be better to have a 2nd collider with a different physics reach.
    \item Presumably many facilities will exist (not just new ones, so 
          balance does not have to be maintained only within a new facility. 
          What HEP can support depends on the size of the limited budget. 
          Given the current limited budget it may be wise to have less 
          redundancy.
    \item Redundancy is a great thing, but at the cost of these detectors, 
          it is not an option that is worth the money it will take away from
          other experiments. 
    \item Not all machine designs are able to reasonably support multiple 
          detectors. However, if two or more detectors are feasible they 
          should be built.
    \item I think its more important that the "next machine" host a large 
          range of experiments, but at the same time the number of 
          collaborations is not as important. The important issue should be 
          to get as much physics as possible out of the next facility. At the 
          same time, astrophysics efforts that have some bearing on what the 
          physics being explored should be hosted, not just "oh we need some 
          astrophysics efforts here too..."
    \item A very interesting fully dedicated project without multi-physics 
          can be very valuable too. There is no general rule.
    \item For an LC, one detector is probably fine. For a hadron machine, 
          two are best.
    \item For any new collider to be built it is absolutely essential to have
          at least two detectors WITH EQUAL STANDING and OPPORTUNITIES. If 
          there would be one then effects below say the 6 s.d. limit cannot 
          be believed. The number of phenomena/reactions studied will be much 
          less diverse. The construction of the detector which will be of the 
          several hundred million dollar size will take much (1.5 - 2 times) 
          longer. One detector only is unhealthy for the field.
    \item Next machine: only one, detectors: at least 2.
    \item In reference to question 11: I would rather see multiple interaction
          regions with significantly different experiments rather than 
          multiple regions with the same type of detectors.
    \item A second collider experiment doesn't have to be completed when the 
          accelerator starts running, but the option to at least add one later
          should be there. Including other physics (e.g. a fixed target 
          experiment and FEL/synchrotron radiation experiment) is important.
    \item Regarding question 11: There is less need for redundancy in e+e- 
          than in hadron collider experiments. The measurements are generally
          cleaner, generally more precise, and generally can be made in many 
          channels in the same detector (as a cross check) and less likely to 
          be created by instrumental effects. It is reasonable for detector 
          emphasis to vary for a high energy e+e- collider, especially where 
          there is a desire to be able to make specific precision tests - by 
          switching for example between Z0 pole measurements and 
          continuum/ threshold discovery measurements ... Unlike hadron 
          colliders, these detectors are generally less costly to build per 
          discovery hence making the construction of two or more distinct 
          detectors, a reasonable course of action. The significantly greater
          baseline cost and complexity of a detector for a hadron machine 
          makes it much more cost effective to build one device and upgrade it
          as the physics and environment becomes better understood. 
    \item I don't think detector redundancy for the sake of redundancy is 
          something our field can easily afford anymore, attractive as it may 
          be. On the other hand, if there is a physics case for two different 
          kinds of beams (e.g. Z pole plus full energy e+e-, or a gamma gamma 
          option), then a second detector may be justified.
    \item The argument about number of detectors, experiment, etc. is quite 
          important, but the REALLY important question is the physics. 
          Independent confirmation of new results is VERY important, but 
          might be achievable within a single experiment. 
    \item With regards to redundant detectors, I believe that they should be 
          built if compatible with the accelerator design. I practice this 
          means I favor multiple IP's at a storage ring, but would not 
          compromise the design of a linear collider to accommodate multiple 
          detectors. 
    \item I assume the next machine will have at least one general purpose
          detector. If there is significant physics that the general purpose 
          detector can not do, but a specialized detector can do at a 
          reasonable cost, this is obviously a good idea. 
    \item HEP lives from the interaction of people and of different 
          experiments. In the extreme limit, having 1 collider with 1 
          experiment for a single purpose (with a rigid hyrachie) would 
          make the "science" that comes out of it not worthwhile. Single
          plans should be scaled that variety fits into the budget. 
    \item Redundancy is a minor reason to have two detectors. The two 
          detectors are not identical ever. 
    \item Two detectors is not an option for a linear collider. Redundancy 
          is important, but I don't see a solution for a linear collider.
    \item If we are to only have one linear collider/VLHC in the future, I 
          think redundancy is very important to give credibility to results. 
    \item Quality is much more important than quantity. Neither the number of 
          detectors nor the number of theorists plays any role. Rather one 
          excellent device operated by careful bright experimentalists and a 
          few outstanding theorists than the unfortunate present situation 
          with too much mediocre people in all these areas.
    \item Given budget constraints, I would pick two different but 
          complimentary programs (e.g. collider and fixed-target) over two 
          large experiments on the same machine looking at the same physics. 
          And that's not just my background speaking ! Redundancy is 
          "somewhat important", but it is also important to not put all eggs 
          in one basket.
    \item The redundancy between experiments (4 at LEP, Belle and BaBar etc) 
          has so far increased the thoroughness of results.
    \item Because of the limited budget, new machines need to have a wide 
          range of experiments. But at the same time, any new 
          measurements/ discoveries need to be backed up.
    \item General purpose detectors, expandable machines.
    \item Greater facilities. Redundancy of detectors. But no redundancy of 
          accelerators in the world. More collaboration between theorists, 
          phenomenologists and experimentalists. Experimentalists should focus
          on physical measurements the phenomenologists suggest them. 
          Experimentalists should also understand how their results are 
          interpreted by phenomenologists (meaning of error for example).
    \item The answers to these questions depend on the resources available. 
          For instance, two detectors are required but only if funds are 
          available.
    \item Even though redundancy is very important, cost may make it 
          impractical.
    \item Having two or more large multi-purpose detectors does not mean that 
          they would do things in a redundant manner.
    \item Past experience with multiple detectors at a single machine has 
          been mixed. The competition both helps and hurts (especially when 
          it becomes all politics). For the money and manpower investment, 
          I'd rather see more variety and less redundancy. Experimenters have 
          become very much more sophisticated in recent years about how they 
          analyze data and how they take steps to avoid fooling themselves,
          so the redundancy is less necessary. Better to focus the resources 
          on making one experiment really good, and having variety in the 
          other experiments/endeavors that are going on.
    \item The cost of detectors should be balanced against the cost of an 
          accelerator. The main argument for multiple detectors is not 
          redundancy per se, but the ability ro make optimal use of the 
          substantial investment in an accelerator
    \item I feel that redundancy and a rich range of experiments and 
          collaborations is essential to the success of the next machine. 
          A good mix of different disciplines is necessary to keep the paths 
          of communication open. Astrophysics and the HEP labs should coexist 
          where they can, but I think that there are some areas where this 
          is not feasible.
    \item Identical detectors are a waste. Redundancy should be considered 
          along with cost, expected physics need for redundancy,
    \item Redundancy between experiments is important, however, there are 
          lessons to be learned from the proliferation of B-factories, which 
          may have been more politically than scientifically motivated. More 
          coordination between labs and institutions could reduce 
          "unneccesary" redundancy, i.e. 3 B-factories in the U.S. alone, while 
          preserving scientifically motivated redundancy, i.e. the ability to 
          experimentally verify measurements.
    \item Your choices are limiting. Redundancy in any important measurement 
          is vital at many levels. This doesn't necessarily imply that two 
          detectors are needed if a single, well conceived and built detector 
          can look at a phenomenon in more than one way.
    \item HEP should focus on a world class facility and eliminate marginal 
          and duplicate projects.

\subsubsection{Maintaining Diversity:}
    \item Breadth and balance are far more important than trying to 
          "pick the winner" for the field as a whole. 
    \item It is important that the next machine be able to host a variety of 
          experiments in order to be most prepared for new physics.
    \item In order to increase the chance of governments funding a large 
          project, it is important to show that it will serve a large 
          community and that it has a broad spectrum of science topics to 
          cover. 
    \item We should concentrate on our own field first and then if other 
          fields can make use of the facilities they should be made welcome 
          to. If money is a problem then the U.S. should stop competing with
          Europe/Japan and agree to work on the major projects together. 
          Competition might be healthy but we're not playing sport. 
    \item Question 9 has may aspects. If you ask whether diversity is 
          important from the political point of view, the answer is a clear 
          "very important". From the scientific point of view it is important 
          that the various fields of basic research are supported, but whether
          or whether not this happens at a future HEP center should be decided
          on the basis of the needs of the researchers. Hence the answer is 
          "very important" if an accelerator (like TESLA) is needed for other 
          disciplines (e.g. as a light source), but one shouldn't define 
          multi-disciplinarity as a primary design goal.
    \item Medical physics research should be valued. Also it would be better 
          to offer significant amount of lectures for the experimentalists who
          are far away from the institutions. 
    \item The LHC and Tevatron have clear goals - find the Higgs and see if 
          SUSY exists in the 100 GeV-1 TeV range. I don't think any such 
          consensus exists for the next generation machines. Therefore, 
          diversity (and new ideas) is important.
    \item Question 10: I feel it is important to expand the definition of HEP 
          to include particle physics that is not done at colliders. Question 
          11: The size of a modern collider experiment 
          somewhat alleviates the fear that a whole experiment will run 
          lock-step into some error. On the other hand, experiments have 
          shown (17 keV neutrino, e.g.) that this sort of lock-step can occur
          even in distinct experiments. So neither is a panacea. Question 12: 
          I'm not sure parity is the right mixture, but some mixture is 
          certainly good. 
    \item Anticipating reduced experimental facilities, it is important to 
          develop and pursue a strategy for maintaining combined nuclear, 
          astro, and HEP physics studies. Perhaps more coordination between 
          these branches/labs would be economic and positive. 
    \item The importance of the collaboration: I believed the new facility 
          should support as many research programs as possible. That not only 
          will make our case stronger before the Congress but also will bring 
          diverse ideas to the field of HEP. Also, the possibilities of getting
          funds from different sources will increase if a very diverse program
          is supported by HEP. 
    \item It is important to try to maintain a marketplace of possibilities  
          so that people have some choice of project to work on. 
    \item We need to focus on the science; much too much emphasis on the 
          sociology these days... A lot of dead wood as well...
    \item I find the balance extremely important. However we should try to 
          convince people and not to force them. 
    \item We should concentrate on a "machine" that gives the widest range 
          of physics as possible. It's time for HEP to move on from one 
          machine=one measurement. Let's look at astronomers for one minute:  
          one machine = hundreds of measurements. That's a different paradigm.
    \item I would hope the next machine is going to be truly versatile, 
          for instance a mumu collider with great neutrino beams as well, etc. 
    \item HEP must maintain its core mission of accelerator based experiments.
          The community has to support "one" new accelerator in the U.S. To 
          obtain this accelerator other projects will have to be eliminated. 
          This is a policy similar to the one adopted by CERN to build the 
          LHC. 
    \item It is crucial that the next big thing not kill the field... either 
          if it fails or if it succeeds. The "base program" must be vital. 
    \item Most of my opinions on these questions are not very emphatic since 
          it is not yet clear to me where the focus of the next machine 
          should be. That will depend on what current and near-future machines
          find. The exception is my answer to 11, since repeatable results 
          are a fundamental part of science, no matter what the field. The 
          problem, of course, is that doubling the number of detectors and 
          broadening the scope of scientific staff support invariably 
          increases the cost. Development of an entire new laboratory, with 
          its associated site acquisition and infrastructure costs, is 
          probably out of the question. Given the demographics in the areas 
          of existing labs, it is probably also out of the question to build 
          an accelerator that would extend much outside of their present 
          boundaries. The next machine will most likely have to fit within
          the real estate we have. This might mean pursuing new accelerator 
          technologies rather than just trying to make existing ones bigger. 
    \item Whilst it is important that we aim to cover a broad range of HEP
          topics it is obviously equally important that the investment in new 
          areas does not detract from the quality of existing programs of 
          research.
    \item Take care of a good pay-off of investments. No unused interaction 
          regions in colliders, multi-physics purposes. Serve a large 
          community, encompassing other fields than only HEP. 
    \item In my opinion it is very important to focus the available resources 
          on "world" projects in order to avoid duplication and maximize the 
          physics return. At the same time, a set of smaller projects, such 
          as astrophysics experiments, should be promoted to complement the
          physical scope, and to bridge the necessarily large gap of 
          'non-physics' periods during the construction of large projects.
    \item Maintain diversity; retain (Bob Wilson's concept of) "nook and 
          cranny" small-scale experiments, in addition to the (unquestionably
          important) major detectors.
    \item HEP should remain focused on high energy particle physics, rather 
          then indulging in interdisciplinary activities. 
    \item We should not spend all available resources on a single project. 
    \item Need to support a diversity of efforts, including flagship 
          facilities at the high energy frontier.
    \item The field has to retain a balance of diversity while maintaining 
          focus on the most important goals. The Higgs search has to be one of
          the highest priorities. Neutrinos is second. Searches for "beyond 
          Standard Model" physics is a third. 
    \item There must be room for the small experiments which are much more 
          beneficial to students than the large experiments. There must be a 
          mix of small and large projects -- HEP should not support only the 
          biggest projects.
    \item A diverse program is essential to the survival of the field. 
          Squeezing out small projects is basically death to innovation and 
          creativity.
    \item I think that also smaller experiments/projects should be supported 
          even in a limited budget scenario.
    \item I believe diversity in our experimental approach to be vital to the 
          health of the field. I find the tendency for the field to coalesce 
          into a handful of extremely large experiments to be very disturbing.
          This trend is particularly damaging for the younger physicists who 
          will struggle to obtain a well rounded education and who will have 
          difficulty in exploring those "off the wall" ideas that have in the 
          past provided some the more exciting breakthroughs.
    \item Given a limited budget HEP should support small groups at the 
          expense of the large projects because the field needs diversity to 
          be prepared to take advantage of opportunities which have not yet 
          presented themselves.
    \item HEP must advance on multiple fronts to search for new (as well as 
          expected physics): (1) highest cm energy possible to search for 
          Higgs and SUSY, (2) neutrino oscillations, (3) Dark Matter, (4) 
          Cosmology and Astrophysics. And finally, "big" as well as "small" 
          experiments.
    \item Have continuous physics results coming out of labs.
    \item Doing good science requires a broad approach and trying to come at 
          the problem from many different angles. Having few collaborations 
          and few kinds of colliders (eg. only hadron colliders) is kind of 
          dangerous. So we have to figure out how keep the field broad despite
          decreasing funds and increasingly expensive experiments. One way is 
          to try and be as multi-purpose as possible: electron machines might 
          also be used as X-ray sources (ala APS or NSLS). Maybe this would 
          allow us to build a photon collider as well. I don't think it's 
          possible to mandate the mixture of theorists, phenomenologists and 
          experimentalists. The best that you could hope to do is encourage 
          balance by choosing how to allocate money.
    \item Right now, nature seems to be telling us to put more effort into 
          neutrinos, while go slow on precision studies of physics for which 
          there is no direct evidence as yet. If the debate over e+e- 
          colliders restricts the opportunity to make new studies of the 
          neutrino sector, it will be sad for the field...
    \item Balancing the overall program will be key to the future survival of 
          HEP. This means not only balancing experiment/ phenomenology/ theory,
          but large and small scale experiments. Large scale 
          experiments/ facilities offer economy of scale, and broad physics 
          reach at the expense of long lead times in construction data taking 
          and analysis. Smaller, more narrowly focused experiments offer a 
          chance to take advanges of rapidly developing trends in HEP research
          which may take decades to pursue in a larger experiment.
    \item I think that the possibility of having more than one detector at 
          the "next machine" should be determined by the budget. I'd be more 
          interested in putting that money into smaller experiments (possible 
          next generation fixed-target expts).
    \item Having a machine that is able to support a diverse range of 
          experiments is important, but I believe that one should not spread 
          one's self too thin; the field needs machines that are high enough 
          energy to get at what we're looking for (alas for the fallen SSC!); 
          in my opinion this should be the primary objective of the field.
    \item Also allow lower energy important experiments in a booster section 
          (or equivalent).
    \item Training the "next generation" of physicists is also important. 
          Training is easier on a small scale than in a huge collaboration. 
          Could one of the experiments be "small" and the other "large"?
    \item HEP needs some multi-billion facilities. But needs in addition 
          multiples of cheaper and 'more exotic/risky' experiments. HEP 
          should realize opportunities in astrophysics.
    \item If HEP can afford several experiments with 500-1000 people budgeted 
          at 100M\$ +, then it should also give priority to smaller (10-50 
          people, 10-50M\$) projects. 
    \item Diversity of different experiments including underground and 
          astrophysics; also R\&D work on development of new detectors and 
          accelerators.
    \item Very important to maintain diversity in areas and not concentrate 
          all funding in one area (e.g. linear collider).
    \item Diversity is very important ... would sacrifice redundancy on a 
          collider machine for different kinds of machines, smaller 
          experiments, etc.
    \item Significant results are beginning to come from particle astrophysics
          and so support should also include funding for such experiments. In 
          addition, it is imperative that we continue with the tradition of 
          many "smaller" experiments addressing very focused physics 
          questions, i.e., g-2, and future experiments which promise high 
          sensitivity tests of the SM (not just a factor of 2 improvement in 
          sensitivity, but ~10X higher sensitivity). Furthermore, we must 
          strengthen our commitment to new technologies, for they are the 
          future of this field. The current growth required to perform high 
          energy tests of the SM is not sustainable, and only through R\&D on 
          new techniques for particle acceleration, and radiation hard 
          technologies for particle detection do we have a hope for the long 
          term future in this field as we know it today.
    \item Given a limited budget, supporting diverse research is much more 
          important than building a "flagship" machine. We are much more 
          likely to learn interesting physics with a diverse program than 
          with a single concentrated and centralized effort. Further, the 
          diversity can help promote physics to the general public.
    \item It looks to me we can now do a "Tesla" type of machine. Let's 
          support it. Also we need to have 5$\%$ of the HEP budget allocated 
          to "small" experiments with great physics importance without  
          having to compete with ATLAS, etc.
    \item Important to have small experiments as well as large ones, so that 
          the young physicist can learn hardware as well as computing. A large
          experiment in Europe and one in the USA/Japan is very important.
    \item The big money is in large experiments. We should stop doing dumb 
          things like CERN to Gran-Sasso, duplicating MINOS. 
    \item A wide array of physics is needed, but not necessarily at the 
          flagship facility. There are still places where relatively small 
          budgets can produce substantial results. 
    \item I think it is crucial that HEP support a range of physics 
          experiments, because we don't know which area will teach us the most
          about getting beyond the Standard Model. 
    \item I think the "next machine" should not be so dominant in importance, 
          I'm in favor of more diverse work like cosmic rays, neutrino beams, 
          heavy ions, and nuclear phenomenology. 
    \item It is wise to bet the future of the field on outcomes that seem 
          very likely, rather than on poorly supported possibilities. LEP 
          learned that only with intense collaborations of experimenters and 
          theorists could they deduce any results of significance. That will 
          be even more true in the future.
    \item The emphasis now has to be on the physics we can do and for the 
          machine builders we need support. The machine physicists are 
          considered second class and that has landed us where we are. We 
          have to make the machine part of the service work to do an 
          experiment. It should just be part of the specialization.

\subsubsection{Balance of Theory and Phenomenology:} 
    \item Phenomenology is under-supported in the U.S. (I am an 
          experimentalist).
    \item Phenomenology is too poorly supported. 
    \item On question 12, the important thing is for the theorists, 
          phenomenologists and experimentalists to work together and 
          communicate effectively, not to have a particular mixture on site at
          one particular machine, though this may help.
    \item It is important to have a mixture, but the market should decide.
    \item I feel strongly that ``theorist'' and ``phenomenologist'' should 
          not be distinct categories, for the same reason that, as I expect, 
          the authors of this survey would not accept ``hardware'' and 
          ``analysis'' experimenters as distinct categories. The best people 
          do both.
    \item Phenomenologists concentrating on theoretical calculations 
          important for experiment (such as QCD processes relevant for 
          Tevatron/LHC) should be hired by experimental collaborations to 
          prevent their knowledge from being lost.
    \item There are too many theorists. There are way too many string 
          theorists. There are not enough phenomenologists.	
    \item Theorists: Phenomenologists : Experimentalists $=$ 3: 1 : 2.
    \item HEP can easily support theorists given a limited budget.
    \item Theorists are cheap relative to experiment, and it would be crazy 
          to cut HEP support to theorists to incrementally increase funding 
          for a facility. 
    \item A healthy balance between experimentalists concerned about physics 
          analysis and detector building is in my view also very important, as
          is a healthy balance between phenomenologists (related to current 
          experimental efforts) and ``pure'' theorists. In particular, the 
          later seems to be grossly off balance at the moment.
    \item There is a severe and very obvious lack of phenomenologists 
          available to work closely with the experimenters in the planning of 
          future experiments and analysis of experimental results.
    \item Do not support string theory.
    \item Being a theorist I do believe that some type of quantum field 
          theory/string theory describes nature at the next generation. 
          However, my opinion is we have not yet learnt to solve 
          non-perturbative aspects of QFT/Strings. Thus it would be important 
          to use the funds that encourage this type of research in theoretical
          physics. The solution to a QFT goes far beyond identifying the 
          symmetries that govern it and being able to perform perturbative 
          calculations at weak couplings. In terms of experimental research I 
          think we should definitely look for signatures of SUSY. 
    \item Now, as we enter a phase of many new accelerator and 
          non-accelerator experiments, and related important astrophysics 
          experiments, it is important to have more experimentalists and 
          more phenomenologists. (We probably have an excess of string 
          theorists now.)
    \item At least in the US, way too much money is given to "string" 
          theorists and not enough to phenomenologists. Europe has a much 
          better balance.
    \item 1) HEP labs should support astrophysics to the level that the 
          experiments impact HEP, or the sites are the best place to perform 
          an experiment. 2) The separation of phenomenologists as separate 
          from theorists is bad for the field.
    \item With regard to 12), I think it is important to get theorists and 
          others who are looking at ideas beyond the programs at an existing 
          machine. Also people who can write up results in a way others can 
          understand and appreciate.
    \item With regard to question 12), the actual "balance" (proportion of 
          each category) needs to be determined appropriately, and this 
          determination is independent of the need for a balance. The mixture 
          is not 1:1:1, but more like 1:1:8. 
    \item It is important to provide adequate funding to the theoretical 
          community, particularly to invest in the computing necessary to 
          carry out lattice QCD calcs, etc.
    \item String theory is largely irrelevant to falsifiable science, and 
          therefore should be funded at a small fraction of its current level.
          In particular, the number of string theory faculty positions is 
          grossly out of proportion to its contributions to our understanding 
          of nature.
    \item Strong theory program is an essential complement to any experimental
          program.
    \item New generation of HEP experiments will require a closer 
          collaboration between phenomenologists and experimentalists. About 
          future accelerators, two e+e- linear colliders would be ideal but 
          if this goes against (politically or financially) the elaboration 
          of a strong research program for the next muon collider, it should 
          be avoided! I also believe that looking for a post LHC hadron 
          collider is premature.
    \item Labs and experiments need theorists and phenomenologists who work 
          directly on the physics issues that are explored by the experiments. 
    \item There are way too few phenomenogists.
    \item I don't think we can mandate a balance of fields, because academic 
          freedom is a very strong reason people pursue this field. However, 
          I think it is a shame that certain types of physicists seem to fall 
          between the cracks, particularly phenomenologists whose work is not 
          flashy but is still crucial to the interpretation of experiment. 
    \item More support for phenomenologists. Too many young theorists end up 
          in string theory while many areas of phenomenology are not 
          adequately covered. This makes it difficult, if not impossible, to 
          correctly interperet experimental results. 
    \item It is very very important that there is a balance between 
          phenomenologists and experimentalists. Of course, phenomenologists 
          should also be talking to theorists, but eventually, theorists will 
          eventually have to come back down to Earth and consider field/string
          theories which can be realistically measured (I should know - I am 
          one !). 
    \item I consider it essential for the experimental HEP community to 
          collaborate closely with theorists and phenomenologists to push the 
          frontiers of research effectively. The mixture, in my opinion, 
          depends on the kind of physics under question. 
    \item Presently, there is a gross imbalance between speculative 
          theory/phenomenology on the one hand and real phenomenology which 
          is critical to extract physics from experimental programs in all 
          colliders in the next couple of decades (cf. speech by Matt 
          Strassler at the Town Meeting). If YPP does not push hard for 
          support of the later, the future of HEP will be bleak. 
    \item For (12) I think the mixture is counter productive. There should 
          naturally be more experimentalists than phenoms than theorists. So 
          it is important and the market should decide.
    \item I think, at the moment, there are lots of theoretical models around 
          but not enough data to test them. 
    \item Not related to budget limitations : communication and work between 
          experimentalists and theorists should be emphasized. 
    \item The experimental community relies upon phenomenology, but there 
          isn't enough support given to phenomenology by the funding agencies 
          and by the national laboratories. This is an issue that needs to be 
          addressed. 
    \item We have a long tradition of a fruitful mix of theorists and 
          experimentalists who understand each other and know where they are 
          needed. Science more or less mandates what that mix should be; it 
          cannot be established by executive fiat.
    \item Question 12: Who would decide what balance is if not the "market"? 
    \item Funding for HEP expt and theory should be allocated within input 
          from both specialties, i.e., expts should be able to encourage 
          useful phenomenology, theorists should be able to have input on 
          which expts are interesting. Current situation with separate 
          review structures leads to a non-optimal mix of both. 
    \item "Balance" should not be promoted through the use of inflexible 
          quotas. The situation at a given institution should be judged by 
          experts on a case-by-case basis.
    \item There is a tendency for students to become theorists. This is 
          probably due to the high status given to theory by the physics 
          community. We should promote programs to drive very well qualified 
          people into experimental physics. Experimental physics is full of 
          interesting and intellectually rewarding challenges.

\subsubsection{Balance with Astrophysics:}
    \item Not all need be invested in machine-based experiments. Correct 
          mandates are a hard thing to anticipate. Flexibility is important 
          and the best physics should drive the choices, not the prior 
          investment decisions...
    \item We have no business supporting astrophysics. I think two detectors 
          are good but not if it chokes off upgrades as we see with CDF/D0. 
          FNAL ought to have forced them to join into one collaboration 
          instead of this ridiculous competition for non-existent resources.
    \item I feel that only a limited number of astrophysics experiments are 
          asking (or are able to ask) the questions that HEP seeks to answer. 
          There is correlation between the fields but the goals are not the 
          same.
    \item Astroparticle physics efforts are a much cheaper way of doing HEP 
          physics in the next generation.
    \item I think that using the same facility for experiments in different 
          fields of physics than HEP, like astrophysics or nuclear physics, 
          is very fruitful and helps sharing costs.
    \item Future of HEP only in combination with astrophysics - we HAVE to 
          merge our efforts.
    \item Don't confuse ``future of HEP'' with ``next machine''. Future of 
          HEP may well be in theory or particle astrophysics.
    \item I think astrophysics is very cool. I have no objection to it being 
          done anywhere. However, it does not seem to me that astrophysics 
          must be supported in HEP facilities if there is some benefit to not 
          doing so.
    \item Next machine should probably also include the $\sim$ \$500M 
          nonaccelerator 
          proposals too.
    \item The astrophysics side gets a large bang for the buck, but 
          traditional HEP accelerator types often discount what can be done 
          with it. Dunno how to change this perception.
    \item Detector technologies and physics becoming much closer in HEP and 
          astrophysics. Very important to have cross-checks of results - 
          keeps people honest and gives competition.
    \item Astrophysics and space-based experiments are good for finding 
          inconsistencies in our understanding of the Universe; they create 
          more unanswered questions than they provide answers to. It is 
          therefore important not to lose focus and continue pushing the 
          energy and luminosity frontiers in collider-based experiments.
    \item It is time to think of space based experiments in collaboration 
          with space agencies, as an option.
    \item Astrophysics is important, but it may not be necessary to do it at 
          the labs. Certainly, it's not necessary to do it at *every* lab.
    \item Much more attention should be given to particle astrophysics. This 
          sort of research probes energy scales and physics which cannot be 
          done at accelerators, attracts wide public interest and support, and
          naturally complements accelerator efforts.
    \item Try to get more out of cosmic rays rather than build colliders. 
    \item I think that HEP should concentrate on supporting experimental and 
          theoretical efforts. Whereas there is a lot of overlap between our 
          field and astrophysics/cosmology. I feel that funding efforts such 
          as the Sloan Digital Sky survey is outside of our remit. Such 
          efforts should be funded from other parts of the budget. Our focus 
          should be on promoting experiments and theory which impact particle 
          physics in a more direct way. 
    \item Hosting astrophysics efforts is one thing, collaborative 
          efforts is quite another. 
    \item I believe that major HEP labs should pursue both their primary 
          accelerator programs, as well as non-accelerator projects. One of 
          the best subfields to pursue is particle astrophysics or high energy
          astrophysics. The tie-in to astrophysical processes which illuminate
          fundamental physics questions is necessary for a proper breadth of 
          perspective at a major lab. 
    \item I'd like to emphasize that HEP efforts must include the fileds of 
          cosmology and astrophysics and non-accelerator experiments. Reliance
          on larger machines will eventually (and soon) end the field of high
          energy physics.
    \item Diverse research is essential as none of us know where the data 
          which leads to the next breakthrough will come from! It might be 
          astrophysics, so we should do what we can to support GLAST etc. but 
          not at the expense of major HEP facilities.
    \item In order to train graduate students/postdocs on particle detectors, 
          astrophysical (or even nuclear) experiments are important.
    \item We should look for more passive experiments. The future of HEP is 
          not in colliders but in innovative combination of astrophysics and 
          passive experiments (like detecting neutrino masses). 

\subsubsection{Survey-Specific Comments on Balance:}
    \item The questions in this section are almost rhetoric questions: the 
          "right" answer is already in the question. Anyway, I agree with 
          the "right" answers...
    \item Answers here require considerable qualification.
    \item The questions are based on the assumption that there will be a next 
          machine. It is also possible that we have used up our credit with 
          the public. If there is no next machine, all detailed questions of 
          relative importance are irrelevant.
    \item Redundancy is a loaded word. One might also say independent 
          verification.
    \item Your way to ask questions is very biasing, I do not know what you 
          will do with the answers but for sure I know what you want people to
          answer. You get an F for fraudulent questioning.
    \item Question 9 is rather obscure? 
    \item I don't understand question 9. All collider experiments provide 
          multi-physics results over a huge range of particle physics 
          subfields. Does this mean that a large facility hosts smaller 
          "other-purpose" experiments? 
    \item Question 11: you are leading the witness ... nobody will answer that 
          redundancy is not important. Still people can think it is too 
          expensive, waste of manpower, etc. 
    \item The answer to question 11 depends on what type of collider. 
    \item Question 12 is meaningless without a definition of "balance." 
          Perhaps directing this question more toward the "promotion of 
          funding for" is more to the point.   
    \item Not sure what you mean by balance, and what balance would be 
          encouraged.
    \item For question 12, it's not clear what the word "balance" mean, from 
          what point of view. 
    \item Again, astrophysics and non-accelerator physics are NOT identical. 
          This part of the survey shows rather incredible ignorance of the 
          field of high energy physics. 
    \item Some of these questions are ill-formed: I don't understand question 
          9 on "multi-physics". What is to be included? Does this mean the 
          machine itself should be multi-disciplinary, such as TESLA? The 
          TESLA proposal to include an X-ray FEL is a very good idea. 
          Question 11: Redundancy is not a good choice of words. One does 
          want complementarity and cross checking. Question 12 is unrelated 
          to the next machine; it is a question to laboratories and the field 
          as a whole.
    \item Options for answers oddly structured, combining issues that need 
          not be coupled the way they are in the selections. This is how 
          surveys "with an agenda" are constructed. What's up here? 9) 
          "diverse (multi-physics)" is ambiguous 11) It is at least "somewhat"
          important for a collider to have $>$ 1 detector, but redundancy is 
          not the only reason. 12) Why can't I select "very important" but 
          "let the market decide". 
\end{itemize}

%% file: comments-globalization.tex
\pagestyle{myheadings}
\markboth{Comments on Globalization}
         {Comments on Globalization}
\subsection{Impact of Globalization}
\begin{itemize}
\subsubsection{Global Collaboration:} 
    \item HEP should support the advancement of the design of a machine 
          (funded internationally) that enables particle physics to be done 
          at the energy frontier.
    \item International collaboration is the key to maintaining a diverse 
          program on a limited budget.
    \item The international HEP community needs to come to a conclusion over 
          physics priorities. We cannot afford to give incoherent information 
          to world governments.
    \item I very much believe in open discussions, followed by determined 
          action, so we can formulate a consensus on how our field can achieve
          public acceptance for scientifically honest and intellectually 
          proper pursuits that mesh well with our international colleagues' 
          activities.    
    \item First, support running approved experiments. Second, try to get a 
          next machine in the U.S. Don't send too much money to Europe 
          without getting much back.
    \item American HEP should be committed to international 
          collaboration/cooperation.
    \item Stress international cooperation; do not demand U.S.-only efforts. 
    \item The world is getting smaller every day with the increase in
          networking, video conferencing etc. If our field is going to become 
          global we must learn to work together from large distances. Part of
          the fun of HEP is working with people from many different places.
          This should not be ignored. 
    \item International collaboration is important to the next series of 
          machines. There is probably only enough money for 1 linear collider 
          and 1 new hadron collider after LHC. A single country cannot do it 
          alone.
    \item Some redundancy is good, but international integration of goals is 
          necessary, in particular in the TESLA and NLC question. It is a 
          waste of money to have two or more redundant accelerator physics 
          programs in different countries. United, the fields stands; divided,
          it falls. Let us not allow our individual pride to override the 
          needs of the field as a whole.
    \item Some international collaboration and less U.S. university politics 
          would increase the available funding (even if you don't receive 
          more). People are too busy making sure their little part of the 
          detector works. HEP needs to change the way collaborations are 
          structured.
    \item Due to budget realities, we need to form REAL international 
          collaborations in order to accomplish the physics.
    \item Science is broad and financial support should be accordingly. Large 
          projects like a new big accelerator requires international (for us 
          at least of European size) support. 
   \item Focus on HEP interest as opposed to national interest. Build the 
         best collider in the country that could do it best. Think 
         international. 
   \item I think the argument that the next big project must be 
        "truly international" is incorrect. We should work to insure that
         U.S. HEP is strong. International collaboration will continue to 
         develop if all collaborators are independently strong. 
     \item Most crucial is to get the lab directors and policy makers thinking
           that global collaborative work is essential.
     \item The idea of having international collaborations is currently taken 
           as a necessity. However, I also think it vital that even given 
           international participation, that detector elements produced 
           should be created by those in the position to best manufacture 
           them, and not distribute by, "oh, well they needed to get some 
           hardware done, so we'll get the 'insert name here' to get these. 
     \item I feel the key benefit of global collaboration in the 
           design/ construction/ operation of HEP facilities is that expertise 
           in these areas will not be limited to the physicists living in the 
           host countries -- in the future one could imagine one country 
           having lots of resources but no expertise, and vice versa. 
      \item Due to the high cost/complexity of HEP research in the future the 
           number of HEP labs will be limited to 2 or 3 globally with specific
           research goals (neutrino physics, very high energy hadron physics).
           The inevitable "global collaboration:" will result in: 1) A 
           multidisciplinary environment with a high level of biasing and 
           company like work style. 2) A shrinking of the physics community 
           3) Lower quality training since hands-on experience for graduate 
           students will be limited. 4) A cutting at the University funds, in 
           favor of the National Labs (to complete the expensive experiment).
     \item Strong U.S. computing leadership is very achievable and very 
           important for the health of the U.S. role in the global HEP 
           enterprise. U.S. computing strength in ATLAS and CMS is currently 
           under-funded and the U.S. leadership role is endangered. 
     \item I am currently working on CMS, and answered the above questions 
           with respect to that R\&D work. CMS and Atlas are good examples of 
           functioning global collaborations. The key to having it work is to 
           not make the mistake we made with the SSC contracts, i.e. requiring
           the money to go back to U.S. corporations if possible. Each country
           involved needs to feel that it has something at stake, including 
           getting contracts for work. 
     \item We have several global collaborations already, and they work. Local 
           institute vs local detector presence is not a new problem, it can 
           be dealt with. 
     \item I think we have ample experience in making large global 
           collaborations work. 
     \item Many of us already work in global collaborations. In OPAL at LEP 
           we have physicists from Europe, Japan, and North America. It is 
           important for physicists to spend a significant amount of time at 
           the center of the collaboration, at the experiment. One can do 
           quite a bit of work using e-mail, the web, video teleconferencing,
           telephone, etc., but significant presence is necessary. I don't 
           see how one can run an experiment or an accelerator without some 
           hands-on presence. 
     \item Non-U.S. HEP physicists in Europe/Asia have much experienced on 
           attending the experiments in U.S. (in the offsite of its own 
           country). 
     \item HEP has always been a global enterprise (e.g., the exchange of 
           information between physicists). The funding tends to be local. 
           While HEP has done much to promote globalization, the interests of 
           governments tend to be more local. 
     \item Global collaboration works already pretty well with physicists, 
           but there is no hope that governments will collaborate globally for
           the length of time needed to carry out the big projects: 10 years 
           to get approval and construct it, 10 more years of running and 
           upgrades based on the original construction, 10 more years of 
           running based on a major upgrade. 
     \item Large collaborations should have itinerating collaboration meetings
           matching the percentage of physicists from each collaborating 
           country rather than the share of investments. Past experience shows
           that the politics of "everybody signs every paper" should be 
           maintained only in case there is a chance for a controversial 
           paper to be published with a minority of signatures. The 
           advancement of physics should not be tied to the opinion of the 
           majority of contributors to large experiments (which will involve 
           in the future large percentages of all active physicists in the 
           field, in some cases). 
     \item It is important that there be several host locations (large 
           experiments) for the global collaborations. It is not good to have 
           one project based at, say CERN, and no other reciprocating projects
           in the U.S. or Asia. 
     \item HEP should become even more a global operation. Continental  
           separation/isolation is not an option. The next step in HEP should
           be planned in an international context, taking into account 
           redundancy vs. complementary experiments on a global level. 
     \item It won't work -- we squabble badly amongst ourselves, and adding 
           politicians to the mix will make things even worse. No one country 
           (among Germany, Switzerland [which really also means Britain and 
           France], the U.S. and Japan) will let their lab close.
     \item When it comes to working on experiments we are internationalists. 
           But when it comes to decisions about the "next big thing" each 
           country wants its own "next big thing". We shall be forced to be 
           more sensible and to pay more than lip service to global decision 
           making. How? I don't really know. But since it is the youth who
           will inherit the future we should help them found a global 
           organization, with a budget and staff and formal links to national
           organizations. This body would provide a formalized forum for 
           "thinking globally". 
     \item 1. What seems to be missing in the present climate is a mechanism 
           for decision making as to proposal by different major labs. In the 
           past, these labs provided a decision making structure for their 
           internal program. Now, some kind of international PAC is needed. 
           The immediate example is the choice between cold and warm rf 
           technology for an e+e- collider. The labs that sponsor each 
           technology seem completely dedicated (as perhaps they should be) 
           to their choice, and will never voluntarily accept the other 
           technology.... 2. "Regional centers" is a possibly contentious 
           topic -- if it is an initial to duplicate many of the facilities 
           of existing national labs at new sites. It doesn't appear that 
           there is enough financial support for HEP in the U.S. for grand 
           expansion in this direction. Rather, the existing labs will more 
           probably evolve towards "regional centers" if they no longer host 
           a major accelerator facility.
     \item Again, this cannot be decided by central command; its anwer must 
           be mandated by the specific activities we will have our review 
           panels recommend for funding, hoping that these recommendations 
           will be accepted by our money tree. 
     \item There will need to be better guidance and leadership than at 
           present, as better management is needed for larger collaborations. 
           A lot of money is wasted by inefficient management at the moment. 
           One major problem will be giving recognition for the people who 
           actually run the detector(s) rather than those who do the physics 
           analyses. At the moment there is little or no reward for detector
           work. I expect this will diminish as the collaborations get larger.
           Similarly, a greater role for technicians, engineers and com. 
           scientists within the collaboration is needed.
     \item Let's for once take an example in multi-nationals. Particularly 
           where it concerns management, physicists are better of *not* to 
           reinvent the wheel. 
      \item It is important to develop U.S. plans in an international context, 
           with international support and cooperation. 
     \item Greater contribution of a global collaboration would be to provide 
           a guaranteed funding profile. Participation in a global 
           collaboration without such a basis will be impossible. 
     \item Global collaboration could be successful if and only if there are 
           efforts from physicists of different countries but similar 
           scientific interests. 
     \item Form small groups of people, 4-6 all interested in working on the 
           same small area. These groups should report progress to the physics
           subject area leaders The subject area leaders in turn reporting to 
           the spokesman. It is rather important that the small group of 4-6 
           all be close enough to talk to each other in person. Good quality 
           software tools, well documented are a must. Real responsibility 
           must be taken for software. 
     \item I can only comment on Fermilab, which is a wonderful place, with 
           many resources... R\&D, computing, and software, can be designed 
           there, for a global machine. 
     \item Pragmatically, global collaborations are desirable, and inevitable.
           National labs without accelerator facilities cost much more per 
           physicist and have much weaker impact on students than university
           labs. Argonne and LBL HEP groups should be absorbed into U. of 
           Chicago and U. Cal Berkeley, and funded per physicist at the level
           of university research groups. 
     \item If U.S. physicists wish to partake in a global collaboration then 
           they need to be prepared to travel overseas, rather than expect 
           foreign collaborators travel to the U.S. 
     \item Global collaborations are bit more complex to keep going than 
           national ones and those a bit worse than regional and on down until
           the best is all in one institution - still, the new detectors are 
           large, the collaborations are large and extended but that is only 
           one more difficulty. If this were easy we wouldn't want to do it.
     \item There are numerous examples of global collaborations that work 
           adequately. However a future enormous collaboration on a future 
           large project will not be a good place to initiate and train young 
           people. Smaller and much faster local projects will be needed as a
           foundation for big international efforts. 
     \item I think the major facilities should be built by international 
           collaborations, but be located in the countries with most developed
           infrastructure, experienced ingeniers and qualified workforce - 
           like U.S., Eorope, Japan. Regional centers, like described above, 
           is a good idea, allowing reduce the need for travel, and easing 
           access for students to such facilities. 
     \item Globalization of HEP is not just a U.S. issue as this section has 
           leant towards. If this 'flagship machine' as you call it is to be 
           built in the states then we need to think long and hard about the
           location of it. 
     \item Clear goals and leadership for the whole collaboration and for each
           regional component. Autonomy (as far as possible) in carrying out 
           each major sub-project. Detector design, development and 
           construction to take place in all regions. [By region I mean 
           Europe, USA, Japan]. Strong involvement of theorists, from design 
           of accelerator through detector to ultimate physics analysis. 
     \item Living and working in Europe makes it difficult to judge the 
           U.S.-situation. Still, competition and collaboration are both very
           essential, as is complementary equipment. I hope that the U.S. 
           (or America as a whole) will get an large accelerator of the future
           (TESLA) generation, but would strongly advocate that it is not a 
           national lab, but rather a type of "world-lab". 
     \item A global collaboration working together is possible. It has been 
           tested and it works! Is the best way for the promotion of 
           science (HEP). 
     \item Global collaborations can work very well. Collaborations definitely
           should be global. 
     \item The U.S. has to develop a "culture" of international 
           collaboration, i.e. to give up the leitmotiv: "We are second to 
           none". 
     \item Globalizaion is a mantra I don't subscribe to; we will have leaders
           that will cause things; the Germans WILL build an NLC because they
           believe in it strongly enough to do it the physicists themselves; 
           internationalization is with us NOW as is; our current generation
           is largely analogus to the previous generation of bubblers; they
           did a lot of the physics and substantially inherited the field but
           rode the coattails of generation 1 who built all our machines and
           started the physics; the bubbler generation did a terrible job of
           running the next initiatives; Isabell, the SSC etc etc... The  
           Japanese, Germans, and even Europeans are making a laughingstock
           of our vaunted "can do" ability look a the initiatives of the last
           15 years..
     \item You will always underestimate the value of a multi-national 
           collaboration (with equal participation among the nations) until
           you are actually in one. 
     \item HEP communities in U.S., Europe and Asia must work out a coherent 
           long term plan that can be supported by the general public and 
           governments. 
     \item I've taken part in quite a few experiments that can be defined as
           "global collaborations" and I've found them very effective. My 
           opinion is that the future of particle physics,in both 
           accelerator-based experiments or astrophysics experiments will
           greatly benefit from the global collaboration approach 
     \item Physicists should start to realize global aspects. I don't care 
           whether a lab or a colleague is in the U.S., Japan, Europe or on 
           the moon. The quality of physics results on this globe is relevant.
           The rest is outdated nationalism. 
     \item I welcome global collaboration working powered by modern 
           communication technologies. 
     \item A global collaboration will only work if the discoveries are 
           shared. LHC will be a test case. If \$500M from the 
           U.S. results only in discoveries by "European scientists" then I 
           think globalization is not going to work. We must also be careful 
           with the results from the Tevatron in Run 2. The good publicity 
           should be spread around to all the participating institutions - not
           just the national labs. 
     \item At networking bandwidth increases global collaboration will become
           easier. My current collaboration, BaBar, is already 
           intercontinental, and I work at my university rather than SLAC. As
           to regional centers, it's really a matter of determining whether 
           it's cheaper to build/maintain them or to fly people to the
           detector. Of course the real experts (those who can fix hardware 
           problems) will always need to be stationed near the detector. With
           regard to national labs, I think there is enough to do in the next
           25 years to keep all of our current labs busy, but see no need (or
           practical possibility) for opening a new one.

\subsubsection{Proximity to Detector:}
     \item It will always be important to spend time at the site of the 
           detector.
     \item The international collaborations based on facilities outside the
           U.S. require large, continuous presence of U.S. personnel at those
           sites. Too many U.S. physicists want to play remotely (i.e., stay 
           at home in the U.S., and participate only via teleconferences or 
           occasional collaboration meetings). This ends up placing all the
           burden on on-site personnel, and destroys the goodwill in the
           collaboration. Instead of having remote-operations facilities in 
           the U.S. for experiments based abroad, it would be much better for 
           U.S.-based personnel to relocate to the site of their experiment,
           and provide remote-conferencing for them to participate in 
           activities back at their home institutions back in the U.S. Put 
           this question the other way -- how comfortable would U.S. 
           collaborators at U.S.-based facilities feel with having their 
           overseas colleagues remain overseas, participating only remotely? 
           This would not really make for a very cohesive collaboration. Same
           principle applies in reverse. DO NOT promote remote operations as 
           a substitute for on-site presence. There is no substitute for 
           human presence on-site. 
     \item I think that being closer to the detector is fundamental and if 
           one has to choose I would in the end choose to be closer to the 
           detector than to the supervisor (provided the supervisor can 
           handled it!). As for the U.S. national labs, I really believe that
           the ones that are in being are necessary for the future of HEP,
           but I also believe that before evaluating whether open or not new
           labs the HEP community as a whole has to be involved: the "next 
           machine" needs a truly international effort and "national" labs 
           might not be the right answer! 
     \item I am definitely in favor of international collaboration on 
           experiments. I think the major element of a students' education 
           that will suffer is lack of proximity to the detector. It is also 
           not clear to me how the experiment will survive with all component 
           experts away from the detector. I think physicists should remain 
           committed to spending some amount of time away from their home 
           institution. 
     \item Hands on hardware experience with the detector or in the case of 
           accelerator physics hands on experience with the different
           components to control the beam is important if the student doesn't
           have a strong experimental background. The idea would be that 
           students receive the adequate education in experimental physics at
           their universities before going to the lab. I believed that global
           collaboration can definitively boost the progress in science. It 
           will allow to have more and reliable results with lower 
           investments. It will improve communication among scientists with 
           positive consequences in their research. 
     \item Hands on experience is important for people in the formative part 
           of their career. I rarely do hardware/ electronics/ DAQ, but I can 
           do
           it if needed. It is the fluency that is important. As a supervisor,
           I find that some students need more direction than others and 
           benefit by my proximity. Other mor independent students thrive at
           the lab. I am not sure I understand the role of a regional center 
           in lieu of a lab. It is not clear that developing the 
           infrastructure for a center at a greenfield site is a wise use of 
           resources. If a linear collider were sited somewhere in asia, I 
           would be more inclined to spend my summer at the lab than a 
           domestic center. Similarly, the center of LHC will be CERN, and 
           that would be were I would spend extended leaves. 
     \item Travel will remain important, but teleconferencing, especially if 
           it can realistically simulate a physical gathering of scientists 
           will help many aspects of the design and analysis of data. I feel
           about labs much the way I feel about detectors, redundancy is 
           important, even if it is only partial. And labs scattered around 
           the world helps to insure this depth. 
     \item If anything I think for young people traveling overseas to work is 
           an advantage. It adds important perspective. It gets more difficult
           as you get older. 
     \item Globalization has an impact on personal lifes. It is no coincidence
           that nobody in our European based group is married or has a family
           and being single is seen as a bonus by the department head. It is 
           devastating to one's life, going beyond the bimonthly problem and 
           taking it to completion: no family possible. Alternatively one 
           cannot participate in physics. 
     \item A collaboration works much more efficiently if everybody is at 
           the location of the experiment. Teleconferences etc. cannot replace
           permanent personal contact. Travel budgets are important. 
     \item Working in or near a lab when learning experimental physics make a 
           big difference ... 
     \item The question of advisor vs. detector proximity depends strongly on 
           the individual, this can't really be answered in general. Labs 
           should be kept open as long as they are doing good physics. If 
           they are not doing good physics they should close.
     \item Students should learn not to be dependent on their advisors, 
           particularly when working in large collaborations. it is also not
           essential to be continually near the detector (in fact, to 
           concentrate on independent analysis it can be better not to be).
     \item Students would have to be resident at the lab, since it is the 
           only way to learn what they need to learn. 
     \item Proximity to the detector is important to the roles that a large 
           number of physicists take on. But it is also very important to 
           have access to a large number of collaborators. In question 16 
           this is largely why I side with detector. 
     \item Hands-on hardware experience is important, but needn't be with the 
           particular detector you're working on. It's not like we all know 
           EVERY bit of our detectors inside out, is it? Or am I in trouble 
           with my supervisor ? 
     \item It's hard to predict the future. Everything depends on the 
           broadband technology and storage breakthrough. Making an "educated"
           decision now is illusory. Look at video conferencing - it's been 
           around for 20 years, and it still can not replace a day worth of 
           presence at the detector site. 
     \item Building of a detector or of an accelerator can be distributed 
           easily. Once it comes to analysis the best place to be is where 
           the detector is. 
     \item Hardware experience: As annoying as I personally find it, it really
           is a good thing re: building character and staying in touch with 
           reality. Advisor vs. detector proximity: Depends on the stage of 
           one's career in which one finds oneself. When trying to learn to 
           swim, advisors are crucial. After some measure of autonomy is 
           gained, some quality time with the detector is probably in order. 
           Use common sense.
     \item It is critical that people working on an experiment work together. 
           Teleconferencing is a pitiful substitute for collering someone in 
           the hall. The usual place for meeting is the central place -- the 
           experiment's site. If the site is overseas, our grad students will
           live there (as usual), but so will the postdocs and untenure - 
           types and their families. Why bother moving back to the States 
           after the experiment is over? 
     \item Broadband is going to help a lot, especially when we have learned 
           to use it better. However, people need to work on their detectors 
           and beams physicists also need to have hands on in many important 
           situations. This means that the foreign travel process has to be 
           adequately funded and the approval process simplified. Maybe this 
           is not a problem outside of the national labs, but it is a major 
           problem within. I don't know what the highest level multinational 
           institutional structure should look like; inevitably it will be 
           formal, remote, and less responsive than desirable. However, at
           the collaboration level existing models will probably work. 
     \item Collaborations need to improve information flow. The technology 
           exists. 
     \item People who use space satellites have been doing this for years, 
           though one group still controls the "experimental apparatus". 
     \item Computer technology allows a true international collaboration for 
           virtually any experimental facility. This should be largely 
           exploited to attract the best talent worldwide for any set of 
           experiments. 
     \item This seems rather obvious. It's possible to operate an apparatus 
           intelligently anywhere in the world, from anywhere in the world 
           with today's technology. In ten years, I can imagine we would no 
           longer actually fly somewhere just to take shifts. 
     \item We already have lots of experience in experimental HEP. Working in
           global collaborations will become easier as communication 
           technology improves. However, there is no substitute for spending 
           significant time on site to develop an understanding of the 
           hardware! 
     \item Question 15: unless you are a theorist, or you plan to continue 
           your career on Wall Street or in the Silicon Valley. Question 16: 
           it is important to be in an intellectually-stimulating environment,
           especially for grad students. Sometimes labs provide that, but 
           quite often universities have an edge. For a student, it is 
           important to be able to see "how the lab works" and have hands-on 
           experience, but it is also important to be exposed to the life 
           outside HEP, especially given uncertain career prospects. 
           Therefore, as an issue of "globalization", it is important for 
           universities to be able to be actively involved in the physics 
           analysis, and not have to go to the lab to run software. 
     \item 1) A global collaboration will only work if the members also have 
           direct in-person contact at reasonably close periodic intervals. 
           Near future technology is incapable of replacing the vital role 
           of personal contact. Communication will deteriorate, and 
           regionalized banding-together will become a major problem if there
           is no central gathering. 
     \item I think globalization of experiments and of machines is of vital 
           importance. While it is still indispensable that at some time in 
           the career one has hands on experience, this is not vital during
           the long time an experiment runs. There working on one continent 
           while the experiment is on another is certainly possible. In fact
           many of us are already doing this quite successfully. 
     \item It will be very tough, considering that the less developed 
           countries will have less sophisticated video facilities, less 
           travel money. Basically, if one wants to be a productive member of 
           a collaboration, then one should be stationed in the first world 
           (that includes Japan). 
     \item At a fundamental level, people need to feel involved and needed. 
           Large collaborations have succeeded so far b/c there has been 
           strong effort to spread them around the world, so most institutions
           are "near" something. Additionally, the construction/maintenance
           tasks have been specifically chosen so that small university groups
           can specifically and markedly contribute. 
     \item Question 16 is a serious problem. Having young people live at the 
           experiment running "open loop" is bad for everyone. On the other 
           hand, doing an experiment remotely is just not feasible; too much 
           of the communication is lost. The supervisor needs to spend more
           time at the experiment and/or a mentor system needs to be 
           established to help the young people stay near the real axis.
           This tends to undermine the supervisor, but that is just life. 
           Question 17: Regional centers as described above really require
           just good network access. This does not necessarily require big
           hardware at all labs. On the other hand, it is not clear from the
           question how large the investment needs to be to have a RC. For 
           example, if the cost is a few \$M, then both LBL and SLAC should 
           have it; if the cost is \$100M, then only one needs to have it. 
           Question 19 is weird. What is the relation between the 
           "importance" of labs vs. reducing the number? We might have only 
           1 lab, but it would be very important. BNL is moving towards 
           nuclear physics, and its HEP is winding down. Where does it fit 
           in this question? 
     \item We need to do as much as possible to keep down the "barrier of 
           entry" into HEP, particularly HEP-experiments. If it is possible to
           do productive work near your home institution we would attract more
           physicists and perhaps more money. I wonder if we're approaching 
           another "split" in HEP: between detector builders and data 
           analyzers. (How do we balance this with the value of hardware 
           experience? I don't know). Or perhaps we should be moving towards 
           making data publicly available. I mean TRULY public, rather than 
           what D-Zero is doing with Quaero -which is a good start. Maybe 
           people can get early access to the data by doing service-work. A 
           collaboration would then publish its data rather than its analyses.
           This would allow real specialization in the process of detector 
           building which could help keep costs and manpower requirements down
           (imagine having a "standard" silicon detector or calorimeter). The 
           danger to this approach, of course, is that the detector builders 
           might loose respect and job-prospects. It might not be seen as 
           "exciting" and interesting. I'm thinking of the example of 
           accelerator physics here. How do we avoid this? I don't know.
     \item It's all about networking and data distribution. As it is a lot of
           people on large collider experiments only know the detector 
           hardware as a picture. For future large experiments access to the 
           data and ability to communicate with remote collaborators is 
           paramount. 
     \item We will always need physical presence - after all we're humans, so 
           are your co-workers, and communication is improved by physical 
           presence. But meetings 3-4x a year are good enough. For the rest 
           of the time a distributed data (most important) and R\&D and 
           control scheme should be sufficient.

\subsubsection{Regional Centers:}
     \item Regional centers may be a good idea even if the next machine is in 
           the U.S. They also afford the opportunity to be located in a place 
           with a better quality of life than FNAL, Sudbury, wherever MINOS 
           is, etc. 
     \item Global collaboration not only on the experiments, but also in 
           construction and running the machine is essential. Regional 
           centers with expertise and educational possibilities for 
           accelerator should be strengthened and have the possibility to 
           participate as full members. 
     \item For collaborating groups with hardware responsibility a regional 
           center can never replace the on-site expert. It can only serve to 
           reduce travel efforts for shift takers. So it should be truly 
           regional, since it almost makes no difference whether you fly 
           transatlantic or coast-to-coast. It should be within a day's drive 
           from any collaborating institution.
     \item Regional centers may save money, but they will NOT have a positive 
           impact on detector/accelerator operations. 
     \item Time lag between regional centers and experiment will be too long.
     \item HEP is a international field as such we must be prepared to work in
           a distributed fashion at local universities and be prepared to 
           travel as required to do the research. Most recruitment takes place
           at universities and regional centers will only weaken them. 
     \item It is certainly useful to have regional centers with ultra-fast 
           communication with experiments; FNAL and Brookhaven will do that 
           for LHC. I don't think a remote control room would help. 
     \item I think all data grids, regional centers, and any other such ideas 
           are wonderful ideas to broaden participation and increase the value
           of the physics. 
     \item National labs are a natural and compelling place to host regional 
           centers, especially wrt to U.S. congressional funding. They have 
           immense infrastructure, local support and existing programs. 
     \item The regional centers may be developed at the existing national 
           labs. 
     \item Hardware experience can be gained on R\&D and construction 
           projects that can be run at Universities. Definition of "regional 
           centers" as remote control rooms is unfortunate. I don't think the 
           remote control rooms are important. Regional centers for data 
           analyses would be useful. 
     \item As the quality of video-conferences or other telepresence 
           facilities improves, they should increasingly be used. There is no
           reason why some of the shift work couldn't be done remotely. 
           However, face-to-face contact and hands-on experience will not
           become less important in the near future, even if they may become 
           somewhat more rare. 
     \item I don't know if these types of "regional centers" can take the 
           place of "taking shift" at the experiment. With the improvements in
           computers and networking, the possibilities of monitoring detector 
           performance can be done from home on your laptop. I don't see how 
           this could replace "taking shift" at the sight. Traveling to take 
           shift is nearly equally onerous whether the experiment is near by 
           or half a world away. 
     \item Regarding question 17 ... I am not sure that the quality of the 
           physics results will improve, since this is dependent on other 
           factors, like accelerator performance, detector performance, and 
           the ability to adequately record interesting physics events. These 
           facilities might not be able to impact these factors, but they 
           would certainly be important for remote users to understand the 
           experiment they are working on at a more fundamental level, which 
           will allow such users to better understand the quality physics 
           results derived from this next machine. 
     \item In order to keep the field exciting, we need to keep pushing the 
           frontier. This means bigger facilities, not more facilities. If it 
           is necessary to close some labs in order to make the largest 
           facilities even bigger than we ought to. On the other hand, if the 
           others can contribute in to the main one, then we should support 
           them. The notion of shifting on a detector when all of the real 
           hardware is at another location is preposterous. You would spend 
           far too much time and money getting the remote system working and 
           keeping it maintained. Also, if the remote center will have the 
           ability to actually do anything, think of the security breaches 
           possible. On a collider experiment like D0 or CDF, it may not be 
           possible to check out the detector itself when the beam is on, but 
           having a shifter who can't even look at the hardware just seems 
           like a poor idea. Shifters should also be able to take advantage of
           beam problems by going in which would be impossible on a remote 
           shift. 
     \item I really don't know how this is going to work. I believe we have to
           move to the model of regional centers, but it is a brave new world 
           and I can't predict how well this will actually work. I do think 
           that while contact with the detector is important, the most 
           important thing is contact with your fellow physicists. It is the 
           successful interaction among people that determines our progress
           more than anything else. In my mind this mandates "centers" of 
           some sort. Right now, those are the big labs, but in the future the
           role could be replaced by regional centers provided they develop
           enough critical mass. Inter-center relationships are likely to be 
           complicated and may develop (unfortunate) political tones. 
           Closeness to the detector and the hardware is important for the
           people with strong hardware interests, but the notion that we all
           need to be close to the detector is based on old ideas that date 
           back to the days when experiments were smaller, cruder, and more 
           tabletop. As for the role of U.S. labs in the next 10-25 years, 
           I'd like very much to see one of them (FNAL!) host the Linear 
           Collider. The others would make great regional centers, plus have
           strong programs of "smaller" scale projects, 
           astro-particle-physics, etc. 
     \item The communication technologies that would be needed for such a 
           thing will not be mature enough to trust in quite a long time. 
           The problem is not the ability to build prototypes, which I'm sure
           we can do --- it will be problems with the very large scale that
           we won't discover until we are already deep into the project. Such 
           experiments would be far more meaningful for computer science and 
           network engineering than for physics (which isn't a bad thing, it 
           just needs to be kept in mind). Besides, when the dust settles and 
           the technology *is* working well, we'll probably find the band of 
           access too narrow for our needs: tinkering is only possible when 
           one is immersed in an environment. 
     \item Clearly defined responsibilities assigned to each participating 
           institution. Clearly laid out metrics for performance. A chain of 
           command with some teeth in it. Without it, regional centers will 
           merely be the same hotbeds of confusion that they are now. Also, 
           individuals should not be allowed to split their time among more
           than one large collaboration ($>$500) and one small one ($<$100). 
           Does anybody seriously believe that people who are members of more 
           than one large collaboration can seriously contribute more than 
           name power to either ? 
     \item Regional centers are VERY good for some things (data crunching) but
           remote control rooms can be bad as young people could end in a 
           "virtual reality physics". Physics is (still) flesh and blood. 
     \item My experience on Auger is that remote stations are very important. 
           It is very expensive to travel to the southern site (Argentina), 
           and it is difficult to spend more than 2 weeks down there due to 
           other commitments at Fermilab. 
     \item I believe that it is extremely important, especially early in ones 
           career, to be present at the experiment/lab. This is in part so 
           that the full experience of working with the apparatus occurs, and 
           in part to broaden knowledge of the field. In this sense, regional 
           centers would be a disaster, since they would remove the incentive 
           to send students abroad and promote provincialism in U.S. high 
           energy physics. 
     \item I'm working in Canada, and I'm not sure what exactly is meant by 
           "regional centers". My answers to question 18 and 19 refer to 
           distribution of particle physics centres in Canada. 
     \item Regional centers would help, if for no other reason than it gets 
           collaborators working in the same place. However, they'd only work 
           if there was already a lot of collaborators who are already working
           at the site of the regional center, and the travel costs to get 
           there were noticeably less than to get to the actual detector site. 
     \item Difficult as an outsider to comment on specifically U.S. problems. 
           The Global Accelerator Network sounds great, detail needs to be 
           investigated to see if it really works. A pilot project would be 
           a good idea before committing ourselves to this for evermore! 
     \item The proposed Global Accelerator Network seems like a good start.
     \item I don't think Global Accelerator Network will work that well.
     \item It will hurt to have people remotely looking at problems. 
           Ownership of the detector/lab will only accrue to the resident team.
           The GAN is a nice sounding idea that will not work.   
     \item Re: regional centers - How to monitor a detector and approve 
           physics results in a 1000-member global collaboration (without 
           using up the whole week in meetings/travel) are questions that 
           need to be addressed. Regional centers will need very good 
           connections to the detector site to do any of the monitoring, DAQ
           etc. tasks listed. Even in the U.S., connections from a university 
           are nowhere near good enough to accomplish these tasks remotely. A
           demonstration that this is feasible and reliable is needed to make
           any informed decision on this. Will funding for travel to/from 
           these regional centers be justifiable? Traveling from a home 
           university to a regional center just to use video-conferencing 
           seems to miss the whole point of video-conferencing. Why not 
           improve links to existing universities and organize/streamline 
           meetings better? To improve the quality of physics results and of  
           the peer review process, especially if attending meetings by video,
           more documentation on analyses and their progress would be very
           helpful: slides on web before meeting, descriptive note distributed
           before meetings, analysis code in a public area specific questions 
           sent to speaker before meeting. Re: National labs: Current national
           labs are funded; it's clear that the funding for building new ones 
           is not there, eg SSC in Texas, NIF over-budget. CERN/FNAL are good 
           examples of what you can do by building on what's already there, 
           rather than starting from scratch, which usually costs a lot more
           in terms of infrastructure and people. Closing some labs? ANL, 
           LBNL, LLNL, BNL, FNAL,... do diverse research. Could this really
           be done at any one university? A list of what we would lose in 
           terms of resources and research projects is necessary to make any
           informed decision on this. 
     \item We should have regional 
           centers which already correspond to national labs which do HEP: 
           BNL, FNAL, SLAC mostly. Other nations may also be included in 
           having regional centers (Japan, Brazil, Canada, India, Russia, 
           China, Australia, S. Africa). 
     \item Work will always be centered physically where detectors and
           accelerators reside. Virtual control rooms (aka. regional centers) 
           are ridiculous. 
     \item There is a role for distributed analysis, but I do not see the 
           need for regional centers to monitor data taking. This does not
           mean that remote access by experts as needed is bad, just that it 
           can be accomplished without the investment in regional centers. As
           far as analysis is concerned, it is distributed now, and it will
           continue to be distributed, whether we like it or not. Again, I do 
           not see the need for regional centers to facilitate this aspect of
           accelerator-based experiments 
     \item About "centers"; quality not quantity. 
     \item I think this is very important, but I would prefer to leave details
           to those experienced in this area. Difficult issues involve money, 
           credit, and politics. 
     \item The existence of regional centers will be a benefit, but it's not 
           clear to me what the cost/benefit ratio will be compared with other
           ways the money might be spent. 
     \item I feel that a hands-on feel for the actual detector is very 
           important in getting a feel for the data being collected and 
           helping in wringing the most out of it. I feel that "regional 
           centers" will help to keep communication between those, with 
           valuable experience, that can't relocate to the lab and those that 
           are local to the lab. I feel that the current U.S. labs can play a 
           large role in this and that they must play an integral role in any
           future machines to be built in the US. to the lab 
     \item A single collaboration could have many (dozens perhaps) centers 
           around the U.S. These would be at universities and labs where many 
           collaborators reside, or where many could easily commute to. 
     \item Focused at a national lab with a few regional centers for Monte 
           Carlo work. Grid connections. 
     \item I think it is impossible to run an accelerator, or a detector, 
           remotely. If a power source dies, someone has to be on site, so 
           that the voltage gets reset and operation resumes. Since there will
           necessarily be staff and scientists on site, I think that regional
           centers (other than the labs) will create "unneccesary" redundancy,
           and result in less efficient data taking and therefore lower 
           quality science. The efficiency of data taking becomes 
           exponentially more important, as the new facilities proposed
           require enormous amounts of power per year- on the order of 
           150 MW/yr. At a cost of 0.6M\$/MW, (from FNAL power consumption 
           and operating budget today), inefficient data taking is an 
           insupportable strain on the HEP budget. 
     \item Regional centers will degrade the quality of beams and analyses, 
           but MIGHT improve the ability of remote physicists to participate.
           I don't think the idea will work very well. 
     \item Global collaboration would have to have region centres but it 
           would still be very important to allow for visits to the real site.
           National labs must continue to supply electronics/detectors 
           development expertise and support smaller experiments. 
     \item I very much encourage the cross-participation in different 
           international projects through the so-called regional centers. This
           should apply to both European participation in U.S. projects and 
           U.S. participation in European projects, as well as to potential
           projects in other countries and to the sharing of responsibilities 
           within the U.S. or Europe. 
     \item Running the experiment should be a local affair done by 
           professional staff. Data-quality monitoring, testing and debugging
           should be possible externally with good networking facilities. 
     \item As I see it, there seems to be a somewhat natural division in terms
           of the functionality envisioned for "regional centers" -- first,
           they would provide control and monitoring systems for remote 
           detectors/accelerators; second, they would provide regional access
           to full (or perhaps partial) datasets from the experiments they are
           controlling/monitoring, along with concomitant computing resources
           needed to adequately exploit the datasets 
     \item I regard the integration of human resources in smaller countries of
           the utmost importance. Countries that do not have the economical 
           resources to contribute money but do have a lot of young physicists
           in need of guidance. This will bring new ideas to the field and can
           easily be achieved using video conferencing technologies and such.
           "Regional Centers" should be cloned also in these places. 
     \item Virtual control rooms are a poor substitute for "hands on" detector
           work. They will help run an experiment, no doubt, but should not be
           used as an excuse to prevent students from traveling to the actual
           experiment. 
     \item We've learned a lot about global collaborations from current 
           large-scale experiments. We should use what we learned there and 
           see where the problems are to be able to ensure some truly global 
           team-work. Regional centers with lots of resources sound like a 
           great step in the right direction. 
     \item It is crucial for the students to be at the experiment. I 
           personally to not believe in regional centers. 
     \item Regional centers are a good idea but the host inatitute will 
           always carry the most weight since where the hardware is that's the 
           place where the action will be. 
     \item It could become chaotic or just be an organizational nightmare if 
           there was more than one of these regional centers. 
     \item It already works in some cases for detectors. It can also work for
           accelerator studies, but with local hardware experts on site. 
     \item BTW, the issues here apply to LHC operations, as well as to future
           facilities. 17) There are other reasons for "regional centers", 
           including greater ease of participation by both senior experienced
           physicists, and young physicists who would benefit greatly from the
           experience. 19) I'd like my answer to be "much the same role" AND 
           "about the current mix of lab sizes and types", BUT "the role of 
           some labs may have to change". 
     \item Regional centers should replace long trips e.g. to Europe due to 
           shifts; a role of regional centers easing analysis is also 
           desirable; personal discussions cannot be replaced by video 
           conferencing; regional centers should work as well for HEP as they
           are for international companies! 
     \item The next machine is already being built in Europe with large U.S.
           participation. Regional centers are probably a good idea but need
           to be tested. I would advocate regional centers plus ability to 
           work from everywhere. Technology will allow it. 
     \item Local centers are probably fine, for data analysis and processing 
           etc. Maybe even for shifts. Nevertheless a shift needs interaction 
           with the hardware and local experts to fix things, so you won't get
           around to have a center among centers. 
     \item As a member of an experiment taking place outside of the United 
           States, I encourage all young members of a collaboration to spend
           as much time at the experiment as possible. Regional Centers are 
           not a good idea. People learn by interacting - and I do not mean 
           by video conferences. It is crucial for people to work at the 
           experimental site as much as possible. 
     \item We have to realize that s/w is a bigger and bigger issue for
           collaborations. This means that efforts in that field should count
           with a significant weight with respect to h/w activities on the
           detector. Being a European working at SLAC I could also see the 
           many advantages of having more remote control options available.
           It does however need to be built in to the experiments right from 
           the beginning.
     \item Regional centres are very important, not only in U.S. but also at
           other places of the world. All high energy physics experiments are
           facing a manpower shortage. Situation can be improved if work is 
           distributed properly all over the world. 
     \item This is the first I've heard mention of the idea of "regional 
           centers," and I'm skeptical. Access to the hardware and 
           face-to-face communication are very important. Access to the 
           detector is probably slightly more important than proximity to the
           advisor for a graduate student, but I believe the difference is 
           small. 
     \item I think the main advantage of "regional centers" is the effect upon
           people's lifestyles and the balance between finding a job and
           needing to commute. I would thus place them near large cities that 
           do not currently host national labs. 
     \item I'm not sure the remote control room itself is a good thing. 
           However, a regional center which brings together scientists from 
           different collaborating institutions is very valuable. This would 
           be a more important aspect of regional centers, in my opinion. You
           might need a virtual control room to give the place focus, I guess. 
     \item Regional centers strike me as an extremely stupid idea. 
     \item In case regional centers take over from the national labs on 
           certain physics programs, the later may be restructured to pursue 
           state-of-the art technologies and possibly house test grounds for
           future experiments. 
     \item Sort of like how an unmanned international spaceprobe is operated. 
     \item I am worried about the decline of hardware/software work at the 
           universities because it reduces the opportunities to excite 
           undergrads about HEP and reduces opportunities for outreach to the 
           public. Also funding grad students/postdocs as sub-contractors to
           the labs on lab-funneled money is worrying because of lack of 
           stability. We should think carefully how to maintain the 
           strength/presence of HEP at the universities while implementing the
           regional centers idea. 
     \item I find it hard to believe that an accelerator or a complete 
           LHC-scale detector can be run remotely. The goal should be to 
           maintain operation of smaller detectors or sub-detectors from 
           remote places. An extremely good computer network is crucial. 
     \item A global collaboration will work best if the focus of collaborative
           technologies is on enfranchising collaborators as well as possible
           at their home institutes, rather than focusing on regional centers. 
     \item I'm not 100$\%$ sure what you envisage here: do you mean the entire 
           experiment run from a single virtual control room, or a dispersed 
           shift crew spread between several such plus the actual control room
           near the experiment? In any case, I'm a sceptic. Shift crews (after
           initial phase) are small, so dispersing seems unhelpful (I'd expect
           average occupancy of one of these remote control rooms would be 
           zero or one). Hence I assume you mean to run the entire experiment 
           from the remote site? Might work, but network links are not 
           infallible, and I cannot see any collaboration accepting that the
           control room could lose contact with the experiment (which, if it 
           is remote, it *will* do at some point). Given that there will 
           probably be some need for physical intervention (at the level of
           removing faulty DAQ modules) and that initial data recording will
           be local (you will not ship the RAW DAQ stream to a site on another
           continent to record it), I just don't see this being very helpful.
           DAQ teams put a lot of stock in reliability, and this proposal 
           could only reduce it. Remote advice, yes, but not remote control. 
           If shift crews were large, with many experts around to monitor (as 
           opposed to control), I could imagine these people being dispersed, 
           but my experience (from PETRA, LEP and PEP-II experiments) is that
           this is not how things will be after the first few months (and in 
           the first months all experts will be on-site with the detector in
           any event). I *do* see a role for regional analysis centres, but  
           not for remote control rooms. Oh, and while I answered "don't have
           a detector" to Q14, that is because my role in BaBar is offline 
           software and ATLAS does not exist yet. I was a DAQ expert on OPAL,
           and am a first-level trigger designer for ATLAS - hence my normal 
           role is in trigger/DAQ (but not with my current live experiment).
           I should perhaps remind you at this point that I'm European, so 
           don't have strong views on the U.S. national labs (personally I'd
           prefer them to stay open, but I don't think the Virtual Control 
           Room is a strong argument if that is the objective). 
     \item Regional centers could make gigantic collaborations tolerable when 
           it come to data taking and analysis. 
     \item More than control rooms, you need places with a critical 
           intellectual mass to understand and advance the physics of the
           experiment, along with data taking. 
     \item Coming together and working together in the main places (e.g. next
           machine) is much more important than outside control centers. 
     \item Comments: 1) experts need to know the hardware precisely (i.e. 
           expertise) so he needs to work with it on place, therefore I
           disfavor regional centers 2) additional national labs (whether in 
           U.S. or Europe or wherever) are only cost intense and do not 
           contribute to new physics more cost-effective is one physics place
           and multipurpose/redundant detectors.
     \item I don't think "regional centres" are important - I don't think we 
           could run the DAQ / control room by video conference! But we will
           need many sites carrying out data analysis/ monte carlo generation,
           much like the SAM model at D0 or the grid for LHC.
     \item The point about "regional centers" is equally valid for 
           collaborators outside the U.S. - if you aren't in the country the
           "next machine" is in, it can become very difficult to do good 
           physics work. The quality of beams and physics at the detector might
           not improve with the existence of reasonably easily accessible 
           "regional centers", but the quality of work done relating to 
           physics done at the detector could be greatly harmed by the 
           non-presence of those trying to be active collaborators. 
     \item National centers are great, but most important would be to have 
           enough funds to ensure frequent travel to experiments. 
     \item My last answer is what I think should happen; not what I believe
           will happen. As for regional centers, one needs to be careful 
           about how such things would be coordinated. While there could be as
           many monitors as desired in principle, in fact there should be only
           one active control room where such things as voltages and detector
           parameters are set. One could envision control rooms in different 
           hemispheres handing off control of the experiment at shift change 
           so that everybody works the day shift (although many prefer 
           evenings). But it can be difficult enough to track what happens 
           shift-to-shift in a single control room where only one group at a 
           time is in charge. Multiple, simultaneously operating control rooms
           is a recipe for confusion, no matter how good video conferencing 
           eventually gets (and it has a long way to go). Such centers might
           provide advice or input to an operational control room at the 
           experiment, but their number should be limited. They would 
           undoubtedly cut down travel expenses in a global collaboration, but
           sometimes there is just no substitute for being there. 
     \item Quarterly face-to-face collaboration meetings; 2-3 regional
           data-taking monitoring centers in the U.S., others in other 
           countries
           at same per-physicist density as in U.S.; hands-on work on hardware 
           for all young experimenters at home and at the detector during 
           early part of career, with periodic returns for detector upgrades; 
           expanded role for video conferencing as this technology matures. 
     \item The regional centers should be at universities as well as at 
           national laboratories students at all levels must be involved 
           research must be recognized by university administrations as an 
           intrinsic part of responsibilities of faculty members 
     \item Virtual control rooms might be sited at Universities. 
     \item Instead of "regional centers" better facilitate international 
           meetings, support abroad trips for the scientific staff etc. 
           Collaborations should not be split in national subsets. 
     \item A new "green-field" site looks awful hard to create in the current
           funding climate. If we pursue this, we have to understand why we
           are creating this new infrastructure and be able to defend it 
           scientifically. 
     \item Although I currently don't have a detector, I did work on D0 at 
           Fermilab as a grad student. In my case, I didn't see the detector 
           for several years after joining the experiment because it was
           commissioned and well into running. Later, I did become involved in
           the upgrade detector R\&D and construction, and given the choice, I 
           was closer to the detector than my advisor. I believe that while 
           regional centers could prove to be useful in the more globalized 
           situation we seem to be entering, there will still need to be a 
           large number of people actually present at the host lab. I don't
           think regional centers can replace the personnel needed at the host
           lab (at least not yet and certainly not during commissioning). I
           also believe there is great value in being physically at the lab
           for some time where the experiment/accelerator is running. I do 
           think the regional centers can minimize the time spent at the lab, 
           and would facilitate bringing new students onto an experiment. I
           also think they would enable people to avoid uprooting an entire 
           family to work on an experiment across the globe. Finally, these
           centers would hopefully enable more people to remain interested
           and active in the fields of high energy and accelerator physics.
           In order to realize these benefits, there is a great deal more 
           work to be done in order to make these centers feasible.! 
     \item We need to concentrate on a few large labs. Wherever the
           next collider will be, people there are able to run it, i.e. U.S. 
           people are not essential. On the other hand, we need 'regional 
           centers' for those involved in machine/experiments (hopefully 
           incl. people around the world) to reduce necessarity of traveling.
           It will/should be a global project. 
     \item The quality of beams and physics may or may not be improved;
           however, the accessibility of the physics throughout the U.S. is
           important to physicists, the larger physics community, and the 
           general public - especially current K-12 students who embody future
           particle physicists. 
     \item I don't think the quality of beams would improve but the quality of
           physics results would, so I have separate answers for those two 
           things. I put yes anyways. Also, I don't see an option stating 
           that the national labs should act as regional centers and then some
           other centers should be created as well in other geographical 
           areas. I would choose this option. 
     \item Regional Centers are important. If the proliferate, they could 
           weaken the role of the national labs. This would not be good. On
           the other hand, the university roles should also not be weakened 
           and are already under stress in the current environment. 
     \item Identify a number of centers (3-5) identify a core of people at 
           each center (15-20) daily involvement in operation and rotating 
           shifts among the centers.
     \item The concept of "regional centers" is flawed. As a faculty member, 
           I would be more than willing to send my students off to work at 
           the lab where our experiment is operating -- but if that's not 
           feasible, the next best thing is to have the students at my 
           university. I don't see that the benefits of regional centers
           comes close to compensating for the diminution of the universities' 
           roles. 
     \item I worry about regional centers for experimental control. Will the
           host lab really allow someone sitting across the ocean to change 
           the trigger for example? Many times, there is no substitute for 
           being at your detector. 
     \item I am a strong supporter of regional centers. The large detectors 
           may be supported by professional staff. Hardware experience is 
           critical and can be gained in R\&D, test beam work, etc so we can
           develop and build new detectors when we need to. We need some kind 
           of national lab structure. Your response choices were limited. I 
           do not really know whether all existing labs should continue but 
           I hope that they will be given a chance to evolve to follow the 
           community's needs.
     \item I doubt if the beams will get any better with remote centers -- 
           you really need someone standing beside the operators and attending
           their meetings to get some action on beams. More or less the same 
           with physics results -- you need someone to go and power-cycle the
           power supply or whatever needs to be done to get the detector going
           again, and that person has to be present at the detector. An 
           experienced team of shift-takers usually outperforms people who 
           come in to do one or two shifts/year in detector reliability. But
           it is important for the physicists to be familiar with their 
           apparatus (and to know when it failed and how, so they can do 
           their work accordingly). CERN collaborations are already global.
           We use the web heavily for documentation, have phone conferences 
           (video conferences are awkward but we are starting to use them),
           but mostly fly people out for collaboration meetings and shifts
           (and even sometimes for very difficult detector problems). E-mail
           is a wonderful tool.

\subsubsection{The Role of National Labs:}
     \item National labs should become "regional centers" also. 
     \item Build new colliders at existing labs - labs without colliders 
           (and without plans to build colliders) should be reduced to 
           "institute" level.
     \item What I really mean is that labs should continue to play a strong 
           role, but that in order to ensure a strong physics program we 
           should not take it as a given that all labs should remain. The 
           emphasis should be on ensuring a strong HEP physics program, not 
           ensuring the survival of labs IF (and only if) they have outlived 
           their usefulness. At the moment, no lab falls into this category. 
     \item Some of the national lab HEP groups are really just massive 
           University groups (e.g. LLNL, ANL, LBL). Other labs should be 
           encouraged more in this direction as the direct accelerator 
           operations roles are eliminated. 
     \item The role of national lab will depend on what they can contribute to
           the discovery of new physics. If there's nothing they can do on 
           this they will have no importance. 
     \item Having a domestic program for smaller experiments is extremely 
           important, and the labs must continue to play a leading role in 
           this. Right now flavor physics is more exciting than its been for 
           decades, and an active flavor physics program (particularly 
           neutrinos) is very desirable in addition to a high energy collider. 
     \item I would like to clarify answers to 18 and 19. For 18, there should 
           be at least one center per experiment. For 19, I believe that 
           national labs will still play a very strong role in the future of 
           HEP on a global scale, but this may require a reduction in the 
           number of these labs being operated. 
     \item The issue of the future of the U.S. labs is a thorny one. At the 
           present level of funding and with the current number of labs, it is
           difficult to nurture development groups of excellence in critical 
           areas (most HEP software comes from CERN, for instance). Fewer labs
           (with the same budget, of course) would allow this. On the other 
           hand, a single mega-lab (like CERN) has its own dangers. The 
           diversity that comes from different labs with different emphasis 
           and different directorates would be lost. For instance, would SLAC 
           objectively consider hadron colliders or Fermilab e+/e- colliders? 
           I am at a loss to make a firm choice between the two directions. 
     \item I think the labs should have an important role, perhaps more 
           important, that doesn't mean there should be more of them. 
     \item Ludicrous to even consider closing any national lab that has a 
           strong physics program - as all do, when you take the time to 
           understand their program and the physics that they do. 
     \item Some national labs should be closed or develop in very different 
           directions. Other labs should continue at the same or greater 
           level.
     \item Close a national lab and/or convert it to an international lab 
           ala CERN. 
     \item Closing some labs is probably the only responsible way to 
           strengthen others. 
     \item Funding is going to be a bigger and bigger issue as we try to get
           more money ... consolidation of labs could potentially save 
           millions a year. Especially since I see us having fewer machines 
           in the future than we do now (one e+e- LC as opposed to 2 high
           energy e+e- machines today) due to funding constraints, it seems
           silly to keep open more labs than we need to to support the  
           machines. 
     \item It would be nice to have less competition *within* labs for new 
           ideas, R\&D, resources, experimental proposals. 
     \item The importance of the U.S. national labs should be driven by the 
           status of HEP research. I don't understand question 19. If there 
           is motivation to do a new experiment and open a new lab for that, 
           let's do it. If there is no motivation, what are we talking about? 
     \item Question 19: I do not understand, though, what closure of a lab 
           means. Argonne and LBL are closed for direct HEP experiments, but 
           they are the host of synchrotron facilities and have strong user 
           groups. LBL also has a strong accelerator R\&D group that is 
           synergistic with the synchrotron radiation lab. In the future, 
           Brookhaven and SLAC might also `close' in this way, but it would 
           be unrealistic and counterproductive to have them drop out of HEP
           and return their full budgets to a common pool. 
     \item Not all labs have to have centers; some could be at universitities.
           But too many dilutes strength. The role should be similar. It is 
           too early to say ``some should be closed''. That is anyway a 
           case-by-case thing for politicians. 
     \item Labs are not only to be seen with respect to the one large major
           machine in the world. There should be a variety of further, 
           somewhat smaller experiments supported by local labs. 
     \item Existing labs should provide support centres for universities, to 
           ease access to machine site - leading machines should be 
           distributed regionally including U.S.            
     \item Labs should be much more closely connected to universities, i.e. 
           students should be encouraged to go to the labs etc. 
     \item I think consolidation of high energy physics into a smaller number
           of labs would be beneficial if the freed resources could be made 
           available for important future projects (such as NLC). 
     \item The labs should be staffed by researchers; how can afford to do 
           research on a long time scale. The University model works well on
           a much smaller time scale, and industry is of not help. 
     \item I hope, you mean "national labs doing HEP". We do not need such 
           centers at LANL, ORNL, NREL, etc. Additionally, LBNL and SLAC 
           probably need only one such center. 
     \item In the future only two labs will continue with high energy physics
           at colliders - CERN with LHC and another with the linear collider. 
     \item National labs are important for large projects to be carried on; 
           it is important to keep the funding of University Research at a 
     \item Some labs should close to allow new ones to start up higher level 
           than now. 
     \item U.S. national labs have provided greater opportunity for 
           international interaction with both experimental and theoretical 
           physicists. It has helped me a lot, and I hope, the U.S. physicists
           too. 
     \item To suggest that some labs should close is not the same as 
           declaring that the labs are crucial. It is quite possible that some
           labs have completed their mission, however important they have been
           or will be in the near future. 
     \item "National labs" doesn't necessarily mean HEP or even physics labs; 
           some shifting, depending on the market, will occur. It should be a 
           combination of what the public finds important and what foresighted
           government knows is best for the nation, for the world, and for 
           human-kind. 
     \item If the budget for the new facility requires that some labs should 
           close that's ok, but we should be careful not to kill all smaller 
           projects that may exist at these labs, like fixed target 
           experiments.
     \item Labs without active accelerators should probably be closed. We 
           cannot afford the luxury of maintaining inactive labs in the 
           present budgetary climate.  
     \item Budget constraints might force the closing of one of the HEP labs,
           but the U.S. program would be much better off if we can keep the 
           current labs. With respect to whether it is more important to be
           near the detector or near the advisor...it depends on what stage 
           of career a student is in. During data-taking it is important to 
           be near the detector. During anaysis, it is more important to be 
           near the advisor. 
     \item I believe that the role of national labs is very important ... BUT 
           ... there may be a need for rationalization. 
     \item If national labs perish and the next machine is outside the U.S., 
           we will need an enormous travel budget to get necessary hands-on 
           experience and take part in an experiment. 
     \item I believe that the national labs will remain useful as staging 
           centers for the future accelerators, but that their role should be
           carefully examined. The dominance of DOE support for the 
           laboratories needs to be better balanced against the University
           program. Much of the leadership and innovation comes from the 
           universities. 
     \item If the model is regional centers in the U.S., then it seems to me 
           that the answer to 19 should be "different role/ same number". The 
           role is no longer to provide beam to the experiments, but rather to
           support them in other ways. "Regional" implies at least several,
           and east/ west/ central is an obvious mix. 
     \item Need to allow a greater variety of work to proceed and be 
           undertaken at the national labs, which are a natural focus for
           multi-institutional collaboration. 
     \item The number of U.S. national labs has already been reduced and
           should not be reduced further. It allows some diversity which
           in research is an important factor. 
     \item Cannot support too many labs. Distribution of resources can reduce
           the number of experiments due to overhead. These labs need to be
           globally distributed. 
     \item Question 19: It depends on the roles of the labs. There are no 
           entitlements, but the field should be sufficiently diverse to have
           at least 2 major labs (and a few smaller labs) doing productive 
           work.

\subsubsection{Survey-Specific Comments on Globalization:}
     \item Again you have poorly worded questions. You assume one detector. 
           You assume a supervisor. 
     \item Geez, do you really think the U.S. are the center of the particle 
           physics world? 
     \item Don't see why U.S.-related topics should dominate. Is U.S. the 
           motherland of HEP or the only provider of HEP-enthusiasm? 
     \item Well, globalization seems to be bit self-centered (on the U.S., or 
           have I gotton something wrong?) 
     \item Why is there so much focus on the U.S.? 
     \item Some questions are U.S. specific.
     \item This seems to be a little U.S. bias/emphasis here, given this is 
           intended for the whole HEP community. 
     \item A little irrelevant unless you're American I feel. 
     \item No opinion about the U.S., I am not in the U.S.
     \item This part of the questionary was dedicated to U.S. people and I am 
           from Europe...(I answered in replacing U.S. by Europe). Global 
           collaboration will be working more and more via video conferencing.
           As far as management is concerned, we should make sure that 
           democratic aspects are respected and that scientific aspects always
           preveal any decision. 
     \item I must admit that I do not know exactly how many national labs 
           you have and how they are distributed, so I cannot answer the last 
           2 questions. I think anyway that having places where a 
           "critical mass" of physicists work together is surely good, even if
           the detector is somewhere else. Nowadays communications around the
           world are very good, but nothing can substitute an old-fashioned 
           discussion face to face for the small groups and especially 
           students.
     \item I can't comment on issues relating to location of regional 
           centers in the U.S., since I am not a U.S. citizen nor do I live 
           or work in the U.S. 
     \item I'm not an U.S. citizen. America is not everything (just a remark). 
     \item Sorry, I'm not familiar with the status of U.S. national labs. 
     \item Why do 17-19 focus on the U.S.? 
     \item Questions 17 and 19 are not U.S. related, they are general 
           questions. Particle physics is an international effort, the U.S. is
           only one of the partners !!!!!!!! 
     \item I don't understand why you asking me so many questions about U.S. 
           It's not the only country which makes Physics in the world and 
           not event the most important for particle Physics. 
     \item Questions 17-19 should be decided by the U.S. physicists 
           themselves. 
     \item ... don't want to comment on U.S. affairs. 
     \item Being working outside U.S., my answers can be ignored. They "mimic" 
           answers reflecting my viewpoint 
     \item Once again, you are taking the very biased (and incorrect) view
           point that the U.S. somehow dominates particle physics. The 
           question you should be asking is "How can the U.S. contribute in a 
           meaningful way to the future of particle physics?". Regional 
           centres are a means by which U.S. physics can CONTRIBUTE to 
           foreign experiments and are therefore important to U.S. budgets, 
           but to no one else. 
     \item I have difficulty answering the questions in this section because: 
           (a) I am an advisor/supervisor (not a student) (b) I work in 
           Europe, and so do not have U.S.-biased view implied by these 
           questions. HEP is a global venture. What is important is that the 
           "next machine" is constructed somewhere. "Where?" is much less
           important. Obviously, for me, it would be more convenient to have 
           it in Europe, but I'll use it wherever it is. I currently do my 
           research at SLAC because that's where the B-factory is. If it were
           in Europe, I'd use it there. 
     \item Half the questions are for U.S. residents only. Strange kind of 
           doing when discussing globalization. All experience shows that 
           techniques like video conf or web based expert treatment is not 
           (NOT!) really working (guess why). I believe that the virtual 
           control room as well as non local experts is complete bullshit. 
           At least, alone rising the question shows full absence of relations
           to reality or newer practice in running an experiment. Sorry! 
     \item What does question 18 mean? 
     \item The questions were geared to students and postdocs. For others the 
           issues are somewhat different and include access to computing and 
           to expertise on how to analyze the data and participate in the 
           physics and collaboration. 
     \item As a tenured European some questions in this section don't apply. 
           On question 14. When OPAL was running I would see it roughly 
           monthly. 
     \item Queston 14: it depends on my responsibility which change in 
           time : daily/monthly etc. 
     \item In question 16 I give my perspective but am, myself, the 
           supervisor. 
     \item Question 16 doesn't make any sense for those of us who are 
           supervisors. I answered in terms of my relation to those who work 
           for me. 
     \item I found that some of the above questions had no answers that 
           totally apply to me. For example, I have no "supervisor." I thought
           this questionnaire was also for senior folk. 
     \item What is the answer to 16 when you are the supervisor? 
     \item You give the choice "N/A- don't have a detector", but not 
           "N/A- don't have an advisor". Much better not having an advisor 
           to not having a detector. 
     \item For answer 16) it depends from the period of the thesis work. One 
           can also travel back and forth. And spend more time near the 
           detector or in home institution depending on the needs. 
     \item My answer for 18 is not accurate. There is no choice for 
           "no opinion" which would be my real answer. 
     \item The answers to 18 and 19 are very "model dependent". 
     \item My answer to question 19 presumes that at least one new frontier 
           facility is built in the U.S. during this period. 
     \item The answer to 19 depends on the precise choices of technology and 
           location. 
     \item The answer to (19) assumes that e.g. CERN will carry the major 
           accelerator. If America hosts the next machine obviously one of the
           national labs should have an enhanced role. 
     \item On 19) The answer here depends on what the future looks like. I 
           would like to see a program diverse enough to support the current 
           labs, but I would not say that the labs are a requirement. In 
           either case I do not think we need *more* labs. 
     \item Question 19 is naively worded. Labs have not closed, but have
           turned into staging areas, in a fruitful way. Others have evolved. 
           Perhaps a national underground lab should open. 
     \item These questions are not well posed. For example, I think question 
           19 that 
           the importance of the labs will be at least as great as now, but 
           that *IN ADDITION* some labs should close. Similarly, the 
           importance and identity of the advisor/supervisor changes as a 
           student's career progresses. 
     \item You didn't have my option about U.S. labs. I think they are very 
           important but I also think we should reduce to one. 
     \item You didn't give enough choices. The labs will not all have frontier
           machines ... of the four now, only one or at most two have 
           frontier machines. This is bound to get worse but labs can still 
           support R\&D on accelerators, on instruments, on detectors for the 
           frontier machines, on precision experiments on the g-2 model, etc.
     \item Alas, multiple choice does not cover the nuances ... these are 
           complex issues. 
     \item These questions cannot be answered in the abstract. The only 
           meaningful answers are those which address specific possibilities. 
     \item Again, limiting in your choices. Regional centers should be 
           considered economically as well. Can we work effectively and more 
           economically with regional centers?
     \item Your possible answers do not cover all cases, which is why I didn't
           answer them. I work within a mile of "my detector", but I last saw
           it a few years ago. I have never "worked on my detector", since I 
           write software. Or does that count? I think hardware experience is 
           very important, yet I have not worked on my current detector's 
           hardware (but did on previous detectors' hardware). I do not really
           have a supervisor in the sense you mean. While I think regional 
           control centers are important, I do not think they will improve 
           beams or physics. Regional centers are primarily important for 
           social purposes: to give a sense of "co-ownership" to institutions
           who will have worked very hard on the facility, but won't be close 
           to it. It is that sense of co-ownership that may result in better
           beams or physics, because of synergistic effects, not because the
           regional centers give access to "better people" or whatever. As to 
           the number of national labs, one would imagine that the regional
           centers would surely be hosted at the national labs, so it is not
           clear how to answer question 19. 
     \item I don't understand your question about frequency of distribution 
           of regional centers ... how often they are founded and closed? 
     \item The questions regarding "regional centers" is misleading. It would
           be natural that the infrastructure of the national labs be used to 
           create the so-called regional centers, since the machine 
           components and bigger detector components will likely be fabricated
           at these central places. It is also the most cost effective method 
           of controlling and monitoring fiscal issues.
     \item The question is not clear. Do the results improve because of RC 
           vs. no RC? or: do the results improve because of RC in the U.S. 
           vs. nothing in the U.S.? And why should the quality of the beam 
           improve?  
     \item Again, your terminology's confusing me a bit. It's not clear to me
           what you mean by "see or work on your detector". This year I've 
           only physically been at FNAL a few times, so I haven't seen the 
           whole detector very often. But, I'm working on a Run IIB Upgrade 
           project, so I work with a certain part of the detector (SVX4 chip)
           every day, at LBL. Having hands-on hardware experience does not 
           demand presence at the accelerator. To me, being at FNAL seems most
           important because it's the central gathering point. It's where the
           greatest number of people are located, and where the highest 
           density of work is done. You can get questions answered quickly, 
           etc. This is, in large part, because the detector's at FNAL. But, 
           if we're taking data smoothly there's no great need to be 
           physically near the detector- except that it's where people tend to
           congregate. Regional centers could provide other high-density 
           gather points. But, would they reduce travel costs/lessen travel 
           headaches significantly? Maybe so. 
     \item My detector is an optical telescope, usually 4-m. I don't have an 
           advisor. No appropriate N/A. 
     \item As a theorist, I cannot answer most of these questions. However, 
           I would like to strongly encourage global collaboration. 
     \item Seems as if you mainly ask experimentalists; that's for me 
           (theoretician) quite disappointing. Furthermore: The questions 
           seem to concern mainly U.S. experimentalists...?
     \item You are missing the point on "next machine". My "next machine" may 
           be neutrinos at the Oak Ridge SNS. Where does that play in the 
           spectrum? Are cosmic rays (supernova neutrinos) a "next machine". 
           By asking these leading questions about a next machine you are 
           pre-supposing an answer and narrowing the range. Regional centers
           just add another level of management, siphoning of resources from 
           labs and universities. 
     \item I wrote these based on the present feeling. In near future, the 
           situation may be significantly changed by the progress of 
           computer-network technology etc. 
     \item (16) Don't have a supervisor. (17) I believe physics quality will 
           improve; however, I am not convinced about beam quality. (18) I 
           don't understand your question. A regional center is some facility,
           room and so on. How does one distribute this? (19) This question is
           ill posed. The importance of national labs depends on what their 
           roles are. Is it (in the context of this survey) where accelerators
           are? Is it where we concentrate computing power for experiments 
           operating at CERN for example? Is it where some unspecified 
           concentration of effort takes place? As an example of the last one,
           Livermore is currently a national lab without its own HEP beams and
           is not a site for LHC experiments. 
     \item Again, the questions seem to relate only to a new LC and leave out
           other fields. Examples are B physics, neutrino physics and 
           astro-particle physics, etc., which are all fields in the midst of
           discoveries and which should also have a future!? 
     \item Your survey ignores the impact of large collaborations on 
           non-accelerator physics. Overlooking this gaff, the above responses
           are still very true. 
\end{itemize}

%% file: comments-outreach.tex
\pagestyle{myheadings}
\markboth{Comments on Outreach}
         {Comments on Outreach}
\subsection{Outreach}
   \begin{itemize}
   \subsubsection{How to Do Better:}
      \item As long as outreach activities (to the general public or to 
            politicians) are seen as not essential for career development, 
            there will not be enough of it. We should try to build a 
            connection to the general public, but at the same time try to do
            this not on the terms set by the industry, i.e. `what products do 
            you develop'. Our interests are not identical to those of IBM, 
            Microsoft or GM and we should emphasize the cultural aspects of 
            what we do a little bit more. After all, the ancient Greeks are 
            not remembered for the amount of olive oil they sold but for their 
            cultural and scientific achievements. That doesn't mean that we 
            shouldn't also take advantage of economic opportunities - if we 
            as a community would only get any royalties every time somebody 
            uses the web, we wouldn't have to apply for funding ever again. 
      \item I think outreach is like pushing on a rope: you don't get anything
            but a local effect. A television series might allow a lasting
            impression, but probably would receive a small viewership anyway.
            I think the only way to improve funding is to demonstrably MAKE 
            MONEY. If you can PROVE that HEP contributes X dollars to the 
            economy for a given investment N years ago, then you have 
            something. Better yet, just have the beneficiaries of the 
            created wealth lobby congress directly themselves. I think we 
            should perform some actual "value mining" to determine what we 
            have that the world wants. Our analysis tools, like ntuples and 
            PAW/root, should be cleaned up and marketed. Instead of paying out
            a load of cash to develop fancy silicon chips, we should develop
            fancy silicon chips and then license the techniques. 
      \item The "meet - a - scientist" program is a good thing and need more
            volunteers.
      \item Try to find a way to explain to the public the purpose of HEP, its
            goals and applications (e.g. developing detectors for HEP we 
            contribute also to the development of detectors for cancer 
            treatment etc). Organize public talks at Universities, advertise 
            these talks with posters, tv, radio, make nice films for CERN or 
            other laboratories, organize visits to schools and universities 
            and tell them about HEP ... 
      \item I believe we could be more active and more honest in our outreach.
            For example, BaBar's recent announcement of a 4+ sigma 
            CP-violation effect in the B system was reported in the press 
            under headlines implying that the matter - antimatter asymmetry of
            the universe was now understood. This is an example where we 
            should have been more clear about the implications of our 
            measurement. The difficulty, of course, is that it's very hard to 
            explain what we do to people without a substantial amount of 
            science education. Long-term success can only be attained by 
            supporting educational efforts in the elementary and secondary 
            schools that communicate the interest we all have in learning more
            about the universe. 
      \item There must be more emphasis upon the synthesis of the different 
            sciences in the public understanding. Connections between physics
            and the other sciences and technologies must be drawn for the
            public in order for the U.S. physics program to maintain 
            relevance. 
      \item More emphasis should be put on finding commercial outlets for 
            HEP, as this will pay for the blue skys research of the future. 
      \item To involve the general public one good idea would be to think at
            our collider detectors as big lotteries. Since gambling is big in
            the U.S. (as everywhere else there is money to spend) we should 
            start funding our experiments by selling events before we collect
            them. A dollar for one event, with variable payoffs if the event
            turns out to be a W candidate, a top candidate, etc., would do 
            the following good things: 1) bring money to HEP from an entirely 
            new source 2) boost people's interest in HEP and allow them to get
            an education in particle physics; 3) media would become interested
            too; People buying an event number would become the "owners" of
            that event once it is collected. If an event results in a paper 
            being published, the owner of the event would get cited in the 
            paper... Outrageous but fascinating, ain't it ? 
      \item Mini-series as on the Discovery Channel.
      \item Outreach does NOT mean constructing a webpage. Actual contact with
            the public should be sought. 
      \item Put a complete directory on the web where one can get an answer
            to any question ever asked about Fermilab.
      \item To answer the public whenever questions are asked every scientist 
            should be willing. But I hate if propaganda gets more important 
            than careful work. 
      \item Operating a good Web site promoting HEP might be good.  
      \item Heck, if I knew a sure fire way to inform the public about what 
            we do, I'd include it. But a big step in the right direction 
            would be to encourage the young and enthusiastic people to not be 
            afraid of people running in terror when you tell them what you do. 
      \item The majority of the public has never heard of a quark or even the
            photon. We need a massive PR effort. 
      \item I think that outreach to the general public be increased greatly. 
      \item We need to cultivate a grass roots interest, as astronomy has 
            done. Then we have a base to turn to influence the political 
            process. What we need is more of what Lederman has been doing at
            Fermilab and in Chicago. We need to attract private funding. We 
            need to get modern physics including particles into the 
            high-school physics curriculum.
     \item The P.R. situation has improved in the last few years, but more
           physicists must get involved. 
     \item The level of awareness of the general public is, in my view, poor.
           It's our responsibility to improve that and in so doing attract 
           public backing for our work. 
     \item Our outreach seems to be directed at getting people to support our
           work. Our obligation to funding agencies, and the public, is to 
           inform and educate regardless of the outcome. 
     \item I don't think we have figured out how to do this effectively, given
           the U.S. (non) culture. In some European countries, such as Italy, 
           there is much more appreciation for science than in the US. 
     \item I fear that without educating the general public and Congress on
           both the basics of HEP and reasons why we this research should be
           done, we will NOT be able secure funding to keep the field of HEP
           as active, diverse, and dynamic as it is today. 
     \item Be honest! Say why we do what we do: because it is exhilerating. Not
           because it is "important". Combating poverty is important. And be
           prepared to answer, honestly, why the public should spend billions
           to learn about the Universe. Emphasize the core goal: learning
           about our place in the Universe. No, it ain't about finding yet 
           another bunch of particles and measuring their esoteric attributes.
           Who cares about that? No, emphasize that we do this because, for 
           whatever reason, we (all of us) want to understand our place in 
           the Universe. Explain to the public that yes our instruments are 
           costly, but very good value for money when compared to many other
           things our governments do, some of which are truly dumb. We've just
           got to be more honest all around. 
     \item I do work on outreach, but somehow we are inspired to do physics
           early on in our lives. It is not a choice as something that drives
           us. People don't want to work hard to understand things that make 
           them feel inadequate. We can only continue to try to rephrase 
           things in language that can be understood but some will never 
           understand. It is critical though that we don't sell our souls 
           (GRID is an example) or try to convince them that the next project 
           will cure their cancer. Hyperbole is useless, honesty in what  
           motivates us must prevail and spin-offs are nice but secondary. 
     \item This is a two way street; you can't reach out to someone who is not
           interested in being contacted. 
      \item HEP tries to answer some of the most basic questions about the 
            universe around us. But it is also a luxury, even in the First
            World. Past questions from Congress include "What use are quarks? 
            Can I eat them for breakfast?" (!) Past answers include "It makes 
            the country worth defending." The culture of stereotyping physics
            and maths as difficult is strange. (So History, English and 
            Spanish are easy then?) We need to reach out more and show people
            how basic physics principles work in their everyday lives: from 
            ice-cubes and air-conditioners to aeroplane wings and rockets to 
            positron emission tomography and MRI. HEP scientists can do this
            and, if they can bring in the advances made as a result of HEP 
            research, both of benefit to society and to the sum of human
            knowledge, so much the better. A society that appreciates the love
            of learning and the benefits it may bring is far more likely to
            fund the next accelerator. Personally, I've found that people are
            quite fascinated by accelerators and detectors and want to know
            how they work. The most unexpected people will present you with a
            HEP article that they clipped out of the local newspaper. In 
            comparison, astronomers have it easy: they have the glossy
            pictures. We have to work a bit harder to explain our pictures. 
            When it comes down to it, a telescope and a particle detector are
            both luxuries compared to clean water, stable power grids, medical
            care, tax relief... 
      \item Our mission in life is to explore the cosmos in every way that 
            our collective imaginations may conceive. Outreach is important 
            partly for funding/survival reasons, but also to simply convey the
            beauty of what we have found (and are looking for) to the general
            public. The truest test of our love of the field is our ability to
            do that effectively. Every physicist on planet Earth must take 
            this responsibility to heart. I think if we all take this
            attitude, then the rest will take care of itself. 
      \item Clearly, we need more. However, there are great people out there 
            doing at least there share on this - working with high schools and
            younger children. Once a month I talk with groups of junior
            high/ high school children to answer questions about the lab. 
            Suggestions for better outreach: 1) HEP needs one person who 
            stands tall and makes public appearances. Someone that people will
            begin to recognize. Where is this person ? NASA has one - hell, 
            the NRA has Charleton Heston. Why can't we get a single, 
            charismatic spokesperson ? Also, we need more PR money. When I 
            turn on the discover channel I should see lots of shows about HEP.
            There should be lots of half-hour shows on this stuff floating  
            around. CNN should have a section for us. The major obstacle in
            this is convincing people we are as interesting as space. To do
            this, we must simplify our language and not forget to dream... 
            and imagine ... NASA is not afraid to show pictures of streets
            lined with coffee shops on the surface of Mars. Why are we afraid 
            to predict things like extra, non-standard model forces giving us
            new energy sources ? No one seems to have the guts to quote 
            anything unless they have a few sigmas behind them... 
      \item Small, interested sections of the public are being reached by 
            various laymens' books, TV and radio programmes, which present
            frontier research (in terms of factual detail) in an exciting, 
            engaging way, but not enough is done to convince the public at 
            large of the need for this research. Once you have convinced the 
            public of this need, it then becomes possible to convince the
            appropriate politicians. Also, we have to demonstrate to the 
            politicians that we are coming up with ingenious and efficient
            ideas for experiments, rather than just throwing money into the
            research. 
      \item Seems to me we are far to "esoteric" in Public view. The 
            transistor was once esoteric too... 
      \item HEP labs should be completely disconnected from classified 
            research. In this way, the public would gain more faith and maybe
            interest in the labs. The public should be allowed to go to the 
            labs and look around (to a certain extent), even without guides. 
      \item I think each person in HEP needs to commit themselves to public 
            outreach. Perhaps once per year, each person should make sure to 
            perform some type of public outreach. I think it is most crucial 
            to reach youths at the early high school level. This can be a 
            guided tour of a lab, a very good talk on particle physics at
            their school, etc. These talks MUST be at the proper level to 
            develop curiosity and interest, and not to show how much you know.
            Also, one should touch upon the technologies which we use every 
            day, such as high speed computing, superconductors, etc. Also, 
            I think we need a budget for public outreach programs. This 
            includes investment in programming on public TV in addition to 
            advertised evening demonstrations/presentations at universities 
            of the "Wonders of Physics". I really enjoyed such a show put on
            by Clint Sprott (at the Univ. of Wisconsin), who set a great
            example. 
      \item I think we need to point out to the general public (more clearly 
            and without any jargon) the benefits that can come from basic 
            research. For instance, I tell people that although the electron 
            was discovered over a 100 years ago, it took many years of 
            research for us to understand its properties. But, if you look 
            around now, almost our entire technology is based on the electron.
            I agree that one should not tie current research to future 
            benefits, but at least we should point out the fact that basic
            research has the potential of completely revolutionizing our way
            of life - whether that is a good thing or a bad thing is a whole
            different issue! 
      \item I think we should all chip in and form a PAC. We need to 
            quantify the economic benefits of our field better (presumably
            with the help of professional economists) and make sure this
            information is widely disseminated. We need to design a really
            addictive video game base on particle physics. 
      \item The field is dominated by the old white boy network. We don't
            have many blacks, other minorities or women in our field. How can 
            we talk about outreach when we fail so miserably at this level? 
            Science is not valued by this society and we are partly
            responsible for it. Our children will pay direly for this failure
            of ours. We need to do more to educate the public, to make the 
            connection of science and society. 
      \item Old men's job. 
      \item I think it is very important to include an inter-disciplinary 
            approach in the planning of future HEP experiments, in the way 
            it is planned for TESLA with the free-electron laser. 
      \item Need to foster programs with engineering schools and computer
            science. 
      \item When the Kameoka data came out there was no place where it was 
            discussed so non experts could understand it. I heard Harry 
            Lipkin give a good talk which did. 
      \item There is a very big difference in the effects of outreach in the 
            different countries. I consider the present contact to the general
            public and the politicians in Germany, the place where I work good.
            The contact to the schools (in particular teachers) and to 
            physicists working in different fields and scientist however 
            should be significantly improved. 
      \item A real catch 22... In Sweden NSF funding is followed by a 
            documentation requirement -- maybe a similar thing for outreach 
            should be considered. 
      \item Non-physicists do not understand what we do at nearly the level
            they should, and this is our fault entirely. HEP asks alot from
            the taxpayer, and the taxpayer should get more in return. Science 
            education is one area where HEP outreach could improve enormously.
            For example, if all HEP graduate students spent 1 day per year
            giving a tour for the general public, or talking with high 
            schoolers, or writing letters to their congressmen, or writing
            letters to their newspaper, the level of education of the general
            public about HEP would increase enormously. We should have an 
            outreach day, where all students, faculty, staff scientists, etc.
            make a coordinated effort to explain what is is they do to the
            taxpayer. Second, a greater degree of centralization would
            facilitate outreach for those who already do it, and make it much 
            easier for those who do not to get involved. A central repository
            of outreach materials, success/failure stories, listings of local
            contacts, e.g. high school teachers who would come for tours or 
            invite classroom talks, would make it much easier to get involved 
            in outreach. 
      \item Senior physicists and well experienced physicists should dedicate 
            more time for outreach and less for scientific policy. Young 
            scientist should mainly focus on research and in designing new 
            challenging projects. Presently (and this is worrying), this is 
            most of the time the other way around. 
      \item Keep better track of HEP alumni -- in fact maybe they should be
            treated more the way universities treat their alumni. This 
            information could be very useful for demonstrating the usefulness 
            of education in the field, and generating support for funding in 
            HEP. 
      \item IF, the HEP community did reach a consensus on what machine to 
            build next, it may not become a reality due to lack of public 
            support. If we spent a fraction of the time we spend solving 
            technical problems, on public outreach and government 
            education/lobbying, we might be successful. 
      \item Sit in the damn boat rather than fighting in shortsighted 
            competitions. 
     \item In my point of view: High energy group is a special group in the 
           physics field. While the other groups, such as, solid physics, 
           material physics, do a lot of practical things, HEP does lot of 
           fundamental study. It appears that HEP does not give good economic
           output. However, it indeed does a lot! the reason is HEP         
           concentrates on those fundamental questions that human being has 
           for this universe. Solving these questions need a great amount of 
           human intelligence. Therefore, having a qualified HEP graduate 
           student means having a man/woman with great intelligence. Of 
           course, not all of them will stay in HEP field. Suppose 50% of 
           them leave to other fields. Definitely, they will develop new 
           brand new stuffs that will somewhat change the style of the current
           world. As an example, I studied HEP for 4 years (another 7 years in
           physics), I found it is very easy for me to understand knowledge in
           other fields, which is a good basis for invention/revolution. And 
           surely a qualified HEP student has accumulate enough inventing 
           abilities when he did his absolute and systematic analysis on this
           complicated universe. I thought of working on some other fields 
           once a while, one example was to work in Gene sequencing field to 
           decode all genes systematically! Hoping/thinking is the way leading
           to invention. Just thinking of some of the modern technology: 
           Internet, CD information storage, Laser technique, they are all 
           directly related to physics. And HEP is a typical and special
           component in physics, just like what nuclear physics did 50 years
           ago. Not only HEP provides economic benefits, it also uniquely 
           provides social benefits when answering those fundamental questions
           human being has for this universe. And this is a very special 
           advantage over other fields. In closing, I am sorry for my poor
           written English if it already made it hard for you to understand 
           this writing. 

      \item Externally changing our approach to the general public with projects like outreach goes some way, however I feel that people would be more interested in the subject if we made changes from within the field. Making our own language more transparent and obvious is one way of doing this as is releasing data to the general public to use. There is still far too much intellectualizing of subjects that could be explained in a much more understandable form (e.g. pages teaching computer skills for physics seem to be written by people who are more interested in showing how clever they are, than in teaching people the skills in the simplest form. It is next to impossible for someone from outside who has no experience of the field to ever get a proper handle on it.). 

      \item We have to be able to explain importance of our research programs for our civilization. It requires very serious thinking and development of strategy how to do it. Implicit assumption that everybody understands how important is science can not work forever. Government and public have to be convinced. On the other side if there will not be long range plan approved, then there is a danger that demographic factor might play significant role and there will be not enough followers for HEP. 

    \item General public is fascinated by macroscopic world, e.g. life, the Universe, arguments for strong and fascinating questions. However elementary particles at all scales, from atoms to quarks and below, are fundamental to understand macroscopic phenomena. This must be stressed with strong cases to reach general public 

    \item Congress/public. Scientists tend to think that laypeople can't/won't understand their work so they spend little time building enthusiasm by explaining it to them. Every day we teach folks who don't understand physics, why is it harder when it comes to the general public or congress? PR is the way of (future) science funding. 

    \item The problem is not specific to HEP. The problem is that there is not sufficient understanding or appreciation of basic scientific research. HEP will not stand or fall on its own. We are part of a larger community. We can probably fall on our own, but we can garner significant additional support only if basic scientific research as a whole enjoys a higher level of support. 

   \item Abate superstitions about HEP, increase public support for HEP applications and importance. 

   \item I'd love to do more outreach. (perhaps more than physics analysis!) We have to sell our products (knowledge + people) to Congress, and other scientists. AND we have to return the public's investment in us by telling them what we've learned. 

    \item It is more important to "educate" to the importance of basic research and its fall out than to press release exaggerated claims to the importance of a result just obtained. i.e. information, not (more or less false) advertisement 

    \item I think large collaborations should publish more often even if it's just updates on what is being worked on. I think general information about the importance of basic research and how its by-products have very applied influence on our lives must be given a lot more readily. 

    \item Outreach is important because we should not take funding for granted. We need to show that we are doing valued work. However, it is important that the outreach be of high intellectual value. I was often appalled with the claims that were made about the benefits that would accrue from the SSC. It is difficult to do a good job of education. We need to get advice from those in the community who are best at the job of communicating science and to learn to do the job well. 

    \item  The NIGHTLINE program on neutrinos was excellent, as was coverage in national magazines. Keep up the contacts, strengthen them by periodic visits, email, phone conversation, etc. and have something to say. 

   \item Perhaps work on a video which surveys HEP/astrophysics work and clearly ties it to the spinoffs which help people. 

   \item Making the work we do as accessible as possible to people working outside of the field is sorely overlooked by too many people. It is my opinion that everyone is responsible in part to make the current research efforts of the field known to the general public is any way possible. Only through increased understanding will we gain increased funding. 

   \item More activity on the www as is happening in astrophysics hands-on-experiments/ activities for general public more math at high school. 

  \item I really think that worldwide the HEP community should make a greater effort to outreach the public making them understand what we do, why we do it, and what comes out of it both in terms of scientific knowledge and technical development and spin-offs. Especially for Funding agencies and politicians a stronger effort has to be made since it looks as the decreased investments in research is a general trend in the world, and HEP especially now needs great joint international efforts. I believe it is an ethical duty to all physicists to devote time to divulgation of her/his research activity to the public as it is a duty to be aware of the possible applications of the techniques she/he's developing. 

   \item In a country like India where there is great need for rural development I am using Monte-Carlo simulations for predicting the reliability of rainwater harvesting system - this is just an example of a meaningful outreach. 
 
   \item Answering the question, "What good is HEP?" is tough to answer. Unfortunately anything that directly or indirectly points to an answer is good. Anything else won't convince the public. 

  \item We are trying to reach the public, but we don't always talk their language. We should try harder to make ourselves understood. With Congress I think we also need to stress additional issues such as visas for physicists from foreign institutions. 

  \item You can never do too much communicating. Much of what we do is excellent, but I think that we still have a major problem in trying to explain WHY we do it, and what the ultimate goal of the 'Theory of Everything' is - this phrase (quite amusing to the particle physic insiders) is meaninglessly arrogant to other scientists, and just meaningless to the general public. We need a better articulated goal.

    \item Subject of high energy physics and its output is hard to understand for common people and politicians. We should try to show people how the big high energy physics experiments have benefited common people. 

    \item Personal contact (including casual conversations) changes attitudes in a big way. If every high school student had a chance to see the big picture in HEP, then even the small amount of coverage HEP news gets in the popular press might be enough to maintain interest. 

    \item 1. Physicists must urgently understand that the language and methods of information distribution in the community and in the rest of the society different and that we must learn also the other language even if the other language might look trivial. Triviality is a tool in promotion. Use it instead of reject it. We know that we can do better so don't think too complicate. 2. We need budgets for promotion, real independent budgets. 3. We must use much more promotion professionals in strong relationship to the community. Physicist with a hand for the public could build the bridge. Encourage them and give them money. You know promotion investigations usually come back but not alone. 

    \item Most of us do this job because it's fun. We need to communicate that spirit of fun to kids, both to promote our public image and to promote kids' interest in learning how things work. 

   \item Get a unified agreement to phase out one lab, in the course of the next few years. Needs real leadership - and strong prodding from young physicists. 

    \item The amount of outreach in USA is fairly adequate compared to that in developing countries, which statistically produce majority of researchers world wide. 

   \item If we decide to build a new large facility then educating the public and the politicians on it's importance will be crucial. I personally think that we emphasize the spin-offs from our research too much in some cases. We should have some goal of promoting fundamental research as being important in its own right. 

  \item Most of the funding or big scale experiments comes from the government. We should do more efforts to attract people from other areas. We have accumulated enough technological and scientific knowledge that can be applied to bring immediate benefit in other areas and to society in general. Society for sure, will give us back all our efforts 

  \item It is almost too late. 

  \item Saturday Morning Physics-style programs or physics road shows to teach high school physics students about HEP/physics in general should be implemented everywhere! This could also help draw more smart young people into physics. 

  \item  We have to understand that, as any competitor for general interest, we have to put more manpower to marketing (outreaching efforts) than to production (research). The ratio of 1:50, as it is currently, has to be changed to 3:2. Nobody can survive who has a domination of production costs. 

  \item Much more important than press releases about work which the general public usually does not understand at all are continuous low-level efforts like days where the labs are open to the public, public seminar series, and co-operations with schools and science teachers. 

   \item I am not at all interested in "outreach" in the sense "convince everyone that what we are doing is wonderful and important and give us more money". I emphatically believe there are things that are far more important and worthy of funding than HEP, and I find some HEP "propaganda" rather overblown (and philosophically illiterate). I do think it is very important though, that we try to explain to the public what we are doing; make the science comprehensible. In my opinion, our contribution to "culture" is as important as anything else. 

   \item First, find out whether anyone outside of HEP really cares about our results and plans. Can you answer the question: "Who is interested in what I do?" If the answer is simply "others doing the same thing", then look around for something else that could be done. Much of the outreach I have seen has focused on telling our message. We need to spend more time listening to what their questions are. 
 
   \item Streamline access to policy-makers and congressional / senate 
            staff. Host items of public interest (e.g. planetarium, public observing telescopes, displays, etc.) at the national labs and encourage creation of tourist-oriented visitor centers.

   \item Figure out how to speak and write for the general public and the press, instead of other scientists 

   \item We need to have a continuous general public education effort, and rely less on "the big discovery", since the latter are extraordinarily rare these days. 

   \item I feel that more effort should be made to inform the general public about the nature and the goals of particle physics. I an convinced that some budget limitations that are being experienced now are to be put in relationship to the fact that the public opinion is skeptical about the fundamental research on particle physics and little keen on pouring money into it. Strength should be put on the technological fallout made possible by R\&D in particle physics. 

   \item There was a recent issue of Time magazine which had an article on the end of the Universe - it spoke about the history of the Universe, including mention of dark energy and the implications leading to a Heat Death prediction. I thought the article was well written and informative. Today, I saw letters to Time in response to that - not only was there much ado over how depressing this outlook is, there wasn't understanding of what the alternative possibilities are expected to be. This, while really a question of cosmology, is to my mind indicative of the problem. We physicists often succeed in showing new discoveries and such in light of previously understood phenomena; but we rarely succeed - and often don't seem to attempt - to show those same discoveries in the context of related research. A search for the Higgs is interesting, but seems a foregone conclusion when there isn't statements made about competing theories. 

   \item Outreach is very important but hard work. Open house every year would be a good start. Invite politicians/ Congress people. 

    \item Tours and open houses of laboratories and experiments are often helpful. People of the general public for the most part really like the "gee-whiz" aspects of our work that we may sometimes take for granted. An article written for a local paper (not a letter to the editor complaining about low funding) or interview with a local journalist about your current project can go a long way to generate local interest and support. Congressional support is important, but ultimately we really do need grass roots support. There will always be some people who feel that basic research is a waste and the labs should be closed down,but while these people are a small minority, they tend to be vocal. Most people haven't thought much about basic science one way or another and we ought to make sure that those of us who support it are heard too. 

   \item Britain has some nice Science Fairs at High schools.

   \item Stress importance of HEP research results and possible impact on everyday life 

   \item Physicists both at labs and universities devote a great fraction of their time and resources to teaching, mentoring, writing textbooks. This should be counted as public outreach. "Outreach" is misleading because it implies an inside and an outside. The truth is, we are all in the same boat. Our research is publicly funded because it is done in the public interest.

    \item Tell the truth about HEP. It is a cultural activity to discover the nature of matter. It is not done to create spin-offs or cure cancer. That type of stuff killed the SSC. 

    \item Astrology is much more widely accepted than the big bang among the public- this is a problem! 

   \item Currently working as lead mentor for a QuarkNet center at my home institute. This is a great program, but only a start. More needs to be done to contact students and the public at all ages. 

   \item There has been visible progress in the HEP outreach effort - programs like Quark Net, special events at Snowmass are good examples. Frequent interactions of faculty with local congressmen etc should be encouraged, increased emphasis on the society benefits from the by-products of the HEP activities: web etc, training of highly employable people ...

  \item Programs like QuarkNet and CROP should remain strong. Outreach programs like that run at Snowmass are excellent and should be encouraged. Whenever a HEP meeting takes place, there should be an outreach activity included. be 

   \item The record of HEP in getting the public, and therefore the lawmakers, interested is abysmal. I believe we need major efforts to overcome this, including perhaps a popular science magazine dedicated solely to HEP (a la Sky and Telescope, Astronomy, Discover, etc.). 

   \item Outreach is the key to get more funding. I try to do it on every possible occasion and would like to encourage young physicist to engage very actively in science fairs, local school gatherings, talking to your local newspaper people, publishing articles in News bulletins of your labs and universities. I am not ashamed to act like a used-car salesman if it brings extra funding and extra public knowledge to our field. This is what NASA does very well. 

   \item More frequently inform administrators and political leaders on science status by direct formal and informal contacts. 

 \item The public will listen to us if we have discoveries. Without new results it becomes harder to get attention and funding. 

  \item I am a tour guide for SLAC; it is my experience that some people come to SLAC as skeptics or cynics about the justification for our work, but after a mere 2 hours of questions and site-seeing, they are overwhelmed by a sense of purpose and importance. It is inspiring to me to know that it is possible to take somebody with some doubt in their minds, and simply by showing them the lab and talking to them about physics they are convinced that things shouldn't be otherwise!

  \item I think we need to think this through from the beginning. We are starting with a product that is complicated and hard to market. We may need to patiently do a huge amount of education. There may be no quick fix.

   \item I think we definitely need to spend more time/effort/money on outreach. It is extremely important to educate the public about our work. This is true both in the interest of gathering future public support and simply to bring the excitement of science to more members of our community. In order to promote this process of outreach, there must be recognition of its importance and of the people involved. The method of funding such work must improve also. For example, currently the DoE does not allow universities to spend DoE money on education or outreach. This makes no sense since they are funding people/programs at educational institutions. Currently their mission statement does not allow for such expenditures; however, this mission statement could be (and should be) changed to allow such activities. We could also take our cue from other scientific institutions who maintain extremely good public and congressional relations. NASA, for example, can! crash Pathfinder into MARS and still get money from Congress to continue their work. And the general public still thinks that NASA is doing extremely exciting and necessary work. Unlike HEP, NASA spends between 2 and 4% of their entire budget on outreach and education. This appears to be a strategy that works for them! Finally, Congress and the funding agencies are what keep us running (or don't), and it is imperative that we maintain a steady and positive influence in their arena. If we ignore this need, we run the risk of letting our field wither away. 

  \item I don't really know what folks in the U.S. are doing these days as far as outreach. However, it is extremely embarrassing to hear of press conferences being called to pre-announce scientific results before scientific peers have even had a chance to see the results. (Cf. CDF, Babar, Pons\&Fleishmann). This practice should be discouraged, via funding-agency pressure if necessary.

  \item Need to provide a broader base for developing ideas of how to involve the general public better in physics goals. 

  \item I'm still in school, so outreach for me isn't an option. Congress as well as the general public are largely in the dark when it comes to science due to how little science is taught in our schools. So I feel much more emphasis on the importance of science should be given in schools (at all levels) such that the next generation of "the general public" will" will know it's importance. 

   \item I used to be a SLAC tour guide -- and found that the people taking the tour where very positively impressed with the activity at the lab. Perhaps such programs should be expanded, perhaps via TV. And sending physicists to schools for brief lectures. 

  \item Outreach already represents a significant consumer of resources. Seems to be "reasonably" successful at funding agency level. If activities are to be increased maybe this should be done by appointing full-time staff to be most effective.

   \subsubsection{Outreach Organization:}
      \item The largest problem is that even with increased ``outreach'', our 
            message gets lost when we don't speak with one voice. Too many 
            divisions or local battles within the field are fought in public
            forums (eg. funding agencies). It's difficult to get a lay-person
            excited if they are frequently hearing a dissenting voice at the
            same time. 
      \item I feel that the HEP community presents too fractured a face to 
            the funding agencies. It is necessary to unit behind something to 
            help in the funding. I feel that the general public is very much 
            out of touch with what we are trying to accomplish. I have 
            offered to donate some of my time in the answering of questions
            from the public that come in through the FNAL webpage. 
      \item Regional outreach organizational might work in the same way as 
            indicated above for research centers. 
      \item Physicists who are not working with a large fraction of the time 
            on communication with the public need guidance on how they can use
            limited time effectively. Some people must devote a large fraction
            of their time to organizing this effort.  
      \item Need for a better control/management over the huge projects our 
            field requires, in order to avoid shameful endings like the SSC...
            Get people to like physics: tough job, but it's the only way. 
      \item It is incredibly hard to explain HEP physics to the general 
            public, and doesn't come naturally. Courses in graduate school or 
            workshops on public outreach would certainly help. This would help
            fund-hunting, too. 
     \item We are badly lacking in leadership, in good communicators spending
           almost full time in Washington and in dealing with the public through
           TV, OP ED articles. We must raise the level of respect for the 
           future Carl Sagans. We should surely join with other disciplines to
           keep a Neal Lane type in Washington working for science. We should
           award a significant prize for the best op ed article in a local 
           newspaper etc. 
     \item We need Carl Sagans and should try to discover, support and 
           reward them. 
      \item Important to focus more outreach on "decision leaders". Outreach 
            to general public is important in principle, but is not a 
            primary path to strong support for funding. 
      \item Perhaps more organized ways of contributing to outreach could 
            help. I think a lot of people are willing to commit some time, 
            but would need to make a large time investment to get anything
            going from scratch. Some way of allowing people to contribute a 
            little here and there could really help, and contributions could
            really add up (a bit the way providing bins so that recycling is
            easy really promotes recycling!) 
      \item We need to develop training material to use for outreach. This 
            provides a consistent and clear message.
      \item We need training programs in outreach. Too many of us really
            don't know how to do it. 
      \item We need more coordination of outreach efforts. We need information
            on how to "plug-in" to programs which work. 
      \item I would do more outreach if I had better guidance as to how to 
            do it well. We should strongly encourage those who have the 
            talent for it.   
      \item Hiring people just to do outreach work would be useful. 
      \item I think people with good communication skills, who have wide 
            general view of HEP, should do more to popularize HEP and physics
            science. 
      \item Outreach should be done by those who have a talent for it. The 
            rest of us should give them support and recognition. 
      \item I already spend some time on outreach activities. My observation 
            has been that many who do little or nothing will do a good job if 
            asked. Part of the activity of outreach activists ought to be in 
            recruiting and organizing capable exponents of the sciences who 
            are not sufficiently self-motivated to seek the opportunities to
            connect. 
      \item It's hard for us as physicists to explain what we do in an 
            accessible way. It would be good to get people who are good at 
            explaining (such as science writers) and really get them 
            informed/involved in the experiments, such that they can use their
            skills but actually have the knowledge of the topic to get things 
            right. 
      \item This cannot only be done by a few "professional" spokespeople. We
            have to get more of our colleagues involved in the various 
            "outreach" activities. 
      \item This is not my forte so would be a waste of resources for me to 
            reach out to the public. Specialists in this area are needed. 
      \item I recognize the importance of these things, but I personally am 
            not that excited by them. But there are many people in our field 
            that are, thank goodness. 
      \item We need to stop talking about outreach and start doing it. People
            want to do it, but don't have the support needed. Efforts should 
            be coordinated and advertised to build support and excitement, and
            so that our culture values and rewards outreach efforts. 
      \item It's not only the question of time to invest. It's (in my view) 
            the task of the "HEP management" (whoever this is in the various 
            Labs and countries) to create the infrastructure for outreach 
            activities (institutionalize public events, visitor days, school 
            and university contacts, etc.). Some such activities are going 
            on, but they are mostly pushed by individuals. 
      \item Rather improve quality/ coordination of outreach efforts than 
            quantity.

\subsubsection{Value of Outreach:}
      \item The problem with outreach is that it isn't part of the job. 
            Faculty don't get promoted because of their work educating the 
            general public. At best, it is a little extra - but if the basic 
            research work isn't enough to stand on its own, outreach won't 
            push one over the edge. And it takes time and talent and, 
            probably, instruction to do well. Question 23 - I put "maybe," but
            the answer really is "yes, if it becomes a real part of my job" 
            and "no, it isn't an official part of my job which is already very
            demanding, and I don't spend enough time with my family now." 
            Note that I do a fair amount of outreach now!     
     \item  Outreach should play a role in deciding tenure/ permanent 
            positions. 
      \item RECOGNITION BY THE LAB/ UNIVERSITY MANAGEMENT: outreach activity 
            should be spelled out on evaluation reports, and be integral part 
            of the career path. 
      \item We should value outreach more in this field! One example is to 
            give physicists an extra vacation day for every two weekend days 
            that they spend doing outreach (such as Saturday morning physics 
            at Fermilab, etc.). Actually, this example could work for any 
            Fermilab employee / user who is able to give tours/ lectures on 
            the weekends. In general, though, we as a field should spend more 
            time and resources educating the non-physicists. 
      \item To increase outreach, management must provide working time for 
            such, for those willing to help. 
      \item There has to be a mandate from the agencies and the 
            university/ academic career structure to recognize and foster 
            these activities. Doing more with less is getting "old" and 
            ineffective. 
     \item  Perhaps mandate some level/ form of outreach for different career 
            levels. 
      \item Outreach is the duty of tenured faculty. 
      \item As a postdoc doing theory I do not have time for anything other 
            than research. The reality is that selection criteria in the field
            depend only on the quality/ amount of research. If I dedicate some
            of my time to outreach (in principle a nice and respectable thing
            to do) then I produce less research = my chances of getting a next
            position/ postdoc decrease rapidly. Clearly if the system does not
            change then I do not see how theory postdocs can be expected to do
            outreach. 
      \item Most people are already overcommitted. To spend substantially more
            time on outreach, they need to have other duties lightened. 
            Universities could lighten teaching loads, labs could lighten 
            other service work. There are only so many hours in a day, if 
            people are going to spend substantially more time on outreach, 
            something else has to give. 
      \item For large experiments, public outreach might be a recognized 
            service task (maybe this is a bad idea, but it's worth
            considering). What else can we do? I don't know. To a certain
            extent, we already know what to do, but we just don't do it.
            What's required is more of a cultural shift within the community
            so that public-outreach is seen as more valuable. As it is,
            outreach is seen as something that you do in your spare time.
      \item I would personally be willing to dedicate more of my time to 
            outreach, if it would be also count as much as writing a scientific
            publications. 
      \item If my career uncertainty were not so strong, I'd be much more 
            willing to do outreach. As it is, I feel such efforts detract 
            from doing things that make me visible to hiring committees.
      \item HEP needs better marketing! But outreach doesn't pay off for 
            one's personal career. 
      \item I already have dedicated much time to Congressional outreach 
            (2 trips with SLUO + UEC to D.C.) but am constantly amazed at 
            attitudes of my colleagues when I tell them of my involvement -- 
            typical responses are "That's not doing physics!". They often
            don't understand the importance of these activities. 
      \item I already do a lot of outreach, but regularly get belittled for 
            that effort by faculty, and ESPECIALLY other graduate students. 
            Physics grad students and postdocs need to get over themselves 
            and realize that those "C students" in the labs we used to teach 
            are now the staffers of Congress determining how much funding we
            get. 
      \item Outreach or die. 
      \item I consider outreach as an integral and important part of the 
            faculty position. I will dedicate my time to it when I become 
            faculty. 
      \item Outreach is important, but young people are not rewarded for this.
            In fact, in can be detrimental to their careers. On the other 
            hand, the older crowd has failed to do a good job. I am in a 
            quandary as to how to deal with this.
      \item I might do some outreach in my Country instead of in U.S., just 
            to have more time at the Lab when I am in the U.S.
      \item Those who do outreach should not be outcasts. This is an 
            essential part of our future. 
      \item We need to create incentives for outreach and obtain an outreach
            line-item on the DOE budget.
      \item Yes, but this effort should be paid as well, since it is an 
            essential part of the task of the physics community. I am 
            astonished that most efforts of U.S. labs into this direction are 
            completely voluntary. In this respect, CERN seems to be far ahead
            when is comes to outside communication.  
      \item Outreach to the public has to be appreciated more. So far, it is
            more considered a personal hobby but it seems not to be adequately
            appreciated within the scientific community. 
      \item Those doing outreach at university don't get the recognition of 
            those doing the research and bringing in the funding money. 
      \item Outreach to public best done by talented individuals - very 
            important, but much HAS BEEN ACHIEVED over last years - I can't 
            do more or quality of my work would reduce - worse end of stick. 
      \item I'm saturated with outreach activities already. 
      \item You didn't ask how much time I spend already. I spend a LOT of 
            time already. 
      \item I already spend an extensive amount of my time on physics 
            popularization. (a lot more than most of my colleagues). 
      \item I already spend a lot of my time engaged in public outreach, and 
            usually enjoy it. Probably that is also a measure of how much one
            should do... 
      \item I already spend a significant amount of time in outreach. Let us
            also not forget that teaching is one of our most substantial forms
            of outreach. When we leave students with the impression in our 
            courses that physics is a dead field concerned with inclined 
            planes we ensure our future...     
      \item My contribution to outreach frequently consists of writing letters
            to Sci American, NY SciTimes, American Scientist, Physics Today, 
            Atlantic Monthly, San Diego Union Tribuen on topics from whether
            Pluto is a planet, the play, Copenhagen, to cosmology. These 
            letters rarely get published! I'm doing as much as I can. 
      \item I admire my colleagues who perform outreach activities. I recognize
            my impatience in these endevours, and find it very hard to 
            participate. Outreach is very important - it is just not something 
            that I'm good at. 
      \item I have volunteered to help out with various outreach programs. 
            When I have worked I have gotten primarily positive feedback,
            and yet they rarely seem to ask twice. Other people have told me
            of similar cases. 
      \item Science seems to be in the lift again. Nevertheless, persuading the public of the interest of physics remains important. However, being a physicist has no status.

\subsubsection{Schools:}
      \item More researchers should spend time doing outreach to local 
            schools. Shape children's attitudes when they are impressionable. 
      \item The American public does not have a good knowledge of what HEP 
            does. This is especially a problem considering that scientific 
            literacy in the public and Congress is fairly low. More public 
            lectures, lab open houses, education- oriented HEP web sites are 
            needed. 
     \item Let's get schools involved. Let's make it a standard to have each
           student visit a physics lab once in her/his lifetime! Funding is 
           important, but I care more about the general public. 
     \item  Reach to the children. 
     \item I think we should help high school teachers to make physics 
           lectures more interesting with videos, public lectures, small 
           experiments, visits in the labs, etc... At least here in Europe, 
           physics is thought pretty boringly, without much attempt to 
           excite the natural curiosity of young people. And of course if
           one starts as a student to consider physics boring and abstruse,
           it is very difficult he/she will change mind later on. 
     \item If we expect funding for our field, we need to do a better job of
           getting "good" and exciting science in the hands and minds of the 
           general public. Students need to develop their enthusiasm and 
           interest at the K-12 level. We need to work with High School 
           Physics/Science teachers. We need to be more pro-active with 
           promoting physics to our congressional representatives. Letters 
           and visits to congressmen. 
      \item Education starts in school. School physics (and science in general)
            education in this country is pathetic, especially for the enormous 
            resources the U.S. has. Talk to any European and they will tell you
            that any time. This should be one of the messages to the Congress,
            and the public at large. Quarknet seems to be a good beginning, 
            but its success remains to be demonstrated. Unfortunately, it takes
            a generation to change public's attitude... 
      \item On QuarkNet I've hired two high school students this summer, 
            and all I hope for is that they tell their friends they worked in 
            the Physics department, did good work and were paid a lot. 
      \item Support high school teachers to do research on HEP experiments in 
            the summers. DoE used to do this, but quit. I worked with one of 
            these teachers and was very impressed with his enthusiasm and his
            contributions to the project. 
      \item Bring in more high school, college students and high school 
            teachers to universities and nat'l labs during the summer.
     \item We have generally been a dismal failure when it comes to getting 
           our case out to the public and to Congress. The SSC debacle is the
           best recent example. It is not clear where to start, but it 
           certainly would not hurt to have a better science educated public 
           in general, and that starts in the public schools and with the 
           public school teachers. 
      \item We need to do outreach right, not necessarily more. For example 
            several hundred thousand people/ year take introductory physics. 
            They are sitting in a room with a physics prof for about 100 
            hours. Do we do anything with this captive audience? 
     \item Outreach needs to cover all age groups. Today's K-12ers can be 
           tomorrow's physicists - but only if we manage to get them 
           interested in physics. Otherwise they miss out on the excitement
           while making lots of money... 
      \item Request to NSF \& DOE to provide funding to minority-serving
            institutions who are trying to establish a HEP research program at
            their institutions. Provide funding for HEP research in South
            Texas, a region with the largest number of Hispanic-American 
            population in the United States. NSF/DOE have traditionally 
            focused on HEP funding to mainly Ph.D granting research 
            institutions, leaving out the non-Ph.D granting minority 
            institutions. This must change.  
      \item Outreach to the public is too important to be left to the HEP 
            field. The physical sciences need a coordinated approach to this 
            issue. My personal opinion is that most of the effort and funds 
            should be expended where it will multiply the effect. That, to me,
            dictates an emphasis on training teachers at the grade through 
            high-school level and less effort on singling out gifted students
            or students from disadvantaged groups.
      \item HEP as well as other great modern physics topics need to be put 
            into the Freshman Physics curriculum, which is where most people 
            in the general public get there life's worth of physics education.
      \item Basic science education in schools must be improved worldwide. If 
            the science teachers are good, the subjects are taught in 
            interesting ways, the public and politicians interest and 
            support will be a free additional benefit. Go to the schools with              the best people! 
      \item Outreach is important and has the greatest impact when targeted
            at educators. If we prepare kids in school for the intellectual 
            and abstract rewards that HEP offers, then and only then will 
            they (and their Congresspeople) be receptive to other forms of 
            outreach. 
      \item An effort in media-ization of particle physics is needed to gain 
            the interest and maybe support of the public. Some popularization 
            at the TV needed maybe. But the most important may be to convince 
            high schools to include a sensibilization to the questions of 
            science in their school programs. 
      \item Outreach means more than asking for more, which it seems is all
            we really do. As far as public education is concerned we should be
            in the schools far more than we are. 
      \item An indirect and very long term way to help guarantee funding is to
            ensure a scientifically literate populace. One place to start is 
            to help educate those who will become primary education teachers 
            to be good at and, hopefully, enjoy science. Such an attitude can 
            help reduce the uncool perception of an interest in science, 
            particularly among young girls. I do not know exactly how to 
            accomplish this, but I suggest looking to other countries, 
            especially Japan and Korea, where the educational systems may not 
            be ideal, but young people are much more receptive to science and 
            there seems to be much less stigma in being scientifically 
            literate. These investigations are not directly a job for 
            physicists, but we should be involved. 
      \item I am a part of educational outreach already. Introducing HEP to 
            the classroom is a high priority of my collegues and me. 
     \item  I already do some outreach --- teaching various physics courses in
            local university's adult education program -- my courses are 
            specifically tailored to adult non-physicists who are interested 
            in learning more about what's going on in modern physics.

\subsubsection{Reaching Funding Agencies:}
      \item The funding outreach has been spotty. Some districts do very 
            well others make no attempts at all. 
      \item I think a wise investment would be a coordinated lobby effort, 
            with liaisons stationed not only in Washington, but also at each
            of the labs. Those stationed at the lab could "rally the troops"
            and have the sole responsibility of organizing support for 
            Congressional petitions and developing a good repoire with local 
            officials. Those stationed in Washington could be our voice. 
      \item We need to work much, much harder on members of Congress. They 
            are appallingly ignorant. We are appallingly inept in our dealings
            with them. Recently HEPAP has done something good, ie develop a 
            "briefing book" on the example of NASA. I think we spend a lot of 
            time and effort in outreach to the public, and I don't know that it
            does much good. We lack pretty pictures; histograms don't excite 
            the public's imagination. I don't know how to solve this problem. 
      \item Hire lobbyist to deal with the legislature. Seek advise from 
            advertising agencies. 
      \item Hire lobbyists in Washington. Individuals should meet congressmen
            more often. 
      \item We need to get far more physicists interested in lobbying the 
            government to support science. In parallel, we need to restore our
            reputation among the more influential sectors of the general 
            public -- this will be very difficult. 
      \item Somehow, we have to get money. 
     \item  Seeing your Congressman is important, but we should take a lesson 
            from the AMA, Gun Lobby, etc. and employ some more professional
            help to get the message across. This could be financed thru the
            large Professional Societies and should be broader than just HEP. 
      \item It seems clear to me that the current Congress has little 
            appreciation for the value of basic science for feeding the 
            country's economic "engine". It is imperative that we work harder
            to make the case for strong support of basic research both to the
            Congress and the general public. Of the G-7, the U.S. is one of the
            lowest per capita investors in basic research and unless we can 
            make progress in influencing Congress of the value of this 
            research to the long term health and competitiveness of our 
            economy, it will be very difficult for us to play a leading role 
            in HEP in the future. We must expand our interactions with members
            of Congress and the public to make them aware of the value of 
            this research and the impact that their budgetary actions have on 
            the continued viability of these programs. 
      \item A great deal of effort has, and is, going into outreach. It is 
            necessary, but I am not sure it is effecting Congress (well the 
            HEP budget has been flat for a decade). Hard go to convince the 
            public and the Congress to support HEP. 
      \item I simply do not know if we as a field are doing enough to 
            "outreach to funding agencies." I know I could do more. 
      \item You can't wait until you need money to start doing outreach. No
            one believes the words out of your mouth when you hand is 
            outstretched at the same time. HEP is a tool of foreign policy. 
            CERN has amply demonstrated the values of science as a way to 
            encourage nations to work cooperatively together for the better 
            good. The U.S. needs to understand this. 
      \item This is a tough problem, given that the funding techniques used 
            by the U.S. changes and the funding of other countries is quite
            different. 
     \item Keeping the public informed about the results of the use of their 
           money for HEP is mandatory, not optional. Physicists should lobby 
           Congress to earmark funds for this purpose, or strongly encourage 
           the funding agencies to do so. The comparison of how well people 
           know about what NASA does with their money to what they know about
           what HEP does with their money is not accidental. It is because 
           High Energy physicists in general do not communicate well, so the
           answer is not to put High Energy physicists on outreach, but to 
           put professionals on the job like NASA does, and pay for it. 
     \item MOVE THE APS MEETING BACK TO WASHINGTON. 
     \item In order to do outreach on the level required to make a significant
           impact the funding structure must be changed to allow resources 
           to support these efforts. 
     \item I don't think that high energy physics has much clout on the 
           political end; as for public interest, it doesn't seem to be very
           high; I think the field makes itself a little bit inaccessible in
           general; clout will come with interest, though I don't have many
           thoughts on how to engender an interest in it, other than 
           continued efforts to inspire human interest in the fundamental
           constituents of the universe.
      \item Have better people in the DOE Office of Science. Rosen \& O'Fallon
            are poor communicators with OMB, Congress and physicists in the 
            field. 
      \item To maintain funding levels we must communicate with the people why
            we should be supported. If the people don't support the science 
            the governments of the world have every reason to continue to 
            apply cuts, slowing existing projects by years. 
      \item Since it's a fact that the government caters to the desires of 
            corporate America far more than to the concerns of the public, we 
            should step up our efforts to get the so-called "power lobby" in 
            our corner. The role of public outreach should focus on attracting
            bright people to our field (and retaining them) in order to ensure
            a healthy future. 
     \item Do not neglect outreach to corporations. 
      \item HEP has failed to answer adequately the question of WHY society 
            should pay for HEP. The reason for this is that no one in HEP has
            learned enough about recent social science research to be able to 
            understand the long term impacts that HEP is likely to have - 
            especially through future career switches by senior physicists 
            now in HEP - and how to explain these impacts to Congress and the 
            public. 
      \item The public would be far more supportive if they knew what we were
            doing. That has to happen before they can be convinced it is worth
            their tax dollars. 
      \item Many of my friends and acquaintances know a little about HEP 
            (usually by association with me) but few folks I meet know what 
            the HEP labs do or why this research is important. We, as 
            physicists, need to sell our field more to the public and garner 
            their support before we can ask Congress to increase support for 
            the field. Particularly now when we anticipate building a new
            facility with an expense of many billion dollars.
     \item I think the government would embrace physicists and their projects 
           more if they included as part of their project outreach to the 
           general public. Politicians often don't understand what the return
           on their investment is, which allows them to rescind their funding
           much more easily. I feel strongly about the commitment we ought 
           to have to educating the general public about what we do. 
     \item Write your Congressman. Write the local paper. Tell them what you
           do and why it is important. Have open houses at your local 
           university. Invite your reps to visit your favorite national lab.
           Above all, realize the public is interested in science and isn't
           stupid. 
     \item I believe that the recent coupling of public outreach to funding 
           is a disaster that needs to be overcome. It is naive to think that
           any election result is likely to be much influenced by the question
           of funding for particle physics. The politicians need to 
           understand the long term benefits of funding basic research. The 
           general public should also be taught this lesson but that is a 
           huge effort which has more to do with the education system than 
           efforts on the part of physicists. 
      \item What is the difference between "not nearly enough" and "far too 
            little"? SSC had a very strong public outreach program in K-12 
            education. This served the role of helping the country as a whole 
            to improve science education, and it also gave name recognition to
            the project. Doing more of this would be good, but DOE has 
            restricted the use of funds making this tricky. Labs might use 
            some limited discretionary funds in this manner, but the scale 
            may be wrong. 
      \item Consider allocating funding explicitly for outreach. In 
            particular, well paid and larger PR groups in the national labs 
            would be a good start. However, do not neglect the important 
            influence of the universities.
      \item Outreach in a sense can never be enough. However a balance has to
            be found for the time spent on these activities. In general 
            however labs / funding agencies have to realize that outreach 
            costs money, and have to be more willing to fund such activities. 
      \item I am not well enough informed. In fundamental research,
            independence from "money givers" is vitally important! 
      \item Society gives us as much funding as it understands of the reasons
            of our research. Thus, if we don't do the job of informing society
            of the importance of what we are doing, we should not be 
            astonished about the lack of funding we receive. 
      \item HEP needs to work out moral evaluations of how much money / 
            effort the public should be asked to donate to the uncovering of 
            given pieces of knowledge. e.g. is the supplementary knowledge 
            that an NLC could provide worth \$5-10 B? Perhaps the public 
            could be involved in the process. 
     \item  We need to come up with a better plan for selling HEP research to
            conservative administrations. We need to showcase not only the 
            'sexy' physics, but also immediate (or near-immediate) 
            applications of developed technologies. We also need to show how 
            the U.S. will lose much more than just the science edge if U.S.
            HEP funding continues to be below the rest of the world community. 
      \item I think many scientists regard the money we get from society as
            something due, and I think it's wrong. Our societies are making 
            not negligible investments in HEP and we should be able to explain
            why (even if it is more a cultural than pragmatic reason). 
      \item There are many claims on public monies and we do not have an 
            entitlement. It is absolutely essential that we explain our claims
            to both the public and the funding agencies. 
      \item The community needs to understand that billion dollar facilities 
            are too expensive to be an entitlement for scientists. Our gloss 
            from the atomic bomb is long gone, and we are viewed as just 
            another constituent group which wants money for its pet projects. 
            We need to get the message across that fundamental research, even 
            if it doesn't make a better mouse trap, is very important to the 
            future health of the scientific and technological base on which 
            our modern prosperity is based. 
     \item Physicists of the present generation tend to feel entitlement. 
           The world owes them a living for their hard work. My generation 
           (over 70) never felt that way and all of us fought to keep the 
           budget up. We must get support from University Presidents as we
           had for FNAL in 1965 and as we had at the start of SSC (but we 
           lost it by the end).
     \item We need to completely eliminate the idea of entitlement so 
           ingrained in the DOE. Both at the agency and grant level, people 
           think that they should receive basically the same funding that they
           got last year. This has got to change. NSF is better in this 
           regard. 
     \item The relationship between globalization and outreach to funding 
           agencies should be carefully considered. The agencies need to be 
           sure they will get something from the investment. 
      \item Two comments: 1.) I do lots of public outreach in addition to 
            working with students. 2.) Although I believe we should keep our
            funding agencies, Congress, etc. well informed, I don't think we 
            should expect too much from this. We tend to be rather arrogant in
            thinking that once people understand what we are doing, they will 
            give us 100$\%$ of their support. This is unlikely in the face of 
            tough decisions that funding agents and politicians must make. 
      \item This has to come from each of us; are we worth the public support?
            I VISITED 19 congressional and senate offices last time in WDC 
            making this case - Orchestrated formated contacts come across as 
            phony at least to me. Do YOU have vision of our role in U.S. 
            society as a scientist and the importance of the scientific 
            enterprise? 
      \item I have absolutely no experience reaching out for money. There are
            people who are very good at this even though they are primarily
            physicists. Maybe these people should get a full-time position 
            seeking funding. 
      \item I frankly cannot judge if enough outreach to funding agencies 
            exists; it is evident though that this outreach is not 
            sufficiently effective. 
      \item Outreach to the general public is by far the most important in the
            long term. A lot more needs to be done to open up and justify 
            future projects. If we have trouble justifying the cost of a 
            future project, then this suggests a failure on our part rather 
            than the ignorance or stupidity of the public. 
      \item It just happens that I am in almost daily contact with 
            Congressional aids. This is not a steady state.
      \item What it takes to reach the funding agencies has not really been 
            done and it relies on a creative idea of a few. A serious 
            systematic study of what it takes to get the level of financing 
            and interest from the Congress does not seem to have been taken 
            yet. The effort of information to the general public is not 
            relying in real knowledge on what it takes to get the general 
            interest and is still driven by a somewhat aloof attitude 
            (big progress has been done though). 
      \item Spend more money on it so we can get on-going national outreach 
            programs.
      \item We need more federal funds spent on science. Present government 
            has squandered the surplus on a huge tax cut for the wealthy 
            rather than investing in the nation's future. This is the 
            fundamental reason U.S. pure science is in decline. That includes
            HEP. 
      \item To maintain a dynamic equilibrium in the field, we need consistent
            and predictable funding. Similarly, if the field is to grow, we 
            need a well-defined funding profile which will allow for realistic
            planning. Even if we are going to lose funding, we need a 
            realistic profile so that we can plan effectively. Unfortunately,
            the U.S. government lacks a history of accurately projecting 
            future funding levels, and seems to have little interest in 
            doing so.

\subsubsection{The NASA Model:}
      \item Learn from NASA; attract topmost people in science to public work. 
      \item NASA and biology is much more adept at getting results (even ones
            that are not truly "new") reported. Even what we consider the most
            mundane HEP result can be presented in a way to the public. The 
            APS should do better in this. 
      \item NASA does a good job at outreach. We should try to emulate them. 
            But it is easy to say that we should do more outreach, and it is 
            even easy to say that I personally would be willing to do more of
            it (although I already spend some of my time on that). There are 
            some obstructions that need to be overcome. One is that DOE 
            specifically prohibits the labs from the type of outreach and PR 
            that NASA has. Another is that there is a general culture in HEP 
            that could almost be called anti-outreach, or at least "outreach 
            is fine, as long as I don't have to do it". This inhibition is 
            often quickly forgotten when it is time to publish one's somewhat
            over-hyped results in the New York Times. 
      \item NASA and the Hubble Space Telescope are good examples of groups 
            that have reached out to the public. It will be necessary for 
            high energy to do this if it wants multi billion dollar support. 
      \item Emphasize what has been learned about "inner space" and tie it 
            with the cosmos. NASA has good public awareness, even though 
            they've had some well-known failures. HEP does not. Astronomy is
            more glamorous because the sky is right up there for all to see. 
            Maybe we ought to hire a Madison Avenue PR firm. 
      \item NASA manages to make much of little, e.g. showing glitzy computer 
            animations of the Ulysses probe flying by the Earth. As scientists
            we tend to shun glitzy presentations, but these romanticized 
            presentations appear to work in inspiring the public's 
            imagination. (e.g. the merit-less International Space Station is
            handily gobbling up a budget of a size comparable to the 
            canceled SSC.). We need another Carl Sagan's "Cosmos" to do for 
            HEP what he did for astronomy. 
      \item NASA has shown over the years how it is possible to convince a 
            broader public how important and in the national/global interest 
            it is that a resourceful civilized nation support basic research. 
            We'll have to learn how to bring our concerns to the public in a 
            way that convinces others that we are talking about the public 
            concern. Nobody enters our field just to "make a living", or to 
            "make money". Let that message sink in, and let us try to 
            convince the public on whose payroll we work, that we are doing it
            in their best (defining understandably what "best" means) 
            interest. This is by no means easy, but it is clearly incumbent 
            on us. 
      \item Get more physics into the news media. (NASA seems to do well 
            there.) 
      \item If you want to see outreach, watch what the astronomers do. 
      \item Astronomers beat HEP physicists hands down when it comes to 
            outreach. What have we done to compete with a Hubble picture? 
      \item Follow the example of astrophysics who are more efficient in 
            their outreach. 
      \item HEP has got much to learn from the outreach work carried out by
            the large astrophysics project laboratories plus NASA etc.. 
     \item I already spend a significant fraction of my time on outreach 
           activities. The big problem is that it is difficult to provide the 
           public with easy to understand visual images that can compete with 
           astronomy. 
      \item I think that this area is probably the one most neglected by the
            community as a whole. We should really make an effort to
            understand issues like" Why can NASA get funding for a \$100B 
            space station, while there is a feeling that a \$5 B project will
            somehow handcuff our field? What is it about the space program 
            that captures peoples interests? Can we capitalize on that 
            curiosity to help promote our field? What needs to be differently? 
            I have several ideas on this subject. First of all, there seems 
            to be a perception that taking some time from research to spend on
            outreach will be detrimental to a young physicists career. We 
            need to understand as a community, the value of outreach, and make
            sure that it is recognized when decisions are made about 
            hiring/tenure/ etc.? We need to try to appeal to the 
            public/ Congress on a more visceral level. It is one thing to 
            explain the importance of basic research. However, these kinds of
            arguments seldom seem to have the same effect as showing someone
            a beautiful picture of a supernova from the HST. More thought
            needs to be put towards understanding how can we best capture
            the beauty and excitement of our field. 
      \item We must figure out how to do outreach to the public as effectively
            as the astrophysics community does. Without the pretty pictures, 
            this is at least difficult and maybe impossible. 
      \item The field of particle physics could become the field of particle 
            astrophysics. The sum of HEP + astrophysics funding + coordinated 
            outreach to the funding agencies could equal more than the current
            base for HEP + astro. 
      \item Above is for astrophysics, which (as was pointed out) "captures 
            the imagination." Astrophysics gets a lot of press, and the 
            astrophysics community seems to have more people willing to 
            interact with the public. I feel it is important that the public 
            get the chance to interact with "real scientists" -- all too often
            we depend on the press to mediate interactions which then become 
            all too far removed from what we are actually doing. Scientists 
            are real people and should be capable of communicating results to 
            the public (where often they are incapable of communicating even 
            with their colleagues). I dare say they even have a responsibility
            to do so.

\subsubsection{Survey-Specific Comments on Outreach:}
      \item More U.S. specific questions. 
      \item Not relevant for foreigners. 
      \item This appears to only be relevant to U.S. physicists. 
      \item Seems like "we" means USA here? 
      \item Answers apply to Europe. 
      \item Again, I can't comment about issues relating to funding from
            the U.S. Congress. 
      \item NA, I am not an American citizen. 
      \item I don't know about the outreach program to funding agencies
            in the U.S... 
      \item My answers apply to Europe/Germany. I don't know the U.S. 
            outreach activities too well! 
      \item Question 21 is naive. Reaching higher levels at DOE and reaching 
            OMB is as important. 
      \item The answer to question 21 requires a level of political 
            understanding that neither I nor, I suspect, 99$\%$ of the physics
            community has. 
     \item I don't know the current level of outreach in order to answer 
           these questions. 
      \item "Not nearly enough" and "far to little" are not exactly distinct. 
     \item "Not nearly enough" and "far too little" mean the same to me! 
            Outreach (both to the decision makers and the general public) is 
            clearly extremely important. However, I've declined to answer this
            section because as a European I have little knowledge of what is 
            done in the U.S. 
      \item I think you need to make a clearer distinction between lobbying
            and education. I'm not sure it's appropriate to lump them together
            as you do here - outreach to funding agencies is really is a 
            political question and should be treated as such. Simply educating
            your congresswoman or -man about your neato-science doesn't 
            translate to \$\$\$, and we shouldn't expect it to. 
   \end{itemize}

%% file: comments-building.tex
\pagestyle{myheadings}
\markboth{Comments on Building the Field}
         {Comments on Building the Field}
\subsection{Building the Field}
   \begin{itemize}
\subsubsection{Timescale for Projects:}
     \item The field is aging and young people are essential for the new 
           ideas and perspectives that they can bring. I am afraid limited 
           budgets the extended time that young scientists spend considering 
           how to secure their next job reduces the overall vitality of the 
           enterprise. 
     \item It's frustrating how long it takes to participate in and get data 
           from an experiment. The scale is now 10 years/ experiment. If 
           you're unfortunate to be near the beginning, you may never get 
           to see any physics data! 
     \item As an undergraduate myself, the most disconcerting aspect of 
           particle physics is the time investment required of the scientists.
           They talk of the VLHC being finished in 2030?!? I'll be fifty then!
           Unless something changes, young people are going to be turned away
           because they don't want to be locked into a path. 
     \item Hey. Everyone knows we can get more cash in other work. We still go
           into this field because we like it. I can trade off pay for doing
           this - it's great. However, the time scale of these large 
           experiments is getting too large to continue putting PhD's through 
           as we are doing. Universities need to think about their 
           requirements for PhD's - otherwise how can anyone think of entering
           things like the LHC right now if they know they probably won't get 
           a university-required publication for 7-8 years ? 
     \item I think the ever increasing size, cost and timescale of experiments
           is driving people away from the field. Unfortunately, this is not
           such an easy thing to change. 
     \item A mixture of astrophysics + accelerator based physics in a single
           field could encourage more young people to enter the field. 
           5yr-time scale astro projects could balance 20-yr time scale new
           accelerator based projects. 
     \item If the field is not making much progress in physics then it will
           not be exciting to young physicists. (Your options for question 28
           could have included "inadequate rate of physics progress".) 
     \item Lack of control -- What do you tell a student now who sees that we 
           are already designing or building machines (LHC, NLC, VLHC) that 
           will span the next 30 or 40 years. If the machines are set, then 
           the questions we wish to ask have already been decided. Yes, you 
           can make an impact 'inside' an already agreed upon direction of
           physics inquiry, but there will be no opportunity during the span
           of an entering student's career to embark on a new track and 
           actually see any result. 
     \item Long project cycles are a big problem. 
     \item In question 28, I worry about the very long time to mount an 
           experiment and to get results, relative to the 5 years a student
           should spend in graduate school. 
     \item No facility! Time scale for results. 
     \item Experiments take too long. 
     \item Bad Morale: Growing size of and timescales for experiments are
           putting larger premium on social-group dynamics that reward those
           who are more politically savvy. In an environment that already 
           makes it very difficult to have some semblance of security, this 
           inevitably lowers morale. Such an environment also suppresses
           original/creative thought that perhaps cannot be expressed well at
           its inception. This then drives a somewhat "monolithic" approach 
           to the science. 
    \item  I think people are also influenced to leave by the lengthy and 
           uncertain timescales for future projects. I doubt that the average 
           time to get a Ph.D. in HEP is significantly different than in
           other areas of physics. But under even the most optimistic
           scenario, a linear collider, for example, is still at least ten 
           years away. 
     \item Management is also a very big issue, and more than salary, it is 
           the image that you need 5-7 years of postdoc (additional servitude)
           before becoming a scientist. It is hard to justify being 34-40
           before starting a tenure track position. 
     \item Long time scale of experiments from conception to first 
           publishable data. 
     \item The pace of discovery in HEP has slowed considerably in the last 
           decade and has probably discouraged young researchers from pursuing
           a career in HEP. 
     \item The time from entering grad school to become a faculty member is a
           way too long. Each step, i.e. grad student $\rightarrow$ postdoc 
           $\rightarrow$ faculty is very important; but the phases are too 
           long. Shortage of permanent job opportunities are to blame. 
     \item The pace of discovery has been too slow. 
     \item Large time scale of HEP projects may influence young physicists
           to leave the field. 
     \item Duration of big experiments is approaching human lifetime. In this
           prospective a HEP experiment is no better than work in a company, 
           with lower salaries and long hours. 
     \item As an "old geezer", I'm not a good reference for these questions. 
           What I find most discouraging for young physicists is the fact
           that the time scales are so long that they cannot see a whole 
           experiment. They gain very little hardware experience, and we may
           end up with a new generation which could not build a new detector
           in 2025 even if given the opportunity. Additionally, the inputs 
           students have in an experiment is very limited. Rather, they have
           a hard time doing anything but work in a corner of the big picture
           and use a large amount of hard- and software which they don't
           understand at a fundamental level. 
     \item Large experiments imply large collaborations and also imply a long
           time between initial proposal and final published results. For
           persons to feel they're contributing, the time has to be less 
           than 5 or 6 yrs. 
     \item Slow rate of innovation plays an important role in many people's 
           disillusionment with the field. 
     \item Another feature that would influence me to leave the field is the
           amount of time it takes to get funding and then to have the 
           experiment completed. 
     \item Although personal reasons drive this the most, the biggest concern
           is that there may be no physics out of a new post-LHC facility for
           25 years. A young person coming out with a PhD in particle physics
           today may wonder if *any* machine will be built after the LHC. The
           community needs to come to a decision about what to do after the 
           LHC on a time scale that is not beyond the career lifetime of a
           new PhD graduate. Postponing the start of construction of the next
           e+e- collider until after the LHC would result in many young people
           leaving the field. 
     \item The length of time in graduate programs is also a negative factor 
           for many students. 
     \item The total amount of time new particle physicists are required to 
           spend in grad school and then on post doc positions is just not 
           justified. It feels as if it was a way for universities to find 
           cheap labor... 
     \item Main concerns: long time needed to graduate and to establish 
           yourself; limited number of permanent positions. We need to speed 
           up promotion of the best young talents in the field, increase
           rotation at the management positions, operate with more realistic 
           schedules. 
     \item The length of time spent in graduate school needs to be minimized, 
           yet the student should still get both hardware and software 
           experience. They need not work on one experiment from start to end.
           Perhaps they can work in accelerator physics for hardware 
           experience and even analyze data from previous experiments for 
           software experience. There is certainly a lot of good data out 
           there that could be analyzed. I think if graduate school is not 
           so long ($\sim$ 5 years or less), then the issue of pay afterwards 
           is not so bad as you start earning money earlier in you life. But
           it must be clear that there are permanent jobs that are obtainable
           on a reasonable time scale. Not being able to get "tenure" until
           one is in their mid to late 30's if pretty crazy. One is typically 
           trying to raise a family at that point and to have so much at risk
           that late in life is not a good thing. Perhaps the whole system of
           tenure needs to be re-evaluated. 
     \item When I think about my future after finishing my last degree the 
           biggest reason for leaving HEP is the collaboration sizes. Another
           factor is that the timescale between experiments is so long.
           Running ATLAS for a bunch of years, and then building a
           super-dooper-NLC just doesn't seem that exciting. By the way, what
           was the deal with questions 25 and 26??? Lifestyle/Hours, 
           possibility of fame, allure of working in a HEP lab??? You've
           got to be kidding, right? 
     \item The biggest problem we face is the length of time required to mount
           \& perform experiments. This, coupled with the need to establish a 
           "physics" reputation in order to get a permenant job, drives many
           talented people from the field. 

\subsubsection{Leadership and Large Collaborations:}
     \item HEP moves too slowly and in the USA, leadership is lacking. 
     \item Field desperately needs more leadership, fewer useless meetings. 
     \item It is important to have scientific leadership which present a 
           balanced and focused physics program for young physicists. 
           Current sociology in HEP is too fad-driven. 
     \item Lack of vision and leadership from senior people, and the 
           reluctance of younger people to assume the role. 
     \item Lack of jobs and weak management hurt the field. 
     \item I believe that the deplorable state of present management is the
           main reason for young physicists to leave the field. The lack of
           professionalism in the working place, as well as the incompetitive
           salaries and lack of growing opportunities are merely consequences.
           The field simply is in dire need of influential people with vision.
           With the decade+ timescales for new experiments, scientific 
           spin-off needs to be implemented by design in every stage of
           development. 
     \item We are losing many of our best young people due to poor management 
           (both in the experiments, and at the national labs). 
     \item Lack of permanent job opportunities is closely connected with
           weak management. 
     \item Some problems: oversized, over managed collaborations with too
           little impact of every person; uncertain future (employment, 
           continued viability of field, openness to new directions). 
     \item There has been too much politics in HEP. Good/best persons often
           are not recognized for their contribution. Many people take credits
           without doing work. Too many people just promote their students, 
           employees and friends.
     \item Regarding young physicists leaving the field: one comes with the 
           idea of working to understand nature's most intimate workings, and
           discover (s)he is the most insignificant gear of an enormous 
           machine whose control is often dictated less by physics motivation 
           than political/ sociological/ budgetarian/ schedule, etc. reasons. 
     \item Most of the problems in question 28 result from the poor 
           management I selected. 
     \item 'Weak and unstructured management' - a new form of international
           management is needed. Otherwise the funding and the plans for 
           extended facilities will be refused by any society. 
     \item The main problem is the existence of an aging generation of 
           physicists with poor managements skills and overcommited to many 
           projects. As a young physicist I hate to see all these professors
           in the mainstream academic institutions to waste millions of 
           tax-dollars to improve their political positions and do 
           unauthorized research. Things can drastically change if: a) vanish
           tenure positions b) hire managers/ physicists to manage money and 
           people c) force university groups to handle graduate students and 
           postdocs as physicists in training not cheap technical manpower d) 
           teaching professors not allowed to be involved in research projects
           while they maintain teaching obligations or administrative tasks in
           their universities. Skillful and talented people are leaving fast 
           while the old guys are proud for eliminating the competition. 
           Finally, young physicists should worry about the problem of 
           thousands of people putting their names in experiments that they 
           do not even know what the initial mean. It should strictly 
           discourage university teams to split peoples time into multiple
           projects. 
     \item The working atmosphere in HEP has changed. With larger 
           collaborations, the amount of "corporate politics" has increased.
           It's not a large group of smart people working together anymore but
           rather may small groups scheming for their particular interest.
           Academic freedom is diminished, it is almost impossible to do 
           research in a non main stream field and succeed. Having to put up
           with this makes "normal jobs" that are also by far better paid look
           more attractive. 
     \item Current generation of HEP'ists slowly adopt to new globalized 
           working conditions. Once we managed, collabs will be virtually
           smaller. So that's fine. BUT as long as people still think that 
           HEP is about bombs we're in trouble. See "outreach to schools". 
     \item My first publication had 6 authors my present publications have 
           over 400 authors; my LHC publications will have over 2000 authors,
           and precious few of them are young physicists. 
     \item Question 28 is a guess (that large collaboration sizes is the
           biggest influence away from HEP), but I hear people talking about
           that the most. There's less glory. But rather than playing that
           aspect down, we should consider it an interesting twist: not only 
           are we studying nature, but we represent an innovation in the way 
           physics is done. Communicating well and sharing work seem almost 
           counter-intuitive to us (our major problem here, myself included, 
           is a tendency to border off a project and take it over completely).
           There are a lot of things that established organizations do much 
           better than us (organization, less complete dependence on any given
           individual, managing personality conflicts, etc.) --- we're still 
           beginning to learn how to merge organization with physics goals.
           And if we can do this on a world- wide scale, physics will be less
           of a beast plodding along accidentally and more directed toward 
           human needs and rational goals. 
     \item The number of people we retain in this field is appropriate. The 
           problem is that we don't retain the best people. One of the
           problems with large collaborations is that it is the politically
           savvy ones who tend to stand out and are rewarded. We keep people
           who give slick talks and lose people who actually make experiments
           run. 
     \item I see people leaving my experiments so that they can have more 
           control over their lives. Being a cog in a large project offers 
           little stimulating professional development, and can leave the most
           talented looking for an area where they can have impact and 
           control. Old guys (like me) should develop, encourage, then empower
           the young. Don't expect them to be junior colleagues for the next
           30 years. 
     \item Large collaboration sizes are a bad thing: since joining D0 (I was 
           on a smaller experiment before) I have been amazed at the amount of
           effort which is spent on management and politics. Not that I think
           D0 is over managed but large collaborations need this overhead and 
           its not really what we are here for. It also makes it far harder to
           affect the course of the experiment in a big way: there is too much
           inertia with large groups of people and you are unable to try out 
           some "neat thing" you just thought of without extensive 
           justification. These "neat things" could lead to important 
           discoveries ... or they could be a complete waste of time but 
           large collaborations are generally less likely to take risks. 
     \item People leave because the factors that brought them to the field are
           no longer there. These I think are scientific curiosity and a sense
           of being able to contribute in a meaningful way. Large
           collaborations deprive people of their ability to feel part of a 
           scientific effort. They rather feel as part of a group in which 
           their contributions lack any individuality and are simple submerged
           into the group effort. 
     \item The greatest concern physicist are facing and the major reason they
           are leaving the field is the general decay of HEP. Lack of
           experiments. Poorly planned experiments, based on political
           decisions and non on physics potential. The consequences of all 
           these (loss of funds, less positions, long time for obtaining 
           results). 
     \item You need to create social organizations with which people can 
           identify. No one bonds to a group of 500 collaborators. The 
           typical ant colony aspect of collaborations in which there are no
           meaningful social organizations below that of the collaboration is
           a real killer in terms of getting people to hitch their wagon to
           the work. Also, the field needs to get over the idea that the only 
           worthwhile life is one in which one is married to the field. You
           need to create an atmosphere in which people believe that it is a
           good experience to have and then to move on. 
     \item I think that large collaboration sizes and length of time spent in 
           grad school both deter pursuit in HEP for some students. I
           understand the appeal of working in a smaller collaboration in a 
           different field. Also, the uncertainty of future funding for 
           projects (as they are often tied to governmental whimsy) creates
           some sense of job insecurity for some students--of course, this is
           unavoidable in most fields of physics. 
    \item  Problems which I have seen hurting the grad students include:
           Isolation: Someone gets stuck in a 'job' which takes years to 
           finish, and does not experience the other aspects (and people) of 
           the field. Politics: In a huge collaboration, there is no 
           connection between service work and research opportunities. I've 
           seen topics divvied up according to the prestige of the advisors 
           rather than the interests and capabilities of the students.
           Prospects: There are only so many tenure-jobs possible in academia 
           without vastly increasing the number of students taking physics. 
           There are only so many 'permanent' jobs at the national labs. The
           experiments require many more people than there are prospects of
           permanent jobs for. Despite the reported charms of being able to
           slip from job to job, those of us with families see problems with
           that lifestyle. No, most of these have not hurt me; I just mention 
           what I've seen that looks serious. 
     \item Young people will leave the field if it persists on emphasizing
           politics, and other extraneous factors, rather than science. 
     \item I think people like being in HEP because of the academic and
           somewhat anarchistic atmosphere. They are willing to get paid less
           in exchange for the increased pleasure of their work. HEP is 
           moving in a less anarchistic management style, and this is causing 
           some of the better people in the field to leave. They think that 
           if they are going to have to deal with a rigid business environment
           anyway, why not go somewhere else and get paid more? 
     \item I think that the size of the experiments is a disaster in that it 
           reflects the overwhelming problems we try to take on -- and do a
           poor job of solving. Much of our physics results actually does not 
           get published because people simply cannot reach the data -- 
           because they are working in an unprofessional manner and are too 
           arrogant to believe that they can be taught better by a 
           non-physicist. Options are to have a better idea or accept that
           things cannot always be hacked together. The lack of 
           professionalism results in a very discouraging and depressing 
           atmosphere. 
     \item Working in a big collaboration, where you don't even know your 
           collaborators. 
     \item Personally I think when we have 2000 people on an experiment, we 
           have too many people. To keep more people the field has to have 
           a clear vision of where it is going and why. I think that is 
           missing right now. 
     \item In large collaborations most young physicist work as part of a
           team doing a tiny fraction of the experiment, so: a) do not have 
           an overall knowledge of the experiment b) do not feel they work 
           as a "scientist" c) there will be a lack of leadership with broad 
           knowledge of experimental HEP in the near future. 
     \item The collaboration size in itself is not a problem. The huge amount
           of time consumed by secondary tasks like computing and 
           organization is annoying. 
     \item If collaborations become too big and the individual has less 
           impact, a compensation in form of early positions and better 
           salary is needed. 
     \item Long-term projects of large collaborations should be cut in smaller
           pieces to give students a chance to making a contribution. 
     \item We need more flexible solutions in terms of positions. I would also
           like to see a much more professional management of the people, for
           instance reward the ones doing an outstanding job (and NOT by 
           giving them even more tasks to do!!!) but also sanction the ones 
           not performing correspondingly! So far, the latter class of people 
           do not directly feel the consequences of inadequate performance, 
           but rather their jobs are given to the ones working very well! 
     \item Poor management comes in a very close second. A more professional 
           environment with fewer egos and less beauracracy would help to 
           drive fewer young physicists into industry. The salaries will 
           never be competitive, so the environment has to make up the 
           shortfall. 
     \item Giving adequate acknowledgement of individual excellence without 
           mixing in politics of institutions, etc. 
     \item Although I do not expect to stay in physics in a direct role, I do
           think it is important to retain most students. Money is an issue,
           and as for HEP, I know many people are concerned about the enormous
           collaborations. A friend was recently making fun of Physics Nobel
           Prizes because there is a perception that HEP Nobels, at least,
           really go to the best manager, rather than the best scientist. 
           Fair or not, this attitude can be corrosive for highly talented
           students who do not have an interest in administration. 
     \item Question 28 is difficult. I have chosen management [at all levels]
           because it's responsible for the other woes. We MUST find a way of
           giving the young scientists more responsibility at an early age -
           I was running my own group two years after my Ph.D. 
     \item Young physicists leave (or are upset at least about) for the 
           unscientific way decisions are taken ("politicking") , and by poor
           management of scientific projects. Kicking and boxing is felt to
           be more important to obtain a good permanent position rather than
           scientific quality and correctness. 

\subsubsection{Salary and Benefits:}
     \item I think salary and the bureaucracy inherent in large collaborations
           are the driving factors in physicist attrition. I don't see any way
           to resolve these issues. 
     \item Salary is no doubt another factor. A software professional earns
           much more than a Physicist, and the nature of work he has to deal 
           with is nowhere as challenging as Physics ... I can say this with 
           conviction because I worked in software for 2 years. 
     \item Postdocs last too long and salaries are too low. If you have a
           family, it is almost irresponsible to stay in the field because
           of such low pay. 
     \item I do not believe a young physicist can support a family on his/her
           own. 
     \item Lack of financial security. 
     \item Postdocs need a 50\% increase in salaries, junior faculty need a
           30\% increase. Senior faculty need a 20\% increase. 
     \item Low pay in academia is a major minus. 
     \item VERY long working hours, stress and lack of financial compensation. 
     \item I think the salary situation is more important than most people 
           give it credit for. I for instance do not know if I can stay in 
           HEP any longer and at the same time have children to support. I 
           may have to sacrifice my interest in academia to be able to 
           support my family. 
     \item The salary issue is a big issue, as the job is also very long
           hours, and few benefits. To encourage people to work in the field, 
           probably the most effective way would be to emphasize the valuable
           experience, and the travel opportunities. To get and retain the 
           appropriate technicians/ engineers, I think that some sort of 
           career structure is the most pressing need. 
     \item A professor working in HEP is synonymous to a poor person! 
           Intellectually is so exciting but it is very difficult to make 
           ends meet. The young people see it immediately and only if they 
           are independently wealthy they don't bring up the money issue. 
     \item With \$100k starting salaries commonplace in high-tech and other 
           industries, who will wait for a department to decide on offering 
           you a postdoc position which needs to be renewed every other year?
           Departments need to commit to longer term contracts with reasonable
           benefits and speed up their hiring process. 
     \item Lucrative dot.com jobs caused talented HEP physicists to leave 
           the field. 
     \item Money will always play a large role in the retention of any field.
           If a person does not have the right personality, they will have a
           tendency to go where the money is. The best way to improve 
           retention is through the recruitments of graduate students and 
           targeting those at the beginning that would be more likely to stay.
           Circumstances can always change this but I don't feel there is
           much more to be done. 
     \item Salaries are no longer competitive, assault on the institution of 
           tenure continues, postdocs who are in their 30's are earning 
           teenagers salaries. 
     \item Let young people know about the goals and tools of HEP physics. 
           Then it's a "personal path" to HEP. To stay in the field, permanent
           jobs and adequate salaries are necessary. This is what concerns me 
           the most. 
     \item Also for question 28, lack of permanent job opportunities may cause
           young physicists leave. 
     \item I can stay in physics and work 12 hours/ day 7 days/ week for a 
           grad student grant. Then I can go into industry and make piles of 
           money for less time, or I can stay in physics as a postdoc and 
           still work 12h/d 7days/w and get less than half the money. I would 
           love to do a postdoc, but I dont want to do it 24/7 and given the 
           lure of industry, it looks like I will leave the field. 
     \item I think you have to accept that the allure of cash will always draw
           away a large number of talented physicists. Future issues - money 
           and the diminishing possibilities of independent work. 
     \item I was attracted to HEP because the people were exciting and 
           intellectually vigorous, and the field seemed fundamental. I think
           young people leave the field because they don't see a clear career
           path line. The opportunities don't look great, and the salaries 
           don't justify a so-so job. Better to have a so-so job in the 
           software industry and make lots of money. I can't blame them. 
     \item Most new PhDs in physics I know have moved away from physics 
           research as a career, either to pursue jobs with a higher starting
           salary or because they were no longer interested in pure research. 
           However, few of them have expressed regret about studying physics.
           As well, even if not all new PhDs stay in physics, there seem to 
           be enough who want to stay to fill the positions available. 
     \item Money. The skills we acquire would earn us serious money in, for 
           example, the computing industry. Given the reality of working in 
           HEP, i.e. sat in front of a computer for vast portions of the day,
           well ... 
     \item Wish I knew ? Increasing the salaries must help. 
     \item Lack of permanent job opportunities coupled with low salary is
           critical combination. CEO change positions often, but have 
           sufficiently large salaries to make it viable. 
     \item The funding for life sciences basic research was doubled over ten
           years. We need to do the same. 
     \item Salary questions. Very hard to get somewhere in large 
           collaborations. Difficult to get tenure and financial support. 
     \item Many go to computing for money. 
     \item When I was starting out, I had confidence that the field would be 
           funded in a steady, rational way. I think the SSC cancellation and
           the scientific ignorance of Congress and the administration 
           trouble young people who balance a HEP/academic career with 
           opportunities in the private sector. 
     \item Salary seems to be the most cited reason for young physicists
           leaving the field. However, I am convinced that this would not 
           count so highly if those physicists had not first become
           disillusioned with the working environment. After all, they must 
           have known at the outset that this career is poorly paid, but still
           made the decision to join. I am in a minority group, being both 
           female and a mature entrant to the field. The issues which have led
           me seriously to consider leaving have been the working conditions
           rather than the substantial salary I could earn outside. The field 
           is most definitely male-dominated and aggressive in nature. 
           Everything is geared to pointless races and competition, both
           intra- and inter-collaboration. I despise the internal politics
           and petty power games, and had expected so much better of an 
           academic environment. In academia, if nowhere else, we should 
           reasonably expect a meritocratic system. 
     \item While lack of a competitive salary might be the main reason for 
           young people to move, I do not think we can compete in that area. 
           Instead, we should aim at the smaller group that leave due to lack 
           of longer term positions or general tiredness of the big 
           collaborations. Those issues can be solved maybe by having longer
           contracts and by giving regions more power in experiments. Many
           people grow tired of the micro-management where all decisions have
           to go around the top of the collaborations. 
     \item You might like the physics, studying long years helps you to keep 
           believing in what your are doing but one day you might be bored 
           eating noodles and live poorly waiting for a post. The modern life
           could give you a better way of living. 

\subsubsection{Attracting Students to HEP:}
     \item HEP is an extremely hard sell to incoming science students. You
           tell them that you can expect to spend 6-8 years in grad school,
           earn little money as a postdoc and then have absolutely no job 
           security. Why not just go straight to industry? Retaining talent
           is a real problem that I worry about. 
     \item It's very hard to attract the best people to a field that is 
           contracting, and that has no clear long-term path. 
     \item I do not believe HEP really has a problem with finding and keeping
           talented people. 
     \item Considering the number of new machines and accessible problems, 
           there may be too many people in the field. 
     \item Some more manifest excitement in the field, particularly evidence 
           for physics beyond the Standard Model (not counting the neutrino
           sector) would do the most to bring the best people in. But we can't
           just order that up, we need some help from Nature. Still, we can
           visit classrooms, and speak at other venues, to try to convey our
           own excitement to members of the general public, and especially 
           to children, who generally have more of that initial spark of 
           curiosity. 
     \item We should not try too hard to attract young physicists and in 
           any case tell them that it has become very difficult for an 
           individual to make a sizable contribution in our field. 
     \item There's no doubt that we are losing talented people to industry. 
           However, given that we still find enough people to fill out 
           thousand-member collaborations, and that faculty openings still
           generate a flood of applicants, I can't argue that the field is 
           under-manned. On the other hand, it's clear that fewer grad
           students are coming into the field than in the past. I'm not so 
           comfortable with the idea of aggressively recruiting more -- that 
           may lead some to make unfounded promises about the future growth/ 
           financial support of the field. Those that are truly interested 
           will find us. 
     \item I think we are training too many people to be high energy 
           physicists. Which means we should reduce the number of large
           experiments we are conducting. 
     \item One thing that seems obvious is the field is over crowded. So 
           increasing the general funding in a way that increases the number
           of experiments and the salaries is very important. The HEP 
           community has recently taken two steps in the opposite direction 
           by shutting down SLD and LEP. There are now very few choices 
           available for graduating physicists. 
     \item Attracting grad-students to HEP means making travel less necessary.
           In an American grad-school you typically spend $\sim$ 2 years 
           taking classes before you start research. During that time you 
           build friendships and maybe find a significant-other. The latter
           is very important, b/c frequently this is the person you'll end up
           marrying. Students do consider this when deciding to go into the 
           field. I certainly considered it, and I know other grad students 
           who decided against joining HEP for these reasons. 
     \item This problem is a critical one for the continued viability of the 
           field. If we are not able to continue to attract talented young
           people into this field, we will not be able to sustain a strong
           program. From my viewpoint there are several problems needing
           attention. First, the need for permanent jobs in the field. As the 
           current faculty at universities approaches retirement, it is
           important that some of these positions go toward hiring the next
           generation of researchers. Second, the uncertainty of our long 
           range planning process and its vulnerability to rapid course 
           changes due to the whim of Congress. When young people see programs
           like the SSC shutdown and NASA funding shrinking, it indicates to
           them the uncertainty of continued funding in these programs and
           they look elsewhere for future employment. We must find a way of
           making science policy in the U.S. that can be reviewed/ approved
           by Congress but that has timelines longer than a few years in order 
           to allow large projects the ability of being completed. Our
           European colleagues have such a system and as a result they have 
           been able to take the lead in building several important HEP
           facilities over the decades of the 80s and 90s. 
     \item The field is interesting and that should be emphasized. Working
           for 40 or so years in most other fields of work is not necessarily 
           attractive, even if the pay might be better. 
     \item Salary and benefits being an attraction to the field? You're 
           kidding, right? 
     \item A bright vision and a happy spirit of (future) accomplishments will
           attract them (especially those who might also be attracted to
           biology, computers, etc.) Permanent jobs are needed to keep them,
           of course. Also substance behind the vision. 
     \item You catch them by giving well structured and ambitious lectures.
           You keep them by helping them produce good results and creating a
           team-like work atmosphere. 
     \item The mixture of freedom, large scale international collaborations, 
           and intellectual challenges is what attracted me most into the 
           field. I think we have to give young people more of a perspective 
           that they can have a career in physics after all. This means more
           permanent jobs have to be funded, and overall more flexibility is
           needed in employment of people. 
     \item Communicate the excitement; ensure solid and diverse post-graduate
           opportunities; be willing to cede control downward to newer folk. 
     \item It would appear that one is getting more results of broad 
           application more readily in areas like genetics engineering and
           molecular biology, also computer science. It is my personal belief 
           that particle physics theories are not all that attractive in 
           their present, very complicated form. Simpler theories are needed.
           Easy to say, hard to do. But these are the lines along which I'm 
           working, in cosmology as well as particle physics. 
     \item Particle physics appeals to people who are driven more to discover 
           new things and to have freedom and variety in their work than to 
           make money. That is probably as it should be. 
     \item To be honest, I have very little concern. I am convinced that
           within 10 years, new fundamental discoveries will be made. If it
           were not to be the case in the coming 15 years, I would certainly 
           leave the field of HEP! What makes a research field interesting, 
           in addition of its intellectual content, is also the degree of 
           exchange and of live inside the community. On that respect, HEP 
           is still in pretty good shape! 
     \item Many interesting options available to young physicists. To attract
           to/retain in HEP must address the work environment: salary, 
           permanent jobs, intellectual challenge, and personal quality of 
           life. Too much time spent away from home base can interfere with
           success at university or other base and also with family life. 
     \item We CANNOT build another cutting edge machine unless we have VERY 
           BRIGHT young HEP physicists working actively on accelerator
           physics. We cannot therefore sustain our field or achieve our
           goals unless we change our attitudes. We must not continue to 
           allow R\&D work to be an automatic career dead end. Furthermore, 
           the best physicists are well rounded people who understand their 
           instrumentation AND are active in data analysis.

\subsubsection{Availability of Permanent Jobs:}
     \item Right now for software developers, the industry provides better 
           opportunities. In the long run, the possibility of a permanent 
           position is the main issue. 
     \item More permanent positions ! 
     \item The only way to encourage talented young people to enter and stay 
           in the field is to have more permanent job opportunities. Period. 
     \item In my experience (35 years) physicists leave HEP when they cannot
           get a decent position. A few leave for a better job (higher salary,
           startup, etc.) On the other hand, it would be a mistake to create a
           large number of permanent jobs ... a budget downturn would stop all
           new construction. 
     \item 1) Lack of permanent jobs, plus strong drain from industry 
           (we are employable) 2) We need to remember HEP training is 
           applicable to a wide variety of career paths, NOT JUST ACADEMICS!!! 
     \item The biggest problem young physicists have with our field is the 
           very uncertain nature of its future coupled with the obvious
           evidence of the lack of growth over the last decade. Uncertainty 
           plus recent stagnation are deadly for people making choices about
           the future. 
     \item At least to me, as a young graduate student, by far the scariest 
           thing about HEP is the apparent shortage of permanent positions. 
           This is fueled by the lack (at least on my part) of a concrete idea
           about how hard it is to get a permanent position. Unfortunately,  
           it seems like to do HEP you have to be at a major university or a 
           national lab - small universities just seem not to be involved or 
           are not welcomed. 
     \item I have seen a significant number of young, extremely talented 
           physicists leave the HEP field because they could not obtain a
           tenure track position. I see this as a response to the narrowing 
           of our field, and as a continuing repercussion of the closing of
           the SSC. I believe this is a difficult problem that goes all the 
           way to Congressional support. 
     \item There are not as many permanent jobs available as the number of
           graduating researchers. So most of them are eventually forced out
           sooner or later. On the other hand considering the kind of training
           they get, they find things easier elsewhere. 
     \item I believe nowadays it looks almost impossible to be an HEP (young) 
           physicist and have a family. This is due to the not so much 
           competitive salary but above all to the lack of permanent job
           opportunities. Nobody can tell what she/he's going to do in a
           couple of years after her/his PhD ... and many people are now
           taking two postdocs ... having at least 5 years after graduation
           to spend around ... without any guarantee of a permanent job. I
           feel it is very frustrating and many people leave because of it.
           Outside HEP after having gained a PhD we get great job opportunity,
           maybe not that permanent but so many and so well paid that the 
           situation doesn't compare with the academic career! Especially
           pursuing the academic one, after a couple of postdocs a physicist 
           is practically out of the market for other jobs because of her/his
           age! This is very bad, nowadays it looks like a faith/vocational 
           choice ... choosing physics no matter what. That has to be changed 
           or young physicist will keep leaving the field. 
     \item The lack of permanent jobs is in my opinion as important as the 
           lack of competitive salary. 
     \item The number of permanent jobs have to be increased. Often young
           people spend hardworking days as graduate students or postdocs and
           still have no promise of a secure future. So, naturally they leave 
           the field. Salary also is an issue, though not so major I think. 
     \item I've been told at every step of the way, from when I decided what
           my interest was as an undergrad that I'd better not get my hopes up
           for actually working in the field. I don't know if its a matter of
           getting people to stay in the field so much as getting people 
           OPPORTUNITIES to stay in the field. 
     \item Lack of permanent positions seems to be the dominant reason for
           leaving the field when it comes to young HEP physicists that have
           proven to be very good physicists. Salary reasons are brought up
           more by "average" candidates. My guess is that more permanent jobs 
           - at the intermediate level between postdoc and Professors - 
           would benefit labs and universities greatly. 
    \item  Permanent job opportunities are extremely important and there are
           many examples of how people stay in the field for decades without
           a stable position. HEP community anyhow pays these people, but 
           nevertheless they do not have a guaranteed future. I think that
           young people have to be convinced that there are places for them 
           and the rest depends on their work. 
     \item Lack of permanent job opportunities IN A PLACE WHERE I WOULD WANT 
           TO LIVE. (In a city-- not in a suburb or the middle of nowhere.) 
     \item My biggest concern is that the outlook for permanent employment in
           HEP in the U.S. sucks. People will stay if there will be permanent
           job opportunities. 
     \item If I leave the field it'll be because I could not get a job. 
     \item I think that the greatest problem facing young physicists, and this
           includes myself, is the apparent lack of long term job
           possibilities in HEP. I really enjoy the work, but for right now I 
           don't see much of a future for myself in HEP, and this is largely 
           due to the fact that funding does not seem great enough to support
           a large influx of new HEP scientists. I am very interested in job 
           security and I don't see that right now. 
     \item The lack of permanent jobs should be related to the need for good 
           teachers in small colleges and universities and need for regional
           centers. 
     \item Permanent jobs have to be given at an earlier age. It is 
           unacceptable to let young people play "Russian Roulette" till the
           age of 40. If at that age a research career doesn't work out, it 
           is too late for changing to industry. Also, accepting the usual 
           combination of mediocre salaries and outrageous working hours 
           requires a lot of idealism. 
     \item The field does not have ANY organizational effort to make sure a
           balanced (wrt hardware, software, management, analysis skills) set 
           of people get a fair shot at the limited set of permanent jobs.
           There is little perception of fairness or good sense in promotion
           results. 
     \item Young physicists leave due to lack of perceived future for the 
           field. 
    \item We need to come up with a long-term vision of the future of the 
          field. Without that, I think that young people will look elsewhere
          for interesting problems to work on. 
    \item More job security and more opportunity to have major roles in large
          collaborations.
    \item I'm actually an astronomer (wanted to do that since I was 5), but 
          work on cosmic rays led to work on more HEP-like topics. So, I 
          sorted of diffused into the field. Lack of permanent jobs, and the 
          corollary that the temporary jobs pay a lot less than similar jobs
          in industry (particularly the computer field) is what has sucked
          away many good people. 
    \item Job opportunities and the future of the field as a whole seems shaky
          at best. 
    \item Too many good, young physicists are not able to find positions in
          Academia or at Physics Labs. If the job market were better, they 
          would be attracted to graduate school and Post Doc positions, but 
          the opportunities appear to be dwindling. It's important to have 
          shorter-term physics projects that can meet the time scale of a 
          graduate students or postdoc. Having a few big experiments with 
          long lead times is self-defeating because it erodes the possibility
          of a graduate student or post-doc to make a noteworthy contribution
          (esp. before the analysis is finished). 
    \item The permanent jobs need to occur after one postdoc not three or
          four. 
    \item More permanent positions, maybe even for people without a Ph.D. 
    \item The job market is a joke especially in Europe! 
    \item One simple important change to ensure young physicists have the 
          possibility of permanent positions is to very strongly encourage 
          retirement of older physicists. Since this cannot be forced in the
          U.S., it must be done by example (e.g., Hans Bethe retired at 67). 
          Those who are or wish to remain active should seek the emeritus
          status on a competitive basis. 
    \item All jobs in theory are occupied by strings guys; they act similar to
          Napoleon's idea " there is only one way to achieve goal: promise
          more and more and do not fullfil them, but instead promise more". 
    \item NSF has effectively terminated university-group Research Scientist 
          positions; this is a disaster, both for the field and for the
          individuals. We need more physicists than can be justified as 
          teaching faculty, and they should be university- connected, with 
          long-term job security. In medicine, engineering, and social 
          sciences, such positions are plentiful and secure; in HEP in 
          particular, with the requirements of long term presence at remote 
          laboratories, such positions are much more relevant and important. 
    \item We need fair access to job opportunities, not just those with 
          connections. 
    \item Many young physicists see how hard it is to get a permanent
          position. When this is balanced against lucrative private sector
          salaries the choice many make is obvious. 
    \item Lack of clear career paths to allow individual young scientists to
          contribute uniquely to the field. 
    \item Again, we need job opportunities to encourage young people to do
          theory in support of experiment. 

\subsubsection{Leaving the Field:}
     \item The issue of most concern to me is how a young physicist and 
           his/her family build a life beyond physics, including when and how
           to leave the field, whether at or after retirement or through an 
           earlier career switch. I raise this issue because I think that the 
           greatest societal contribution of HEP is provided by highly 
           accomplished HEP physicists who switch careers to address major
           societal issues (even if only part time, as in the case of Sid 
           Drell and his work on arms control issues.) 
     \item It must be stressed that EXPERIMENTAL HEP is a fantastic training
           ground for lots of jobs. I have had students start up programs in 
           medical physics reactor safety medical imaging business consulting
           ALL do well. Their training in HEP has taught them to think 
           fundamentally, deeply, work in a group, think about the latest
           technology AND have fun. Those who are lucky enough to get 
           rewarding jobs in the field have other rewards.
     \item The question should be how we can encourage young people to do work
           in HEP and then leave the field? HEP experiments need large numbers
           of grad students and postdocs to successfully build and run an 
           experiment, but there are not enough permanent positions for them
           all. The field needs to specifically address this fact so that 
           young physicists aren't simply used as cheap labor. 
     \item We need to restructure / reface the Physics Institution. There is 
           STILL an overwhelming mentality that doing physics is elite, and 
           that applying the skills elsewhere is a betrayal. No one likes
           that. By accepting and training people in the field and then 
           LETTING THEM GO, we form an immediate base of ambassadors, 
           sympathetic supporters, etc. in all professions! 
     \item Don't scorn those who leave after a PhD. Use them as a resource. 
           If students realize that a PhD in HEP can lead to a good job, in
           HEP or not, then they will consider it a viable option. The 
           physics, intellectual atmosphere, and challenging problems that
           exist in this field can captivate students, but they must feel 
           that all the work will pay off with a job. 
     \item The jobs in the field simply cannot support all the young people 
           who enter grad school. Thus, we should not write off all young 
           people who leave HEP as "failures". Trying to keep more physicists
           than there are jobs makes conditions worse for those who stay. 
     \item Why are young physicists leaving? A lack of a coherent vision in
           the HEP community, and a sense of abandonment of public interest 
           after the SSC. 
     \item One frustrating thing about the field is all the whining that has
           been going on since the cancellation of the SSC. We need to get 
           over this and move on.
     \item I'm switching to medical physics industry after I graduate. I 
           have lost interest in HEP and in being a post-doc. 
     \item HEP in the U.S. took a shot in the head when the SSC was
           terminated in the fall of '93. The future is very bleak beyond the 
           LHC. Post-LHC, there is (in my opinion) no sensible physics 
           justification (viewed in terms of cost to build) for a 
           U.S.-based NLC or muon-collider. By the time such facilities would 
           be built and made operational here in the U.S. very little new 
           information would be gained. Good young people will only stay in a
           field if it has a viable future. This requires a viable/ sensible 
           project on a sensible time scale and adequate funding. 
     \item We should encourage talented young people to do graduate work in 
           HEP that has the potential for applications outside the field, so 
           they are ready to continue in HEP or, given the realities of the 
           field, find employment elsewhere. There is no reason to feel that 
           someone who has done PhD quality work, and has been published in
           PRL or Phys Letters, and then gets a job in industry is a failure. 
           That person is in fact a great success. 
     \item I would not wish to continue in high energy physics beyond a 
           doctoral degree because of the atrocious social atmosphere/
           interaction I have found (as a female) in both the universities and
           at the national labs. I can't imagine spending my career in such 
           cold and distant/preoccupied places. 
     \item I know that I got into the field because of my interest in studying
           nature. As for why I am still here, my thinking varies from time to
           time. At this point it would seem a terrible waste to leave before
           I finish my degree. After that, I will reevaluate my position. Many
           of those influences you mentioned might cause me to leave the 
           field, or I might stay. At times I have felt that I am exactly
           where I want to be doing exactly what I would choose. Other times, 
           I wonder why I am wasting my life in a field where two or three 
           projects would seem to be a career. That "feature of the field" may
           be the worst. I am on D0 now, and if I stay until it finishes I 
           would be here until I am at least 35. If I then join another 
           collaboration from the beginning and stay until the end (something 
           I would like to do, getting in on the ground floor), then I would 
           probably be ready to retire when it finished. This is NOT 
           appealing. 
     \item I did leave the field. I am working in HEP related computing field
           now. I left because there was no direction to what I was doing and
           because there was little or no opportunity for me as a postdoc in
           a large collaboration to forge my own direction. 
    \item  I think the lack of a clear long-term vision for High Energy Phyics
           makes the likelihood of future contributions based on an "interest
           in physics" discouraging to many talented young people. I think 
           that young people leaving HEP is not in itself a bad thing. It has 
           happened for years, and in fact a Ph.D. in HEP is good training 
           for many complex challenges in life. But I think that more of the
           most talented young people in HEP are leaving than before. 
     \item I think the NUMBER of young people retained in the field is 
           probably OK. My disappointment is how often the best people leave
           and the less good ones stay. 
     \item If they are truly excited about the field, they will make every
           effort to stay. The graduate training system faces difficulties,
           however, in light of the long lead times of experiments, the size
           of collaborations, the reduced number of experiments, and so on. 
     \item I tend to think people leave physics because of a mismatch between
           their abilities and physics opportunities. 
     \item Many of the brightest people I've known entering HEP have left for
           different reasons. 1. Inability to obtain a permanent position. 2.
           Better salary in industry 3. Not happy with direction of field. The
           first two reasons above are related and constitute the largest 
           category of people I know leaving the field. Most of my colleagues 
           have worked in experimental HEP developing detectors and ensuring
           that quality data is being taken when the detector is running. In
           these cases, lack of recognition of their contribution to the 
           experiment came because they could not demonstrate "enough physics"
           done during their postdoc, tenure evaluation period. Given a larger
           salary and reduced hours at work in industry help attract the 
           bright hardworking folks away. Item 3 above is also correlated with
           item 2. The people I know who've left for this reason have done so 
           because they feel that they had little opportunity to do truely 
           independent research. Specifically, with the advent of VERY LARGE
           collaborations and experiments, they often felt stifled by the 
           committees which often run the development of a detector or 
           pursuing a particular physics analysis. Again, a larger salary and
           fewer hours at work in industry help attract these people away. 
     \item This is fairly complicated. I have seen many younger physicists 
           leave over a 20-year career. Some I thought were not talented 
           enough to make the professor cut, saw it coming, and left. Others
           could have gotten jobs at national labs in computing, for example,
           and since they could make easily twice as much in industry for much
           the same work, left for those reasons. I doubt I would go into HEP
           now as a new student because the collaborations are too large, the
           time to get results are too long, and the specialization is too
           great. I joined a collaboration of a dozen at the proposal writing 
           stage and got my PhD on the analysis. I did every single part of
           the experiment: beam-design, detector-building, trigger electronics,
           DAQ, MC, analysis, and wrote the final paper. I went into HEP
           because it was demanding, forefront physics which would exercise 
           all my talents and intellect. I could be thinking about the most
           important problems in nature while also figuring out how to design
           phototube bases and model electormagnetic showers. How can CDF/D0
           compete with that? Why would anyone want to be one of 1000 people 
           looking for the Higgs? The sense of continuity and ownership isn't
           there as it used to be. I would probably do experimental 
           astrophysics or even molecular biology if I were in my twenties 
           again. 
     \item The distinct possibility that there will be no (or a limited 
           number) of new facilities or experiments to attract the best minds
           to the field. 
     \item I don't know that we should encourage them. with no machine in
           the U.S. in the foreseeable future, we should leave HEP to the 
           Europeans. Plus, the salary stinks. I stay in the field due to 
           inertia ;) 
     \item Salary scales vary widely between countries. U.S. postdoc is 
           pretty good compared to the UK's \$22,000, which is graded by age,
           not experience! Lack of permanent job opportunities - how many HEP
           professorships are there out there? There are far more postdocs. 
           What happens when you don't get a permanent job and are in your 
           mid-thirties? Length of time to complete an analysis is very long.
           Huge collaboration size is also an issue. It takes all these people
           to make the detector work but there are far fewer analyses. Lack of
           professionalism can definitely put people off and overpower their 
           interest in physics/ science/ nature. Quality of analyses/ 
           results/ peer review/ meeting presentations could all do with 
           improvement. Note that it is not necessarily the best people who 
           stay in the field - a lot of them move on because they are good ... 
     \item The worst thing about the field is that there isn't enough room for
           people who just want to do science for a living. My perception is 
           that I have to have a particular career track in order to say in. 
           Question: Is there some way I can make a stable living by doing 
           science and not worrying about grant-writing and surviving
           university/ lab politics? I think that a great deal of potential is
           lost because we have a very limited view of what a scientific 
           career should be. Alternative career models need to be considered.
           YPP is in a unique position to be a "think tank" on this very 
           important issue. 

   \item I think young people leave the field primarily due to a failure to obtain degrees and secured positions prior to age ~30. However, it is not necessarily a bad thing that there is some talent and dedication required and some winnowing of the field even during the graduate/post-graduate periods. 

    \item It's not clear to me that the most talented ones don't stay. They can't all stay, there isn't room it would be exponential growth which is unsustainable. The negative attitudes and wining of people making it unpleasant is the kind of thing that almost drove me away. 

    \item I think the key to keep people in the field is to keep building better machines so that their are continual opportunities to make new discoveries in move the field forward; experimentation is our lifeblood. 

    \item No fundamental discoveries being done in the U.S. and all basic research being done abroad e.g. CERN. 

   \item 1) lack of permanant positions (in particular in Europe) 2) Small salary that can be compensed only by academic freedom or intellectual atmosphere. If "politics" is too important than it would be very hard to recruit new physicists. 

    \item Some of the best young HEP physicists I have known left the field because their work was not appreciated and they were not given appropriate credit for what they did, i.e., some senior person took credit and then did not further their career appropriately. 

   \item Lack of intellectual challenge, that is to say "fun".

   \item I have stayed because I don't know what else to do. People are leaving because that have figured that out. Our hours suck, our pay sucks, and what are the rewards? No social life, no family, and no recognition for a job well done. 

   \item  Not enough interesting topics are the feature of the field which I think might most influence young physicists to leave the field 

   \item I think many of the factors mentioned in item 28 contribute to people leaving HEP. 

   \item Strong diversified scope of research without multidisciplinary boundaries; possibility to develop and pursue new experimental ideas 

   \item Lack of interesting topics and long times between experiments. 

   \item Make sure that there is variety in our field. If every project becomes a mega-project our field will die. Relatively small, quirky, experiments must be part of the mix. We should see HEP as a broader endeavor; HEP should wholeheartedly embrace particle astrophysics - this is our future. My main concern is the mega-collaboration and the inadvertent squeezing out of the tangential thinker, the maverick, the young person who simply insists on thinking differently from the rest. I also worry that the "rewards" system is too ad hoc and prone to the "favorite son" syndrome. 

   \item To answer, one must do relevant research, i.e., go ask a sensible cross section of young people. 

    \item The field in the U.S. will wither when we no longer have the highest energy machine in the world. 

   \item Most SHOULD leave! They get a great education, only a few should stay anyway. Exp. Particle Physics is providing a very broad and rare education in all fields. 

   \item Although all factors listed in 28 are valid, I believe the overriding force is that a good fraction talented people avoid physics because more primal reasons. Physics is "not cool." Judging from Hollywood, we are assumed to be geeky, poor, lacking in social skills, badly dressed, etc. Computer science has a tremendous advantage in that it is perceived in terms of fancy cars, mansions, and retiring at 35. I don't think there is a fix. This materialism will continue until a major shift in society occurs. 

   \item Many of the listed issue are equally important, especially salary, lack of permanent jobs. And lack of enthusiasm from "older" physicists. 

   \item People leave because the infrequency of seminal results coupled with the difficulty and time scale for experiments. 

   \item The big problems are permanent positions and -- in the recent past -- totally unreasonable salaries in the computing industry. There would be a lot more permanent positions in universities if the old guys would retire at a reasonable age. 

  \item Lack of first rate frontier facility. 

  \item People leave for many reasons. Some go for money, some get frustrated for not being able to find a job. Both are reflection of the general public's and funding agencies attitude towards basic science. Outreach is important ! 

   \item Where the jobs are, people will apply. No matter what you do, only a self-destructive idiot or a rich fool would sacrifice five to ten years of his life with no prospects to stay in the field, and only slightly better prospects to leave. 

   \item The role played by the schools is very important; more contacts should be done there, with the pupils and with the physics teachers. And in addition, that you do not get any professional help in planning your career. 

   \item In questions 25 and 26, interest isn't strong enough. Fascination is more accurate. This is what we must convey, without looking out-of-touch. In question 28, there are many such features. A combination of uncompetitive salaries and fierce competition for few places is probably a contributing factor for many. And it's not necessarily the competition for permanent positions. Many people want this at some stage in their career, but for some younger researchers, a series of short-term places would be preferable, if only these places were easier to come by. Finally, a lot of people who display a lot of talent in their early years have sort of ended up doing this by default and leave partly because they are just getting bored - this is something we probably can't do much about. 

    \item The answers to 25 and 26 are really that it has been tremendous fun. I work on great problems with really smart people. For 27, it seems the field is not attracting the best people. This seems a larger problem than retaining people who are not sufficiently motivated or do not enjoy physics research. More and better undergraduate programs would help. For 28, it seems that some collaborations are so large that it might be easy to lose the big picture and get discouraged. 

   \item  I strongly feel that the lack of permanent jobs and the relatively poor salary do great damage to the field. It also appears to be forgotten how much hardware skills are required to continue to build the best possible detectors and accelerators. This has been to my detriment in the field. I worked hard to make a name for myself in building hardware. My work ethics and poor financial compensation greatly contributed to the demise of my marriage. Later, when my marriage fell apart, so did my career. Afterwards, I had great difficulty in finding employment in the field. I sincerely doubt that I am the only person to which this has happened. It is a severe problem in the field and needs to be addressed. 

   \item By offering new and interesting opportunities in the field and competitive salary, of course. 

   \item A perception rightly or wrongly that the field is closing down. 

   \item Question 28: I think young physicists leave the field because of lack of opportunities of all kinds. The future is cloudy at best. The collaborations are too big and it is difficult to make a name for oneself. The best people are often not rewarded for the best work, rather everything is politics. 

   \item Since I can remember, some left some stayed. Even the fraction has not changed ... from my thesis experiment (1969-72, about 2/12 students remain in the field. Competition to stay is a must... but I think that money is a miserable motivator. 

  \item Lack of posts - time scale of a single experiment - lack of 'fantasy' permitted in large collaboration. 

   \item This is obviously a very difficult question. I think alot of U.S. students don't like the idea of having to relocate to a national lab, and thus it is sometimes difficult to attract U.S. students into HEP. I think we have to convey the excitement of the field and explain the big picture. Give a sense of history of particle physics, and where there place is. Explain that it is an opportunity to do fundamental research while also learning about hardware, software and physics. Particle physics will not be for everyone, but it prepares you for doing other things at well. We should point to statistics of where HEP people go after they finish their graduate work. I believe the employability and jobs are very good after graduate work in HEP. And, if they are dedicated, they can stay in the field and devote their life to research, or perhaps get a faculty position. 

    \item I see little evidence that talented young physicists who are motivated to work on physics are leaving the field. In point of fact, there seems to be a lack of qualified talent. 

   \item  The fraction of physics PhD's which stay in physics is roughly unchanged since the 50's. I think, however, that the reason they leave may have shifted. Post doc salaries have declined (w/rt inflation), and the typical post doc period has increased. This means they spend more time at a lower salary. Given the current and historical uncertainty in finding a tenure track position the trade off is less worth it now than previously. 

   \item I left HEP because there is no room to get involved in short timely experiments and a big "hunting for the Higgs" experiments. The funding situations forces one to a all-or-none participation. 

   \item  Most important point driving young people away: perception that the field has NO FUTURE, and that the existing power hierarchy is destroying HEP. 

   \item The young talented people should have ability to participate actively in physics research and have opportunity to use their creativity in much better and more organized environment. 

   \item Uncertainty in the future of the field is the largest deterrent in my opinion. 

   \item Lack of competitive salary, lack of permanent positions and workload one is expected to do go together. The combination makes for very unattractive long-term perspectives. Allure of working at HEP lab, the intellectual and, not to forget, the international and open atmosphere in HEP are the strong arguments in favor of staying in field. 

    \item Many things influence young people to leave: 1. too much work, too little pay 2. too large experiments 3. forced to live in places not of their choosing 4. macho culture. Why was number 4 not on the list? It's the number one reason for WOMEN to quit. 

   \item Lack of opportunities for individual achievement. Dearth of exciting challenges for individuals. 

    \item I think that the combination of accessible science and the structure of the community make working in HEP substantially less attractive than it was in the 1970s. 

    \item In my view the limited perspective we are able to offer to younger excellent scientists in particle physics is the biggest problem of the future of the field. 

    \item Need to assure them (and all of us) of the long term future of the field. Note that one would expect a substantial fraction of the HEP students to go onto work in other fields. 

   \item We [I] have been encouraged to work on hardware because thats "how I will get a job", but the truth is that its the physics that get people jobs. Collaboration spokespeople should do more to encourage support for the people who make detector work. 

   \item Lack of competitive salary, scarcity of professional positions (academic and at nat'l labs). 

   \item Give them a decent working environment. This means changing the culture ... I believe this is particularly applicable to young women. I've stayed in the field because I can work my life round it due to my particular personal circumstances ... 

    \item We will have to give young people the context in which they can develop their best skills in an agreeable collegiate atmosphere, where excellence is remunerated by recognition and a competitive salary/set of working conditions. We have to recognize that money-oriented people will likely leave he field, but even they will have long-term interests in what they left behind; they are our outside support in many ways- even if their interest is in designing and producing products that find application in our research and have to be acquired from them. 

   \item  Don't put up labs in the middle of nowhere, we Europeans cannot withstand country life. Payment is always an issue. 

   \item Both salary and the lack of permanent job positions are forcing many talented people to leave the postdocs are lower in standard, and the best graduates don't stay on to do PhDs. Long durations abroad can put off those with young families from continuing. No travel is not an option for both career progression and financially. 

   \item The long hours are not the main problem - that's mostly the same anywhere if you want to have a career. The combination of lots of travel and the total absence of economic security that postdocs enjoy however would not be acceptable in any other profession. It speaks for the idealism of our community, but not for our brains. At the moment one can encourage young people to do graduate work in HEP also because they will easily find a well paying job afterwards in the industry when they want to settle down and have a family. But we can't really offer them much inside HEP. As a result some crucial expertise e.g. in computing is becoming rare in HEP. I believe that more permanent positions for researchers below the faculty level should be created. 

    \item HEP is partly lost the flavor of a fundamental science, to my opinion. It's treated "as any other job" currently, which is not true. The candidates should start being collected and educated at schools rather then universities/colleges. And yes, too many talented young physicists leave to IT-related companies due to much higher salaries. This of course partly kills HEP... 

   \item Add question 28: the ratio (salary*safety of adequate future)/(collaboration size) is too small, which is a hint towards higher acceptance of our work by the public. The large collaborations result in (too) hard work for the better people to drag the slow ones with. 

    \item People would stay more if we focused more on the exciting prospects for finding and exploring new aspects of why the world is the way it is, and the role they could play in that. 

   \item There are several issues. Large collaboration sizes and lengthy projects are two. Other problems now are the lack of consensus and direction for the field, mostly due to cost, I think. We really need a long-range plan to keep the young people interested and make it clear that there is a future for this field. 

    \item Actually I think it is the combination of all the reasons you list in question 28 which compelled so many of my contemporaries to leave. What concerns me the most is that more and more physicists have spouses that work and they want a family. The old way of doing HEP is not compatible with this choice and making this choice should not preclude a good physicist from doing HEP. 

   \item The problem is really a combination of 3 factors you mention: lack of permanent job opportunities, lack of a competitive salary and the length spent in graduate school. I seen many of my peers leave school because they were offered 3 times as much as what their salary would be 5 years from now. In a society like the US, consumism oriented, it's clear that you were not going to retain any student at all. 

   \item Have in each HEP laboratory an adequate number of short-term fellowship positions for young post-graduates and enhance the mobility between HEP laboratories and national research institutions (like INFN in Italy) and Universities. Funding of initial research positions must be partly transferred from national to international HEP laboratories to have better initial selection in a field where saturation has been reached but the risk is not enough young and talented peoples replace a relatively old population. 

   \item The 21st century is not supposed to be the century of physics. 
 
   \item The fact that getting funding for future research is getting harder. Fundamental research is not seen by a priority of the government or the public. 

   \item Too much emphasize is placed on political connections rather than raw ability. We *must* help promising young people to be noticed. We *must* promote good science. 

   \item Streamline the process from graduate work to full- time positions in HEP, accept more BS/BA employees where higher degrees are not required (software, hardware), offer more training. 

   \item Question 25 - This "allure" happened because I grew up near FNAL and started working there as an undergrad. 26 - I sometimes think it was just luck. I like what I do (both day to day and the overall goal) and the atmosphere (working with intelligent people and the academic environment), but as to why I'm a "success" and others who seem to me to be much more talented are gone - luck. Question 28 - Or just a perceived lack of permanent job opportunities. There has been a big exodus of talented people - and it has fed upon itself. But I don't think people leave because of just one thing. I'd put length of time in school, management (we were trained to be physicists, not people managers!), large collaboration size, and travel (although younger people don't worry about this as much until they have a family) high on my list. 

   \item In the set of answers to p. 28 I think you don't have one, which may be the most important - the lack of public interest to the field. For young people it is very important to feel that they are doing something which is strongly needed for the humanity. I choose the salary, because salary is also the indicator of how society rates your job. 

   \item Change the "weeding out", "only the best" attitude towards people - there is no reliable measure of such a thing and this attitude turns off students, post-docs and young faculty. 

   \item I guess grad students (but even more post-docs) are way overworked while being underpaid. There are not too many permanent research positions and of course even less professorships. Considering the chances many young physicists find in industry, consulting etc. they take the positions that allow them more regular hours, better pay and career chances. The wonderful thing that HEP has to offer though is the way in which individuals get to identify themselves with an experiment or collaboration. It seems more meaningful and rewarding to work in HEP than in industry partly because HEP offers a non commercialized environment... 

   \item Additional answer to question 28: Besides the difficulty in finding permanent positions, the lack of clear direction for the HEP in the U.S. could be another factor driving young people seeking jobs in other fields. 
 
   \item Uncertain future. Long time to build machines. Large experiments. 

   \item  In my experience young people who leave the field do so because they have just not found they "fit in". This is for a number of reasons. Anything from salary, to poor management, to lack of interest in the "whole" field (hardware, data reduction, politics, funding, etc.) 

   \item Loss of interest/Too much pressure for the money. If students were allowed to work on important analysis and if the work progressed and moved on I think it would make a difference. 

   \item Most have extremely good software ability, or electronics training, but lack the training to build detectors, big or small. 

   \item Need better HEP funding. Many people left the field when the SSC was closed. 

   \item  Lack of future projects to participate in. We have priced ourselves out of the market unless we can convince society there is benefit in multi-billion dollar HEP projects. 

   \item In the recent past we've seen too many good people leave the field, not just because of negative aspects listed here, but also because of extremely attractive opportunities elsewhere. 

   \item We need to accept that people want to "have a life" but still be part of HEP. The field tends to consume and then disillusion young people. 

   \item For question 28 I feel that I could have almost checked every box. A question like this is never black or white but many people seem to leave the field to pursue a more lucrative career (and one that they often find less interesting). If there were more stability, security, performance related benefits and better structure I would imagine many well qualified people who left the field may not have originally done so. 

   \item  It's hard to say what most contributes to young, talented physicists leaving the field of HEP. The amount of travel required is a huge factor, in my experience. For me personally, it is the biggest reason I am disenchanted with the field. Large collaboration sizes and the long, long, long time-line associated with the foreseeable projects are also a huge factor. Lack of pay is another factor, though substantially less important, as far as I am concerned. Time \& distance are the major problems with HEP today. 

   \item  The reason that I might leave the field is that I have interests other than physics, but physics requires such dedication that I don't have time for other interests. A career in physics, though fascinating, does not seem likely to change the world. I personally would be more enthusiastic about a career in physics (or science in general) if I felt that the scientific community took a greater interest in societal issues. Suppose I could be a scientist *and* a social worker - then I could sleep at night. Seriously. 

   \item Most HEP grads these days have extremely marketable skills in a number of high-tech related fields. In my own personal experience, everyone I know who has recently left the field has stepped into jobs with double or triple my post-doc salary (and I'm considered well paid for a post-doc!). Combine this with the present lack of prospects for permanent jobs in the field and it makes me wonder why ANYONE stays in particle physics. In the past, I think there used to be perks (social status, academic work environment etc.) associated with academia, but this is now largely gone, and in a large collaboration environment, you might as well be working for IBM and making six-figures instead. 

   \item There must be a long term future of frontier research to keep young people in the field. 

  \item Answer given to question 28 above refers to working in Accelerator Physics. Too much current research is oriented to specific existing or proposed machines, and not enough is devoted to basic research. 

   \item Are we finding any new physics? What if we find a Higgs thing... are we done then? Out of work? 

    \item There are many very interesting and rewarding fields for a talented physicist to work in today. Competition from other fields has tended to move people out of the field. In good economic times we may see a resurgence. 

   \item We need to develop respect for a wide variety of talents and accomplishments for the field to succeed. Too many of the essential tasks to not receive sufficient recognition to permit a self-respecting physicists (or engineers) to expect his family to subsist on current wages with current recognition. 

   \item Why young physicists leave: often it's the ones that are not number 1 in academics or subjected to a "bad" advisor that quickly disappear, though the lure of money in industry is probably the straw that broke the camel's back. Some people are not the best, but many of them are excellent creative folk that just got unlucky in a lab or on some tests... and fizzle out over several years. Is there an easy implementable answer? No. 

   \item  Question 28: a combination of a) lack of competitive salary b) lack of permanent job opportunities c) weak or unstructured management (worse in the case of large collaborations). 

   \item I guess it's a mixture of lack of competitive salary, lack of permanent job opportunities (depends on the subfield, of course) ... 

   \item Oh boy - I could fill several pages on all the bad things about HEP. The good thing, which kept me going, was the quasi-spiritual point that this is something that is unnecessary to our existence, and is therefore something noble, untainted even. Despite the obvious, and considerable, spin offs to the rest of society, this is something we should not be afraid to point out again and again. The above point about lack of permanent job opportunities only vaguely touches on the nub of the problem. In reality, nobody has a permanent job anymore - it is the absence of choice to stay at one place for more than the usual 2/3 years, even though it is more than likely you're still communicating and working with the same people. Most depressing is the fact that the people who succeed in HEP these days are no longer the most talented (ability is more or less a given) but rather the most stubborn, and even political (in the academic sense). Do we really want these individuals to be running Physics in the future ? 

   \item HEP is a little remote from everyday problems that challenge young people. We hide behind Cosmology (The Big Bang), since that seems the most exciting aspect of our work. We have lost sight of the importance of understanding the structure and interactions among fundamental particles. The theoretical ideas behind particle physics (Higgs mechanism, super symmetry) are difficult to grasp. Permanent positions are scarce, and yet the really exciting work is often done by post docs. 

    \item The problem is other way round: it is not the question how to encourage young people to stay, the question is what are the chances to stay. In some sense, HEP community is manipulating the young people - most of them (us) are rather idealistic and prepared to work much harder than they are really paid for - but then, when they are 40, only a small fraction could stay doing physics and the rest should go... To be clear, I think that it is not responsible to take much more graduate students or PostDocs than there will be available permanent job opportunities in the future! (P.S. If you think that is normal that people with PhD in physics mountain software or install local networks, then think how often have you seen medical doctor instead of veterinary taking care of cows.) 

   \item  Offer good perspectives in terms of salaries and job security; more permanent positions instead of 2 year (or even less) contracts. 

  \item Not enough new and unanticipated results. 

   \item There are too many people seeking faculty jobs for the system to support. Meanwhile, there is a severe manpower shortage. At this stage, additional personnel would do more for the money to advance the field than additional hardware. More post-docs staying for longer is not the answer. 

   \item I think we are not attracting as many of the really talented young scientists to the field as we used to. I think this is because other fields (computers, biology, astro) are looking more attractive these days. The best way to counteract this is to get the field moving again with some major discoveries. 

   \item At times, there doesn't seem to be much respect for young physicists in this field. I.e. at the BaBar conference at SLAC at the end of June, a younger physicist giving a presentation would be corrected in a condescending manner by the older physicists when the younger person would make an error or an apparent error. I have not experienced this behavior directed towards me, but I have seen it happen enough to know that this could be a turnoff for some young physicists in this field. 

   \item The mixture of the lack of permanent job opportunities, lack of competitive salary, and especially the fact that HEP doesn't attract the most brilliant people any more which is a down going circle. 

   \item To let people stay we urgently need: better salaries (especially money compensation for overload MUST come) permanent jobs must be NORMAL (people still can change if they like) environment must be much more professional management must be much more professional AND finally for all of you who still didn't get it: The man power situation in the big experiments is a joke! The global situation in a lot of experiments leads necessarily to a complete burn out of the encouraged people after approx. 5 years. We have it in our experiment and it is simple the truth. And I know a minimum of 5 new experiments at which it will be the same. Another thing is that the leading HEP heads seems to have to have not at all understood that next or new generation experiments not only force some orders of magnitudes more computer power or other technical stuff but also orders of magnitudes of better organization and better structures. In my opinion we need COMPLETE new forms of collaboration organization. AND THERE MUST BE A QUALITY RATING SYSTEM FOR THE MANAGEMENT with personal success responsibility. 

   \item I worry about the uncertain future of the field as a whole and about the necessity to do a nation-wide job search, commute, etc. If you offered me a job at Fermilab (within the current range of salaries/benefits) and promised me the lab would exist and do interesting physics forever, of course I'd take it. But the funding for a particular lab, or for the whole field, may dry up, and there's no guarantee that accelerator-based HEP will always be an interesting field in which to work (i.e. both interesting day-to-day work and interesting scientific results). 

   \item  Large investment (all types:: time gov-money etc) and little physics gain. 

   \item Each individual is merely a chain in an assembly line - little chance to see the whole picture of physics. 

   \item The most destructive factor would be the closing of all major accelerator facilities in the U.S. 

   \item  Lack of competitive salary is only one of a vast array of items that needs to be addressed if HEP is to retain more young people. Everything on your list is important. We have got to wake up and act. 

   \item Intellectual excitement waning and lack of individual to be able to have a major impact upon the direction of research. 

   \item I don't think people leave the field for any one single reason. I think they leave the field (if they do) because they can no longer function in the environment, which has decidedly not improved with the increase in the size of the collaborations. I think the work environment has become more competitive, more burocratic, yet with typically weak and often unstructured management, and in the end people get disillusioned and wonder why they should work for such a low salary. I always used to say that physics is 95% hard labor and 5% fun, but the 5% fun made it worth it. I think the 5% is tending more and more to zero. Travel and length of time in grad school are not important, although grad students should be treated (and paid) a lot better than they are. 

   \item  Other fields now have just as challenging problems as HEP. And, some of these fields have a shorter turnaround time than in ours. This makes these fields more attractive to some of the younger scientists. 

   \item We should give younger people more responsibility. I am seeing more young people giving plenaries at conferences; this is good! 

   \item Another problem to keep some people in the field is the difficulty to combine private and professional life. An example is the rumor that tenure track positions and divorce are highly correlated (and I unfortunately know examples of it). I guess this is a problem in many other professional areas, especially in the U.S., but it's clearly not positive. It is by the way saying that you didn't even list this in your list. 

   \item Lack of optimistic perspectives for jobs, projects, and the physics. "Horganism" has infected high energy physics. The illness was coined by Daniel Kleppner (MIT) after Horgans book entitled "The end of science". It signifies the tiring of old physicist at the end of their careers. Particle physics has lost its optimism and its ability to work in the full breadth of the field. The concentration of all resources on the Higgs, the neutrino mass and CP violation alienates young physicists. These are all sensed as problems for the few big shots and little is left for the young to their own research. Why is particle spectroscopy or QCD related physics considered finished? 

   \item Question 26: the main reason I have stayed so far is really that the alternatives that appeal to me (philosophy, voluntary work) would give me an even more uncertain job situation and no or low salary. Still, I intend to leave physics unless I get a more permanent job in a couple of years. I believe the postdoc round is killing people -- it does not allow a proper life outside physics. As a consequence, it is not the "best", but the most obsessed physicists that stay in the field. People with a more "rounded" outlook are discouraged. It is also extremely difficult to have a family. It is necessary to reorganize the "career path" so that people do not have to sacrifice their whole life every two or three years. Allowing people to stay longer in one place would be the most important thing. 

    \item We retain perhaps 1/3, 1/4? likely about right; I'd drive a taxi to do this and roughly expect that of others these are the people society should support however shabbily ; many don't share that... leave because they don't find a sinecure and need that; etc. etc. or all those other reasons that sometimes work out... some find something that they really need to do or that needs them enough inspires them... I've always assumed I'd follow my greatest interest it has never left this stream of physics I can imagine a number of ideas I might have or might follow if I had a great idea or opportunity... A great part of our societal worth is the people we spin off who find a greater outside role than follow the path of our mentors to the grave... This is still my path... Intellectual transfer really happens through people transfer; Harold Varmus remarked we gotta keep physics going if we are going to have progress in biology/medicine etc. Physics perhaps particle/astrophysics is the Plane Geometry of our scientific time 

   \item Funding uncertainty leads many promising young physicists to find alternative careers. The U.S. does a very poor job in attracting and keeping the brightest and best for this reason. 

   \item Lack of excitement in the field may also contribute to people leaving the field. 

   \item The field (at least in the U.S.) has taken on the aura of hanging on by its fingernails in recent years. Even good, motivated people spend a long time with little or no job security, leaving them ripe for the picking. 

   \item For question 28, I'd actually say that a mixture of poor salary and lack of permanent job opportunities is the problem - very hard to say which is bigger issue. 

   \item I don't think there is enough communication between Physics and Computer Science at Universities. Many HEP questions (how to store so much data, how to look through data, triggering, etc) are interesting computing problems, that might better be handled by someone trained in CS, not physics. Certainly, these jobs should be open to physicists with interest and experience, but there should be more of an attempt to interest CS undergrads, grads, and postdocs in the work we are doing. The same goes for engineering and detector design, though I see that happening now much more already. 

   \item The essential answer for 28) may be lack of chances/places where younger people meet articles or other things related to HEP. If many young talents are willing to come to HEP, the market will be expanded. 

   \item Stability and existing long term future of the field is extremely important to attract young talents to join this field. This field is in great risk to become a dead branch of science such as nuclear physics. 

   \item Many physicist leave or don't even join HEP because they have the feeling that almost everything has been done and that the new detectors are only built to rule out theories that are made to justify the detectors... 

   \item A discovery / deviation from the Standard Model / some completely not understood physics effect would certainly make the field more attractive ! (at least to experimentalists). 

   \item Particle Physics is also an excellent school for young physicists that aim at becoming familiar with a technique and then, if necessary, to move out of the field of pure research to take a job in one of the highly diversified fields that the training in particle physics opens up. 

   \item Lack of appreciation of their work. Physics is more than the (self)glorification of a few. 

  \item After LHC there is no long term future, except astro-particle physics. 

  \item Allow them to produce something in any step of the young physicist occupational career, especially support them in following their own ideas and dreams. 

   \item One should encourage young people to work in HEP, I think then there will be no need to hold them... they stay, I hope! 

   \item That's a really good question. 

   \item There is not enough opportunity for young exp. physicists to dream up an experiment and do it. 

   \item I guess the problem is a combination of a competitive salary, permanent job opportunities, and the possibility to have a responsible and interesting position within a collaboration that still allows creativity. 

   \item Good projects will attract good people. 

   \item 1) large collaboration, establishment and the success of the standard model are not appealing. The latter we have to live with for now, the first two we should think about. 

  \item Comment to 28: Lack of possibility to be creative and propose individual project is a reason that young people leave the field. However unavoidable at the present stage of HEP.

   \item Young physicists will leave the field if reasonable career paths at adequate compensation are not available to them. 

   \item Academic freedom/tenure is the correct motivator. 

   \item Staying in the field 100\% of the time should not be the HEP goal. Physicist are also trained to contribute to the scientific advances in all fields. No more than 50\% should stay in the research field, otherwise we have failed. 
   \item There is no single answer to the question of why young people leave the field after investing several years of training. Job opportunities is one. Sense that personal intellectual contributions are difficult to make is another. Lack of recognition is a third. 

   \item I would add to my question 28 answer also: - lack of competitive salary - length of time spent in graduate school - weak or unstructured management. The issue which concerns me the most about the future of the field is the lack of a major U.S. lab on the horizon. This will strongly curtail any new interest in HEP as a career. 

   \item The field has become too competitive and ruthless. The competition is often based not solely on the intellectual capabilities, but on other considerations, with the latter being weighted higher and higher. Lack of social life for many of our colleagues and a very long delay between the idea and physics results in the experimental science is another driving factor. Inflation of theory surely drives many young theorists away from this field. This all is going to change in the next 10 years, with a plethora of new discoveries coming from the Tevatron, LHC, precision measurements, and the future machine we are going to build. If we have guts to put ourselves behind such a machine, which is clearly a linear collider somewhere in the world. 

   \item The interest in physics and the Academic life style which, I think, is connected with creativity needed in the domain are both reasons for having stayed in the field. The large collaborations size might be a factor too. It does not allow the right amount of contact and exposure. As for the conjunction between the length of time spent in grad school and the lack of competitive salary, it might influence young peoples to leave the field, but I think that it touch the more materialistic part of the young peoples who are not as committed to science? 

   \item We have a problem of not having enough new experimental results. Presumably this will improve when the LHC turns on. 

    \item HEP is dying - too few opportunities for new discoveries. Large experiments controlled by few people at few large National Laboratories. No room for innovation and excitement of discovery to excite young people. Young people come from universities - and universities seem to have no role in the future of HEP. 

    \item Talented young people should have more chances to travel and work with distinguished scientists. Also competitive salary have to be provided. 

   \item Unless financial support of the field increases, we should not encourage people with other interests to stay in the field. Enough people will stay without added encouragement to meet the diminished needs of the smaller-than-present effort this country will support. 

    \item Raise salaries, increase benefits, create more jobs, reach out to the public, create role models. 

   \item  People who like physics, and HEP in particular, will enter the field and stay regardless of the obstacles/incentives provided. 

    \item We don't work for high pay. But we should offer the chance of permanent positions and stability for younger physicists. 

    \item  Danger that field might be nearing an end may encourage young physicists to leave. 

    \item The inability of the HEP field to focus on single major issues or to rank within the several topics addressed by the field, has lead to a severe 'balkanization' of the field. This in turn has lead to a loss of funding and secure positions for even the best young physicists. 

   \item  Several people I know have complained about elitism of peers, especially more senior established physicists. 

   \item It seems rather hit and miss depending on who you end up with as advisor/supervisor or even surrogate supervisor. I've known several really bright students who have left the field after their PhD's because of disillusionment with their advisors/training given. Similarly I've known several bright post-docs unable to get permenant positions and then they leave - normally to higher paid jobs in industry. There needs to be more of a career structure and more standardization as to the level of support given by graduate advisors - its at that stage that it seems make or break. 

    \item I think that question 28 has many correct answers; it varies by person. I know individuals who are so fed up dealing with the overwhelming hubris of physicists that they wish they'd never entered the fields. I know just as many for whom salary or job security was an issue (either due to personal finance or family issues, or because they could do the same research in industry and make 4x the money). However, I also feel that behind all of this is a general malaise -- that physics just isn't going to places where we are able to answer many of the questions raised in the last 100 years. 

   \item I don't see huge numbers of the most talented physicists leaving the field after grad school -- I've seen a few, but not an overwhelming number. Most of the physicists who have left the field have talents which are best utilized elsewhere. I see lack of recruitment of new students as a much more serious issue -- this is a big problem and I don't have any really easy answers. Having lots of neat new results (which we do right now) and a reasonably attractive future (which we don't) are critical. 

    \item When I joined the field you could feel the passion that physicists brought to work every day. We are no longer that kind of field. We need to get that back. the problems you mention are generally real and we need as a field to address them. We have much more flexibility than one imagines. Many of the bad things we've done to ourselves. 

   \item Question 28 : I suppose it is really a mixture of reasons -- no competitive salary for relatively very hard work, no future in the sense of getting a permanent position, no stability (moving! , theorists do not even know a year in advance where to ) and security . A comment on the length of the grad school - don't think this is really an issue. It takes only three years in the UK to get a PhD, yet many people are still leaving the field because of the above reasons 

    \item The strongest advice I can give (and I'm backed up by Nobel laureate Steven Chu) is to pick your advisor/group rather than a specific area. The most common source of disillusionment I've seen and experienced is mismanagement, which can produce feelings ranging from "I don't have any time to myself" to "I don't get any guidance" to "I'm not learning anything by doing this" to "I want to graduate someday, dammit" and a bunch of others besides. It's even more important in large collaborations or collaborations spread thin. Advisors need to realize that their students are not only vital resources to the field, but actual people with their own goals and needs. If young physicists don't find a challenging and constructive work environment in academia or government research, they will undoubtedly leave to look for it in industry or commerce. 

   \item Lack of apparent progress on central questions most influences young physicists to leave the field. There is a perception of little intellectual excitement in mainstream HEP. Of course, that perception is only partially accurate. The field needs to maintain diversity. 

   \item  I believe the good times for particle physics are over. Society will not be willing to support particle physics at the levels we have enjoyed in the Cold War era. Also, we seem to have run out of ideas to advance our knowledge with affordable experiments that can be carried out by small groups of people. These facts turn away the most talented young people, and I cannot blame them. If I were a young graduate student today, I would not choose particle physics as a career myself, but rather a field where I felt I could have a personal impact in the foreseeable future. 

   \item I think the work in this field is extremely interesting and challenging. I also appreciate the academic environment, the freedom found within the field, the multi-national interaction, the intellectual atmosphere, the opportunities to do great things, and the ability to travel to interesting places. Finally, I love teaching students about science (and physics in particular), and I am hopeful that I will eventually be able to do so at a university. However, I do believe that the lack of competitive salary and benefits (especially after one spends 6+ years as a graduate student on near-poverty wages) will cause people to leave HEP. This combined with the lack of permanent positions available puts industry jobs in a rather good light. Working as a postdoc does not even remotely guarantee job security. This without beginning to acquire much in the way of a retirement fund. It seems many in this field assume that the work itself overcomes any practical monetary issues. While I believe the work is fascinating (and have chosen to stay and work toward the possibility of a permanent position in the field), I also think we need more recognition of (and action on) practical issues like salary, benefits, and retirement packages. Without this, we will always have people leaving the field. Now, that said, I think that an education in HEP does prepare one a great deal for a job in industry as well as academics. Companies want to hire smart people who have shown the ability to learn quickly and who can solve problems (of whatever sort that industry has). I don't think this message has been adequately delivered to undergraduates. I think most undergrads believe a degree in physics necessarily leads to academic work. This is not the only career path available to someone in physics, and should not be the only possibility (we should not limit the options available to HEP grads), and we need to spread the word. This would bring more students into the field, and once they are here, they will see the benefits of this kind of work. They will also see the problems, but more students in the field means that more people are available to stay. 

  \item Depersonalization. We can attract and keep people if we can give them significant roles which make them feel special/responsible/important. The larger and larger experiments make this difficult unless care is taken to place and support good people. 

   \item Why do people leave the field? Easy: they get a crappy salary (after studying for 8 years or so) and they get fired after 2-3 years (i.e. NO job security). 

   \item Two things: 1. You should title the field Elementary Particle Physics and not HEP. 2. HEP (at accelerators) has not been exciting enough to continue to attract as many people as it has in the past. This could change. Also the size of collaborations and the remoteness of the experiments are a problem. 

   \item Where are the jobs going to be in the future? The field looks like it is shrinking. This more than anything else drains talented people away into more financially stable/rewarding fields. 

   \item There have to be jobs available before this is an issue. Ideally, the scheme of having to do multiple post-doc jobs each without tenure, of short duration should be revised. This encourages people to carry out research which produces visible results regardless of their genuine merit, rather than carrying out true research which may not give a positive result to put on a CV. Salary issues vary dramatically from country to country so are not such a strong issue. 

   \item Although pretty much everyone in HEP could make much more money in a whole list of fields. That's science. But many people who would have gone into HEP regardless of the money don't because there are so few jobs. I think it's as bad as trying to make it into professional sports. For the little money though I see why so many people don't even bother. 

   \item I see not a bright future of High energy physics I think next century will be on astrophyics and medical. I feel it is too late to change the field Physicist are too arrogant. they thing the politician and public are fool/stupid, I think it is otherwise. 

   \item Hard to attract people when permanent job prospects are so dismal. "Perpetual postdoc syndrome" is probably main thing scaring me away from pursuing HEP as a career (after grad school). 

   \item Other areas of physics have very compelling interests and I believe the HEP may have difficulty competing with them in several areas: technical \& scientific merit as well as funding and future employment opportunities. (these areas can be as diverse as the development of quantum computing to protein crytsalography). 

   \item I'm thrilled that you offered the answers I chose in questions 25 and 26. In fact, it is true that HEP satisfies a spiritual urge in me, and solves the same difficulties many academics experience in finding a place in the church/ mosque/ synagogue. It's a strong reason why I got into the field and why I have remained. The spirit of Academic inquiry is another. It's an environment and lifestyle I don't expect to enjoy anywhere else. The principle "do no harm" is one I feel is at least satisfied in our discipline. That said, I am leaving the field in September for industry. It has been a very difficult decision, and one which I'd hoped I'd never have to make. The political process necessary to market oneself to get the next job, I find distasteful. It's also a slow and daunting procedure, for which my family and myself no longer have the patience. It seems true that politicking and becoming visible generally counts for more than does doing good physics, necessarily. It has been hard for me in some cases to see my peers enjoy successful career trajectories, and to see particular models emulated, when in fact my career so far, in which I believe I have shown I am every bit or more the physicist of my tenure-track colleagues is not rewarded. I don't believe I'm the only disgruntled person in this situation. It might not be unwise for the YPP to investigate this sentiment among young physicists, as it could be more common than imagined. It's also true the travel is a limitation for me and my family, though I don't claim it's necessarily wise to try to restrict travel in our field. It is necessarily a component of HEP. I don't mean to grouse too vigorously. It may be true that I have recently taken stock of just How badly I want to work for what I once thought I wanted. 

\subsubsection{Survey-Specific Comments on Building:}
   \item Tick marks would be better in this section. It is not so easy to 
         reduce the answer to one point. 
   \item You really need a rank order here, because a number of things 
         contribute. Second would be large collaborations. 
   \item Question 28 it is too simplistic. I actually think many of those 
         things will be equally weighted. So I would answer all. 
   \item Questions 25, 26 and 28 there are a number of answers and its not 
         easy to identify the 'most'. 
   \item This is a terrible piece of the survey. The answers are much richer
         than are allowed here. 
   \item I am unhappy to answer question 28 the way I did, but it is my honest
         opinion. I definitely think that lack of permanent job opportunities 
         and bad management/ organization are large contributers as well. 
   \item I am not happy with response to 28. Intellectual excitement, joy of 
         discovery, even if it isn't your own ... is what will keep people 
         in ... lack of excitement, not salary, within reason, will drive them
         out. We need to provide stimulating activity while waiting for the 
         new machines, as outlined above. 
   \item Would have been good to allow for multiple selections in 25, 26, 28. 
         Several of the replies are almost equally valid. 
   \item Question 26: You've omitted some important options: Inertia and fear 
         of change. Also, some of these questions do not have a single answer. 
   \item You should provide the option to select multiple choices in question
         25. For me additional very strong motivations were: - hardware work
         - strong mentor. 
   \item Again, single answers often don't say it all. In some of the above,
         I would pick several of the reasons and add more of my own. I am not 
         a young physicist, so I don't know what most motivates young people 
         these days. But like the other issues, I think it is more than one 
         thing. For example, it could also be that other areas of science may
         seem more interesting. Many have said that the 20th century was the 
         physics century, 21st will be that of biology. There may be some 
         truth to that. If so, a smart young science-oriented person may want
         to contribute to learning about the human brain, or cure some 
         disease, or whatever. 
   \item Multiple answers to 25 \& 26 might be useful-- e.g. love of hardware 
         on top of interest in physics. 
   \item I wish you had given the ability to choose more than one option for 
         25 and 26. For me, "most" depends on the day of the week.... 
   \item It would be interesting to compare what people think the primary 
         reason for leaving is for others vs. for themselves. 
   \item Well, what answers did you expect? 
   \item What's "adequate numbers of talented physicists in HEP" ??? How can 
         anyone define that ??? About "issues facing young physicists" - 
         THERE ARE NO JOBS ! 
\end{itemize}

%% file: comments-physics.tex
\pagestyle{myheadings}
\markboth{Comments on Physics}
         {Comments on Physics}
\subsection{Physics}
\begin{itemize}
   \subsubsection{Most Compelling Physics:}

  \item I think that one of the most important problems for the next 10-25 years will be an understanding of quantum gravity (from either string theory or some novel approach). In phenomenology it would be interesting to see the signs of unification and supersymmetry. 

  \item String theory is the most interesting topic for the future. Surprisingly, it's not one of the options for the "most compelling physics". 

  \item I think we're on course to do the necessary Higgs/SUSY searches. I hope in the next 10-20 years that the big machines don't preclude doing interesting cosmological observations. 

  \item Once we've learned a bit more about masses the next field must be astro physics. Prepare! 

  \item The most important physics for the field is string theory. 

  \item I want to pursue the energy frontier (which in many ways includes Higgs/EW etc). Lets build the VLHC. 

  \item Search for non-Standard Model physics. 

  \item None of the fields you mentioned are interesting alone. For example, studying any of them at a hadron collider without studying QCD is a foolish quest because every interaction starts between quarks and gluons. The interesting question is how does this all fit together. We need these divisions because without them no problem would be surmountable, but to focus on one to the detriment of the others would seem utterly foolish. Who knows where the next breakthrough will lie? The only real question is: Does the Standard Model tell the whole story, and if not what is the whole story? 

  \item Most important is new energy frontier colliders to give us a big discovery window for exotic processes. 

  \item  A better understanding of QCD and strong interactions is fundamental for extract crucial information from experiments and therefore explain the mystery of EW symmetry breaking. Moreover the next generation of colliders may find new strong interactions and we will need do be able to deal with them. More funding to lattice QCD is a must. 

  \item  There are a number of questions as to the completeness of the Standard Model beyond finding the Higgs. These questions must be pursued at the highest energy. There is almost certainly a new vista to be seen at the TeV scale and it cannot be exploited with LHC. The momentum of progress is much more difficult to sustain now, but it is important. These investigations will require a new machine. The facility could be built in the U.S. based on (currently stagnating) accelerator research at the U.S. labs. However, the facility need only be in the U.S., if the political winds are favorable, both for the construction and for the operation. 

  \item So questions 30 and 31 are very speculative. I think the most important physics for the field over the next 25 years will be looking for quark/lepton-substructure. Over the next 10 years or so it will be EWSB, but hopefully by 2010 we'll have a much better handle on it than we currently do. Those are the things that I'm most interested in precisely b/c I think they are the most important. Or rather, I think that these are the most likely places to look for new physics and push back the frontiers of the field. Neutrino physics is also a very good candidate for this, but neutrino experiments are very hard to do and frequently very narrow. However, we need to be open to the possibility of something completely unexpected. It strikes me that we're at a stage right now where we don't know what the right questions to ask. And not only that, but we may well not yet have asked these questions at all. A major new facility does need to be built in the U.S. if only b/c without such a beast, the U.S. HEP program will likely wither away. We need to maintain our momentum. 

  \item Physics beyond the SM. You mentioned exotic particles searches which is different. 

   \item SUSY (if it exists!) since this would be a huge extension to the SM and for once will be something we experimentalists can measure first without the theorists spoiling the fun by telling us the answers first! If the Tevatron and LHC don't see SUSY then we will need a larger facility - but a proton based one ... e+e- will be useful if we find SUSY. 

  \item I find biophysics the most compelling in the next 25 years 

  \item Even with enough funding for the next collider the field will not really sustain itself in the long term unless there is a major breakthrough in the accelerator domain. Thus the most important physics for the field is accelerator physics. 	

  \item To questions 30 and 31: *new physics* (given that we find something new) if no significant deviations are observed experimentally, or if theoretically no progress is made beyond the Standard Model, everything becomes boring.

  \item Also known as fractal space-time theory, scale relativity physics was born ten years ago but is still unrecognized though its several verified predictions (mostly in astronomy and astrophysics at present). It has also a bunch of theoretical predictions in HEP, among them Higgs mass, coupling constant values, etc... 

  \item 1. An important near/midterm physics opportunity is exploration of the MNS matrix in the lepton sector. This requires upgrades so large as in effect to be major new facilities. 2. In the range of 10-25 years I very much hope that the most important physics is that beyond the present standard model -- based on discoveries made during the next decade. If we're still doing the physics of the 1980's over and over, we're in big trouble.

  \item I am more interested in studying the fundamental symmetries (in addition to CP) in the various interactions (strong, weak, EM, etc). This will be my primary research focus in the next few years. As for the field, much of the accelerator based effort is directed at 3 topics: 1. CP violation (Kaons, B system) 2. Neutrino Oscillations/mass 3. Higgs/Electro-weak symmetry breaking. These fields have adequate/over representation while I see that a true understanding of QCD in the boundary region between perturbative/ non - perturbative regime is still in its infancy. Theoretical work is now beginning to yield predictions (lattice, chiral perturbation) in many areas. Much of the experimental work in these areas can be accomplished at energies within reach of existing accelerators. 

  \item I'm afraid I still do not want to consider astrophysics as part of "the field," even though it has some of the most compelling questions to answer in the next 10-25yrs. In fact, I would have probably put down "cosmological constant" for question 30 if I were tenured and could choose what I wanted to work on. I think it is important to nurture diversity in the field. I am concerned about putting all of our resources into the "next machine," because it does mean that we will not be able to fund interesting, small experiments. I think it is less important where the next machine is built (other than for my own personal well-being). I would prefer not to spend my professional life traveling, but I will do it if the next machine is built abroad.

  \item With a limited time remaining I find fundamental searches most appealing i.e. neutron oscillations and dark matter searches. 

  \item EWK symmetry breaking is clearly top of the list ... except that we may be seeing the tip of an exciting iceburg with neutrino oscillations ... so to flag this I put neutrino physics. In any case if oscillations provide enough fuel to make significant progress on why there are 3 flavors, baryogenesis, etc then this is of equal importance to understanding EWK symmetry breaking. 

  \item  Many fields may be interesting including but not limited to QCD, neutrino physics, EW symmetry breaking. I suspect that all of the comments will be important over the next 25 years.

  \item In the near long-term (!) we need to survey, in depth, the 100 - 1000 GeV range and answer the questions we currently have. Then we need a machine to answer the questions we'll be asking after that. Where these machines will physically be is going to become increasingly irrelevant.

  \item I think that there are a number of questions in astrophysics have attracted the interest of physicists in the recent past. I expect this trend will continue as the time scale and cost of new HEP facilities continues to grow. 

   \item Choosing one topic for the field for the next 10-25 years is not right approach. All the fields need to be supported but not all at the same level at the same time. No one subfield should be supported at more than 50-60% of the fields resources (not counting additional construction money). Otherwise the field will lose its intellectual vibrancy and will be short changing itself for the longer term - the next 15-50 years. 

  \item The field needs to continue to work at the highest energies and to search for what happens at those highest energies. 

  \item We've almost reached the point of confirming one of just a couple plausible scenarios to wrap up what we know at these energy scales. I don't think that will exhaust what HEP can do but we will have to start getting more creative-and reward some amount of risk-taking. 

  \item Although I find neutrino physics personally interesting, we must explore the possibility for new physics at higher center-of-mass energies. 

  \item Getting beyond the standard model is probably the most important area of research for the next few decades. I am unclear as to the definition of major. If you mean 100's of millions then I think it is inevitable. If you mean billions of dollars then I have reservations ... we should not put all our eggs in one basket.

  \item My personal goals for the next ten years lie at the interface of QCD and flavor and CP violation in the quark sector. Ten years hence, there will be puzzles (from LHC data) in EWSB so around then, my focus may switch. 

  \item Compelling and important physics: experimental searches of new phenomena related to extra dimensions, searches for baryon/lepton number violation; CP, CPT violation, flavor violation, neutrino factories, underground experiment, cosmological experiments, muon collider. 

  \item Note that "the physics I find most compelling" may not be the physics I would want to work on. 

  \item I'd like to think I am a physicist, not "nuclear" or "particle" or "electron" or "proton" or EW/Higgs physicist. You options target 5 year time scale, not 10-25. Understanding how the universe works is what the fundamental science is all about. Sometimes this includes solid state or atomic physics ! Right now we are just trying to peek outside the Standards Model window. What comes next is not entirely clear, and that's the beauty of it. Unfortunately we have to commit funds 10-15 years before a major facility can be built -- but that only means that we have to be flexible and best equipped to be able to answer questions that suddenly pop up. 

  \item My choice of question 30) really depends on what will be found in the other fields. If supersymetry is found I would surly join the researches there, but meanwhile we should not forget other topics which can be, as in the QCD case, long time work. I do not know what will become the most important. It's a surprise. We have to keep our eyes open in all fields not to bias the researches. 

  \item There are still many open questions on heavy quarks spectroscopy. With a negligible amount of money (with respect the huge experiments) a much better understanding can be reached. 

   \item Personally, as a theorist, unification topics have always engrossed me. At the moment, string theory is the hot topic. In terms of experiments, those I would be most interested in are searches for specific signatures of supersymmetry, unification beyond the Standard Model or duality. As for the most important, I don't think it's possible to say anything is the most important. As scientists, we can't be satisfied while there are still any unknowns, so it is important to keep attention on all areas. 

  \item Clearly the area of EW symmetry breaking is the most pressing area in the field. However many other questions are closely related to this -- so searching for new physics like SUSY, doing precision measurement in the EW sector etc are all equally important. However they are essentially driven by the first question.

  \item The problem of EWK symmetry breaking and the origin of mass is the most pressing one in our field, and arguably one of the most important in all of science. What's exciting to me is that within the next ten years we should know a great deal about the answer, thanks to the Tevatron and the LHC. But the LHC can't do it all, and for this reason I believe that an e+e- linear collider is the necessary and correct next step to advance the science. It's desirable that such a facility be built in the US, but the main thing is to build it, with whatever technology, SOMEWHERE, starting SOON. 

  \item It's all important. 

  \item All parts of physics are important, but the most fundamental thing is work on the unification of knowledge (not only unification of forces). 

  \item I worked for 5 years in HEP (with quite good success). But now I prefer to work in a subject which has more direct consequences for mankind. I am fascinated about the possibilities of FELs.

   \item  Questions30 and 31: it is really difficult choice in 10-25 year progress in cosmology/ astrophysics/ cosmic rays/ can not be achieved without the research at the accelerator experiments, in my opinion. Let's say that at Snowmass meeting some experiment will report discovery of some supersymetric particlesor establish masses of neutrinos or discover new families or ... impact on other fields is at once. 

  \item The problems I work on involve several of the topics you listed in 30), and I honestly don't what will be the most important physics for the field over the nest 10-25 years. However, I am troubled by the fact that despite the enormous efforts over the past half century I've been in the field, one can still not answer the question, of the mass of the electron, what fraction is due to its interaction with the quantized electromagnetic field? Why do the masses of some of the elementary particles exhibit such simple regularities? The fact that in molecular biology we learn that molecules store information, pass that on to other molecules, which use the information to construct other molecules, including molecules that repair the information-storing molecule is so fundamental that we have to wonder whether the elementary particles also exhibit similar properties, and how would we determine this by suitable experiments? 

  \item The Higgs field, SUSY, CP violation are fundamental properties in Physics. We indeed are living in exciting times and the future promises to be even more exciting (especially if even one the above mentioned phenomena turn out to be true). 

  \item Questions 30-31: All of the above. I am excited by a broad range of physics. I think the field has many avenues to explore, and I will be happy to work on any of them. 33) It's important that the next major facility be in the US, but I wouldn't risk the long term health and solidarity of the field, or our relations with Congress, over this issue. If the LC is built in Europe, we will collaborate on it, and will also build many other smaller experiments in the US, from which we will extract rich physics. 

  \item Dark matter, dark energy. Matter seems to be of minor importance. Supersymmetry must be established (or eliminated).

  \item The most important physics in the next 10-25 years may well be something totally unexpected. High energy frontier facilities have the best chance at this, but this physics may come from elsewhere and we need to maintain diversity in experiments and facilities. 

  \item  I am most interested in the development of mathematically coherent and easily explainable pictures of nature. When physicists give explanations that don't make sense, or only make sense in the context of a wealth of unquantifiable intuition and experience, the recipient of the explanation is not richer, they are more confused. The view of physicists as know-it-alls will in the end be harmful to the field. 

  \item Personally I would be most interested in the discovery/study of charm mixing. Babar/Belle data will likely suffice. 

  \item We have to be able to react to the most recent information obtained. Understanding Electroweak Symmetry Breaking is clearly the most urgent problem before us. What we see at the Tevatron next week may point the way - or we may have to interpret other new data in a way that indicates our most promising next steps. Open-minded curiosity will have to set the stage. 

  \item In reference to question 31, I think the most important experiments are those which will indicate beyond-the-standard model physics. This may require a higher energy machine, better machines for precision physics, or both. 

  \item  The most interesting physics fot the field probably will be astrophysics as a whole, including dark matter, inflation, MWBR, neutrinos. 

  \item Question 30: at the same level, I would also select "astroparticle physics" (which is not offered - cosmic rays is not the same). re 31: I consider it most important that activities are supported which allow for progress in all these fields. In my view, it would be the end of HEP if we disregard fundamental questions to concentrate on one or a few key points. 

  \item EW symmetry breaking, neutrino physics and dark matter are certainly most important to the field, in particular I believe that strong interaction phenomena (similar to QCD) play a major role in the EW symmetry breaking. Also, QCD plays a strong role in understanding results from the new accelerators. Hence I think that while experiments should be focussed around the above phenomena theoretical research should be more diversified than just "beyond the standard model" phenomenology" or just "superstrings". Accelerator wise we need the Large Hadron Collider for discovery of new particles, a linear electron-antielectron accelerator (or alternatively a circular muon collider) for high precision physics experiments and a neutrino factory long baseline experiment. In addition, astro-particle physics detectors should attract more attention too. 

   \item Everyone seems to think the most important physics is Higgs. In my perverse way of thinking I'd love it if we had no clue right now what will truly be important in the next few decades. 

  \item Owing to the nature of the work I have been doing, I am getting more drawn to computational physics. All the physics fields are important, but perhaps emphasis on the fields that can be used to generate publicity should be strongly encouraged, as this would lead to greater public awareness of HEP and greater support for funding. Given the way the Net/Grid is expanding, a new facility can be put virtually anywhere 

  \item Even when MINOS and OPERA are done, we won't have a clear picture of the masses and other characteristics of the neutrinos. It's important that we get to know as many things as we can of this mysterious particle that might hold the key to more clues.

  \item I do not know what will be important 15 years from now, and I wager that no one else does either. Consider two plausible alternatives: * LHC finds, and NLC confirms, MSSM. Dark matter is identified. * Neither Higgs nor SUSY is seen below 1 TeV. Dark matter remains mysterious. Our plans in 2016 will be vastly different in these two cases. 

  \item  Need a "new machine" to go beyond the Standard Model. Hopefully clarify the nature of the "ultimate theory".

  \item There needs to be fascinating new physics to stabilize the Higgs scale, possibly supersymmetry. The goal of HEP in the next 20 years should be to completely quantify all the particles and interactions at the TeV scale. The LHC alone cannot do this, because an e+e- machine is a necessary complement for precise measurements. Since the LHC is already in Europe, it is desirable to have the e+e- machine in America. A nice reversal of the 1990's :) 

  \item For CP violation in neutrino physics (not the only important part of the issue of course) it is probably necessary on physics grounds that the experiment span continents. 

  \item It's really impossible to say what will be the "most compelling" field of HEP 25 years from now. It is true that a great deal of strong interaction physics, non-perturbative QCD, remains unexplored even at current collider energies. Cosmic ray experiments suggest that there is indeed interesting physics and probably surprises here that merit study. 

  \item  I think the most compelling physics will have to do with hints towards the high energy theory of which the Standard Model is a low energy effective theory, such as SUSY.

  \item Personally compelling physics: genuinely new, discovery physics For the field: EW symmetry breaking must be attacked early in this time frame; the longer time scale is appropriate for the exploration of the energy frontier

  \item I am very interested in some specific physics issues for the next 10 years, and they will be addressed in large measure with current or planned facilities. But what comes next will depend on what we find in the next few years. That doesn't mean, however, that I wouldn't plan the next accelerator based on our best current guesses about what we will need next since we can't afford to wait ten years to find out what will pan out. We have to take some risks and make some gambles just to keep experts busy and making steady progress. 

  \item It is all interesting depending on what we find out as time goes along! 
  \item I include SUSY searches as part of EW symmetry breaking. I would not classify these with less motivated random "exotic" particles. If SUSY is discovered it will probably require a new machine to fully explore the spectrum and get at the underlying theory. This need not necessarily be in the U.S. if it is a truly international machine and people can fully participate through regional centers.

  \item I think that the most important and compelling goals of the next 25 years is the the precision testing of the standard model at high energies ( higgs search/ discovery and developing in parallel the capability to test and discriminate non-standard model processes. In parallel we should be pursuing precision tests at lower energies (cp violation, rare decays etc) attempting to reach all the standard model limits for these processes. 

  \item We need to look for the unexpected, not just do the standard model testing and Higgs searches.

  \item By exotic particle searches, I am referring to SUSY particles. I feel that the most important physics for HEP over the next 10-25 years primarily concerns: Higgs physics, cosmological constant, and exotic particle / SUSY searches. 

  \item Supersymmetric particles, dark matter, the Higgs field and extensions of the standard model are clearly the focus of the future. At the same time, the challenge to understand the structure of the visible matter surrounding us (protons and neutrons) from first (QCD) principles is still very exciting (and not solved).

  \item All of these physics topics are important and compelling. But my bet is that, since we think the standard model is not the correct/complete theory of EW symmetry breaking, the most important thing is to get evidence of what lies beyond the standard model in this sector. It could help point to the solutions of some of the other problems listed, such as CP violation, dark matter etc. for the continued health of the field internationally the U.S. should take on the responsibility of hosting the next facility. The commitment would, however, somehow have to be in the form of an international treaty that is not easily broken by congress. It would be a disaster if an approved project with 50% non-U.S. contribution is half completed and then unilaterally killed by the host government. 

  \item Astro/ particle is probably as important as Higgs, etc. 

  \item Politically, we need a major discovery at the energy frontier, to guide theorists, allow experimentalists to target experiments, and most importantly, to generate sufficient public interest to justify funding future experiments. 

  \item If I knew what physics would be most interesting in the next 10-25 years I'd drop out of physics. 

  \item What will be most important? It's a crap shoot where the next discovery will come. I would bet AGAINST the QCD and heavy ion/quark gluon plasma fields. 

  \item My answer to 30 and 31 is a combination of QCD interactions, EW symmetry breaking, CP violation, and hadronic physics. I think we need to use all of those to better define our understanding of QCD, to find out where it breaks down and if there is a "better" version of it. Why is it apparently OK to renormalize? etc. 

  \item We need more testable alternatives to the standard model, particularly in the Higgs sector. It is also possible that string/M/ brane theory may provide new ideas.

  \item The most important physics for 10 - 25 years from now is the discovery we haven't necessarily dreamed of yet. This implies that we build the energy frontier machine -- and only one qualifies -- the VLHC. It should be built in the U.S. because the superior site if available. To describe this as an exotic particle search is not unreasonable but probably too restrictive to be appropriate.

  \item By picking one, I have to not pick the others. I don't care what (I'm in accelerators), but look for something. I think it's wise to trade off with Europe every 5 years with the largest and/or newest accelerator.... Of course that means planning for post-lhc now! At the same time, not collaborating (eg, the U.S. at cern) is childish. Am I asking for too much money?

  \item 1) the four related 'super questions' are (a) the origin of mass (b) the origin of the generations (c)the origin of CP-violation and (d) the nature of the Dark Matter. 2) the priority is for a linear e+e- collider, preferably with an upper energy limit without major engineering upgrade of at least 1 TeV, and preferably 1.5TeV. Should LMA be the correct solar neutrino solution, the next highest priority is a neutrino factory. 

  \item Question 32: which one, 30 or 31 ? W.r.t. 30 the answer is yes: It is amazing to see how little is spent (on the theory side) on dedicated machines for lattice QCD -- our lack of precise knowledge of strong interaction effects precludes, at this time, any definite conclusion from the BNL g-2 experiment; the frequently quoted 2 sigma incompatibility with SM physics is fake, theoretical uncertainties are grossly underestimated, it seems to me only lattice QCD can help.

  \item I regret, my choice is not unique, but the form doesn't allow to select more than one topics. I would have like a choice of "beyond the SM", which encompasses several topics in your list. 

  \item The field lacks direction. Unfortunately I have no clue either where it should be going. 

  \item Neutrino physics is still quite an open field. We understand so little about it. It requires many accelerator, cosmic ray and reactor experiments. The accelerator program should not be limited to long baseline efforts, but should include measurements important for reducing backgrounds in cosmic ray and long baseline experiments, as well as searches for new physics with high flux, narrow band beams. 

  \item It's clear that the EW symmetry breaking mechanism is the next frontier for probing the standard model. Much of the program will be done at LHC, but a linear collider may also be needed to sort out the details, especially if the Higgs sector is more complex than that of the minimal standard model. The new facility should be built wherever possible, where "possible" has both technical and political components. If we insist that it be in the U.S. or nowhere, the risk of not having a new machine increases.

  \item QCD is the big theoretical challenge, and I don't see it going away. But I am an experimentalist, and I see several experimental challenges to be of equal importance: certainly QCD and Higgs physics; but also cleaning up CP violation and the CKM matrix, neutrino physics, and probably a few issues we haven't thought of yet. 

  \item I don't think anyone can say which physics is the most important for the field, I would say that we should continue working in a variety of new places for the best chance of finding what's behind the standard model. It might be neutrino physics, it might be the energy frontier, it might be precision "low energy" measurements. We need to keep forging ahead in all of these areas. I certainly don't think all of the new facilities should be in the U.S.  but one big facility should be here, as well as a "base program" of smaller but equally important experiments.

  \item I think all are compelling. Many people will argue that their subject is the most compelling. I am indifferent at this time. 
  \item Many of the topics are vital to the field, not a single one. 

  \item The most important question for High Energy Physics is what is beyond the Standard Model. Whereas no one can say for sure where this is likely to be found, understanding the Higgs sector and looking for new particles at the highest mass scales achievable are excellent places to look.

  \item Looking back the history, a lot of discoveries were not foreseen, personally, any effect to understand the nature are interesting, and may end up with a break through. 

   \item Question 30 : I find it hard to choose one. this is going to change over time, depending on the results of the ongoing experiments, so I guess I don't want to answer this question. Ditto 31. 33- while I'd like to say this is true (i.e., it isn't important that it's in the U.S.) I fear that politically it *is* important. However, I feel this is an attitude that needs to change. I don't think that fewer (or less high-quality) people would go into physics if the experiments were, say , in Europe or Asia. 

  \item B factories and the next generation experiments are expected to solve most of the long standing puzzles of CP violation. So now it's high time the SM is put to its ultimate acid test, i.e. the test of the Higgs phenomena. 

  \item I think it is short sited to ask us to chose what "the most" important physics is. We should have a diverse program and seriously question any program that is so expensive we can only have one focus.

  \item I'm interested in many things myself, so I just picked one for question 30. As to question 31, it depends on what you think the field is. If you think it includes astrophysics and cosmology, I think that the next most important results will probably come from astrophysics. But in order to answer questions 32 and 33 the way I think you mean them, I chose the subject that requires a major new e+e- facility, because I think that is what is needed to keep accelerator-based particle physics a part of "the field": if we don't build one soon, there may not be such a facility for a very long time (never say never), and "the field" will comprise only astrophysics. I think it is not really that important that the new facility is in the US, but if we want to convince the funding agencies and congress that we are serious when we say that a particular facility is just what we need, then it would surely look strange if we didn't at least prefer the machine to be in the U.S. 

   \item All are interesting and important; however, too much effort may be directed at a fashionable problem to the detriment of others which seem less urgent. 

   \item If the Higgs is the only new particle found in the next 10 years (no susi-particles etc.) it might be a possible death sentence for HEP but certainly hard to sell new even more expensive projects to to the funding agencies. 

  \item I am personally most interested in the structure of nuclei. This field will be strengthened by the advent of radioactive beam facilities. New data will be a challenge for theory on one side and experiments dealing with low intensity beams. 

  \item The most important field is the testing of the SM. If we find new particles (like SUSY) the most important thing would be the study of its properties otherwise it would be the precion test of the SM, e.g. Higgs parameters electroweak parameters and the unitarity triangle.

  \item Important physics: experiments that impact the whole range of astrophysics and cosmology: solar physics, nucleosynthesis, supernovas, dark matter, dark energy (quintessence) and more Major new facility: Two are needed (1) Major Underground Lab. The U.S. needs an equivalent (improvement) of Kamioka mine and Grand Sasso labs. (2) High flux neutrino source in the 10-50MeV range (decay at rest) for studies related to solar and supernova neutrinos.

  \item I include SUSY in EW symmetry breaking when answering 30/31. Actually further unification, gravity etc. should also be on the list (our theoretical colleagues always link SUSY to that anyway). As for location, no, I think the quality of the facility is more important than location (except not dumb locations like Waxahache, Texas...). 

  \item For "other" I think that physics beyond the standard model is important. At the present time I would emphasize string theory (or, as it is sometimes called, the theory formerly known as strings), but I would not like to restrict myself to one particular approach for the future. 

  \item The fundamental nature of space-time, the vacuum, and quantum gravity might be of interest. 
  \item Specify a single physics topic in not appropriate. The goal of our research should be always pursue new discoveries and also to enhance our understanding the physics laws that governing our universe. 

  \item Variety of research is important.

  \item Compelling physics, that is the question. Personally I do not find any of the physics topics currently on the plate so compelling to justify the huge efforts required to explore them. But unfortunately I do not have the magic wand of a solution to this question. 

  \item More attention should be paid to recent developments in alternative accelerator schemes, in particular based on high-power lasers.
 
  \item I think, if one decides to spend more efforts on cosmic ray physics, this should be done at established labs (since the expertise is available there already) 

  \item Personally I find fundamental interactions and symmetries most interesting. For the field interdisciplinary work and openness to new idea will be crucial. The field would be dead, if I could say now what will be important in 25 years. If that an be done, we better stop right now and enjoy fishing or so.

  \item The most important point is that a number of fundamental experiments be pursued. These range from searching for electric dipole moments of the neutron, electron, muon and atoms, to studying the nature of EW symmetry breaking. There has to be a balance of topics under study, since nature has often given us surprises, rather than what we were sure we would find. 

  \item In my opinion there isn't "the field" but rather a wealth of things to do: Higgs/EW symmetry breaking CP violation Neutrino physics cosmological constant/dark matter. 

  \item Question 30: would also add dark matter searches as example for fundamental questions attracting me. Question 31: for me no difference to 30): stop working in a few years. Question 33: could well be outside U.S., if U.S. contributes still.

  \item  If I knew the answer to 31, I would lose interest in the field. 

  \item I think theories like supersymetry with no current evidence, while they should be investigated, are taking up a disproportionate amount of research. 

  \item 25 years: too long timescale. 

  \item I think that all are choices are important for the field in the next 10-25 years and that the U.S. should lead the way. 

  \item I believe we will learn the most by pushing to higher energies.Push to higher energies. The particle physics/cosmology connection is the aspect I find most exciting. 

  \item It is essential for the field to start seeing answers to the mass problem. Not just observing the Higgs, but actually getting a glimpse of a deeper structure beyond the Standard Model

  \item I am amazed that one of the most important physics direction of the next 10-25 years: physics that stabilizes EWSB scale at ~1 TeV, has not even made into this survey. Clearly, faculty should do a better job teaching young physicists what's of crucial importance in this field! (This neither Higgs physics/EWSB or Exotic particle searches; this is physics BEYOND 1 TeV!) 

  \item  I think that it is important that we maintain a diverse physics program, this includes physics at lower energies for QCD, B physics and CP violation, and physics at the energy frontier for EW symmetry breaking, new particle searches, as well as non accelerator based experiments for neutrino physics (for example). I don't really know which of these will be most compelling for me ten years from now. 

   \item It's not possible to pick just one area as being most important for the field. Instead I would pick three-- CP violation, electroweak/Higgs, and neutrinos. For CP violation a new facility is not needed. For neutrinos, a new facility would be needed. For EW/Higgs, its not yet clear if a new facility is needed beyond the LHC. 

  \item About 31): there is a whole bunch of physics subjects which are important for the field. Answer to basical questions : e.g. CP violations , neutrino physics, cosmological constants/dark matter. Once answered those questions will open completely new fields for research, and other basical questions will rise and open new domains of physics. Nailing down the standard model and giving and answer about the Higgs is equally important if found in the energies reached by RunII or LHC, it can open the door to new domains of research as well. However if not reached at that energy level it can drag physics (forever) in directions where it should not go. 

  \item Important physics may not be located in one specific field. It is important to be competent in several topics.

  \item I would most like to see the continuation of particle physics as a whole. This will allow for the discovery of new physics over the next 10-25 years. The type is not as important to me as it would be to a physicist who has devoted the past 20 years (or whatever) to a particular subdivision of particle physics. 

  \item All of the above plus extra/hidden dimensions. The next 10 years will help reveal/uncover new phenomena that will help drive the physics opportunities of the next machine. 

  \item EW symmetry breaking, neutrino physics, astro-particle physics and cosmology are equally important topics. The discoveries which will be made will decide which one is most important. 

  \item We know here is new physics, physics beyond the existing models, today, in the lepton sector. You don't have to build a new machine to find that new physics. We need new facilities and perhaps new beams to address that physics. But at a small fraction of the cost for a new VLHC or NLC. 

  \item While we are pushing hard to probe dark corners of the Standard Model, nothing breaks the mold like a new particle. To accomplish this it seems reasonable to argue that accelerator physics has to be pushed to smaller, more powerful technologies. Instead of turning a blind ear to those who have novel or "impossible" ideas about new accelerator technologies, maybe its time to toss a little venture capital their way. All of this begs for a new facility, either in the long or short term, at which such R\&D can be nurtured. 

  \item I selected CP violation but would say that I was interested in flavor physics -- especially why there are three generations. The U.S. needs to have a large facility -- as the world's richest country it should be willing to invest in this kind of research, which generates fundamental knowledge.

  \item Most compelling fields -- black hole physics, quantum gravity, string theory. I am intrigued that some issues in these fields may even be addressable in near-future high energy machines such as a linear collider. 

  \item There's plenty of great stuff going on. I personally find myself attracted to precision (and thereby, as it happens, mass production) astronomy such as SDSS, SNAP, LSST and a whole bunch of other four-letter acronyms. But astrophysics in general is the way forward, and specifically, I feel the future will belong to those who come up with increasingly clever ways to let nature do the accelerating. Cosmic ray detectors, neutrino telescopes, and the like could provide a great deal of information on new physics (esp. hadronic interactions) at absurdly high energies. 

  \item I find any physics that addresses the questions - why is there mass - why are there 3 families - why is there more matter than anti-matter compelling. These are questions that are independent of field. 

  \item I think that neutrino physics is becoming much more interesting and important, and I think it is not receiving the recognition it deserves. However, when I say this science requires a new facility, I do not mean on the same scale as the linear collider (for example). The advancement of neutrino physics can happen much cheaper, and potentially in parallel to other work in the field. Now, eventually, if one needs to build a full neutrino factory or muon collider, it will be expensive. I also weigh the last answer against what I see happening world-wide, meaning that this path would enable us to maintain useful action in the U.S. while participating in a facility elsewhere. 

  \item In the short term, I think neutrino physics will still teach us a great deal, and I have less faith in accelerator results than non-accelerator ones. Longer term, cosmology (dark matter, possibly dark energy, etc.) will be the important area.

  \item I thinks the biggest thing in Physics will be moving beyond the standard model and then trying to incorporate gravity. And since the U.S. lead the world in particle physics research the entire length of the previous century, I feel we should take our time and see what comes out of the LHC. 

  \item The Higgs sector is likely to be our best window to what's "out there" beyond the SM, if anything. Some Higgs boson is almost certain to be found at the LHC or the TeVatron, but may be poorly measured. SUSY particles may be there and will be the best long-term bet for HEP if they are there. If the Higgs comes in too light or too heavy, then something is "out there" and we should look for it. The LHC can do much of the discovery work, but precision measurement may require an e+e- linear collider. With good global participation, it does not matter where the laboratory is. With poor funding ability in some countries, the U.S. may be financially the most capable of hosting the project. (this may change as some countries' economies improve and they become more willing to finance HEP, or less, who can tell what the politics will do?) 

  \item Depends on what is found: if progress on EWSB is made "soon" the solution may/will probably raise new important questions. Almost certainly continuing development of major facilities needed to answer these. Q.33: Politically probably yes, to retain the strength of commitment of U.S. physicists and funding agencies.

   \subsubsection{Does the Physics Require a New Machine?:}

     \item LHC is being built. That is what we need to do it. 
     \item LHC may find the Higgs, but how much about it will they be able to 
           measure. Vacuum for U.S. based physics after FNAL Tevatron closes? 
     \item The LHC will address most of the next round questions. 
     \item New experimental ways of probing the Symmetry breaking mechanism 
           are required other than just building bigger and bigger colliders.
           Ingenuity and new ideas are strongly necessary, especially due to
           the shrinking budgets for coming years... 
     \item It is precisely because nobody can answer the question about which
           experiments are going to provide the next big breakthrough that we
           need progress on many fronts. We need a LC, but also a neutrino 
           factory. We also need the U.S. to be strongly involved, and that
           probably means (for political reasons) that one of the new large 
           facilities should be sited in the States. 
     \item The primary scientific justification for the SSC was the 
           elucidation of the mechanism(s) responsible for EWK symmetry 
           breaking - the origin of mass. The LHC will (hopefully) address
           this, but may not, because it may be too low in energy. Post LHC, 
           it makes no sense for the U.S. to build an NLC or a muon collider,
           because by the time they become operational, there will be little
           to add to whatever is known from LHC experiments, which will have 
           been running for some significant number of years. In my opinion, 
           the only sensible thing now for the USA to do is to build a p-p 
           accelerator with energy $>>$ LHC. 
    \item  If a new neutrino facility is built in the next 10 years, it would
           be very wise to have available beams from all stages (low to high 
           energy). 
     \item There should be a reasonable mix of non-accelerator and accelerator
           experiments. Neutrino mixing, Higgs, and supersymmetry are the most
           important physics goals over the next few years. There will still 
           be many unanswered physics questions after LHC and FNAL Run II are
           finished. We will need new machines to explore the energy frontier. 
     \item We may learn quite a bit from the Tevatron Run II and the LHC. This
           has to be seriously understood in planning a new facility in the
           area of EW symmetry breaking, for example. The U.S. does need a new
           high energy facility, but we have ourselves to blame for the
           current situation. We didn't we rationally plan for the SSC to be
           an international facility at Fermilab and then plan for a $\sim$ 2 
           TeV e+e- linear collider to come on afterwards? Instead we act like
           a bunch of little kids, fighting each other, and the DoE has made 
           serious mistakes. The DoE should not be trying to manage the field.
           They are far from being wise leaders in high energy physics. 
     \item The present proposals for some of the next machines are evolutions
           based on present technologies. The muon beams, TESLA and the low 
           field VLHC are exceptions. The first offers the possibility of 
           something completely new, the second uses a more efficient
           technology, and the third is the best way to get back to the energy
           frontier. Simply evolving present technologies - apparently beyond
           their scope - by feeding them more and more money is not a 
           responsible way to spend the taxpayers' money. We need to support 
           new ideas and implement the new concepts, or our machines will
           always be costing more and more without becoming more efficient. 
    \item  The opportunity of the underground laboratory (given its scope
           and cost) must be pursued independent of the collider decisions. 
     \item I feel that the next machine needs to be an electron collider to 
           study the EW sector along with the LHC. I feel that a VLHC is 
           necessary to study energies beyond LHC and carry the physics
           further. I feel that it makes the most sense to build the VLHC in
           the U.S., based at FNAL, since much of the infrastructure exists.
           The electron-electron lab can be built in Germany while the studies
           and planning is being completed for the VLHC. 
     \item My suggestion of accelerator physics is born of desperation as I
           don't think we presently have a fiscally viable new facility on 
           the horizon. The hope would be that accelerator R\&D might produce
           a cheaper alternative. 
     \item Current proposed new accelerators are unrealistically expensive. 
           We need smarter, more versatile in physics potential, cheaper
           machines if we ever expect to get approval from the government 
           to build a $>$ 1B\$ facility. 
     \item It is clear that the LHC will not provide all the answers we are 
           seeking but should help us toward the next step for both experiment
           and theory. It is important to complement the science to be done at
           the LHC with another type of machine that will provide further 
           insight into the foundations of the standard model. 
     \item If "major" means so expensive that all the rest will be down sized,
           we better put more effort in acceleration technique so that higher
           energies can be reached with different kind of machines. In the 
           meanwhile there is PLENTY of physics at lower energy that can be 
           exploited with less expensive new facilities or upgrades of 
           existing ones. 
     \item I think we should get moving on VLHC to cross a new energy 
           frontier, or else go get real jobs in San Jose, leave behind a few
           hundred people to finish up ongoing work on neutrino properties and
           EWSB, and hope our friends at CalTech see gravity waves. 
     \item HEP has a long record to shut down facilities much to early - LEP
           being the last of the list. 
     \item There are already major new facilities either planned or in
           operation both for QGP and cosmological constant (CMB). I do not 
           know if any major new facilities will be needed beyond those 
           already on the table -- probably not. 
     \item For Question 31/32, I think that the LHC at CERN should find 
           something new. A new experiment, probably an e+/e- collider, will
           be required for precise measurements of this new physics in 10-15
           years or so. 
     \item If there is to be a future in accelerator based HEP then we must
           invest in accelerator R\&D to identify and develop affordable new
           technologies. 
     \item Although I would like the next facility to be a neutrino factory, 
           I am resigned to the fact that the momentum in the community is 
           behind a linear collider and I think there are good reasons for 
           this decision. Given that, I will support the best design and 
           fastest timescale to build a linear collider (TESLA) and push 
           for R\&D for a neutrino factory. 
  \item Should try to put more effort into R\&D, both accelerator and detector related. This will make it possible to make a better case. Also it should be made very clear to powers-that-be (politicians) that R\&D is an integral (and essential) part of our activities, and that special funding should be provided for this effort.

\subsubsection{Location of a Frontier Facility:}
  \item Stop being cry-babies about the U.S. losing their leading role in HEP,
        and start thinking what you (the U.S.) can do for world HEP. 
  \item I would only support a new machine at Fermilab. I do not want to go
        through the same procedure used for the SSC. Without these 2
        constraints I do not support the construction of a new machine in 
        the U.S. 
  \item I don't believe the U.S. can maintain a vigorous program without 
        a frontier facility. 
  \item I'd like to see a U.S. national underground science lab built very
        soon. 
   \item The U.S should have a major underground experimental facility for 
         HEP. 
   \item U.S. National Underground Laboratory is necessary. Please, no 
         Homestake. Who is going to work in South Dakota? 
  \item It may not be important for physics that the facility be in the U.S., 
        but it's important for U.S. physics. 
  \item It's not important to HEP in general that it be in the U.S. The 
        Europeans can do the discoveries. but, it is important to U.S. HEP. 
  \item Incremental upgrade of Fermilab will deliver the neutrino factory and 
        maintain a healthy HEP in the U.S. Going for the linear collider (it 
        looks like the TESLA option is going to win out) will merely delay the
        acquisition of a neutrino factory, since it looks as though TESLA will
        go to DESY (as it should). DESY did the R\&D! 
  \item A new facility in the U.S. is crucial to keep young physicists from 
        leaving the field. 
  \item A new facility in the U.S. does NOT mean it has to be a "U.S. lab!" 
  \item From a science point of view, it doesn't matter where the new machine
        is located. From a political point of view, however, thinking that it 
        doesn't matter is worse than asking for too much money. 
  \item Preferably the facilities should be distributed equally throughout the
        world. But this should not be used as an argument that the LC should
        be build in the U.S. 
  \item It needs to be in space! 
  \item Several new facilities are needed. It is important that one be in 
        the U.S., but I don't think it is important (from a physics point 
        of view) which one. 
   \item To survive, we have to get past the competition to host machines in 
         the sense that Congress decides it isn't worth supporting HEP unless 
         the facility is local. The space station is not in any state of the 
         USA. 
  \item The U.S. needs no less than to have the highest energy machine in the 
        world. A sub-TeV lepton collider will not provide that. 
  \item I have decided to do physics in Europe because, after the SSC, I 
        believed that the U.S. would not be able to pull it together. The NLS
        business is evidence that I am right. But the U.S. must have a program
        or the European program will flounder. This is clear from the time 
        slippage of the LHC without an existing SSC: recall SSC was scheduled 
        to run in 1999 and lhc in 1998. We now see LHC in 2006? 7? 8? 
  \item I would prefer a linear collider in the U.S., but feel that it is 
        more important that we have one somewhere. 
  \item I would only support a new machine at Fermilab. I do not want to go 
        through the same procedure used for the SSC. Without these 2 
        constraints I do not support the construction of a new machine in 
        the U.S.
     \item I believe it is a major mistake for the U.S. not to have a major 
           accelerator in place during the reign of the LHC. The HEP community
           here will be marginalized in important ways. We can partly offset 
           this by as-strong-as-possible participation in LHC and other 
           international objectives. 
     \item U.S. people should get over the fact that we can't have everything 
           here. Other countries, while not having the same money as us, when 
           combined can do very well as collaborations. It is also good that 
           there is international pressure on individual governments to 
           continue support. The U.S. cannot guarantee that its government 
           will always be HEP-friendly, but an international collaboration 
           always has some HEP-friendly members at any given time. Therefore, 
           we should get used to the idea that the next big project could be 
           anywhere in the world. Of course, being from the U.S., I would like
           it here. But there are other factors against the U.S. - It is so 
           damned hard for foreigners to get visas to stay here. It is scary 
           for some to leave the U.S. because they don't know if they will be 
           allowed back in. The lack of respect by this country for scientists
           from other nations is alarming. 
     \item The location of the next machine should not depend at all on 
           national pride. Much more important is that it be located in a 
           truly international lab, as CERN is (I am not saying it has to be 
           at CERN...). 
     \item Just get some funding and build one *good* accelerator with some 
           decent (at least 2) detectors on it. Location doesn't matter, 
           joining forces does. 
     \item It is very important to have some major facility operating in each 
           region. For Fermilab, the most promising suggestion I have seen is
           by Peter McIntyre, to increase the Tevatron beam energy to 3 TeV. 
     \item HEP will die in a continent that does not have a machine (well, 
           there will be a few small university groups).
     \item I don't think the performance of the machine will depend
           on its location, given that the basic resources, such as constant 
           electrical power, are guaranteed. There are some obvious problems
           at SLAC with the brownouts, and the phones not working ! Seems
           like it would be easier to run SLAC in Mexico ... 
     \item I think that a new e+e- collider built soon would benefit the 
           field. I'd prefer it be built in the U.S. so that I could more
           easily work there and for the interest in physics it would bring 
           to the U.S. However, if the U.S. doesn't want to pay so much, we 
           should still certainly support a linear collider wherever else
           it may be built (e.g. Germany). 
  \item There is a natural tension between being a good internationalist and 
        being a U.S. PHYSICIST. You read it in the various Gilman etc. 
        reports which instruct us to be international in our planning but to
        be sure that we maintain U.S. leadership. I believe these are not 
        incompatible. The TEVATRON has given U.S. leadership for 20 or so 
        years.
        I believe it makes sense for the U.S. to continue this with VLHC.
        High energy in the domain of multi hundred TEV will expose new
        phenomena rivalling anything in our history. That is the usefulness of
        the astrophysicists and even they may be as children, throwing pebbles
        into the surf of a vast ocean of ignorance. In some sense the Higgs,
        supersymmetry are "tired" discoveries. But what the hell is dark 
        energy? 
  \item The facility should be the best for physics in whatever site gives 
        the best chance of worldwide funding and participation. 
  \item To clarify, it is even more important that the U.S. try for the next
        than it succeed. Not trying is tantamount to abdicating a major 
        league participation. 
  \item The U.S. needs a frontier machine after the LHC starts up in 2007 
        (more likely 2009!). It is not necessary for the next machine 
        (probably a linear collider) to be built in the U.S., but planning for
        the machine after that is essential, and R\&D must start immediately. 
  \item A facility is better than no facility, no matter where it's built. 
  \item Location in the U.S. would give a significant boost to the U.S. 
        program; for this, location is important. If, on the other hand, 
        putting it in the U.S. means it doesn't get built, this would be a 
        tragedy. 
  \item We need to have the Tesla LC design built near Fermilab. 
  \item It is preferred that new facility be in the U.S. A new facility abroad
        (i.e., no new major ones in U.S.) requires substantial reorganization
        of U.S. field; not impossible, but there will be resistance. 
  \item I am a U.S. citizen, and want my country to be the leader. 
  \item As a European I think it's important that a new facility should be 
        in Europe. 
  \item Again, I am from Europe and I think that it is important that a new 
        facility be in Europe (for Europeans) and in the U.S. (for Americans),
        etc. Of course, this would not prevent strong exchanges between the 
        various continents. 
  \item Every country wants it to be in their country. We need a better 
        argument ... 
  \item The most important property of the new facility is that it be 
        economically viable. While it would be better for U.S. physics if it
        were in the U.S., one should not rule out a foreign site if it appears
        that this is the only way to make the machine happen. 
   \item Priority is that a next facility a e-e collider in the O(1TeV) energy
         range is built somewhere. Location is of second importance. 
   \item We have to give up that science will die if the U.S. doesn't have a 
         facility. Of course, as weather research has learned, if you are 
         willing to form collaborations broadly, your research will die. In 
         their case, computers did them in. 
   \item The new facility should have international character and could be 
         located anywhere in the world where there is reasonable 
         infrastructure. 
   \item The location of a new e+e- collider is not as important to me as
         it used to be. 
   \item It appears to me that we should have a new e+e- collider somewhere. 
         Surely it is best for U.S. physicists if it is in the U.S., but the
         location is a matter for negotiation.
   \item The physics reach should be the primary objective. The location 
         should be decided based on logistic and funding possibilities rather 
         than national arguments. 
   \item Being in Europe I naturally find it more compelling to have the NLC 
         in Europe. I also happen to believe that TESLA is the best concept
         around, therefore I think that TESLA at DESY would be the best thing
         to happen with TESLA at Fermilab still being better than any other
         NLC somewhere else. 
   \item With the current political climate in the USA I find it very hard to
         believe a truly new project has any chance of being funded. 
   \item Good physics can be done anywhere in the world, if the resources are
         available. Whether the U.S. government will provide good funding for
         the field if the next accelerator is built in Europe or Japan rather 
         than in the U.S. is an open question. It may be possible if there is 
         a consensus for a plan which includes a facility in the U.S. 
         subsequently with a European/ Japanese commitment to participate. 
   \item The only reason I suggest that we need a facility in the States is 
         because of reality. The university system is not yet prepared to
         offer graduate students overseas access. If it were just the science 
         that we needed, we could stick the new machine ANYWHERE. But we also
         use the field for training, and for that we need people to have 
         access to the project. 
   \item It is most important to get the new machine. If that means building
         it in Germany, so be it. I would be willing to take part. Whether my 
         present employer would be willing is a different question. A linear
         collider is an essential tool for pursuing the physics. It would be 
         nice, but not necessary for that machine to be in the U.S. For the 
         health of the field IN the U.S. it would be very good if it were
         built here. Otherwise there will be a decay of the infrastructure to
         remain viable in the long run. If the machine is build abroad, the 
         U.S. must be an active participant. We must also learn from the SSC, 
         for which the U.S. was not really serious about it being an 
         international machine until it was in trouble. I see many signs of
         this recurring. 
   \item I think a new facility has to be built in the U.S. It is unreasonable
         to imagine that the support will be there to train enough graduate 
         students in facilities in Europe. Or for more than a few senior 
         people to spend a reasonable amount of time there. 
   \item Prior to WWII, most of forefront physics was centered in Europe. 
         America was known for its engineering might. After WWII, elementary
         particle physics was rewarded with several facilities in the U.S. 
         because of the critical role they played in the development of the 
         atomic bomb. They were recognized as a national resource that needed 
         to be maintained and developed. As we have moved away from cold war, 
         so to has America moved away from supporting basic science in favor 
         of applied physics with shorter term pay back. This has lead to a) 
         a drain of graduate students and postdocs in the physical sciences 
         into professions that have strong industrial support (i.e. dot com 
         programming and networking, chip design). More senior persons in the 
         field are leaving U.S. facilities to work in Europe (i.e. formally 
         LEP, now LHC) which will capture the energy frontier since the death
         of the SSC. 
   \item The U.S. will have a larger role if it is in the U.S. 
   \item We don't need everything here, but we do need something here. 
   \item It is important for the field to move ahead with the appropriate new 
         facility, anywhere. It is also important that the U.S. have forefront
         facility, and this is the only reason to say in Q33 that the
         facility should be in the U.S. The U.S. should play a major role, 
         wherever it is built. 
   \item Any new lab doesn't necessarily have to be in the U.S. Why not build
         it somewhere else (other than of course money) ? 
   \item Physics is worldwide the same. I don't see any point in insisting 
        that future experiments have to be conducted in the U.S. 
   \item I believe we have to talk international now and the U.S. has
         to truly join the effort of anybody else, Japan, Europe, Asia, 
         Australia etc. etc. to build the "next machines"! 
   \item U.S. is currently far behind Japan and Europe as far as the facility
         to carry out numerical calculations in lattice QCD is concerned. 
   \item New facility should be somewhere with some existing infrastructure 
         for the type of accelerator chosen to minimize the cost. 
   \item It is vital that the USA remain a major player in HEP and that many 
         of the brightest students are attracted into the field. That will 
         only happen if there is a laboratory with the resources to undertake 
         the most important research. In the next couple of decades that means
         a linear collider. 
   \item It's important that the U.S. HEP community survives and thrives. If 
         this means they need a facility in the U.S. then my answer to 33 is 
         "yes" - but as an outsider I don't know. Maybe international 
         facilities based wherever are as good/better (the UK works like 
         this). 
   \item The U.S. certainly needs to remain a major player in the field, but 
         it is also important that Europe and Asia maintain strong programs. 
         So, while I certainly think the U.S. should be preparing for a new
         facility, it need not necessarily be the NEXT facility. We could 
         build the next-to-next facility, for example. 
   \item If the next lepton collider and LHC are both in Europe, what major 
         facilities would the U.S. have probing the energy frontier? 
   \item I am European, so I would actually prefer that large facility to be 
         in Europe ;) 
   \item It is obvious that any new, large machine will necessarily be an 
         international facility, but the U.S. needs a new facility to keep 
         domestic science strong (perhaps make it stronger) and to keep some
         of the younger people who are leaving the field. 
   \item The credibility gap engendered by the SSC debacle still exists.
         Large project management in the U.S. has not improved much since. 
         Even though I am U.S.-based, for me this risk weakens the case for
         the next machine being the "U.S.'s turn" enough that I answered 'no'
         to 33. 
   \item I don't believe that all the world's HEP facilities need be in the 
         U.S. 
   \item The Tevatron and LHC will hopefully cover much of the EW symmetry 
         breaking mechanism, but probably a linear collider will be needed to
         do this accurately. Given that there may only be one such linear 
         collider in the world, I don't see any particular reason why it 
         should be in the USA. 
   \item The importance of placing a new detector in the U.S., in my mind, is 
         that of continuing to attract people to the fields of physics in the
         future in America. The young adults just entering college for the
         first time are the ones who need to have a solid reason to expect
         work in physics, and their youth and childhood should include 
         exposure to HEP as something they could do when they grow up - it is 
         often seen too much as an endeavor which only some very few can 
         participate in. 
   \item The location does not matter much. Groups working in large 
         collaborations are already used to being located far from the 
         experiment itself (Tevatron and LHC for example). Even if the new 
         facility is not in the U.S., U.S. institutions can still have a very
         important participation in it (cf LHC). 
   \item If a new accelerator is not built in the U.S. in the intermediate 
         (10-25 yrs) future, the field here risks serious stagnation. The 
         senior scientists will travel abroad to the experiments and important
         conferences and travel funds will be insufficient for postdocs and 
         students. The field will be very unattractive to many bright young 
         people. 
   \item Why do you want to know if it should be in the U.S.? Maybe people 
         feel it should be elsewhere. Anyway, to be convincing I think we
         should work on our physics motivation and than deal with politics. 
   \item Naturally, the next big machine should be in USA after Europe got
         the LHC. 
   \item I think a warm technology e+e- linear collider should be built at 
         Fermilab, with a gamma-gamma option vigorously pursued. 
   \item It is not important for the quality of the SCIENCE that the next 
         accelerator be in the U.S. However, if the next accelerator is 
         elsewhere, the U.S. will cease to play a leading role. This may be 
         inevitable, but will be unfortunate for the careers of high energy
         physicists in the U.S. 
   \item There are countries other than the USA. 
   \item The major reason I think it's important to put the facility in the 
         U.S. is personal. I would rather do work at a lab in the U.S., so 
         that travel would still be a N-day thing rather than N-week thing. 
   \item A linear collider in Germany would cost the U.S. 1B\$. If built in 
         the U.S. it would require 1/2 of the cost and, using U.S. accounting,
         6 B\$. So we pay 3 B\$. Saving of 2 B\$ if done in Germany. 
   \item I strongly believe that U.S. should maintain leading role in high 
         energy physics. If no new major HEP facility is built in U.S. soon, 
         that will be hardly possible. 
   \item Michael Witherell made a convincing case for the LC, although I don't
         think it necessarily should be at Fermilab, though I agree with him 
         that it is important for American science that it be in the U.S. 
   \item It is important to take back the lead in energy frontier machines and
         therefore to build an e+e- machine that complements and expands on 
         LHC sensitivity to new physics. It is important that this machine be 
         in the U.S., for obvious reasons. 
   \item The U.S. should try to take the lead on the next big step beyond the 
         standard model, not the incremental approach of the next linear 
         collider. 
   \item It seems like TESLA is a great plan, and that DESY really has their
         act together. But, it also seems important for U.S. involvement (and 
         funding - Congress seems to like spending continously, rather than 
         projects interleaved with times requiring no structural investment) 
         that it be sited in the U.S. 
   \item The most important thing is the intellectual input into the subject.
         This requires bright students and postdocs. They are attracted by 
         research facilities that enable them to learn fast in a dynamic and 
         challenging atmosphere, and contribute to the subject. The facilities
         need to be accessible and not a continent away. 
   \item To keep the field of HEP alive in the U.S. it is obvious to me (even 
         as a European working in Europe) that the aim should be for a LC in 
         the U.S. To reach the end of the Tevatron without the next big step 
         in the U.S. planned would risk to destroy HEP in the U.S. 
  \item  Build VLHC in the U.S. please !!!!!!!!!!!!
  \item Concensus seems to be that no country, including U.S., is going to go
        it alone for future big accelerators. Important that we plan for the 
        possibility that it will be a long time before the U.S. has the 
        frontier machine here. 
  \item Unless the next machine is in U.S., the U.S. field will collapse (by 
        30-50$\%$). If this happens, HEP worldwide will similarly shrink, 
        although more gradually (perhaps). This will threaten whatever new 
        machine is under construction, and the SSC proved that big new 
        accelerator projects can be canceled once support for them diminishes 
        beyond a certain point. (Actually, this principle was established 
        earlier by the fate of ISABELLE, but not sufficiently understood by 
        enough people at that time.) Ergo, the next mmajor machine MUST be 
        built in the U.S. if there is going to be a next machine. This 
        principle supercedes any other in deciding what machine to build and 
        where it is to be sited. 
  \item If reduced financial support in the U.S. continues, the best option 
        may be TESLA in Germany. There may not be Congressional support for 
        any large accelerator project in the U.S. in the forseeable future. 
  \item If a LC is not to be built in the U.S. soon, then it becomes important 
        to support TESLA in Germany. 
  \item Looking back at the decisions concerning projects and its seeming 
        financial guarantee made in the U.S. in the last few years, I would 
        rather pick an option anywhere outside the U.S.! 
  \item I would choose an e+e- collider. I believe we should bid to host the 
        machine in the U.S. but be willing to enthusiastically support it if 
        it is sited elsewhere. 
    \item Linear collider programme is an essential next step to complement 
          LEP. It would help world HEP if the collider were sited in the USA 
          in order to bring forth the next generation of particle physicists. 
          It would be a terrible loss to world HEP if the USA did not host a 
          major new project in the next decade. 
    \item I believe it is fundamental that the U.S. always host a frontier 
          machine...either we should have the frontier linear collider or 
          hadron collider, but if both are outside the U.S., U.S. particle 
          physics will die.
    \item The U.S. should aim to maintain at least two accelerator labs. 
          We have had BNL, CESR, Fermilab and SLAC and that has been healthy.
    \item Some large physics experiments in Australia or Spain would be nice. 
          Those would be interesting countries to live in. 
    \item As written earlier, the geographical location is a non-issue. 
    \item Benefit NOT being in the U.S. due to funding problems in the U.S. 
          SSC is an example why not to have the next facility in the U.S. 
    \item I spent decades working at CERN, so a facility doesn't have to be
          in the U.S. to be useful. On the other hand there must be some
          facility in the U.S. or the field will simply die. With a career 
          choice of "this hole in the ground in Switzerland" or "that hole 
          in the ground in Switzerland", the field is dead. 
    \item Question 33: regional centers is an excellent idea to keep more 
          countries involved, so that it no longer becomes an issue of
          nationalism (which is too often how major programs are sold to the
          public!). Of course, I would like to see the next big initiative 
          at U.S./FNAL, however it is more important that we make a decision 
          that will move the field forward. 
    \item Higgs is important only for the morale of the field. In some ways 
          we'd be better not to find it, so we could start some new 
          theoretical work. The facility needs to be in the U.S., again,
          for morale reasons. 
    \item The next discovery is equally as likely to come in any area. If a 
          large new facility does not exist in the U.S. by the end of the LHC,
          then the U.S. will become a third-world nation in fundamental 
          physics. The sacrifice of any major field of research hurts all 
          science. 
    \item Location of TESLA is unimportant. Building a linear collider should 
          be top priority regardless of location. 
    \item No machine in the U.S., have proven not to give reliable schedules. 
    \item It's just a wish - I don't expect it to materialize. We will be 
          damn lucky to ever see another facility in the U.S. 
   \item Let physics decide what is needed. Then build what is needed (no
         matter where, the point is that the best machine is to be built, 
         again no matter where). 
   \item In my view the location of the next machine is not as important as 
         the question if it will be build or not. 
   \item The big worry is that if the U.S. does not host a major new
         accelerator over the next 10 years, we will lose precious know-how
         and our technology base. That will also contribute to the attrition
         already taking place, especially with regards to the young people 
         leaving the field after their PhD or a few years of postdoctoral 
         work. But, by far most important, is for HEP to come out with a
         clear plan for the future that we all should support. If that means 
         that my preferred option is not selected, that's fine as long as
         there is a sense that the physics at a new facility is exciting and 
         we will learn a lot from it. 
   \item The U.S. should strongly support simultaneous world-class e+e- 
         and hadron facilities -- one of these two facilities should be 
         in the United States with strong international participation, the 
         other should not be in the U.S. but it should have substantial 
         involvement and investment by the U.S. as a fully committed partner.
  \item  I don't see that the post-TESLA world by definition has to be in the 
         U.S. It could be anywhere ! 
   \item It is very important for the continued health of our field over the
         next decade or more to have some project in the U.S. which is in a
         position to take a leading role in hep research. I am not sure
         whether this is an NLC type machine or something else. This will 
         depend heavily on the proposed energy of the machine. To make this 
         project really worthwhile will require a center-of-mass energy of 
         $\sim$ 0.5-1 TeV, and at the present time I understand that this is
         higher than what is being targeted for at the present time. Further,
         given the timelines for other big projects like VLHC or Muon
         Colliders, we need something in the near term to keep our programs/ 
         facilities/ personnel in tact until such time as one of these big 
         projects can be designed and funded! What that near term program is
         is difficult to say. There are proposals to upgrade the energy of
         the Tevatron ring to allow for a competitive pbar-p program running
         in parallel with the startup of the LHC. This would make sense if it
         could be launched soon and have beams in a few years. With an upgrade
         of the magnets to 12 T, the energy of the beams could be tripled and
         the CM energy raised to 6 TeV, which would give this machine a chance
         in competing with the higher energy LHC pp machine. My feeling is 
         that we need such a shorter-term project if we are to be able to
         make it through to the next level of new machines like the VLHC, etc. 
   \item The field needs to spend a larger percentage of its resources on
         R\&D for accelerators and other new technologies. This is even more
         important as the next-next machine will be even harder to build. This
         shift in resources will delay the next new machine being completed by
         a few years but is important for keeping the field alive beyond the 
         current generation of physicists.

\subsubsection{Survey-Specific Comments on Physics:}
  \item Well, to question 30 and 31 I would have added neutrino physics, 
        one choice is not enough. 
  \item The list of topics to choose from is very uninspiring. None of them 
        are particularly important or compelling. There should be something 
        along the lines of seeking the unexpected. For example, we need more 
        young physicists to fervently believe that the standard model is poor 
        and inadequate and have a physics plan to research this particular 
        topic. 
  \item I think your lists are too restrictive! What we desperately need right
        now is to map out the structure. Is SUSY there? What is the dark 
        matter and energy? Extra dimensions? These general questions are much 
        more important than the width of the Higgs or precision measurements! 
  \item Most of the items on your list are important and compelling. I hope 
        we can have efforts in all those areas. 
  \item All of the science topics given in the list are important, so my 
        answer is - all of the above. We will do U.S. particle physics 
        considerable harm if we try to specify one aspect of the field at 
        the expense of others. 
  \item Question 31 is misleading: Flavor physics and Higgs Physics might be 
        two distinct approach of the same problem. GUT might become a 
        reality $\rightarrow$ must do both! 
  \item Again, a bit U.S.-centred. 
  \item Is this now a young physicists panel or a young U.S. physicists panel? 
  \item "REQUIRE a major new facility" -- is an inappropriate phrasing ... 
        "important" to whom -- me, the field, the U.S.? 
  \item I find question 31 very ambiguous. I do however think that a new
        facility is totally necessary, probably an e+e- collider. 
  \item This is a naive and old-fashioned list. I cannot choose one of them, 
        and I hope most active people cannot. What are "exotic particles"? 
  \item Question 32: new with respect to what? is LHC seen as new here? 
        Then yes. If not, then, maybe. 
  \item Question 30: The list of options you give tells me how bankrupt the 
        thinking in the field is. Would you think finding a 2nd cosmological 
        relic exciting? Question 31: The "field" for is investigations into 
        the fundamentals of how nature works... the "what's it all about?" 
        questions. For me the big one is a looking out from our lab's to see 
        how the processes we've uncovered at accelerators are manifested in
        nature. 
  \item Question 32 is poorly written. Does "selected" refer to question 31 
        or 30? 
  \item I think this is poorly phrased: instead there are perhaps two time 
        periods to consider ... one at 5-15 years and another at 15-25 (30?). 
        It seems to me there are likely different physics goals and machines
        involved in these two periods. 
  \item I find questions 30 and 31 unuseful. There are a number of important
        subfields, and picking out only one does not fit with taking a broad 
        view of our field. 
  \item You are missing supersymmetry in this section! 
  \item You should have allowed a ranking of the importance of the physics
        fields with weighting. There are three or four fields that are 
        important to the future of fundamental (or particle physics) one of 
        these is cosmology (broader definition of cosmological constant),
        symmetry breaking, accelerator physics. Others such as cosmic ray and
        neutrino physics while topical are unlikely to produce a major 
        advance in understanding frontier physics. 
  \item I found your options TOO LIMITED for a 10-25 year from now window ...
        These will be radically different in that era. I am driven by personal
        ideas angles on how to approach some aspect presently involved in CP 
        physics, QCD, Gravitation wave physics, think hard about EW sym Higgs
        notions; I want physics and intellectual reach in a new facility
        beyond the Regge Poles of our time... 
  \item This section is ill-conceived. Clearly, any thoughtful physicist would
        consider the full scope of HEP challenging and interesting. 
  \item Why this last stupid question? I repeat: U.S. is not the only country
        in the world!! 
  \item I am not sure that selecting one from the available options given in 
        this section is particularly meaningful. 
  \item Questions 30, 31 -- don't think these questions should ask for only 
        one answer. Compelling and important have many meaning, e.g. 
        "important" as "necessary" is QCD, "important" as "essential for the 
        future understanding of HEP physics" is Higgs physics/EW sym.
        (as well as QCD). Most challenging? Less HEP-mainstream: neutrinos, 
        cosmic rays and cosmological constant. 
  \item This list had a large hole. It left out nucleon lifetime and other 
        high-mass interaction signatures. 
\end{itemize}

%% file: comments-pick.tex
\pagestyle{myheadings}
\markboth{Comments on Picking a Plan/Finding Consensus}
         {Comments on Picking a Plan/Finding Consensus}
\subsection{Picking a Plan and \\ Finding Consensus}
 \begin{itemize}
   \subsubsection{Selecting the Next Machine:}
    \item You assume a "next machine", but is it really the right path? 
          Maybe we are just traveling on the paved road for lack of new ideas. 

    \item We very much need to prioritize on what is THE "next machine" we
          want to build and promote that with the backing of the whole 
          community. We can turn the focus on another facility after this 
          is approved to make progress step by step. Trying to "push" 
          everything at once without priorities is a signal to the funding 
          agencies not to fund any of them. 

    \item Snowmass is a rubber-stamping exercise that is meant to support the 
          decision already made by lab directors to build NLC. This is done 
          in the absence of any other good ideas, particularly on how to build
          accelerators. It ignores the fact that not everyone in the field can
          be on a BaBar sized experiment with LHC size costs. I voted with my 
          feet and moved to Europe to do physics in the next phase. There will
          be no physics to do (and no money left) in the U.S.

    \item I believe that, owing to the progressively shrinking HEP budget, 
          the focus in the accelerator-based HEP should be on a limited 
          number of projects worldwide, among which mu+mu- ought to be given 
          the priority. As an advice to young researchers, I am afraid that 
          the job openings in HEP won't be that numerous during the few 
          coming years. 

    \item Support antiproton physics.

    \item I believe the only presently sellable machine is the proton driver.
          Since we haven't discussed why our dream machine, the SSC, failed 
          to be completed, how can we assume we won't fail again?

    \item I opt for the VLHC hadron collider at the present time as that is 
          the only "next machine" that has a "guaranteed" potential for 
          discoveries and reasonable time scale.

    \item I believe that e+e- linear collider should be the next machine 
          and to be built at Fermilab site.

    \item Brite neutrino source.

    \item An e+ e- linear collider at cm energy of 500 GeV coming on in 2012 
          is obviously not at the energy frontier since LHC will have already 
          been running at a higher equivalent energy since 2006. It is only 
          a "clean-up" machine.

    \item Focus on ONE big machine for entire world.

    \item The highest c.m. interaction energy observable in the laboratory is 
          the main value, to be recognized worldwide and pursued with any 
          limited budget.

    \item I am in the opinion that we should build an e+e- complement to the 
          LHC ... i.e., a 1 TeV e+e- machine.

    \item Ultimately physics output is very important. So, theorists, 
          phenomenologists and experimentalists should jointly decide which 
          experiment is most important and build a facility according to that. 
    \item The "next machine" should be a Linear Collider. The important next 
          questions are: where ? and which technology ? 

  \item I like the idea of the first choice because I don't know if the muon collider or the VLHC would be better and it would be nice to have the e+e- here in the U.S. But since we need world money, we got to go with world consensus on the location. 

  \item I think we should support TESLA in Germany, with increased high-gradient accelerator R\&D support here. Not sure about muon collider. 

  \item If HEPAP makes a recommendation different than you would prefer, will you support the HEPAP decision? At this point I think that consensus to back any of the several very good options is more important than which one is actually chosen. 

  \item  A Neutrino Factory should be built when/ if possible. 

  \item A new linear lepton collider would be beneficial to the U.S. However it almost seems too late for it to happen. If TESLA becomes reality in Hamburg, the U.S. community should strongly push for a muon collider. In the meantime we need a decent underground lab, e.g. San Jacinto Peak. The future will see more non-accelerator experiments and we don't want to miss that flight either. 

  \item TESLA in Germany soon, and the next generation linear collider (CLIC-like) or muon collider somewhere. 

  \item I think there should be a facility in U.S. that will start to operate approximately when LHC program comes to an end. I don't think the argument "since LHC is a pp-machine we should build an e+e- machine" is right (i.e. VLHC is a good option too). As for TESLA in Germany -- isn't that a decided matter already? 

  \item TESLA in Germany soon, but VLHC and neutrino factory are not mutually exclusive (neutrino factory is much cheaper, and goes to say BNL).

  \item In terms of cost and range of physics, I see no benefit to another linear collider. I think that either the VLHC or a muon collider in the U.S. are the best ideas, the VLHC being the most plausible. 

  \item The location is not important to me. Do the physics anywhere. 

  \item Wait for LHC and neutrino oscillations experiments results. Then decide on LC, VLHC, neutrino factories and mu colliders. Expand R\&D on all of the above, trying to reduce costs and maintain diversity. Build a U.S. underground lab now. 

  \item A muon collider would be comparable to TESLA, with the additional possibility of high intensity neutrino studies. I think a machine like this would be useful AFTER a new discovery machine like VLHC.

   \item A new facility should be carefully designed to have potential to evolve, or flexibility- which I think limits the appeal of the muon collider. Heavy investment should be justified by long term pay out as well as academically rich opportunities. This "richness" may be a mixture of science and non-science opportunities. The HST, for example, is good for students, physicists, and the general public, too! 

  \item While I understand the point of choosing options a point needs to be made. A U.S. choice against a new e+e- collider located in the U.S. may have dire consequence for all other efforts. Politicians like the argument, "If our international colleagues did not think it was worth doing why should we spend our money on a project they considered not worth the investment." I could expend much more verbiage on this topic.

  \item I fully support an NLC/TESLA/CLIC somewhere, as well as LSC and VLHC. 

  \item I believe the choice of the next machine can not be made on a pure logical / mathematical basis. My opinion is based on my physics, politics and "reality" positions.

  \item I think it's important to have (at least) one large facility in Europe and (at least) one in the US. I can't decide between a VLHC and a Muon Collider for the U.S., but it is important to build TESLA soon. 

  \item I favor moving ahead briskly with a e+e- LC; if that means "cold" TESLA-like technology, then fine. It does not have to be in the US, but it does have to have reasonable "expandability" in energy. I am not sure what "reasonable" is and hope that Snowmass can better determine this. 

  \item I think the VLHC is the most important thing we can strive for. This is the energy frontier and I find it very exciting. I would rather pursue this than the NLC if I had to choose only one, but reality is such that I think the NLC has to be build first or we wont be able to form an international consensus. Any new machines (NLC, VLHC) will have to be built by an international collaboration because of the cost (although I note with some interest that the expansion to O'Hare airport will cost the U.S. taxpayer six billion dollars, and no-one is batting an eye...if we could attack the VLHC with the same enthusiasm that we had for the moon shot, for example, we could do it ourselves). We need to get the public excited about this kind of physics, and for this I think the VLHC is a much easier sell that the NLC (everyone can understand the concept of searching for something where no-one has looked before, its much harder to explain to people why precision measurements of theoretical quantities is exciting (something I have been trying to do for years). Bottom line: I think the NLC will do important physics that needs to be done, but the VLHC will be much more exciting. I think it could revive the sense of excitement that drew me into HEP and put us back on the map of "exciting science" as far as the general public is concerned. 

  \item TESLA is leading LC technology today and should be built ASAP. I'm not sure that VLHC makes sense at all, but muon machines might open up new physics. 

  \item My own current opinion is that if the collective particle physics community decides to build a new machine it should probably be TESLA, since the NLC faces some really difficult problems that in the short term are difficult to see the solution to. Personally, I don't really care where it gets built. However, for a next generation machine, to really probe the energy frontier, I think we should be thinking muon collider, for the future. 

  \item We should propose an e+e- linear collider in the U.S. at Fermilab ... take a global view of HEP ... decide between this machine and TESLA (perhaps a new concept comes out of this) .... construct 1 machine.

  \item Question 42: TESLA where-ever soon, upgraded to accelerate muons in a multi-TeV muon collider; followed by a facility with a 200-400 TeV VLHC in an adjacent tunnel to a 100 TeV muon collider, and also with a 140 TeV mu-p collider. Eventually a 1 PeV linear muon collider if this turns out to be feasible. 

   \item Let's get on with it. 

  \item Question 41 is "definitely no". A muon collider sounds really cool (low synchrotron radiation, clean collisions, high energies). I have no idea, however, how the muons could be relativistic enough to be "stored". The value of a VLHC depends on what discoveries are made at the LHC. 

  \item A neutrino factory in the U.S. would also help !!!! 

  \item A large facility for the study of rare cosmic ray physics or neutrino physics that may open a window on new physics at high energies. 

  \item Too old to have a word on that (66). 

  \item From a physics point of view: Is the Europeans want to build TESLA the U.S. should join the effort, but US should also push a new machine at the energy frontier. I would like to see a muon collider but I do not know how far away this is. VLHC could be an alternative. If the Europeans cannot bould TESLA without U.S. funding than forget about it and make sure plenty of money into research for the next 10 years. It is clear that TESLA is much better than NLC therefore I do not consider the latter an option. >From the politics point of view: Here location become important! Europe is becoming leader in particle accelerators. A particle accelerator should be seen as a microscope and therefore this technology is crucial for the economical development of the U.S. It is also crucial for the image of U.S. If we know for sure that Europeans cannot build TESLA than U.S should push a new frontier machine in U.S. and forget LC. If the Europeans can build it anyway than the U.S. should make an effort to have it here. Particle accelerator and frontier physics are much more important to the U.S. technological superiority (and therefore to economy) than the Missile Defense project is. I would like to see this stated clearly to the politicians. 

  \item There certainly should no be duplicative facilities. The planning and the construction should be done globally -- but with a lead country etc. It is very important that the requisite R\&D be done on the other options. The other options appear to lack the credibility of feasibility and cost that the linear colliders have. We must look past the current problems and questions, progress can be expected in the intervening decade (or more) before a new facility will become available. Our goal should be to explore the new energy regime as carefully and completely as possible. 

  \item I don't think we should push for a facility that we are unlikely to get funding for. 

  \item We do need an electron collider b/c of the importance of a broad-based approach. And we need a VLHC to maintain the HEP program in the U.S. This is a difficult question, b/c we don't really know what we're looking for. On the other hand, we can't really afford to wait and we don't have the luxury of being as broad as would be ideal. CAN we do the TESLA/VLHC approach or is even that too expensive? 

  \item Assuming there are a few billion dollars available to build the next machine, we must make sure that this machine will be able to adequately address questions relative to physics beyond the SM (whatever it is). Having little or no information on what is beyond the SM, it is hard to build a machine that we know will be able to address the relevant physics. It may be necessary for demographic or political reasons to go ahead with the next machine without knowing what lies beyond the SM, but as I said, the physics case is not convincing. 

  \item With TESLA now being built in Germany we have, with the LHC, I believe all we need for the immediate future in terms of 'main stream' HEP. A neutrino factory sounds very interesting as does a muon collider and I would like to know more about both and would be interested in seeing them developed. A VLHC is a definite longer term accelerator but we need to see the results from the LHC first and give the theorists a chance to chew on them for a while to have an idea of what to look for at even higher energies. I think what we really need though is a better way of accelerating particles: accelerators (and hence collaborations) are becoming too big. 

  \item No collider physics anymore. 

  \item My opinions are current options are most strongly influenced by current level of funding. Now is not the time to consider new facilities. Now is the time to working on improving current (or currently being built) facilities and research programs. 

  \item My choice for HEP would be: Energy upgrade for an existing hadron collider (VLHC) and continuous research for a muon storage ring/collider. NOT an e+e- collider since it cannot discover something that LHC cannot find. 

  \item I think it is important to keep a facility running in all major labs. The U.S. with Tevatron starting up just now and BaBar only on for a short time with possible upgrades coming, do not need a new facility soon while DESY does. Building the next facility after the linear collider in the U.S. is the best option for me. 

  \item A muon collider should give much cleaner events than the VLHC; however, I would also support the VLHC. (Fermilab would be a great place for either one.) I don't see a need for a post-TESLA linear collider. 

  \item This will be the outcome of Snowmass ... anyone want to take me up on a bet? 

  \item Muon storage rings will provide means to probe intriguing aspects of current Physics and probably unveil new ones. Therefore, although a lot of technological obstacles are in the way, it seems fundamental to build a muon storage ring, with higher priority than an LHC upgrade, I would say. 

  \item I think TESLA is the best choice for the next e+e- machine. I would like to see it built in the U.S. because that is where I live and work - a purely selfish consideration.

  \item My comments are biased by working on the VLHC design study for the last few months. But I have a much firmer opinion now that a superferric magnet VLHC is technically feasible. TESLA appears to be closer to being ready than NLC. I believe there are still lots of technical challenges facing a muon collider, maybe a neutrino factory is easier. 

  \item I do not think the Germans will let TSLA go easily to the U.S. Why should they? If the U.S. had developed TESLA, would we let it go? Would we have built the TEVATRON in Hamburg, after making the magnets work?! 

  \item The choices are strongly coupled to (geo) political situation. What country can/ is willing to host what accelerator. What is the impact of a particular choice on the other accelerators, i.e. does one particular choice then exclude other accelerators? 

  \item Given that Congress is preoccupied with other matters the likelihood of a new project in the near future, started by 2005 is small. Better to have a more complete picture of what to build and HOW before bugging the Congress for the money. 

  \item TESLA at Fermilab now, with a neutrino factory at CERN.

  \item My answer to (42) is based on what I would like to have happen. Realistically, however, I would say that a LC is more likely to be accepted politically for construction within the U.S. The proposed 1st stage VLHC is too similar to the SSC. I'm afraid that Congress has enough of a memory to realize this and would not support it.

  \item The neutrino factory/muon collider is necessary for the field. Mapping out the MNS matrix is just as important as the CKM matrix. I also think linear colliders are not the right machine to build from a physics standpoint. The case is weak and the payoff is possibly zero. When you don't know what you will find you should build a hadron collider at the energy frontier -- LC's are then the best thing for followup. I have read many of the documents for the LC's and the physics case is wishful thinking at best. Furthermore, e+e- machines don't have much of a future because of the technical problems. If we knew how to build a muon collider now I would support starting that as my first option. It would be intellectually new and challenging and MC's are the beginning of a road, not the end. What we should do in the U.S. is go for a muon collider + VLHC and let the Europeans build TESLA. The problem is that we essentially "skip a step" and I'm sure DOE + Witherell are concerned about whether that will kill U.S. HEP. Rational people can disagree on how much to weight that vs. the second-class physics case for TESLA. The NLC is a joke, the technical plans make zero sense -- at least TESLA will work.

  \item TESLA or NLC soon; Major underground lab in the U.S. Involvement in astrophysical studies using satellites and ground-based facilities.

  \item It seems clear to me that the maturity of the physics goals and the technology strongly indicates a linear collider as the next step.

  \item A LC is probably necessary. However, even if built in Hamburg, it won't come on line till 2012 (50 fm-1). This is 5 years after the LHC will be running. Is the Higgs width going to be the cutting edge question then? It may be important, untangling SUSY stuff may be interesting and necessary. However, I think that it would be a disaster commit the long range HEP program in the U.S. to the construction of an LC in this country. The physics is limited and is that of measurement rather than mapping out the grand structure. 

  \item TESLA in the U.S. soon. 

  \item I would emphasize "soon." TESLA (not TELAS!) can be built now; CLIC is far in the future, and the NLC still requires more R\&D. 

  \item The above choice is really rather arbitrary as far as physics goes. The important thing is that we have an LC somewhere ASAP and that the U.S. remain major players in HEP. I have invested considerable effort in TESLA in Germany, hence the above answer, but would happily have TESLA (or NLC, CLIC is a generation down the line) somewhere if the above two conditions are fulfilled in that case, but not with TESLA in Germany. 

  \item We should pick anything that people will fund. Of course, right now nobody is funding anything. 

  \item 400 GeV circular e+e- machine in the U.S., followed by a (probably European) $>$ 1 TeV e+e- linear collider and a (probably American) VLHC. 

  \item Upgrade the Tevatron Collider to 3 + 3 TeV (i.e., build the Tripler), then proceed to build a site-filler ring at Fermilab to accommodate the then best-available magnet technology. This is more sensible than to propose a \$9B project on a new site that will likely not be funded and will most likely be strongly opposed by the public. This field of study needs to gain some common sense and come to realize that there is much to be gained below the energy frontier.

  \item TESLA in Germany. FORGET the NLC!!!! Putting our "eggs" in the NLC basket is the least interesting thing (with current proposed parameters) that we could possibly do. Continued work on VLHC and the neutrino factory/ muon collider (with R\&D in the U.S. for now). Because the timescales for "turn-on" are somewhat different, I believe that the HEP community as a whole can maintain VLHC and Muon programs, but we need to allow for the possibility that both of those programs may not be U.S.-based. 

  \item I work on CDF. I am also working on ATLAS, only because the SSC was trashed out 8 years ago by the U.S. Congress (but we (HEP) deserved it, since the SSC was badly mis-managed by DoE, URA and at locally at Wachahatchie, Tx). There are many in the international HEP community interested in building an NLC. Only one such machine should be built. It would likely get built far sooner if it was built in Europe than here. Post-LHC, I seriously doubt there will be any significant physics discovered at the NLC that wouldn't already be long known from LHC experiments. For similar reasons, building a muon collider in the U.S. is stupid. By the time it becomes operational for the first time, it will be extremely unlikely to compete with physics results coming from the LHC experiments. The only viable long-term plan for honorable U.S.-based physics, I see is a p-p type machine with energy $>>$ than the LHC. Anything else is/will be back seat to Europe. 

  \item I am uncertain what the best move is. If I had to choose ... I guess I would go with TESLA in Germany and a muon collider/storage ring soon. 

  \item With the present price tag, we cannot afford any of these machines.

  \item Regarding 42, one might want to cede the muon storage ring and associated R\&D to another region. 

  \item I think the start of R\&D for the next major (post-LHC) collider should be starting very, very soon. I do NOT think that this collider should necessarily be in the U.S. The future of HEP is a lot more important than in what country it's going to happen.

  \item The problem we face is the following: an e+e- machine at 500GeV or so has so little reach and is so expensive that it's usefulness can be questioned. The VLHC can't be seriously designed until the LHC tells us where the next scale of physics is expected to be, and so can't be started until e.g. 2009, and finished perhaps in 2020. By that time all young high energy physicists will have left for more exciting fields! Finally a muon collider is a good idea (since its reach is high) but noone knows how to build one. Thus we face a serious problem that seems to have no easy solution. 

  \item TESLA in Germany soon is the option I choose, independent of whether there is a VLHC or a Muon machine afterwards. That question is probably the subject of another survey ;-) 

  \item Not very convinced of answer 42). Picking a plan should be done with the maximum consensus hopefully without any other criteria than physics. 

  \item TESLA in Germany, with small U.S. financial impact. Stop the SLAC-FNAL competition and get on with (probably) an e+e- linear collider. Forget the muon collider. 

  \item I think the TESLA should be built. We need to work on the problems of the Muon colliders, neutrino factories. The VHLC will never be built. 

  \item TESLA in the U.S. soon (no other LC option). 

  \item I have actually participated in work on each of these options. These are not the only options. It's important that we identify the scientific imperatives that face the field, and pursue each of these, not just a select few. I see no need for a linear collider. The muon collider or a VLHC may be relevant in the future. However, a great deal can be accomplished with existing facilities, and with cosmic rays. We need more creativity, and less dependence on large, expensive facilities. 

  \item I have been around for many years. The lab where I did my PhD thesis closed just as I finished. I worked on an SSC experiment for many years and am now in an LHC experiment. Mostly I have done e+e- physics, at three different labs in two different countries. My own experience, plus discussions with others, has most affected my current opinions. I do not think that any 500 GeV e+e- will be a forefront machine. We already have the LHC coming on in ~2006. After that has been running for a few years, we will know more. I think we can plan for a 500 GeV e+e- machine, and that we should choose the technology (TESLA vs. X-band) as soon as possible. We have wasted a tremendous amount of money on NLC R\&D and we are still not ready - obviously bad management or the wrong idea - stop throwing more money at it. Since the 500 GeV e+ e- linear collider is not a forefront machine, it will be very difficult to convince any government to spend \$6B. The only possibility is a full international collaboration building the accelerator. Best to make Fermilab an international lab. Even this is just "bells and whistles" - but there is a constituency in Europe (not in the U.S.). We have to convince them to come to the U.S. - not so easy. The NLC people are not the leaders in the highest energy e+e- colliders - there are North Americans who are/were leaders in the LEP2 searches - why are they not active for the NLC? - must be a reason! We need to continue to do serious R\&D towards neutrino factory/muon collider and VLHC - this could lead to a future forefront machine in the U.S. The neutrino factory/muon collider offers a very rich program with physics at every stage - this should be seriously pursued!! 

  \item The physics requires an e+e- soon. TESLA is a superior accelerator to NLC for up to 1 TeV. Getting the U.S. back on the energy frontier requires a VLHC at Fermilab. And CLIC will be CERN's post LHC project. KEK starts now on a neutrino factory based on JHP with detectors in Asia, Europe and/or the U.S. 

  \item TESLA soon in Germany with greater effort in U.S. put towards Muon Collider and VLHC-and perhaps more into novel acceleration techniques. 

  \item I think that TESLA can "clean up" the discoveries made by the LHC, much the same way that LEP did after the SPpS experiments at CERN. To quickly explore the next frontier, one needs to explore at higher energies than the LHC. I strongly believe that if the United States wants to have a prominent role in HEP, it must have the VLHC built here and invite international collaborations to form the experiments. Whoever builds the facility, will be perceived as the principal player and leader in the field, whether we (physicists) like it or not. I realize that this statement is more political than scientific, but ultimately, that will be the driving argument that will move our congressional representatives to build the VLHC built in the U.S.

  \item Given the immense amounts of money we'll have to ask for to build ANY machines, and the fact that any precision measurement machine (LC for instance) should be tuned to exploring the physics that is discovered by LHC, it seems ludicrous to me to start building a machine that may not be able to measure the discoveries! The proper course from a science standpoint is to figure out where the new physics is with a discovery machine (LHC) and only THEN to build a precision machine (an LC). Simultaneous with construction of that machine, we should start ramping up for the NEXT discovery machine. And so on and so forth. It makes no sense to spend money on a machine when you don't know anything about the physics it will be studying. 

  \item I have looked at the LC very carefully. The case is not perfect, but still strong. It seems to me to be the best combo of scientific value, technological readiness, and political feasibility. Given that it should (and probably will) be done, I would like to U.S. to lead instead of follow. 

  \item A new LC (TESLA,NLC or JLC) somewhere in the world soon; continue research for VLHC. This option should have been given, I think. 

  \item TESLA soon, reserve judgment otherwise. 

  \item VLHC at Fermilab should be built next.

  \item  A joint Europe + Japan + U.S. effort to build the NLC in the U.S. would be ideal, but I don't think you can convince the Europeans to commit money when the U.S. Congress could cancel the project at any time. 

  \item TESLA anywhere in the world ASAP. Continuous research in all other fields. No decision for a new machine after first LC before first results from LHC. 

  \item I learned about the NLC, TESLA, VLHC, and muon collider from many different sources. However none of this information changed what my first 2 opinions were. Which are A) now is not a good time to build a big machine here in the 
U.S., and B) that the ideal machine that we should ultimately build (many years from now) is a neutrino factory/muon collider. 

  \item This is a tough call. I am not enthusiastic about the funding prospects, so it is hard to be serious about the choices. We need to go after realistic targets, but in trying to choose based on the expected physics results, keep in mind that none of the Nobel Prizes that resulted from any of the existing facilities were even vaguely hinted at in the original proposals to build those facilities! This is what I call the "Columbus Effect" -- i.e. tell me first what you will find and then I will tell you whether or not you get the jewels... Columbus got the jewels by promising Japan... 

  \item One major machine of each type in the world; more efficient international cooperation; machine priority: 1. TESLA in Germany 2. neutrino/muon factory in U.S. or in Europe or in Asia 3. VLHC in U.S. or in Europe 4. muon collider in U.S. or in Europe. 

  \item I believe we should be choosing between scenarios, NOT simply a specific machine. I believe the U.S. choice risks being suicidal in the absence of parallel discussions in Europe/Japan (i.e. despite the lip service to internationalism, this process is clearly national. while that is naturally a Congressional perspective our deliberations need not be so restricted). 

  \item Response to question 42 involves a guess about funding. 

  \item Three feasible projects, three regions: how complicated can this be? 

  \item Definitely research now. TESLA in Germany makes sense since it fits their program well. I think that there are a lot of open questions on Muon collider/neutrino factory but it has a very attractive feature of offering physics at various points. Here I think that we should implement physicists doing accelerator and particle physics at the same time and with that get rid of the 2 cultures. I dont see how NLC fits into the U.S. program of tevatron/babar followed by LHC. 

  \item This is very difficult because this is really more of a political decision than a scientific decision. The linear collider will go -- if it goes at all -- to the country most willing to put up the money. I think that TESLA in Germany is much more likely than an LC in either the U.S. or Japan. One of the reasons for this is the fact that DESY has brought the synchrotron radiation community into the picture in a major way. We need political allies which we do not have in the U.S. -- remember the opposition of our condensed matter colleagues to the SSC. Do you think that their reaction to a linear collider will be any different? If so, why? 

  \item In the next 10 years the muon collider may succeed in solving its many problems but at this stage, we know how to build an VLHC. We need to work on cost reduction and on public acceptance. 

  \item A linear collider somewhere, and continued R\&D in both VLHC and neutrino factory/muon collider. 

  \item Bill Foster summed it up rather nicely today. Alas, he has a stake in VLHC ! But facts are facts. 

  \item If Europe is willing to fund TESLA they should. There is guaranteed interesting physics in the SM to observe. If the U.S. is willing to fund any of the options, it should do that. If Japan wants JLC it should find the funding. All options are good, and the more we learn in the next decade, the more cost-effectively we can choose. We must have enough other options to maintain the size of the HEP community. 

  \item  TESLA in Germany soon and some other hadronic machine later on somewhere (CERN?) 

  \item I would improve several fields of the HEP, not only the very high energy. I find interesting the asymmetric colliders.

  \item TESLA in Germany seems the most likely way to get a 500 GeV facility soon (which I find very important). A muon storage ring is very attractive, though very challenging. It might be that VLHC is after all a technically more realistic scenario. More research I think should be done before finalizing this step -- but I am convinced we know enough to go ahead with a TESLA like facility. 

  \item Question 35: All of the above is essential! I picked one answer more or less randomly. I know a bit more on neutrino factory, since I joined that effort since its inception, however, we still have quite a bit of work to do to optimize the cost vs physics. I was not charged to do extensive reviews of these project, and I am not informed enough to make a decision. However, let me point out that all Collab. have still extensive R\&D to do before they can spend wisely a few billion dollars... 

  \item Unless the linear collider can be made to be near 1.5 TeV so that it complements the LHC, I would advocate going directly to the VLHC. A super LEP in the VLHC tunnel may be a better way to do the physics of the NLC (e.g. a Higgs factory). In any case, a 240km tunnel for \$1B would give at least three options for the future; low-field and high-field pp, and e+e-.

  \item My choice is TESLA SOMEWHERE SOON. Where depends on the best likelihood of adequate funding. 

  \item  Hold for the muon collider, focus on LHC and neitrino physics and RD for muon colliders. 

  \item An LC would be useful to supplement the LHC physics program, but the cost of a LC is too great at this time to justify building a new machine that is only supplemental -- that is, until we know more about any new physics that may appear at the Tevatron or LHC, we are not justified in asking ordinary people to spend their money on a new machine. I also worry that its energy is too low to cover all the new physics that may be found at the LHC. Politics and sociology seem to be the major reasons for starting now, not physics. 

  \item I picked a new LC (TESLA or NLC; CLIC is not a feasible option for 10-15 years) in the U.S. because it would be most personally attractive to me (not as much travel, etc.) But the most important thing is to build it SOMEWHERE. I've come to this conclusion through almost all of the routes mentioned in the question above... reading, talking, listening, working, thinking. To the extent I think there is a consensus in our field, it's not "LC good, VLHC bad". Rather, I think it's "LC now, VLHC later". I also think that "LC never" means "VLHC never", "neutrino factory never", "future of our field never". We simply cannot defer addressing important questions for another 10-15 years; people will leave, universities and labs will wither, public support for us will erode. In order to make the LC happen, the project will have to be realized in a truly international way. Failure to do this was one of the major mistakes of the SSC. Look at the consequences... if the SSC had worked out, we would *already* have found the Higgs, maybe SUSY, extra dimensions, etc. Instead, the best the VLHC people can do is offer us the SSC in 2020, the obvious difference being that the LHC will have been running for 14 years already, which makes the value added by a 40 TeV machine much less apparent. The real usefulness of the VLHC lies in the high-field 200 TeV extension, which may happen around 2030. So in my opinion the VLHC is out of the question as a realistic near-term option. Machines happen when three factors come together: 1) A compelling physics case; 2) A critical mass of worldwide interest, necessary to garner financial support; and 3) accelerator technology capable of doing the job. Only the LC has all three of these in place at this moment. WE can help make it happen. Let's do it!!! 

  \item My only reason for not picking the muon collider as the next U.S. machine is that the research is a long way from complete. We can begin to build the VLHC within the next few years so as to have it within 10-15 years. By the time that it is up and running, we should be building a muon collider. I believe that there is every reason to believe that we will always need discovery machines.

  \item TESLA seems to be attractive but I am for a more conventional e+e- collider (keep it simple). 

  \item That is the wrong way to look at it. I buy Wagner's thesis that all potential players should bid on whatever is on the table. The chances (based mainly on politics) that any work out is small enough that we should maximize the OR of the probabilities at each opportunity. (I think that there is only one on the table at the moment. It might not have been my first choice all things being equal, but they are not.) 

  \item - muon storage rings of interest and deserve R\&D - muon colliders a dream .... - VLHC deserves real work as one day it will be needed. 

  \item I don't know enough to make a reliable judgment; however reflecting on Thursday night's forum, I worry that if one doesn't build something, one could lose the interest of students. But it should be something that leads to important discovery, otherwise that will also be a turn-off. Very novel accelerator design, as well as other directions in which to go, could help a lot here.

  \item TESLA at Fermilab, with strong ($>50\%$) European influence: in this way, European physics (phenomenology) has the chance to become more popular in the U.S., where fashion plays a major role.

  \item I think that we should proceed with the TESLA design. I think probably Germany is the best location to get to the physics in a timely manner. However, I am not clear on what the next facility should be for the U.S. I think that one solid possibility would be to emphasize several smaller projects (e.g. SuperBeam at either BNL or FNAL, SNAP, Underground Facilility-- with nucleon decay experiments, perhaps a double beta decay expt.), along with extensive R\&D, which along with the above projects, and a globalized TESLA, should help to preserve the infrastructure of the labs. 

  \item I think TESLA is the obvious choice, but I don't have a strong opinion on where it should be based. Regarding the technologies beyond TESLA, it is hard to tell, since most of them still need a few years R\&D work. 

  \item I would advocate a TESLA-style e+ e- collider at Fermilab, knowing that the odds of raising the money to do it are very slim. 

  \item We need to stage the muon neutrino factory: a proton driver, muon cooling and finally muon storage ring. Several top of the line experiments will be performed at the end of each staging, like the muon EDM at the end of the muon cooling stage. By the end of the muon storage ring stage we will know whether we can build a muon collider or not. 

  \item 35) Fifty years of experience in this and related fields. There is, at this meeting, an attempt to stampede the herd into pressing for a 1.5 TeV linear collider. This might, if successful, wreck this field. 

  \item I think that Germany should host the next electron collider but not based on the TESLA model. Rather a more easily expandable model should be used to extend the possible lifetime of the machine. I feel that the VLHC should be built in the U.S. using the existing infrastructure at FNAL. I think that the problems with clean beam inherent in a Muon Collider need to be studied much more and on a time scale beyond what is most easily feasible at this time. However, it would be a good candidate to replace the electron machine in the future. 

  \item 200 TeV is a very nice number. Most young people enter HEP for its potential of discovering something new about our understanding of nature. The VLHC is the machine to fulfill those aspirations. shooting for a discovery machine, over what is essentially a measurement machine, is more akin to the dreams that got us into the field in the first place. 

  \item The choices presented above are too restrictive. If Germany can swing the funding for TESLA, they should do it, otherwise we should wait for TeV II and LHC to tell us whether it is needed and what the energy should be. We should build VLHC after LHC results become available. We should continue muon R\&D with goal of proposing neutrino factory once the design is more solid and better optimized. Linear Collider in U.S. would be a 20 to 30 year trap that would doom us to precision measurements and delay our regaining the energy frontier. 

  \item TESLA in Germany soon, CLIC (e+e- multi-TeV) in U.S. or Europe. 

  \item Three strikes, and we're out. We missed on Isabel, SSC. I don't know what the answer is, but if we miss one more time, I'm afraid it's the end of our field. Any machine we build must have a good discovery/cost ratio. We can't bet our future on a single model (eg., mSUSY, ...). I'm also worried about betting our future on assuming Congress will give us 10+ years of funding without changing its mind.

  \item Ideally, we would already have been constructing a LC as LEP is being torn down, to complement the physics we'll be doing at the LHC. For as much political and social reasons, the next "big" machine needs to be in the U.S. 

  \item I consider first option as the best one for U.S., but I chose third as a most realistic and not bad for U.S. too. Unfortunately, U.S. lost leadership in "collider physics", by loosing many talented "phenomenologists" in recent 5-10 years (by giving jobs to string guys). The current situation shows that USA is far behind Germany in " physics study" (I participated last 10 years in LC communities in Germany/USA). I am sure next all major discoveries will be made in Germany in the next 10-20 years. U.S. became an "astrophysics/ NASA" - nation (which is not bad at all). Tax payers in U.S. will buy strings "science fiction" only -- and that is a tradegy of U.S. physics. Even leaders of U.S. physics community can not find courage to say that "string theory" should be considered as mathematics and it is not related to physics. Only heroic Veltamn says " .. this stupid and senseless string theory " (final talk on "SPACE AND TIME ..." conference in Ann Arbor). Fortunately, there is Europe! Go Germany, and build LC-TESLA. Go U.S., and build stringy "faculty positions"....

  \item Given that we double funding over ten years (which we need to ask for, not for us, but for the kids in high school now), we should build LC and neutrino factory, either in U.S., Europe, or Japan, with international participation. 

  \item TESLA in Germany with major U.S. participation and a very long term plan for another machine in the U.S. would be the second option. 

  \item I think the field needs at least an e+e- collider, and probably a larger hadron machine several years down the road. I would like to see HEPAP make a strong statement that we need to start on these projects immediately. Stagnation is death! 

  \item Importance is that a e+ e- ala TESLA, at present the only technology ripe for early start of construction is built with a time-overlap with the LHC. It will be the LHC and e+ e- collider complementary which in my view view will advance HEP most in the near future.

  \item A lepton collider will be needed as Tevatron II comes to an end, whatever the results are. Given the funding climates in Europe and the U.S. a linear collider is more likely to get built in Germany - particularly if it stops, in my opinion, wasting so much money on HERA which is not going to tell us anything new. 

  \item Question 38 would be more amusing if "too much" were one of the options. In general, I'd say that I don't know enough about anything. 

  \item On question 42: the best option is to develop an international plan; TESLA in Germany and VLHC in U.S. is likely to have international appeal. However, other options may arise and the U.S. should work with international laboratories to choose the best option. 

  \item Cancel the e+e- machine, support LHC as much as possible, go to muon collider as quick as possible 

  \item TESLA is the closest to a machine I would consider worthwhile, however, the c.o.m. energy is too low. If TESLA had a 1.5 TeV c.o.m. energy that was tunable by a factor of 10 I would be all for it. This would be an energy frontier machine with discovery potential that I could support. Otherwise, I think it defies common sense to build a precision machine (e+e-) before we know where the new physics is. Muon Collider and Neutrino factory are nowhere near technically ready to be considered in the category of near term (~next 10 yrs) machines, and I think it makes sense to wait for LHC results to indicate where the next frontier energy scale is before building a VLHC. 

  \item It's mostly driven by what time scales are achievable. But we may have to face only having one working machine at a time, probably alternating continents. Meanwhile, I will look at astro/ underground/ cosmic for shorter timescales. 

  \item IMHO, construction of VLHC and the muon machine in the U.S. are not mutually exclusive. 

  \item Having soon a linear collider in U.S. (SLAC) and Europe (DESY), running possibly during LHC. Focus in parallel on the development of a muon collider (with its associate neutrino and muon factories) both in Europe (CERN) and U.S. (FERMILAB) and maybe also Japan. Think of a VLHC after LHC preliminary results and also after precise electroweak results at linear colliders). 

  \item All choices face daunting political challenges. We need leaders who can unite the field and present a coherent strategy to governments and funding agencies. 

  \item TESLA in Germany offers a very good opportunity to forge the first real international collaboration. It also fits the available HEP budget in the near term. We really need to increase R\&D in VLHC and get on the stick with public outreach. 

 \item 1-1.5 TeV, high luminosity linear collider in the U.S. - superconducting or not is not an issue. It's the energy and luminosity that matter. I think the 500 GeV case is a bit overstated.

  \item The muon collider option has received too little attention, while the neutrino factory has received too much. At the moment, we don't really know if a detector can do pattern recognition at a muon collider -- I don't think a neutrino factory is worth building if the muon collider that is unfeasible! So can it work or not? I don't buy into the idea that the field will die if we don't choose an option right now today. I think we will know more in a relatively short time and I don't want to bet all my money until I see my last card. The LC options won't go away, in my opinion, and the VLHC and MC options become more mature. What's a few years to wait to make sure we get the right machine for the right price? Let's see if that Higgs is really where the "true believers" say it is. 

  \item CLIC in the U.S. makes no sense. there should also be more research into muon colliders, both in Europe and the U.S. 

  \item A linear collider very soon anywhere in the world. After first LHC data decide upon the next step(s) i.e. LC upgrade/VLHC/muC 

  \item There is a clear need to come to the realization that we have to choose our next facility in this country. Our international collaborative stance is well documented by our LHC participation. The Electron Linear Collider is technically well enough advanced to propose it to the Government. It is complementary to the LHC. Other projects should receive further interest and should be funded enough to keep them as future options.

  \item Question 41: guess I know enough to make up my personal choice, but certainly not enough to advocate a particular future vision in the HEP community (which requires a lot of political, sociological and infrastructural thought). 

  \item TESLA in Europe (looks like Germany) NLC in the U.S. as well possibly a bit later VLHC at CERN after this. 

  \item One might prepare for TESLA type machine in U.S., Germany or Japan. I would postpone a definite decision until results from LHC are available. 

  \item In my view, given the prospect of LHC at CERN, the combination of a next linear collider with a neutrino factory would be the ideal combination for the next 15 years. In order to realize these projects without major delays within the available resources, the combination of TESLA at DESY and a muon storage ring (eventual muon collider) in the U.S. would seem to me to be an ideal combination. If realistic, the option of a linear collider in the U.S. (with DESY as a regional center) and a muon storage ring at CERN should be seriously considered as an alternative. VLHC is an option which should be pursued on a longer timescale. 

  \item Research on a VLHC should be continued but a decision prior to first LHC results would be premature. Hence, concentration on an experiment complementary to the LHC is adequate. One such exciting alternative would be a muon collider but it could probably not be realized on a time scale less than 10 years. As for linear electron colliders: my personal preference is the TESLA concept 

   \item I think that also TESLA in Germany soon, with to goal of an eventual muon storage ring/muon collider in Europe is a valid option. 

  \item TESLA includes more options than competing designs (FEL, fixed target, photon beams) and is technologically far sexier. A muon collider (likely to be located at Fermilab) would be an exciting option for the post LHC time. The tendency of internationalization in our field should be continued, with TESLA being a `world laboratory' and not a German or even European facility. BTW, the same should apply to any future large scale facility in the U.S., which includes that the nature of National Labs in the U.S. and the control by the DOE also should change (by about the time the muon collider comes along). 

  \item I think one of the main point to keep the field alive and moving is not to let too much time pass between different project. So my opinion is that an LC (TESLA) should be built somewhere as soon as possible to complement LHC. Where it will be built is less important, even if I think Germany is a good place. 

  \item TESLA soon, muon storage ring in Europe (Europe, because only Europe will guarantee that the construction process will be stopped like SSC)

  \item TESLA/NLC might well be the last machine that the HEP community can build in the foreseeable future, maybe we should know some results from LHC before finally deciding how to spend our money. 

  \item International collaboration is a must! 

  \item TESLA in Germany soon with future neutrino study facilities anywhere.

  \item The linear colliders (TESLA, NLC) currently under consideration are not discovery machines. In terms of discovery reach, a 500 GeV e+e- collider is about 1/3 of LHC. The case for a LC is based on Higgs and other new phenomena (such as Supersymmetry). Unfortunately, neither has been seen. Therefore, it is important that we wait for Tevatron/LHC results. 

  \item  The rest of the world (Asia and Europe) is converging on an e+e- collider. It's about time that the U.S. should start uniting behind the LC project and be a truly global player.

  \item Europe financial effort on LHC is already large. A new e+e- collider is the only realistic and true complement to LHC for the next 15 years. 

  \item TESLA in Germany, a possible next next generation collider based on the results of TESLA and LHC. 

   \item The answer to question 42 requires a lot of political maneuvering (ie, is not determined solely by science). 

  \item  I believe the U.S. economy can and should support two machines in the U.S. We should build an e+e- machine now and a post-LHC VLHC. We should continue research on other options, like a muon collider, as well. 

  \item It will be important to have an e+e- complement to the LHC, and I think it is important for the image of U.S. science that not all HEP take place in Europe for the next 20 years. 

  \item But the best current option is TESLA in Germany, with VLHC in the U.S. 

  \item I certainly agree that the e+e- collider program is timely and very important (as a LEP physicist over the past 15 years). But, to me, the high-energy frontier is still the most exciting and intriguing. 

  \item A linear collider soon, it doesn't matter where but considerations should be made to the future funding scenario. A consensus from the worldwide community should be a top priority. Similarly for future generations of machines, VLHC, neutrino factories, muon colliders, etc.

  \item I am very discouraged by HEP. I wish there was a better exit path. All that talk about good training for other fields was bull! 

  \item It's not clear to me what the successor to TESLA (NLC) will be (VLHC or muon collider). This will be affected by technical possibilities. I don't think we can judge this conclusively at this time. I think it is *very* important that the U.S. HEP community can form a consensus on what they want. Comparing TESLA and NLC it seems clear (at the moment) that Wagner is doing a much better job of planing, promoting, and pushing than anybody in the U.S. 

  \item We have, and have had for a number of years, a major opportunity at Fermilab. We have not taken it seriously. 

  \item Although I have some familiarity with all of the proposed machines, and with the physics issues, I do not know anywhere enough to form a sound judgment. Furthermore, I would guess that very few people in the field have a sufficient overview of the physics and detailed understanding of the options to make an educated decision. Asking people at Snowmass, or other respondents, to express an opinion is potentially dangerous; it reminds me of student council elections in high school. They were popularity contests only peripherally related to any substance. Most of the people at Snowmass are experts in their own brands of physics, and highly partisan. Most of the working groups bring together experts in one area of physics/ experimental technique/ technology. Relatively few people are spending most of their time educating themselves in areas where they are not experts. So most of us are not developing the expertise and knowledge to make the critical decisions. You should be very wary of the results from this survey, especially those which are opinions about what should be done next. 

  \item I think that there is too much focus on just increasing energy. I believe the field needs to think about more creative ways to investigate phenomena. The limits of economics seems to imply that eventually we won't just be able to build a bigger collider. Also it seems that many of the interesting theories such as string theory will most likely be forever out of our energy range for direct experimentation. We need to come up with new ways to test these very high energy theories.

  \item Question 36: I assume the question is related to the state of knowledge NEEDED now. I do not know a lot about the muon collider, but it is quite sufficient to know that is is not a serious option on the current time frame. VLHC is a messier case. It is certainly not a viable option in the 10 year frame; it is sensible in the 25 year frame. 

  \item Building an e+e- machine in the U.S. should be of the highest priority, followed by neutrino physics and particle astrophysics. VLHC and muon collider are at present fanciful, and should not divert resources. 

  \item TESLA soon, because the field is behind in having a new facility on the horizon. A muon storage ring/collider next because it seems to hold the most promise for a variety of physics topics. And because a VLHC seems too much like the SSC for political reasons! 

  \item Tough choice - there are technical problems everywhere, but the big hurdles are going to be political and financial - not clear that the world is ready for this, whether or not we are. 

  \item The cost, effort and time frame are still too large to justify another LHC \$-class facility. We need to be cleverer at a time of fiscal conservatism, and power, CO2 concerns. 

  \item A future machine at the energy frontier will not be ours unless we can restore funding and growth to U.S. HEP commensurate with our ambitions. 

  \item There is strong public support for TESLA in Germany. TESLA seems to be a solid design and well advanced. 

   \item Start soon several upgrade programs; do strong research in accelerator physics; build soon a prototype NLC collider able to produce several Giga-Z and wait for the LHC (or CDF) to tell if there is an Higgs how we now believe it should be, how many of them, which mass(es), etc before building the "major" next facility. 

  \item  Non-accelerator experiments are more interesting to me at present. 

  \item I think it is clear that a new LC soon will help us a lot. It should definitely be built at an existing site to make use of existing infrastructure etc. Whether the site should be in Europe or U.S., I can't judge really. For longer range plans I think we need more research and more information before we can really evaluate ... (At least I do).

  \item It is important to strengthen the neutrino research in the U.S.
 
  \item Neutrino source in the U.S. with underground lab for experiment. 

  \item We really need to access the highest possible energies. 

  \item We need to assess the readiness of each proposal, recognizing the work accomplished and remaining R\&D needed. The countries should commit before site selection with an agreement on the procedure for selecting a site. The host country should be asked to pay a larger portion of cost for the privilege. 

  \item Muon collider is not yet ready for serious consideration, though it may be 5 years hence. Neutrino Factory as interim project at Fermilab is worthwhile as a filler if possible. 

  \item In my opinion, the linear collider is a wrong path to persue.

  \item I don't think that a linear e+e- collider is the right choice for the HEP future, because: - its physics is too much depending on LHC outcome. Any reasonable government would ask for waiting LHC results; - it is almost a one experiment machine, which is bad for sure. 

  \item I don't think the puzzle of electroweak symmetry breaking is simple enough that we will completely unravel it either by pure thought, or by the LHC alone. We need the complementarity provided by a lepton collider, and the e+e- linear collider is the only proposal which can be carried out in the time interval we need it, the next ten years. Therefore it should be our field's highest priority. We should not, of course, foreclose research into other types of accelerators, since further developments in accelerator science will be required to go beyond the LHC/LC era. (CLIC is one such option; I don't think research on it is at the stage where it can be built "soon", though it may make an excellent "afterburner" or higher gradient version of a TESLA or NLC, exploiting the same tunnel.)

  \item My primary emphasis would be on NLC and muon storage ring for neutrino physics, with equal emphasis on each. Whichever technology is ready sooner should be the choice for the next facility, but without diminishing effort on the other. 

  \item Its stupid to consider VLHC before the LHC even turns on. The world only needs one e+e- collider, and whether it is in the U.S. or Germany SHOULD NOT MATTER. The TESLA design is not ideal in my opinion, but we should definitely not have both NLC and TESLA unless the baseline NLC design energy significantly exceeds the maximum TESLA energy, and they definitely should not run at the same time. 

  \item Fermilab needs a big project soon after 2006/7 but VLHC is not yet mature/well defined enough, VLHC may also not be interesting enough in the starting 40 TeV range => only e+e- remains, it is that simple... 

  \item I would support U.S. participation in JLC or TESLA as part as an overall plan that included a future facility in the U.S. I am more comfortable with a NLC or JLC option. We didn't need 2 B factories, PEP-II and KEKB, and it would be a disaster to repeat it with 2 or 3 linear colliders. 

  \item My answer to question Item 42 above is based on 3 assumptions: 1.The German Government will proceed with TESLA in a timely matter. 2. Compared to raising funding a U.S. NLC, it will not be too hard to persuade the U.S. Government to provide some U.S. \$ for TESLA. 3. The overall good of fostering science, no matter where and how it occurs, outweighs the good of fostering U.S. science. 
  \item I'm young yet, and feel that there are some significant advances in accelerator physics on the horizon that could let us leapfrog to high energy accelerators without the enormous size which can put off not only the general public, but scientists who might work on such accelerators. That said, we need to maintain enough experiments at any given time to keep the population of trained HEP Physicists up. I think a very good argument can be made for building a new collider here, or being sure the U.S. works in a major way with TESLA or CERN on these grounds alone.

  \item Do we need VLHC anywhere in the world? We should look at micro-, nano- scale to find a bigger picture. Big doesn't solve everything. 

  \item I would amplify on my response, but I said "definitely no" to my any credibility of my response. 

  \item I would choose a proton-antiproton machine at intermediate energies to study all aspects of low energy QCD and the origin of mass. 

  \item We have to minimize the cost and maximize the utility for other (nonHEP) physicists. We should also make sure the budget for the new machine does not kill smaller HEP research eg fixed target and astro-particle experiments. 

  \item 1) A linear e+e- collider [~1TeV] somewhere, some technology 2) A neutrino factory 3) R\&D on CLIC, muon colliders, VLHC Time scales: LC needed operating around 2010-2012 NF needed operating on similar timescale, may be later multi TeV lepton collider unlikely before 2015-2020 VLHC? unlikely before 2020 [no real physics case sustainable until 2007-2010, except 'higher energy, possible surprises'].

  \item  Muon collider is the highest priority. 

  \item TESLA in Germany or elsewhere soon, otherwise R\&D but no decisions to actually build another machine until first data from TESLA (and possibly LHC) allow to target the scale (and, hopefully, the most promising channels) where new physics shows up more precisely.

  \item All the mega-projects should only be pursued, IF there is enough money left for small machines to do precision measurements and keep the community active. Students need data, the field needs data. A muon collider is a wonderful dream, unfortunately the time is not ripe. I do not believe it can be done right now. Simply blowing up current ideas and technology should never be done after LHC. Actually LHC is in my opinion the wrong way.

  \item There is a strong (though not airtight!) case for a linear collider already, and TESLA is the most advanced design. So, if we need to begin the campaign for a new machine now, TESLA is the best option. My personal "best choice" would be TESLA soon (either in U.S. or Germany) with continued research into VLHC and perhaps muon colliders.

  \item I do not see that VLHC and a muon collider both being built in the U.S. are incompatible -- the muon collider might be built in such a way as to "slip under the radar" of the major-project mentality. It is important that both be built, and at least one be built in the U.S. Perhaps the best method (in a consciousness- raising mindset) is to temporarily allow the leading U.S. facilities to slip off the frontier, and then panic the Congress ala Sputnik: "We're falling behind! The U.S. needs to reassert leadership!" 

  \item I think that we should encourage Germany to build TESLA, and reserve judgment on what to do in the U.S. for a few more years. It's hard to believe that if the "US doesn't make a bid to build TESLA, it won't get built". On the other hand, there seem to be more folks interested in building TESLA abroad than here. One thing I don't understand, though, is how many non-FNAL folks want to build or do physics at a VLHC. 

  \item I like the idea of a VLHC in the U.S., because we have the real estate for such a machine, while other countries may not. However, I am ambivalent about the idea of "staging" such a machine, because I am not sure one can make a case for 40 TeV. (Incidentally, if the U.S. administration has any memory, they will think that this is just the SSC again, and it will not fly. That is another reason to just go straight to 100 or 200 or whatever.) 

  \item Answer to 42 could have been just as easily TESLA in Germany with muon collider in U.S. 

  \item Regarding question 42), I think the VLHC is more likely to work and produce good physics. The location of an e+e- collider preceding a higher energy hadron or muon collider is not crucial. 

  \item The important thing is to build a Linear Collider somewhere in the world. A good VLHC should be developed. (40 TeV, L=10$^{34}$ parameters of SSC is not a good option for VLHC). 

  \item 10 years ago, while I was a student, I was surprised to hear that U.S. was starting SSC, and Europe was planing LHC. In my mind I could accept that only if we expected one project would fail, otherwise, we should aim at one project, putting all effort together. Years later, this became true, U.S. joined LHC project, and dropped SSC, after a lot of lost in human sources and money. In the future, I hope all physicists in the world can join their force together, and don't limit them self by their citizenship of a certain country. 

   \item 42: A TESLA-like machine somewhere, not necessarily in Germany or in the U.S. Your options all tied TESLA to Germany or weren't specific enough about the choice of technology. 

  \item TESLA-in-Germany is not so much a preference as an acceptance of the likely coupled with the need to then plan for the next facility. 

  \item The most important ingredient needed, to formulate a policy for the future, is an understanding, among the lab directors and the HEP community, that short-term compromises and sacrifices are needed, to assure a long-term future.

  \item Any of the options 2,3 or 4 or combinations of them would work. Nothing wrong in pursuing the TESLA in Germany and think of next collider facilities in the U.S. We can also make the judgment later when time is more appropriate and we know more about how to achieve those goals (muon collider etc.)

  \item It is completely clear (an objective truth!) that what we need is a new e+e- collider a la NLC. We can no longer afford to do a lot of "consensus building", yet, we must have a consensus on that. I would go so far as to say that the consensus is more important than what particular facility, and a consensus that doesn't have too many nay-sayers causing trouble down the road. But as I said, it is clear what the facility should be. 

  \item  A TESLA like facility (e+e- linear collider) somewhere in the world soon. The community must then decide, based on LHC data and data from current accelerator upgrade programs, what the longer term goals of the field should be. 

  \item I can't say which item has most affected my opinion. I know little about the specifics of the proposals -- when my experiment ends I leave the field (see above) so I haven't made an effort to keep up-to-date. But in terms of physics reach I support a TESLA like proposal with e-e e-gamma, gamma-gamma and synchrotron radiation. 

  \item  TESLA in GERMANY soon. A muon or VLHC collider anywhere in the world after LHC has been in operation for a few years.

  \item In your wording... VLHC starting now This is what we do well and know how to do... at whatever level TESLA or JLC/NLC as possible ...Can't be our whole field And the proponents gotta go to work ON THE MACHINE not the detectors ( a las SSC ) till 0 - 2 years any idiots can built detectors the machines are everything.

  \item An e+e- collider is essential as a partner to LHC. It is clear that such a collider is on the near horizon. VLHCand muon colliders are much further in the future. Considerable technology development is needed for these to be successful, and their physics programs need better definition. 

  \item I do neutrinos but most people don't. I think we should go with the physics most people find interesting, even if it slows me down a bit. 

  \item  VLHC, let the e+e- options duke it out between themselves. 

  \item I think it would be nice to work on something besides TESLA or VLHC. Sure, somebody has to do it, but don't expect me to stand up and cheer for either, since they will come at the cost of other options.

  \item Why list CLIC as an option for "soon"? CLIC is a post-LHC option, surely? In any case, I'd advocate a new e+e- machine on a short timescale as the priority (which is then a choice between TESLA and NLC). I'm absolutely neutral on location. I would prefer to reserve judgment on what should come after that. 

  \item Do everything. Lobby to get the money we need to do it. Consolidating is severely hurting the field in my opinion. 

  \item I do not buy the complementarity sales pitch as the history is there for all to read. Plus the enormous cost of an e+e- collider is not justified on physics grounds. I opt for a very large tunnel populated with an e+e- collider initially and the VLHC following 10 years of LHC operations. This scenario satisfies a multitude of physics and physics community goals. 

  \item First, make shorter e+e- LCs at two or more places. After the qualification of them by experiments, extend one of this to the higher energies. 

  \item TESLA in Hamburg does not allow the investment in the infra structure (tunnel) to be in 20 years for an upgrade at higher frequency. As far as we know today! The consistency of the soil north of Hamburg might not even be good enough for TESLA parameters. We nee a "Lehman" review for TESLA. 

  \item I support to start focus the future effort in the U.S. on VLHC design. If German government approves TESLA, the U.S. HEP community should support in exchange for their support for the VLHC in U.S. 

  \item After having stopped SSC and joining LHC the U.S. are well advised to go for the next big machine and not wait for the overnext, which might take very long to come. 

  \item TESLA in Germany soon, a muon collider wherever it is favorable to build it (why the hell should it be in the U.S.?). 

   \item A muon collider would be extremely nice to have, but I don't think it will become operational before I retire.. 

  \item TESLA in Germany. Makes no sense to build such a machine with little involvement of the inventors and spiritual leaders. Further there should be somewhere a big proton machine developed. Forget the U.S. nationalism. Build it where it can be build best. Maybe CERN. Then decide what you need at a time: low energy muon machine, Muon collider, neutrino factory, spallation source, ... 

 \item Leave out TESLA, go directly for a muon collider. Also: any new accelerator should be a BIG step. It should be an international collaboration. U.S. versus Europe type of thinking has to be abandoned. 

  \item TESLA in Germany, but in any case a ep and eA linear accelerator (ELFE @ DESY).

  \item As I mentioned in the previous section, the LHC will need to be backed up eventually, but we should wait to have some results before designing a new e+/e- machine such that it can be optimized to the new interesting physics/energy. Mean while, other searches should be conducted such as on the dark matter... 

  \item JLC in Japan and possibly muon collider/VLHC in U.S.

  \item I believe that we do not know enough now to build anything except TESLA, which the Germans have brought quite far. I believe that it is too soon to try to get an international mega-project in the U.S. To tie the future of the linear collider to the U.S. political process at this time is foolish. It will delay things by years! The SSC is still strongly in the minds of Congress, and with the tax cut it will take several years to set things back in balance. We should support the Germans on this one, and do the multi-year ground work for a future **international** project in the U.S. later. 

  \item TESLA in my eyes is a very good machine but can not cover all available physics, so pp physics too is a topic to be covered. In my opinion e+e- and pp complement each other well enough to support both.

  \item Start building a staged neutrino factory NOW. Proton driver/super beam technology is as ready as anything else. 

  \item We should take the fastest and most technologically advanced track to get us the physics regardless of machine location. Because of this, I think TESLA is the best option for a linear collider. They are technologically and politically farther along than the NLC collaboration. There is no reason not to contribute to their effort. 

  \item Obviously the U.S. would like a new facility (post Tevatron). CERN will be busy with the LHC, and I think TELSA is the best candidate for a linear collider. So a VLHC seems to be the best option. Unfortunately I do not know much about neutrino factories or muon colliders. 

  \item A linear collider should be built as soon as possible (the location does not matter), but only at the condition that an upgrade to a multi TeV machine (2.5 TeV or so) is feasible both technically and financially. Although a 500 GeV linear collider is certainly useful for EW (and possible SM Higgs) precision measurement, the energy required to study possible new physics will only be known after several years of running at LHC. Neither TESLA or NLC can guarantee such an upgrade right now thus R\&D should continue.

  \item 1) The next machine should allow a variety of experiments. E.g. should not make neutrino physics impossible. Even though the main direction is/should be EW symmetry breaking. 2) The decision should be based on scientific arguments. 3) The new machine has only to start up after LHC is done. To have again experiments stopped in favor of a new machine is a disaster (LEP/LHC) and only hard to explain to the public. At the moment the community is anyway moving from hot spot to hot spot. The near startup of a new machine leaves the old experiments deserted without fully exploiting there capabilities. The decision should be based on results gathered by Tevatron or even LHC. Then the necessary energy range is known(?). The project should be expandable. Until a final decision is taken R\&D should continue for all possibilities. A start up time of 2020 is not too bad. I would think that the U.S. would deserve the next machine. But the labs can learn from CERN how to build an international organizational structure. 
  \item Working at intermediate energy physics accelerators I think TESLA as a multi-physics machine is highly desirable now but I cannot judge which other machine should be installed in the U.S. as a post-LHC machine.

  \item I do not know enough technically to choose between TESLA and and x-band e+e- machine, but I know enough to understand that we need to build an e+e- collider and we need to start immediately. VLHC, neutrino factory and muon collider should all be studied, but those that imply that any of these options are viable in the next 10-15 years are misleading us. An e+e- machine does not preclude any of those machines. 

  \item A neutrino factory might be an affordable domestic project in the relatively short term.

  \item A 500 GeV NLC will have less energy than LHC and is probably not justifiable. Muon collider - we do not know how to build. VLHC needs a new method as SSC has shown us not to be workable. 

  \item A thorough physics program of the LHC + long baseline neutrino experiments now. An e+e- machine soon (US), followed by CLIC + muon collider at CERN later. 

  \item Bring back the SSC. 

  \item I am for an e+e- collider somewhere in the world. The Germans/Europeans cannot afford TESLA by themselves, and I seriously doubt that American funding agencies will commit enough of a contribution for yet another machine not on American soil. On the flip side, if a non-TESLA option is built in North America, I doubt whether the Europeans will then commit enough funds. It seems one of the few "workable" (politically) options is to build a TESLA type design LC at Fermilab. However, I remain concerned about the present top design energy of TESLA of 800 GeV. We need more headroom in energy expandability (i.e., to 1.0-1.5 TeV) to ensure the physics.

  \item I do think that we need a LC soon, but I don't know where it should be located. There are compelling arguments for several locations. I also think that we should plan for a VLHC in the post LHC era now. And continued R\&D into muon colliders. I also think that a neutrino factory should be considered. Finally, let's not neglect the low energy machines for beauty and charm physics.

  \item TESLA is a well worked/shaped project and fills a gap between accelerators and physics projects. It was conceived in Germany , it might as well be built there and the efforts and man power should be turned in the U.S. to VLHC with some participation (if possible) to TESLA. The physics prospects does not justify another LC. The experience needed in accelerators to make the gap for VLHC might be gained from TESLA and the other existing projects. As far as physics and detector developments, U.S. should participate to all the ongoing projects those in America and abroad. 

  \item Also need to pursue a near-term program of diverse physics at the national labs and universities. 

  \item Rebuild the university programs and look for new ideas while we wait for results form Fermilab and CERN. Pursue R\&D on the muon collider, VLHC and other new technologies. 

  \item  In order to balance the field it would probably be the best to build TESLA at Fermilab. Building it in Germany, however, could be realized on a shorter time scale (which I think is very important) and chances for getting major funding from the German government seem to be better at the moment than from the U.S. government. 

  \item This option: - politically more feasible; - it gives optimal distribution of basic facilities around the world. 

  \item TESLA in U.S. or Germany, as soon as possible. Continued R\&D in VLHC and neutrino factory/ muon collider. 

  \item As noted above, I am not as concerned as to the type of facility. I do believe that research needs to continue with or without a facility to continue pushing the frontiers with the latest technology. Personally, I have worked with hadron colliders so I would like to see more work in that area. In the larger picture, continuation in some form is the most important piece of the puzzle. 

  \item An unspecified LC (i.e. TESLA, NLC, etc) in an unspecified location (i.e. Germany, U.S., Japan, etc), with continued studies on other facilities without specifying what they are or where they should be situated. The available choices (except "other") couple the near term and the long term in ways that do not always make sense. They also couple physics driven choices (e.g. LC) and politically driven choices (such as location) in ways that are not universally accepted. I suggest that they be split apart if there are future surveys. 

  \item A second choice would be TESLA in Germany with a vigorous program of smaller efforts (e.g. neutrino) in the U.S. 

  \item Furthermore, I favor the warm technology, as I think it's the most promising and I think reach in energy is the most important thing in the end.

  \item A combination of many of these items rather than simply one alone. 

  \item There should be a viable physics program in all regions of the (particle physics) world. There should also be a viable physics program in B physics, neutrino physics and astro-particle physics. A decision on a next LC must keep these two points in mind. 

  \item This is a somewhat curious question -- yes, I'd like an accelerator in the U.S., but it's ultimately a political decision, and not ours. If Congress decides we *won't* build one, I would put all my weight behind TESLA. I will send in more comments separately, thanks. 

  \item I think TESLA should be given the green light/funding to go ahead in Germany ASAP. The U.S. should stop its R\&D budget on this and look towards the next generation. Maybe best to wait till after the LHC turns on to see whether the way forward is VLHC or muon collider. But whatever, we must ensure that the lessons from SSC are learn't. 

  \item Because I am distracted by my duties to the BaBar collaboration and by my responsibilities within my group structure, it is difficult for me to form any opinion about any one technology or site due to the simple lack of time to conduct the research. As a physicist, I would be a fool to claim I could form any opinion about any of the proposed projects. 

  \item I want to see the most productive global physics program evolve. This to me means that there isn't any duplication of the largest accelerators. Most important, though, is for us to take advantage of all possible funding streams. If the German government is really willing to put up \$2B towards TESLA, the debate should be over and the U.S. should take them up on the offer and participate while taking the bulk of the U.S. program money and spending it on other efforts. If we push ahead for a U.S. LC so hard that the German project fails, then we will have kissed off a huge influx of new money which will then not go into particle physics -- and the world program will be poorer. 

  \item If you count responses, I have to say mine should count as 1/2. I am conflicted. Whatever we do, it will take 15 years and I strongly want a strong domestic accelerator based HEP program while we are constructing it. 

  \item I do not care where the collider is, I would like it to be built and physics be done there. It should be a linear collider first, then ( without a doubt!) a VLHC-type machine. 

  \item See above and my answers in part VI. I'm basically an astronomer; I haven't even taken a course on Standard Model particle physics. So I don't feel qualified to say which accelerators are and are not good for the rest of the field; my comments in part VI above are based only on the suspicion that engineering difficulties will eventually prevent accelerators from getting to energies where some fantastically interesting things are happening (esp. quantum gravity). I encourage particle physicists to explain to me, as a non-expert, the advantages of their particular brand of collider and why I should fund it with my tax dollars. 

  \item Although it was not an option, I don't care where the LC is built and would like to see continued research in VLHC and muons. 

  \item I am very skeptical that the taxpayers will be willing to come up with the amounts needed to build any of the proposed machines, especially if the scientific community is not unanimous in its support. 

  \item It has become clear to me that TESLA is far more advanced in technology, outreach, budget, and politics than either the NLC or CLIC. It also seems quite likely that TESLA will be built by Germany. And I don't think it makes much sense to try and build TESLA in the U.S. What German in his/her right mind would want such hard-won technological advances/innovations to become a reality in another country?! While I don't advocate that we pilfer their technology, and I don't think we should build NLC (it's not remotely ready yet..nor has site preparation been done in earnest), I do think we should participate in TESLA if the Germans build it. If they decide against the project, then perhaps we should try to build it here..having given them the chance first. I also think building two linear colliders in the world would be a mistake. So, assuming that Germany will build TESLA and we will participate, we still need forefront activity in HEP in the U.S. This is why I chose the path to a muon storage ring. This path offers much diversity and the ability to step toward the final goal in specific, physics-rich, well-defined, not-exorbitantly-priced stages. A proton driver as the first stage would provide a great deal to the neutrino physics community (as well as the hadron collider folks if built at FNAL). It could also provide a lead-in to the VLHC if that path were eventually determined to be the best for the field. 

  \item So much depends on the unknown funding and political climate, that the choice is not simple, and may have little to do with what I consider the most desirable machine, the NLC. 

  \item No new facility. 

  \item I came to Snowmass ready to support a linear collider with the TESLA design in Germany with continued research in VLHC and muon collider. After attending the consensus building presentations I find that I can NOT support a linear collider because the physics is not there. The linear collider has the same discovery potential as the LHC and it comes 5-8 years later. The statement was made that the LHC will discover things and the linear collider will be needed to understand them. I can not believe that we will sit around for 8 years with discoveries that we can not understand. It would probably take much less money and time to fund theorists to figure it out! I find giga Z physics rather boring. If there is nothing new in the reach of the LC then it will have been a terrible waste of money - at the expense of many small experiments. This is not a risk I want to take. I think small experiments which cover a wide range of topics are the best way to keep a healthy field and to keep young people interested.

  \item Linear collider in JAPAN. The U.S. is too politically unstable to entrust with such a large project. (Remember the SSC!) 

  \item 42) GSI-proposal of an European heavy-ion-facility. 

  \item A new e+e- machine somewhere ASAP. With continued research for the next step. 

  \item TESLA in Germany is also a good idea (but without a new project in the U.S., the U.S. laboratory and university programs will shrink. -- TESLA in the U.S. may cause the German program to shrink, but perhaps less so due to the close proximity of the LHC).

\subsubsection{Decide to Wait:}
  \item Wait for first results at the LHC (Higgs + SUSY), THEN decide if we 
        need a VLHC (only if s/th unexpected comes from LHC). Muon ring (don't
        underestimate R\&D + costs!) and TESLA is too much for a 30 year 
        period, but we need TESLA ASAP (for Higgs + SUSY precision). Cut that
        "in the U.S." crap - who cares, R\&D is spread over the world already,
        so jobs are not the issue. "If we fund, we want it here" is 
        ridiculous, too: Physics is decentralized, and so should its labs be.
        It's participation that counts. Nobody can buy physics. 
  \item There could be unsuspected surprises in the next decades that change 
        future priorities. 
  \item Given the time lag between experiments and to keep many aspects of 
        HEP viable, it is clear that one would like to begin construction of a
        new facility in the next few years. However, I personally do not know 
        which type of new facility would offer the best chance for discovery 
        or illuminating an existing problem in HEP. In the time between now 
        and the first data coming from the LHC, it is not clear to me that 
        the parameters of a new facility can be adequately determined to best 
        optimize the physics output of the facility. I have listened to 
        arguments from proponents of each facility and have trouble comparing 
        the physics reach of each facility on the same footing. What is clear
        to me now is that without consensus of the whole field and attending
        support/ mandate from the public, an aborted attempt to construct a  
        new facility will likely mean accelerator based HEP will suffer an
        unrecoverable contraction worldwide. 
  \item Wait a while after hearing results from CDF and B-factory and even
        LHC if necessary. It's good time to decide after 2005, I believe. 
  \item With the LHC starting in a few years, it is a bit early to be 
        lobbying about it yet. Thinking about it, yes. 
  \item Will the findings from LHC affect the choice? 
  \item Seeing what Run II and LHC tell us is important before commiting to
        the next big machine.
  \item After LHC, the field may take a direction that we cannot guess at 
        this time. An even higher energy of say 10 times more may not be 
        very productive. 
  \item LC physics is not compelling for such a big machine unless relevant
        discoveries are made in Run II or at the LHC. 
  \item I think the field needs to focus on determining if there are one or 
        more Higgs bosons. This may be done (or started) in the FNAL Run II 
        program but certainly will be carried out at LHC. I think it is a bit
        premature to think about the "next" accelerator although the advent 
        of LHC brings the question to the forefront in the U.S. If anything,
        the question underscores the tragedy of the SSC! 
  \item I believe that we need the results from at least Run 2 at Fermilab
        before we can make an intelligent choice for our next option. 
  \item I think before we seriously start talking about a neutrino factory or 
        VLHC, it would be good to finish and get some physics results from the
        neutrino oscillation experiments that are now in the stage of 
        construction and from LHC-a,b. TESLA is good, but we need a new
        facility in the USA. So for me this is just a choice of NLC versus 
        muon collider. I do not know enough about muon colliders to estimate
        the potential of such a machine. But NLC would be undoubtedly very 
        useful. 
  \item I would not proceed on any expensive new accelerator without doing the
        following .. 1) TESLA/NLC should wait till we know exactly where the
        Higgs is. 2) The logistics of establishing truly international 
        collaborations should be worked out at the governmental level. By this
        I mean that an organization should be established which will be funded
        through international treaties and that would have the ability to
        manage the construction of new facilities whenever and wherever they
        are required. 
  \item The LHC results are very important to decide for the future. 
  \item Don't think that waiting a few years and then joining/building major 
        facility is possible -- what will young particle physicists do if we
        are going to wait 5 years to spend 2 years deciding to spend 8-10 
        years building collider? "Wait 5 years" is tantamount to eliminating
        HEP as a major component of U.S. science program. Would rather decide 
        honestly that we are going to wind down than pretend that we are
        going to wait. 
  \item Perhaps we should wait for the physical results of approved, but not 
        yet started projects to see which way to go. 
  \item Catch-22: If you don't start now, you won't be done "in time" but
        right now you don't know enough to know what would be best. 
  \item 1. The environment for international cooperation needed for projects
        of the magnitude of an e+e- collider or a VLHC does not seem to be in
        place. Therefore, impractical to embark on such projects now -- since
        it appears that no single country can fund these devices. 2. The 
        physics case for precision studies of electroweak symmetry breaking is
        not yet strong enough to start construction. As an experimentalist, I
        think there should be more tangible evidence for new physics before 
        one contemplates high-precision measurements. For example, 
        astrophysical detection of WIMP's whose character is supersymmetric 
        would be a tremendous boost for the case for an e+e- collider. HEP 
        should be doing much more to support astroparticle searches that would
        have a critical bearing on our future accelerator program. 3. In the
        mean time, nature seems to be providing us with very strong 
        encouragement to study neutrino physics in much greater detail -- with
        particle accelerators. While the total cost for a complete program is
        fairly high, we are fortunate that neutrino physics can be approached 
        by incremental investments. Therefore, I recommend giving higher
        priority to facilities to study the neutrino sector this decade. 
  \item If I were the person who makes the final decision, I would spend many 
        months learning the issues before deciding. This is a serious question
        that can't be answered by a gut feeling. 
  \item Despite all the excitement, it is much too early to worry about what 
        I'll do next. The best options will/ may get ignored - the politics of
        the field will probably override other considerations, so it's more 
        efficient to do my current research and await a decision. 
  \item While the desire to build a new machine is in keeping with a long and 
        enormously successful program of exploration, the costs, both human and
        financial, have risen to a new scale. I am not personally convinced 
        that going ahead without a better idea of the physics to be probed is
        appropriate. The concern for a U.S. facility, as opposed to a truly 
        international one, I have little patience for. Yes, it is personally 
        difficult to work in Europe (for example), but I think this is a small
        sacrifice to make if we want our fellow citizens to come up with
        billions of dollars in research support. 
  \item I'd like to say reserve judgment but I also think the next machine 
        should be built in the U.S. to ensure that there is a strong HEP 
        community in the U.S. Without the next machine in the U.S. I believe 
        that the slow demise of HEP (at least in the U.S.) is inevitable. 
  \item I think that research into all of the above options can continue 
        however until the results of Tevatron Run 2, LHC and the B factories 
        are well understood it seems a little early to make a final decision. 
        I also just make the point that all of the above options only offer 
        a 'final' decision as being built in the U.S. There are other 
        countries in the world and there is no real reason why a new 
        experiment has to be built in the states. 
  \item Recent RF breakdown problems encountered at the NLCTA and at CLIC 
        cloud the choice between normal and super-conducting collider options.
        More work is clearly needed. TESLA currently appears quite viable, but
        upgrading much beyond 1 TeV is not at all viable, so it appears to be 
        a short-term solution with no good upgrade. The muon collider remains 
        at least 30 years in the future. With the big tax cut and a 
        science-hostile president, the funding climate for a big machine is 
        lousy. We should use the remaining 3.5 years to do basic accelerator 
        technology development and make a vigorous pitch to the next 
        administration. If the Germans forge ahead with TESLA in the interim, 
        fine. The need for a serious machine reaching beyond 1 TeV will 
        remain. Make no judgment this year-- the options are not completely 
        understood. 
  \item Unfortunately, politics will play a major role so we also have to 
        consider what is possible and what is safe for the field. If we 
        spend a lot of money on something which proves to be uninteresting, 
        then we are in real trouble. I do not think it would be wise to push 
        for an early LC in the U.S. I said VLHC but we should reconsider in 
        a few years. 
  \item HEP faces the problem that the length of time to create a new facility
        is much longer than the characteristic time for the extension of
        knowledge. Hence, one is reduced to guessing and gut biases; the 
        standard of knowing what physics you explicitly hope to learn driving 
        design choices cannot accommodate the present rate with which HEP
        physicists want to build new machines. 
  \item The answer - fair amount, but need to know more reflects my opinion 
        of the state of understanding of the community - we don't know enough 
        to make a firm positive decision for these options. 
  \item HEP-community comes into the same waters which led to the extinction 
        of dinosaurs 67 My ago: too big and too little flexibility. We should 
        wait for a few years with our next decision; TESLA is coming too early
        (for a number of reasons). Nevertheless, it has to be decided now (or 
        soon) in order to keep the field going. The projects are getting too 
        big for the community. 
  \item It is too soon to tell. Need much more careful studies about physics 
        objectives. 
  \item I have a popular level exposure to the issues pertaining to future 
        accelerators. I would like to reserve my judgment for future. 
  \item It is very difficult to make decisions for a future period of 20-25 
        years. For me physics problems have to be attacked from many sides. I 
        am for broad science. With limited national budgets big science 
        projects e.g. accelerators must be supported by the international 
        community. 
  \item It is the duty of the community to make full usage of LHC/ Tevatron 
        physics before building the next machine. The day we know the Higgs 
        mass we can design the proper apparatus, before the risk of a machine 
        slightly too small is too big. See PETRA. 
  \item The field should decide and converge on a machine to support now, not 
        reserve judgment for several years. 
  \item There is a lot of things which has not been understood yet about the 
        matter and I don't know if after the LHC another big project will be a
        key is this understanding. That's a lot of money so we need a 
        discovery to push a little bit higher ... otherwise nothing will come 
        from that. 
  \item We choose the next machine to find the answer, but we don't yet know 
        the question and we can't afford to ask the wrong one. I think we need
        to wait at least for preliminary results from Fermilab's Run II before
        deciding on the next machine to pursue. It would be nice to have LHC 
        results as well, but the lead time needed for building a machine might
        not permit this luxury. 
  \item Either VLHC or muon collider depending on results from LHC. 
  \item Based on physics motivations let Tevatron and LHC run for a while and
        then decide. In the mean time R\&D for accelerators and detection
        methods should continue. 
  \item I would choose very differently depending on what is found at the LHC
        and the Tevatron. The problem is how to ensure that we have enough 
        trained physicists around for the next step, and this probably means 
        building some facility even if it is not the "right one". 
  \item Let the LHC find what it is supposed to find, take info from 
        non-accelerator experiments and take a decision combining all HEP 
        knowledge and not just part of it. 
  \item We need more accelerator research. 
  \item Reserving judgment would kill our field. People and expertise would 
        get lost! 
  \item I would pick either to wait until the LHC has been running for a 
        few years or start now to design a very high energy pp machine. 
  \item This is the don't make a choice solution. It is the one of those to 
        pick but I think it is unfundable. I believe that we'll see major 
        facilities in Japan and Europe and we'll miss out following this 
        track. That would be a shame. We need to focus. NLC has obvious issues
        (i.e. what if there's no SM $\sim$120GeV Higgs?) that we'd know the 
        answer to before we run but not before we make a fast-track funding 
        pitch. I think this is a poor idea. Wait for 5 years and do R\&D. Then
        build it when there is a mandate. Support TESLA if it gets funded in 
        the meantime. VLHC is too far to commit before the LHC results are 
        digested. Neutrinos are the only sure, staged program with a physics  
        mandate. Too bad there is a smaller user base. 
    \item Support advanced accelerator R\&D. Simply building bigger machines 
          from extant technologies is not practical much beyond the 0.5 TeV   
          scale. Plan for the future by developing plasma, laser, high 
          frequency RF, superconducting RF acceleration technologies, and 
          in the process of pursuing these technologies you will attract 
          eager young minds.
   \item Mark Twain once said something like "it is hard to predict, 
          especially the future". Because we do not know what will happen 
          (or not happen!) at the LHC, nor at Run 2, the next machine is a 
          big risk. For this reason, diversity is important for this machine, 
          sort of like hedging your bets or diversifying your stocks. On the 
          other hand, it has to always be the best guess at physics at the 
          moment, and in the absence of any brilliance (such as for SPEAR or 
          SuperK, for examples), it is usually the highest energy and/or 
          luminosity.
          important question.
    \item The worldwide HEP community may be able to persuade the world's 
          taxpayers to fund one new large facility. We should therefore avoid 
          a rush to judgment about what such a facility should be. We do not 
          know what is around the "energy corner". It would be a better use 
          of public money to spend it on truly advanced R\&D towards a future 
          facility, and defer a decision about what to build until about 2005.
          By then we should have a better idea about which way Nature's wind 
          is blowing.
    \item If the "next machine" cannot be done properly, it should be 
          postponed and instead one should build a several tau/charm, b and Z 
          factories for precision measurements. If everything goes into 1 or 
          2 super-collaborations, science dies. We can see the start of that 
          at CERN. 
    \item I am not convinced that there must be a "next machine". "Big 
          science" may not be the best use of resources, especially if the
          scientific payoff is very uncertain (as seems to be the case 
          currently).

\subsubsection{Survey-Specific Comments on Picking a Plan:}
  \item See my comments to the HEPAP subpanel: follow the link on: \\ 
        \verb+http://hep.uchicago.edu/~rosner/+ 
  \item This is not a high energy physics survey, this is a high energy 
        experiment survey. On that issue I would rather not express my 
        opinion in print. How would the experimentalists like it if I made a 
        survey asking, Q: what do you think about the importance of orbifolds
        and orientifolds in future N theory research? a) ... b) 
  \item I thing that your survey is not very well prepared: First you are 
        concentrating to much on the U.S. and then your answer structure is
        not very well suited for your questions most the time. You are mainly
        asking for yes or no / black or white, but sometimes this is not
        possible and more options of the supplied answers are important. 
  \item See text of part VI. You did not seem to have other in mind but 
        building a new machine in the U.S. when you created this survey. 
  \item This form seems to be U.S.-biased. 
  \item Why something has to be in the U.S.? I find these questions very
        nationalistic, should decisions in science not being fact based. 
  \item Sorry, this is a bad survey. It doesn't allow for any of the 
        richness of the relevant discussion ... (Which is not to say that I 
        discourage you from attempting to address the issues ...) 
  \item You're talking to the wrong guy for this section. 
  \item You don't seriously believe that the selections you present begin to 
        cover the realistic, or even probable range of options. I find it 
        difficult to believe that a precision machine such as a LC is going to
        warrant the projected cost without decimating the field. I assert that
        this section is so biased that no reasonable conclusions can be drawn 
        from it. 
  \item The choices for question 42 are too limited. I would prefer: a new 
        e+e- collider (TESLA, NLC) in the World soon, with continued research 
        in VLHC and muon storage ring/ collider technologies. At least one of
        these new facilities should be in the U.S. 
  \item Why only mention the U.S. and Germany? 
  \item An option for QCD physics not mentioned above: ep (TESLA on HERA). 
  \item Expand options to include fixed-target experiments. 
  \item Question 35 is strange -- it is physics issues that has most affected
        my views, but that is not a choice. This is the only scientifically 
        justifiable choice, and as well the only one that could work in the 
        real world of funding and politics -- i.e. the answer to 42. 
  \item I find these options somewhat odd, as a physicist, I am only 
        interested in the physics. Personally I do not care where! If the U.S.
        have enough money, well then let them build anything they want :-) a 
        muon storage ring would be nice anywhere! 
  \item I wish I had time to read about and consider these weighty issues. 
  \item Nice job with this survey. How can I get more involved? 
  \item Well I don't know enough about these subjects at the moment and my
        excuse is that I have to finish my PhD in a year after only starting 
        2 years ago. 
  \item Shouldn't the response to question 42 be disregarded if the person's 
        response to question 41 was "unsure", "probably no" or 
        "definitely no"? 
  \item Question 42: Again, the coupling of multiple issues in the selections 
        available is inappropriate, coupling issues that are only linked by 
        assumptions made by those who wrote this survey. Further comments: a) 
        TESLA (not TELAS) and NLC are comparable in time scale and cost, but 
        differ in other ways. CLIC is much farther off in time, and should not
        have been coupled with either NLC or TESLA. b) Opinion regarding 
        importance of siting the facility in the U.S. addressed in previous 
        comments. 
  \item Ridiculous choices! I don't care where the next linear collider is 
        being built! This focus of the U.S. physics community on itself sucks 
        and is counterproductive for the worldwide advance of High energy 
        physics. Looking at the available techniques only TESLA appears to be 
        far enough advanced for a construction start soon. In addition, paired
        with a FEL it is by far the more convincing concept. I think given the
        technical challenges and the costs of even more future colliders, we 
        have to wait and see how the presently built machines, LHC and a 
        linear collider perform before we can ask any people to burn another 
        5 to 10 billions of dollars on another project. 
  \item As before, I cannot pick from these U.S.-centric options. I would 
        like a high-energy e+e- collider somewhere, soon. At present, TESLA 
        looks like the most feasible design. In Germany would be convenient 
        for me, but somewhere else (eg. U.S.) would be great too. Beyond that,
        a neutrino factory leading muon collider would be attractive. 
        (I don't care where.)
  \item The choices you offer in (42) are much too biased towards the U.S. 
        Please consider that HEP is an international field! 
  \item Everything too biased towards the U.S., humph!
  \item O.k., perhaps because I am German ... - but please, why must 
        everything be in the U.S.???
  \item I continue not understanding your question always about U.S.: many 
        other countries exist, many other countries study particle physics as 
        well as U.S. Have you ever heard about Europe, for example? 
  \item This part of the survey is quite biased towards a linear collider. 
        Please learn a bit more about the field and do some sitting and 
        thinking yourself before presenting a survey like this. 
  \item There is no compelling physics justification for any of the other 
        options listed with the exception of studying dark energy/cosmological
        constant. 
  \item Snowmass and this questionnaire are concentrating due to circumstances
        on a major flagship project. The intellectual health and progress in 
        our particular field depends more on diversity and intellectual 
        ferment than on monolithic projects. A comprehensive science policy 
        for particle physics must encourage diversity in research projects as 
        well as address the energy frontier. 
  \item Your list is incomplete, and most of the items are not under one's 
        control. TESLA  has made the case, and we will have to wait a year to 
        see whether Germany could build it or whether the U.S. can "buy" it. 
        The next machine after linear collider should be dictated by the 
        physics we learn in the next 10-15 years at the Tevatron, LHC, and 
        hopefully LC. Picking the after-LC machine now is gambling - you might
        end up right between two cactuses in the energy desert. We do not need
        a macho "get me to the highest energy possible" machine. We need a 
        machine at the energy where the exciting physics is destined to 
        happen, namely the 1 TeV precision machine, as we already building a 
        TeV-class hadron machine. 
  \item It is very biased to couple NLC and VLHC in one item in the previous 
        question. I am a lot more convinced about NLC than I am about VLHC. 
  \item Question 42: Your options are too narrow. A more likely scenario is 
        that we push for a LC to be built somewhere in the world with U.S. 
        participation. We obviously favor having it done in the U.S., but are 
        willing to support it in Europe or Japan if that is the way that it 
        works out. VLHC is also very important to keep alive and research for 
        such a machine is crucial. The neutrino factory is important and 
        should be supported, but at a lesser level than VLHC. The muon 
        collider needs a lot of more research and science justification. 
  \item Not sure I like/understand the wording of question 42. 
  \item I think this survey is a little biased. I hope this is only because 
        the results may be used to lobby U.S. funding agencies. Perhaps 
        inappropriate to say it, but "global" does not have to mean "in the 
        USA". 
  \item 1) The list of question 42 is too short and too restricted 2) HEP 
        will only have a future if national interests are not completely 
        defining the solutions 3) A neutrino factory in Europe would be fine 
        by several reasons. 
  \item Question 35: No space for `all of the above'? Question 36 : Your 
        survey is designed to make someone who has studied these options 
        systematically and seriously seem immodest. I would be surprised if 
        most people who have studied all of these options tell you so. 
        Question 36 : Had I designed the survey, TELSA/NLC would get one 
        line and higher energy e+e- another. This more correctly reflects 
        the current choice, both the technological and the political question.

\end{itemize}